\documentclass[aps,pra,reprint,superscriptaddress,twocolumn]{revtex4-2}
\usepackage{graphicx}
\usepackage{color}
\usepackage[dvipsnames]{xcolor}
\definecolor{DarkGreen}{rgb}{0.0, 0.5, 0.0}

\usepackage[T1]{fontenc}
\usepackage{textcomp}
\usepackage{bm}
\usepackage{amsmath,amssymb,amsthm}
\usepackage{comment}
\usepackage{grffile}
\usepackage{cancel}
\usepackage{algpseudocode}

\usepackage[
  colorlinks=true,
  citecolor=DarkGreen,
  linkcolor=blue,
  urlcolor=blue,
  linktocpage=true
]{hyperref}

\newcommand{\cket}[1]{\left|#1\right\rangle}
\newcommand{\bra}[1]{\left\langle#1\right|}

\begin{document}


\title{Magnetic-field control of interactions in alkaline-earth Rydberg atoms and applications to {\it XXZ} models}

\author{Masaya Kunimi}
\email{kunimi@rs.tus.ac.jp}
\affiliation{Department of Physics, Tokyo University of Science, 1-3 Kagurazaka, Tokyo 162-8601,  Japan}
\author{Takafumi Tomita}
\email{tomita@ims.ac.jp}
\affiliation{Department of Photo-Molecular Science, Institute for Molecular Science,
National Institutes of Natural Sciences, 38 Nishigo-Naka, Myodaiji, Okazaki, Aichi 444-8585, Japan}
\affiliation{SOKENDAI (The Graduate University for Advanced Studies), 38 Nishigo-Naka, Myodaiji, Okazaki, Aichi 444-8585, Japan}


\date{\today}

\begin{abstract}
We study the magnetic-field dependence of the interactions between two alkaline-earth(-like) Rydberg atoms, ${}^{88}$Sr and ${}^{174}$Yb. Considering the pair of Rydberg states $|ns,{}^3S_1,m_J\rangle$ and $|(n+1)s,{}^3S_1,m_J\rangle$, we show that the effective Hamiltonian takes the form of an {\it XXZ}-type quantum spin model, as in the alkali-atom case [M. Kunimi and T. Tomita, Phys. Rev. A {\bf 112}, L051301 (2025)]. We find that the behavior of the anisotropy parameter for ${}^{174}$Yb at zero magnetic field is significantly different from that for other atomic species. This behavior arises from the interplay of strong spin-orbit coupling and the resulting multichannel redistribution of F\"orster defects in ${}^{174}$Yb. We systematically calculate the interaction parameters of the {\it XXZ} model in the presence of a magnetic field and show that they can be tuned by the field. As applications to quantum many-body problems, we investigate one-dimensional systems in the large-anisotropy regime and show that the folded {\it XXZ} model can be realized in ${}^{174}$Yb systems without fine-tuning of the field. We also investigate two-dimensional square-lattice systems and show that a supersolid phase can emerge in the ground state at the mean-field level.
\end{abstract}
\maketitle
\section{Introduction}\label{sec:Introduction}
Quantum simulation with highly controllable quantum systems provides a powerful approach to exploring quantum many-body phenomena. A variety of platforms for quantum simulation have been developed, including ultracold atoms~\cite{bloch2008many,gross2017quantum,schafer2020tools}, ultracold molecules~\cite{carr2009cold,cornish2024quantum}, trapped ions~\cite{blatt2012quantum,monroe2021programmable}, Rydberg atoms~\cite{Browaeys2020,morgado2021quantum}, and superconducting qubits~\cite{wendin2017quantum,kjaergaard2020superconducting}. In particular, Rydberg-atom quantum simulators have advanced rapidly, driven by recent progress in quantum control and engineering. These simulators have enabled experimental studies of various quantum many-body phenomena, including quantum many-body scars~\cite{Bernien2017}, topological edge states~\cite{Leseleuc2019}, quantum spin liquids~\cite{Semeghini2021}, and spontaneous continuous symmetry breaking~\cite{Chen2023}.

To perform quantum simulations, one needs to engineer the desired system Hamiltonian, a process often referred to as Hamiltonian engineering. Rydberg-atom platforms offer a high degree of programmability: static and/or time-dependent external fields can be applied, and anisotropy of interactions can be tailored through the choice of Rydberg states, driving schemes, and the geometry of the atomic array. In such platforms, standard spin-1/2 models have been realized, including the Ising~\cite{Labuhn2016,zeiher2016many,Zeiher2017,Bernien2017,Leseleuc2018,Lienhard2018,Guardado2018,Keesling2019,tamura2020analysis,Kim2020quantum_Ising,Borish2020transverse,Song2021quantum_simulation,Ebadi2021,Semeghini2021,Bluvstein2021,Scholl2021,Hollerith2022,Zhao2023,Bharti2023,Franz2024,Bharti2024,kim2024realization,Zhao2025,manovitz2025quantum,gonzalez2025observation,zhang2025observation,Dag2025emergent,zhang2025probing,datla2026statistical,geim2026engineering,leclerc2026one}, {\it XY}~\cite{Ravets2014,Barredo2015coherent,Ravets2015,Orioli2018,Leseleuc2019,lippe2021experimental,Chew2022,Franz2022a,Chen2023,Franz2024,Bornet2023,Bornet2024,Emperauger2025,Emperauger2025benchmarking,chen2025spectroscopy,emperauger2025probing,hornung2025observation,chen2026observing,bornet2026dirac}, and {\it XXZ}~\cite{Signoles2021,Geier2021,Scholl2022,Franz2022a,Steinert2023,Franz2024,Emperauger2025benchmarking} models. Beyond these standard cases, higher-spin and more exotic models, including $S=1$ spin models~\cite{mogerle2025spin,Qiao2025,qiao2025kinetically,martin2026measuring} and the Kitaev honeycomb model~\cite{evered2025probing}, have also been realized~\cite{Lienhard2020,Kanungo2022}.  For models that have not yet been realized experimentally, extensive theoretical proposals have been made~\cite{wu2022manipulating,weber2022experimentally,yang2022quantum_Hall,zhao2023fractional,nishad2023quantum,kuznetsova2023engineering,chen2024proposal,Valencia-Tortora2024rydberg,Kunimi2024,yoshida2024proposal,kuji2025proposal,tian2025engineering,nill2025resonant,shah2025quantum,mukherjee2026quantum,Samajdar2026ThreeBodyRydberg,SotoGarcia2026LongLivedRevivals}.

Here, we focus on spin-1/2 {\it XXZ} models, which are widely studied across a variety of fields. 
In our previous work~\cite{kunimi2025proposal}, we showed that the anisotropy parameter $\delta$ in the {\it XXZ} model can be tuned using static magnetic fields. By appropriately choosing the Rydberg states and tuning the magnetic field, we further demonstrated that the isotropic Heisenberg point $(\delta=1)$ can be made experimentally accessible.  We note that Ref.~\cite{kunimi2025proposal} focused on alkali Rydberg atoms such as Li, Na, K, Rb, and Cs.

Recently, alkaline-earth(-like) Rydberg atoms have attracted increasing attention in neutral-atom quantum computing, owing to their favorable properties such as a rich set of isotopes, ultranarrow optical transitions, controllable ion-core excitation, and optical trapping of Rydberg states~\cite{norcia2019seconds,madjarov2020high,young2020half,burgers2022controlling,wilson2022trapping,wu2022erasure,jenkins2022ytterbium,ma2022universal,lis2023midcircuit,ma2023high,cao2024multi,shaw2024benchmarking,finkelstein2024universal,Muniz2025high}. For high-quality quantum simulation and quantum computation, it is essential to characterize alkaline-earth Rydberg states, because Rydberg-Rydberg interactions are used for engineering many-body Hamiltonians and two-qubit gates. In contrast to alkali atoms, alkaline-earth(-like) Rydberg states require a multichannel description: the Rydberg-electron wave function is influenced by an ionic-core structure and can be described using multichannel quantum defect theory (MQDT)~\cite{seaton1966quantum,fano1970quantum,lu1971spectroscopy,lee1973spectroscopy,seaton1983quantum,cooke1985multichannel,potvliege2024mqdtfit} rather than the single-channel theory~\cite{seaton1958quantum}. Recent theoretical~\cite{vaillant2014multichannel,robicheaux2018theory,robicheaux2019calculations,hummel2024engineering} and experimental~\cite{ding2018spectroscopy,lehec2018laser,okuno2022high,peper2025spectroscopy,kuroda2025microwave} advances have substantially improved MQDT-based models, paving the way for quantitative calculations of interaction strengths between alkaline-earth Rydberg atoms~\cite{Weber2017,Sibalic2017,robertson2021arc,mogerle2026accurate}.

In this paper, we extend our previous work~\cite{kunimi2025proposal} to alkaline-earth(-like) Rydberg atoms. We focus on ${}^{88}$Sr and ${}^{174}$Yb, which are bosonic isotopes without nuclear spin. Specifically, we consider a pair of Rydberg states, $|ns,{}^3S_1,m_J\rangle$ and $|(n+1)s,{}^3S_1,m_J\rangle$, where $n$ is the principal quantum number and $m_J$ is the magnetic quantum number, and map them onto an effective spin-1/2 degree of freedom. As in our previous analysis of alkali atoms, we obtain the {\it XXZ} model as the effective Hamiltonian. We find that the behavior of the anisotropy parameter in ${}^{174}$Yb is significantly different from that in ${}^{88}$Sr and in alkali atoms. In particular, the anisotropy parameter reaches $|\delta| \gtrsim 10$ without fine-tuning of the field. We also find that ${}^{174}$Yb exhibits a F\"orster resonance around $n=83$ at low magnetic fields ($\approx 0.1~{\rm G}$). As applications of these properties, we show that the folded {\it XXZ} model~\cite{zadnik2021folded,zadnik2021folded2,pozsgay2021integrable} can be realized in one-dimensional chains of ${}^{174}$Yb Rydberg atoms and that a supersolid phase of hard-core bosons can emerge in a two-dimensional array~\cite{dang2008vacancy,danshita2010critical,Capogrosso-Sansone2010quantum,yamamoto2012quantum}.

This paper is organized as follows. In Sec.~\ref{sec:Methods}, we present the method used to derive the effective quantum spin Hamiltonian. In Sec.~\ref{subsec:absence_of_magnetic_field}, we present the interaction parameters in the absence of a magnetic field and discuss the difference between ${}^{88}$Sr and ${}^{174}$Yb. In Sec.~\ref{subsec:physical_origin}, we discuss the physical origin of the large anisotropy parameter in ${}^{174}$Yb. In Sec.~\ref{subsec:presence_of_magnetic_field}, we present the corresponding results in the presence of a magnetic field. In Sec.~\ref{subsec:application_to_one-dimensional}, we derive the effective Hamiltonian for one-dimensional systems in the large-$|\delta|$ regime. In Sec.~\ref{subsec:application_to_two-dimensional_square_lattice}, we investigate the ground-state phase diagram of a two-dimensional square-lattice system and discuss the possibility of a supersolid phase. In Sec.~\ref{sec:Summary}, we summarize our results. Technical details are provided in the appendixes.

\section{Methods}\label{sec:Methods}
In this section, we explain how the effective {\it XXZ} model is derived for Rydberg atoms and how the interaction parameters in the Hamiltonian can be tuned. Our method is based on standard second-order perturbation theory~\cite{Bijnen_PhD_thesis,Whitlock2017,Kunimi2024,kunimi2025proposal,Wadenpfuhl2025unraveling,dobrzyniecki2025tunable}. For completeness, we review the framework used in our previous work and its application to the present system.

\subsection{Single-atom Hamiltonian}\label{subsec:single_atom_hamiltonian}
Here, we consider a single Rydberg atom in the presence of a magnetic field applied along the $z$ axis. The single-atom Hamiltonian is defined by
\begin{align}
\hat{H}_{\rm single} &\equiv \hat{H}_{\rm Ryd} + \hat{H}_{\rm mag},
\label{eq:definition_of_single_atom_Hamiltonian_under_magnetic_field}
\end{align}
where $\hat{H}_{\rm Ryd}$ is the field-free Hamiltonian of a single Rydberg atom, and $\hat{H}_{\rm mag}$ describes the interaction with the magnetic field. The eigenstates of the field-free Hamiltonian are denoted by $|nl, {}^{2S+1}L_{J},m_J\rangle$, where $n$ is the principal quantum number of the Rydberg electron, $l$ is the orbital angular momentum of the Rydberg electron, $S$ is the total spin, $L$ is the total orbital angular momentum, $J$ is the total angular momentum, and $m_J$ is the magnetic quantum number. In this paper, we do not consider ion-core excitations. Thus, the orbital angular momentum of the ion core is zero, and the total orbital angular momentum of the atom coincides with that of the Rydberg electron, i.e., $L=l$. Since the label $l$ is redundant in this case, we omit it hereafter and denote the Rydberg state by $|n,{}^{2S+1}L_J,m_J\rangle$. This state is hereafter referred to as a bare state. The corresponding eigenenergy is given by
\begin{align}
E=I-\frac{h c R_M}{\nu^2},
\label{eq:eigenenergy_of_alkali-earth_Rydberg}
\end{align}
where $I$ is the ionization threshold for the singly excited Rydberg series, $h$ is the Planck constant, $c$ is the speed of light, $R_M\equiv R_{\infty}/(1+m_{\rm e}/M)$ is the mass-reduced Rydberg constant, $R_{\infty}$ is the Rydberg constant, $m_{\rm e}$ is the electron mass, $M$ is the atomic mass, $\nu \equiv n - \delta_{\rm eff}$ is the effective principal quantum number, and $\delta_{\rm eff}$ is the effective quantum defect. 

We next consider the effects of a magnetic field. For a magnetic field $\bm{B}$, the Hamiltonian $\hat{H}_{\rm mag}$ can be written as~\cite{hummel2024engineering,peper2025spectroscopy}
\begin{align}
\hat{H}_{\rm mag}
&\equiv \hat{H}_{\rm Z}+\hat{H}_{\rm D}
\equiv -\hat{\bm{\mu}}\cdot\bm{B}+\frac{1}{8m_{\rm e}}(\hat{\bm{d}}\times\bm{B})^2,
\label{eq:Hamiltonian_magnetic_field}
\\
\hat{\bm{\mu}}
&\equiv -\mu_{\rm B}[\hat{\bm{l}}_{\rm c}+\hat{\bm{l}}_{\rm r}+g_s(\hat{\bm{s}}_{\rm c}+\hat{\bm{s}}_{\rm r})],
\label{eq:definition_of_hat_mu}
\end{align}
where $\mu_{\rm B}$ is the Bohr magneton, $g_s$ is the electron $g$ factor, $\hat{\bm{l}}_{\rm c,r}$ are the orbital angular momentum operators for the core and Rydberg electrons, $\hat{\bm{s}}_{\rm c,r}$ are the corresponding spin operators, and $\hat{\bm{d}}$ is the electric dipole operator. The first and second terms in $\hat{H}_{\rm mag}$ represent the Zeeman term and the diamagnetic term, respectively. For $\bm{B}=B\bm{e}_z$, where $\bm{e}_z$ is the unit vector along the $z$ direction, the diamagnetic term reduces to~\cite{hummel2024engineering,peper2025spectroscopy}
\begin{align}
\hat{H}_{\rm D}
=
\frac{e^2B^2}{12m_{\rm e}}\sqrt{4\pi}\hat{\bm{r}}^2\left[Y_{00}(\hat{\bm{r}})-\frac{1}{\sqrt{5}}Y_{20}(\hat{\bm{r}})\right],
\label{eq:expresssion_Diamagnetic_term}
\end{align}
where $e~(>0)$ is the elementary charge, $\hat{\bm{r}}$ is the electron position operator, and $Y_{lm}(\cdot)$ denotes the spherical harmonics. Here, we denote the eigenstates of the single-atom Hamiltonian in the presence of the magnetic field by $|\widetilde{n,{}^{2S+1}L_J,m_J}\rangle$, and refer to them as dressed states. The presence of the $Y_{20}$ term in $\hat{H}_{\rm D}$ implies that a dressed Rydberg state is given by a superposition of bare states with the same parity. In particular, dressed $S$ and $P$ states are formed by superpositions of $S$ and $D$ bare states, and of $P$ and $F$ bare states, respectively.

\subsection{Interaction Hamiltonian}\label{subsec:Interaction_hamiltonian}
Here, we consider the interaction between two Rydberg atoms. In this work, we retain the dipole-dipole interaction as the leading term in the multipole expansion. 

We assume that one atom is located at the origin, while the other is located at $\bm{R} \equiv R(\sin\theta\cos\varphi\bm{e}_x + \sin\theta\sin\varphi\bm{e}_y + \cos\theta\bm{e}_z)$, where $R$ is the interatomic distance, $\bm{e}_{x,y}$ are the unit vectors along the $x$ and $y$ directions, and $\theta$ and $\varphi$ are the polar and azimuthal angles, respectively.

The dipole-dipole interaction Hamiltonian $\hat{V}_{\rm dd}$ is given by~\cite{Ravets2015}
\begin{align}
\hat{V}_{\rm dd}&\equiv \frac{1}{4\pi\epsilon_0}\frac{\hat{\bm{d}}_1\cdot\hat{\bm{d}}_2-3(\hat{\bm{d}}_1\cdot\tilde{\bm{R}})(\hat{\bm{d}}_2\cdot\tilde{\bm{R}})}{R^3}\notag \\
&\equiv \hat{V}_1+\hat{V}_2+\hat{V}_3,\label{eq:definition_of_dipole-dipole_interaction}\\
\hat{V}_1&\equiv \frac{1-3\cos^2\theta}{4\pi\epsilon_0R^3}\left[\frac{1}{2}(\hat{d}_1^+\hat{d}_2^-+\hat{d}_1^-\hat{d}_2^+)+\hat{d}_1^0\hat{d}_2^0\right],\label{eq:definition_of_V1_dipole_dipole}\\
\hat{V}_2&\equiv \frac{(3/\sqrt{2})\sin\theta\cos\theta}{4\pi\epsilon_0R^3}[e^{-i\varphi}(\hat{d}_1^+\hat{d}_2^0+\hat{d}_1^0\hat{d}_2^+)\notag \\
&\hspace{8.0em}-e^{+i\varphi}(\hat{d}_1^-\hat{d}_2^0+\hat{d}_1^0\hat{d}_2^-)]\notag \\
&\equiv \hat{V}_2^++\hat{V}_2^-,\label{eq:definition_of_V2_dipole_dipole}\\
\hat{V}_3&\equiv -\frac{(3/2)\sin^2\theta}{4\pi\epsilon_0R^3}\left(e^{-2i\varphi}\hat{d}_1^+\hat{d}_2^++e^{+2i\varphi}\hat{d}_1^-\hat{d}_2^-\right)\notag \\
&\equiv \hat{V}_3^++\hat{V}_3^-,\label{eq:definition_of_V3_dipole_dipole}
\end{align}
where $\epsilon_0$ is the electric constant, $\tilde{\bm{R}}\equiv \bm{R}/R$, $\hat{d}_i^{\mu}~(\mu=x,y,z; i=1,2)$ is the dipole operator of the $i$th atom, $\hat{d}_i^{\pm}\equiv \mp(\hat{d}_i^x\pm i\hat{d}_i^y)/\sqrt{2}$, and $\hat{d}_i^0\equiv \hat{d}_i^z$. The terms $\hat{V}_{1}$, $\hat{V}_2$, and $\hat{V}_3$ change the total magnetization $M_{\rm tot}\equiv m_{J_1}+m_{J_2}$ by $0$, $\pm1$, and $\pm2$, respectively. For $\hat{V}_2$ and $\hat{V}_3$, we also introduce the operators $\hat{V}_{2,3}^{\pm}$, which change the total magnetization by $\pm 1$ and $\pm 2$, respectively. Their Hermitian conjugates satisfy $(\hat{V}_{2,3}^{\pm})^{\dagger}=\hat{V}_{2,3}^{\mp}$.

\subsection{Perturbation theory}\label{subsec:perturbation_theory}
In this subsection, we derive the effective Hamiltonian using second-order perturbation theory. Based on the results of the previous subsections, the two-atom Hamiltonian is given by
\begin{align}
\hat{H}_{\rm two} \equiv \hat{H}_{\rm single} \otimes \hat{1}_2 + \hat{1}_1 \otimes \hat{H}_{\rm single} + \hat{V}_{\rm dd},
\label{eq:definition_of_two_atom_Hamiltonian}
\end{align}
where $\hat{1}_j$ $(j=1,2)$ denotes the identity operator for the $j$th atom. We consider a pair of dressed states, $|\widetilde{n,{}^3S_1,m_J}\rangle$ and $|\widetilde{n+1,{}^3S_1,m_J}\rangle$, and identify them with the spin-1/2 basis states as $\cket{\uparrow} \equiv |\widetilde{n+1,{}^3S_1,m_J}\rangle$ and $\cket{\downarrow} \equiv |\widetilde{n,{}^3S_1,m_J}\rangle$. We denote the corresponding eigenenergies by $E_{\uparrow}$ and $E_{\downarrow}$, respectively. 

We define the target subspace as $\mathcal{H}_P \equiv {\rm Span}\{\cket{\uparrow\uparrow}, \cket{\uparrow\downarrow}, \cket{\downarrow\uparrow}, \cket{\downarrow\downarrow}\}$. The corresponding unperturbed pair energies are $2E_{\uparrow}$, $E_{\uparrow}+E_{\downarrow}$, and $2E_{\downarrow}$. We then apply nondegenerate perturbation theory separately to $\cket{\uparrow\uparrow}$ and $\cket{\downarrow\downarrow}$, and degenerate perturbation theory within the subspace spanned by $\{\cket{\uparrow\downarrow}, \cket{\downarrow\uparrow}\}$. The resulting effective Hamiltonian is given by~\cite{Bijnen_PhD_thesis,Whitlock2017,Kunimi2024,Wadenpfuhl2025unraveling,kunimi2025proposal}
\begin{align}
\hat{H}_{\rm eff}&\equiv\hat{H}_{\rm eff}^{(0)}+\hat{H}_{\rm eff}^{(2)},\label{eq:definition_of_effective_Hamiltonian_total}\\
\hat{H}_{\rm eff}^{(0)}&\equiv 2E_{\uparrow}\cket{\uparrow\uparrow}\bra{\uparrow\uparrow}+(E_{\uparrow}+E_{\downarrow})(\cket{\uparrow\downarrow}\bra{\uparrow\downarrow}+\cket{\downarrow\uparrow}\bra{\downarrow\uparrow})\notag \\
&+2E_{\downarrow}\cket{\downarrow\downarrow}\bra{\downarrow\downarrow},\label{eq:definition_of_effective_Hamiltonian_zeroth_order}\\
\hat{H}_{\rm eff}^{(2)}&\equiv J_{\uparrow\uparrow}\cket{\uparrow\uparrow}\bra{\uparrow\uparrow}+J_{\downarrow\downarrow}\cket{\downarrow\downarrow}\bra{\downarrow\downarrow}\notag \\
&+J_{\uparrow\downarrow}(\cket{\uparrow\downarrow}\bra{\uparrow\downarrow}+\cket{\downarrow\uparrow}\bra{\downarrow\uparrow})\notag \\
&+\frac{J}{2}(\cket{\uparrow\downarrow}\bra{\downarrow\uparrow}+\cket{\downarrow\uparrow}\bra{\uparrow\downarrow}),\label{eq:definition_of_effective_Hamiltonian_second_order}\\
J_{\alpha\beta}&\equiv -\sum_{\bm{n}}\frac{|\bra{\bm{n}}\hat{V}_{\rm dd}\cket{\alpha\beta}|^2}{\Delta E_{\rm F}(\bm{n},\alpha,\beta)}\equiv \frac{hC_6^{\alpha\beta}}{R^6},\quad \alpha,\beta=\uparrow,\downarrow,\label{eq:definition_of_J_and_C6_diagonal_part}\\
\frac{J}{2}&\equiv -\sum_{\bm{n}}\frac{\bra{\uparrow\downarrow}\hat{V}_{\rm dd}\cket{\bm{n}}\bra{\bm{n}}\hat{V}_{\rm dd}\cket{\downarrow\uparrow}}{\Delta E_{\rm F}(\bm{n},\uparrow,\downarrow)}\equiv \frac{hC_6}{R^6},\label{eq:definition_of_J_and_C6_off-diagonal_part}
\end{align}
where $\hat{H}_{\rm eff}^{(0)}$ and $\hat{H}_{\rm eff}^{(2)}$ are the zeroth- and second-order effective Hamiltonians, respectively; $J_{\alpha\beta}$ and $J$ denote the strengths of the diagonal and off-diagonal van der Waals interactions, respectively; $C_6^{\alpha\beta}$ and $C_6$ are the corresponding van der Waals coefficients; $\bm{n}\equiv (\bm{n}_1,\bm{n}_2)$ labels the intermediate pair states; $\bm{n}_j$ is the label of the intermediate state of the $j$th atom; $\Delta E_{\rm F}(\bm{n},\alpha,\beta)\equiv E_{\bm{n}_1}+E_{\bm{n}_2}-E_{\alpha}-E_{\beta}$ is the F\"orster defect; and $E_{\bm{n}_j}$ is the energy of the intermediate state of the $j$th atom. In this case, the intermediate pair states are $P$ states due to the selection rule of the dipole operators. Due to this, the first-order effective Hamiltonian vanishes. 

We introduce the spin-1/2 operators as follows: $\hat{S}_j^+\equiv \cket{\uparrow_j}\bra{\downarrow_j}$, $\hat{S}_j^-\equiv (\hat{S}_j^+)^{\dagger}$, $\hat{S}_j^x\equiv (\hat{S}_j^++\hat{S}_j^-)/2$, $\hat{S}_j^y\equiv (\hat{S}_j^+-\hat{S}_j^-)/(2i)$, and $\hat{S}_j^z\equiv (\cket{\uparrow_j}\bra{\uparrow_j}-\cket{\downarrow_j}\bra{\downarrow_j})/2$. Using these operators, the effective Hamiltonian can be rewritten as
\begin{align}
\hat{H}_{\rm eff}&=\left(E_{\uparrow}-E_{\downarrow}+\frac{J_{\uparrow\uparrow}-J_{\downarrow\downarrow}}{2}\right)(\hat{S}_1^z+\hat{S}_2^z)\notag \\
&+J(\hat{S}_1^x\hat{S}_2^x+\hat{S}_1^y\hat{S}_2^y+\delta\hat{S}_1^z\hat{S}_2^z)\notag \\
&+E_{\uparrow}+E_{\downarrow}+\frac{1}{4}(J_{\uparrow\uparrow}+J_{\downarrow\downarrow}+2J_{\uparrow\downarrow}),
\label{eq:effective_Hamiltonian_spin_representation}\\
\delta&\equiv \frac{J_{\uparrow\uparrow}+J_{\downarrow\downarrow}-2J_{\uparrow\downarrow}}{J}=\frac{C_6^{\uparrow\uparrow}+C_{6}^{\downarrow\downarrow}-2C_{6}^{\uparrow\downarrow}}{2C_6},
\label{eq:definition_of_anistropy_parmater}
\end{align}
where $\delta$ denotes the anisotropy parameter. Therefore, as in the case of alkali atoms~\cite{kunimi2025proposal}, we obtain the effective {\it XXZ} model. Here, we remark on the property of the interaction strength $J$. Although the matrix elements $\bra{\bm{n}}\hat{V}_{\rm dd}\cket{\downarrow\uparrow}$ and $\bra{\uparrow\downarrow}\hat{V}_{\rm dd}\cket{\bm{n}}$ can in general be complex for $\theta \neq 0$, owing to the factors $e^{\pm i\varphi}$ and $e^{\pm 2i\varphi}$ in Eqs.~(\ref{eq:definition_of_V2_dipole_dipole}) and (\ref{eq:definition_of_V3_dipole_dipole}), the interaction strength $J$ remains real. This follows from the fact that the initial and final states have the same magnetic quantum number, together with the selection rules of the dipole operators. The only nonvanishing contributions are of the form $\bra{\uparrow\downarrow}\hat{V}_2^{\pm}\cket{\bm{n}}\bra{\bm{n}}\hat{V}_2^{\mp}\cket{\downarrow\uparrow}$ and $\bra{\uparrow\downarrow}\hat{V}_3^{\pm}\cket{\bm{n}}\bra{\bm{n}}\hat{V}_3^{\mp}\cket{\downarrow\uparrow}$, for which the phase factors $e^{\pm i\varphi}$ and $e^{\pm 2i\varphi}$ always cancel. As a consequence, the $\varphi$-dependent phase factors drop out, and the effective Hamiltonian becomes independent of $\varphi$.

Here, we discuss the validity of the perturbative approach. For perturbation theory to be valid, the ratio of the matrix element to the F\"orster defect must be sufficiently small. Previous work~\cite{Bijnen_PhD_thesis} introduced a critical radius $R_{\rm c}$ defined by
\begin{align}
R_{\rm c}^3&\equiv \max_{\alpha,\beta}(R_{\rm c}^{\alpha,\beta})^3
\equiv \max_{\bm{n},\alpha,\beta}\left|\frac{\bra{\bm{n}}\hat{V}_{\rm dd}\cket{\alpha\beta}}{\Delta E_{\rm F}(\bm{n},\alpha,\beta)}\right|R^3.
\label{eq:definition_of_Rc}
\end{align}
The condition for the validity of the perturbation theory is given by $R^3 \gg R_{\rm c}^3$~\footnote{As discussed in Ref.~\cite{kunimi2025proposal}, we can estimate the contribution of the next-to-leading-order correction. Because of the dipole selection rules, this correction arises at fourth order in perturbation theory. The ratio of the fourth-order correction to the second-order term scales as $(R_{\rm c}/R)^6$. For $R=2R_{\rm c}$, this ratio is $1/64\simeq 1.6\%$, indicating that the fourth-order correction is only of order $1\%$ of the second-order contribution.}. In this paper, we evaluate $R_{\rm c}$ using bare states in the numerator of Eq.~(\ref{eq:definition_of_Rc}), while the F\"orster defect is evaluated from the energies in the presence of the magnetic field. This approximation is adopted to reduce the computational cost, and we find that it does not significantly affect the results except close to the F\"orster resonance points.

\subsection{Numerical calculations}\label{subsec:numerical_calculations}

Here, we explain how numerical calculations are performed to evaluate the parameters of the effective Hamiltonian; see also Ref.~\cite{kunimi2025proposal}. First, we calculate the single-atom wave functions in the presence of a magnetic field (dressed states) using the {\it PairInteraction} software~\cite{Weber2017,hummel2024engineering,mogerle2026accurate}. We then identify a dressed state $|\widetilde{n,{}^{2S+1}L_J,m_J}\rangle$ as the eigenstate that has the largest overlap with the corresponding bare state $|n,{}^{2S+1}L_J,m_J\rangle$. In most cases, a single bare state gives the dominant contribution to the overlap. However, in some parameter regions, we find that several bare states have almost the same overlap with a given eigenstate in the presence of the magnetic field due to accidental degeneracies. For example, for ${}^{88}$Sr at $B=163.9~{\rm G}$, we find that $|\langle 51,{}^3S_1,m_J=1|\widetilde{51,{}^{3}S_1,m_J=1}\rangle|^2 \simeq 0.503$ and $|\langle 50,{}^1D_2,m_J=1|\widetilde{51,{}^{3}S_1,m_J=1}\rangle|^2 \simeq 0.497$, indicating strong mixing between these two bare states. In such cases, we do not evaluate the interaction parameters.

After identifying the dressed states, we evaluate the $C_6$ coefficients defined in Eqs.~(\ref{eq:definition_of_J_and_C6_diagonal_part}) and (\ref{eq:definition_of_J_and_C6_off-diagonal_part}). To evaluate the sums in these equations, we introduce artificial cutoff parameters, since the number of Rydberg states is countably infinite. In single-atom calculations, we include the Rydberg states $|\widetilde{n, {}^{2S+1}L_J,m_J}\rangle$ with principal quantum numbers in the range $[n-\Delta n,n+1+\Delta n]$ and orbital angular momenta up to $L_{\rm max}=F$. For the summations defining the interaction strengths in Eqs.~(\ref{eq:definition_of_J_and_C6_diagonal_part}) and (\ref{eq:definition_of_J_and_C6_off-diagonal_part}), we include only the intermediate states satisfying the condition $-\Delta E+E_{\downarrow\downarrow}\le E_{\bm{n}}\le E_{\uparrow\uparrow}+\Delta E$, where $\Delta E\equiv E_{\uparrow\uparrow}-E_{\downarrow\downarrow}$ denotes the energy difference between the pair states $\cket{\uparrow\uparrow}$ and $\cket{\downarrow\downarrow}$. In the following, we present the results for $\Delta n=3$. We have confirmed this by comparing with calculations performed using the larger cutoffs $\Delta E=2(E_{\uparrow\uparrow}-E_{\downarrow\downarrow})$, $\Delta n=4$, and $L_{\rm max}=G$, and found that the anisotropy parameter $\delta$ is accurate to approximately two digits.

At the end of this section, we briefly discuss why we do not use electric fields. Previous works have shown that the interaction between Rydberg atoms can be tuned by applying electric fields~\cite{Leseleuc2018,jiao2022electric}. Because the electric-field term breaks the spatial inversion symmetry, a dressed $S$ state is generally given by a superposition of $S$ and $P$ states. This $S$-$P$ mixing leads to nonzero dipole matrix elements between dressed $S$ states, which are forbidden by the dipole selection rules for the bare states. As a result, the dipole-dipole interaction appears already at first order in perturbation theory~\cite{Bijnen_PhD_thesis}. This modifies the functional form of the effective {\it XXZ} Hamiltonian. On the other hand, a magnetic field yields dressed $S$ states with $S$-$D$ mixing, which do not induce dipole-dipole interactions. Consequently, the functional form of the {\it XXZ} Hamiltonian remains unchanged under the magnetic field. Therefore, we employ magnetic fields, which tune only the parameters of the {\it XXZ} model.

\section{Results}\label{sec:Results}
In this section, we present the results of our perturbative calculations for the interaction parameters of the alkaline-earth(-like) Rydberg atoms in the absence and presence of a magnetic field in Secs.~\ref{subsec:absence_of_magnetic_field} and \ref{subsec:presence_of_magnetic_field}, respectively. In Sec.~\ref{subsec:physical_origin}, we discuss the physical origin of the large anisotropy parameter in ${}^{174}$Yb at zero magnetic field. In Secs.~\ref{subsec:application_to_one-dimensional} and \ref{subsec:application_to_two-dimensional_square_lattice}, we discuss applications of these results to many-body problems. Hereafter, we assume that the two Rydberg atoms lie in the $xy$ plane, i.e., $\theta=\pi/2$. 

\subsection{In the absence of a magnetic field}\label{subsec:absence_of_magnetic_field}
\begin{figure}[t]
\centering
\includegraphics[width=8.6cm,clip]{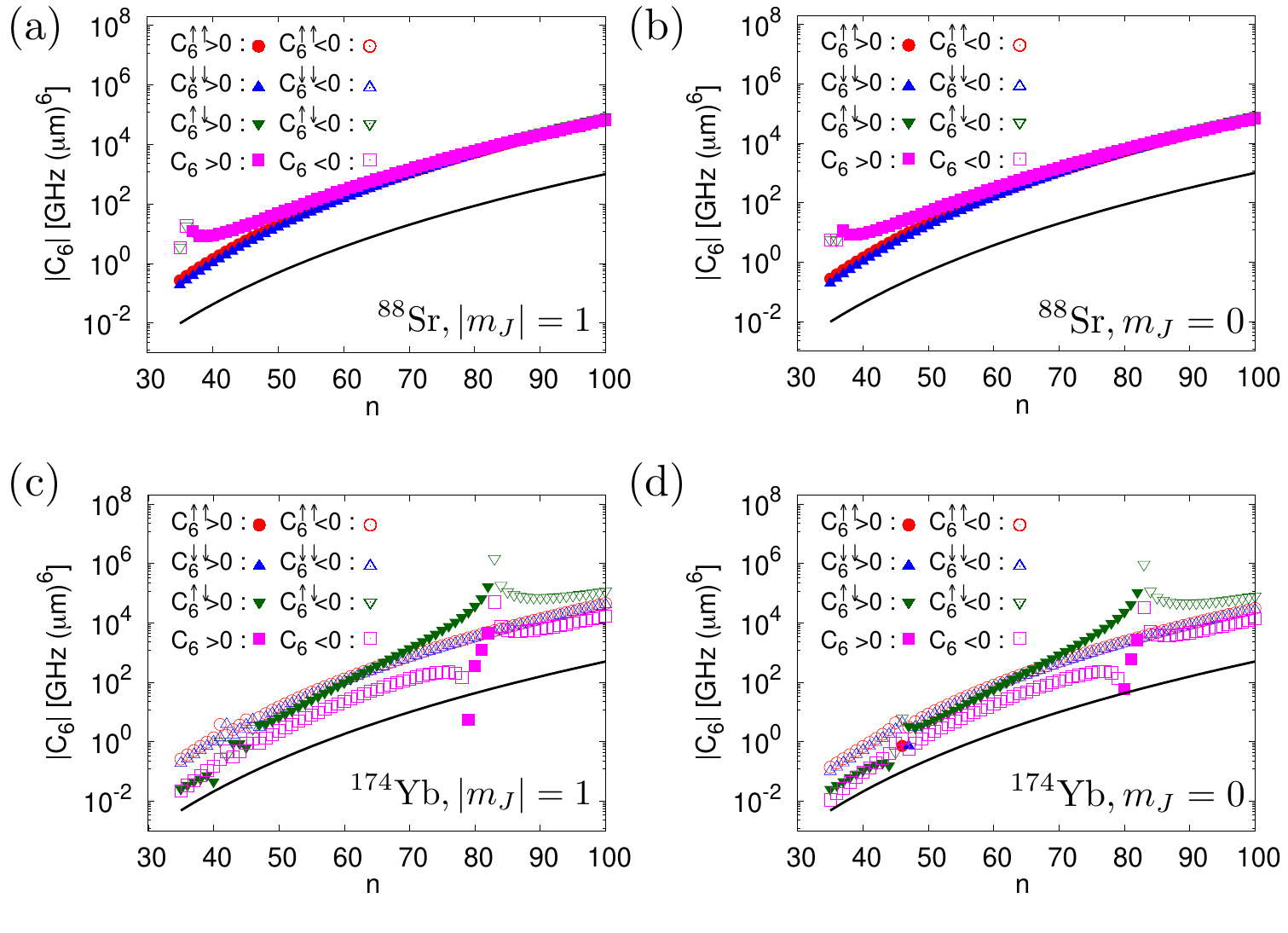}
\caption{Principal quantum number dependence of the $C_6$ coefficients in the absence of the magnetic field for (a) ${}^{88}$Sr with $|m_J|=1$, (b) ${}^{88}$Sr with $m_J=0$, (c) ${}^{174}$Yb with $|m_J|=1$, and (d) ${}^{174}$Yb with $m_J=0$. The solid black line represents the $n^{11}$ scaling as a guide to the eye.
}
\label{fig:C6_in_the_absence_of_magnetic_field}
\vspace{-0.75em}
\end{figure}

Here, we present the results for the interaction parameters in the absence of a magnetic field. Figure~\ref{fig:C6_in_the_absence_of_magnetic_field} shows the principal quantum number dependence of the $C_6$ coefficients for ${}^{88}$Sr with $|m_J|=1$ and $m_J=0$ [Figs.~\ref{fig:C6_in_the_absence_of_magnetic_field}(a) and \ref{fig:C6_in_the_absence_of_magnetic_field}(b)], and for ${}^{174}$Yb with $|m_J|=1$ and $m_J=0$ [Figs.~\ref{fig:C6_in_the_absence_of_magnetic_field}(c) and \ref{fig:C6_in_the_absence_of_magnetic_field}(d)]. In the absence of a magnetic field, the cases with $m_J=\pm1$ yield identical results owing to time-reversal symmetry. 

For ${}^{88}$Sr, we find that all the $C_6$ coefficients depend smoothly on the principal quantum number and take almost the same values at large $n$. At low $n$, however, $C_6^{\uparrow\downarrow}$ and $C_6$ clearly deviate from the other $C_6$ coefficients. This behavior originates from the F\"orster resonances $|36, {}^3S_1,m_J=\pm1\rangle|37, {}^3S_1,m_J=\pm1\rangle \leftrightarrow |36,{}^3P_2,m_J=\pm2\rangle^{\otimes 2}$ and $|36, {}^3S_1,m_J=0\rangle|37, {}^3S_1,m_J=0\rangle \leftrightarrow |36,{}^3P_2,m_J=\pm1\rangle^{\otimes 2}$. The corresponding F\"orster defect is $\Delta E_{\rm F}/h\simeq 26.2~{\rm MHz}$. On the other hand, for ${}^{174}$Yb, $C_6^{\uparrow\downarrow}$ and $C_6$ clearly deviate from the remaining $C_6$ coefficients, and the difference between $C_6^{\uparrow\downarrow}$ and $C_6$ is also significant. This behavior originates from the F\"orster resonances $|83,{}^3S_1,m_J=\pm1\rangle|84,{}^3S_1,m_J=\pm1\rangle \leftrightarrow |84,{}^3P_2,m_J=\pm2\rangle|82,{}^3P_2,m_J=\pm2\rangle$ and $|83,{}^3S_1,m_J=0\rangle|84,{}^3S_1,m_J=0\rangle \leftrightarrow |84,{}^3P_2,m_J=\pm1\rangle|82,{}^3P_2,m_J=\pm1\rangle$. The corresponding F\"orster defect is extremely small, $\Delta E_{\rm F}/h\simeq 436~{\rm kHz}$. As a result, the critical radius becomes $R_{\rm c}=37.7~\mu{\rm m}$ for $m_J=\pm1$ and $R_{\rm c}=29.9~\mu{\rm m}$ for $m_J=0$. These large $R_{\rm c}$ values suggest that the perturbative treatment at $n=83$ is not appropriate under current experimental conditions, because the experimentally accessible system size is at most about $\sim 1~{\rm mm}\times 1~{\rm mm}$~\cite{manetsch2025tweezer}. In many-body systems, this strongly restricts the number of atoms that can be accommodated. We discuss a treatment beyond the two-level approximation in Sec.~\ref{subsec:presence_of_magnetic_field}.

\begin{figure}[t]
\centering
\includegraphics[width=8.6cm,clip]{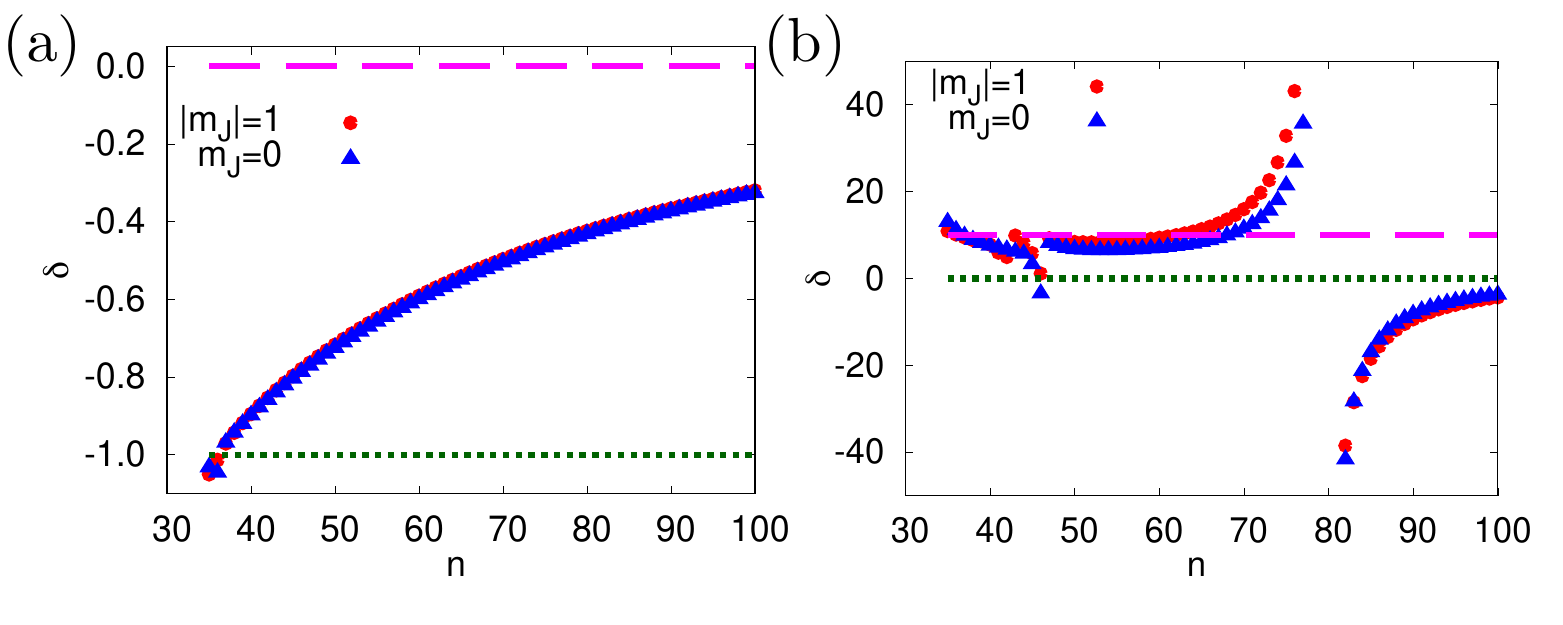}
\caption{Principal quantum number dependence of the anisotropy parameter $\delta$ in the absence of the magnetic field. (a) ${}^{88}$Sr. The dotted green and dashed magenta lines represent $\delta=-1$ and $\delta=0$, respectively. (b) ${}^{174}$Yb. The dotted green and dashed magenta lines represent $\delta=0$ and $\delta=10$, respectively.
}
\label{fig:delta_in_the_absence_of_magnetic_field}
\vspace{-0.75em}
\end{figure}

Figure~\ref{fig:delta_in_the_absence_of_magnetic_field} shows the principal quantum number dependence of the anisotropy parameter $\delta$. For ${}^{88}$Sr, the anisotropy parameter tends to satisfy $|\delta|<1$. The origin of this behavior is that all the $C_6$ coefficients take almost the same values at large $n$, as mentioned above. This tendency is similar to that in alkali atoms~\cite{Whitlock2017,kunimi2025proposal}. 

For ${}^{174}$Yb, the behavior of the anisotropy parameter is significantly different from that in the other atomic species. We find that $|\delta|\gtrsim 10$ over a wide parameter range. This behavior originates from the inequality $|C_6^{\uparrow\downarrow}|>|C_6|$, as shown in Figs.~\ref{fig:C6_in_the_absence_of_magnetic_field}(c) and \ref{fig:C6_in_the_absence_of_magnetic_field}(d). For alkali atoms, realizing a large anisotropy parameter $|\delta|\gtrsim 1$ requires fine-tuning of the magnetic field close to a F\"orster resonance~\cite{kunimi2025proposal}, which makes experimental realization difficult. By contrast, in the ${}^{174}$Yb case, large values of $|\delta|$ can be realized without such fine-tuning. This feature makes ${}^{174}$Yb a promising platform for exploring the large-$|\delta|$ regime. We discuss applications of this property to many-body problems in Secs.~\ref{subsec:application_to_one-dimensional} and \ref{subsec:application_to_two-dimensional_square_lattice}.

\subsection{Physical origin of large $|\delta|$ for ${}^{174}$Yb}\label{subsec:physical_origin}

In this subsection, we discuss why the behavior of the anisotropy parameter in ${}^{174}$Yb is significantly different from that in the other atomic species. The mechanism for the large $|\delta|$ in ${}^{174}$Yb can be summarized as follows. Strong spin-orbit coupling induces singlet-triplet mixing and, together with the multichannel character of the intermediate $P$ states, increases the number of dipole-coupled intermediate channels. It also produces a large splitting of the $P$ manifold. These effects redistribute the F\"orster defects over a broad energy range. As a result, near-resonant intermediate states with both positive and negative F\"orster defects contribute to the van der Waals coefficients, leading to sign-changing perturbative contributions. Combined with the Cauchy-Schwarz trend discussed in Appendix~A, this yields $|C_6^{\uparrow\downarrow}|>|C_6|$ and hence a large $|\delta|$.

Here, we discuss the roles of spin-orbit coupling. We first consider the differences in the intermediate $P$ states between ${}^{88}$Sr and ${}^{174}$Yb. According to the dipole-selection rules, the states $|n', {}^3P_J,m_J'\rangle$ $(J=0,1,2)$ are coupled to $|n, {}^3S_1,m_J\rangle$ by the dipole operator. In the case of ${}^{174}$Yb, in addition to these states, $|n',{}^1P_1,m_J'\rangle$ is also coupled to $|n, {}^3S_1,m_J\rangle$ by the dipole operator. This arises from singlet-triplet mixing induced by strong spin-orbit coupling. This effect is incorporated into recent MQDT models~\cite{peper2025spectroscopy,kuroda2025microwave}. Consequently, ${}^{174}$Yb has a larger number of possible intermediate states than ${}^{88}$Sr.

\begin{figure}[t]
\centering
\includegraphics[width=8.6cm,clip]{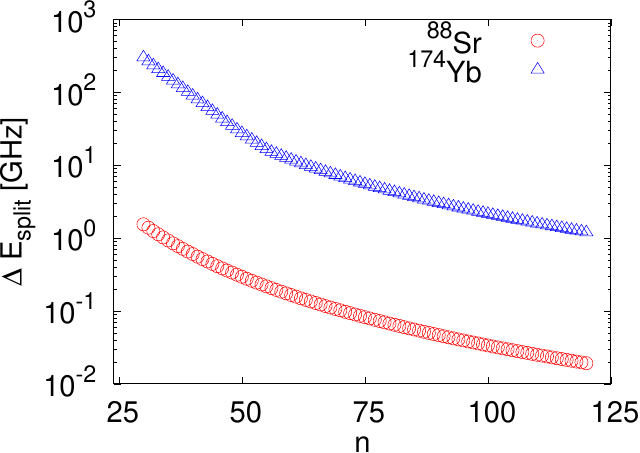}
\caption{Principal-quantum-number dependence of the energy splittings of the intermediate $P$ states. The energy splittings are evaluated as the maximum energy difference among the ${}^3P_J$ states for ${}^{88}$Sr, and among the ${}^3P_J$ and ${}^1P_1$ states for ${}^{174}$Yb. The maximum energy splittings of ${}^{88}$Sr are given by $E_{{}^3P_2}-E_{{}^3P_0}$. The maximum energy splittings of ${}^{174}$Yb are given by $E_{{}^3P_1}-E_{{}^1P_1}$ for $n\le 55$ and $E_{{}^3P_0}-E_{{}^1P_1}$ for $n\ge 56$.
}
\label{fig:energy_splitting_sr_and_yb_in_the_absence_of_magnetic_field}
\vspace{-0.75em}
\end{figure}

The second role of strong spin-orbit coupling is to produce large energy splittings within the intermediate $P$ manifold. Figure~\ref{fig:energy_splitting_sr_and_yb_in_the_absence_of_magnetic_field} shows the energy splittings of the $P$ manifold $\Delta E_{\rm split}$ for ${}^{88}$Sr and ${}^{174}$Yb. The energy splittings are evaluated as the maximum energy difference among the ${}^3P_J$ states for ${}^{88}$Sr, and among the ${}^3P_J$ and ${}^1P_1$ states for ${}^{174}$Yb. This result shows that the energy splitting in ${}^{174}$Yb is about two orders of magnitude larger than that in ${}^{88}$Sr. To clarify the effects of the large energy splitting, we introduce the energy difference of an intermediate $P$ state from the average energy, defined by $\delta E_{\bm{n}_j} \equiv E_{\bm{n}_j}-E_{\bm{n}_j}^0$, where $E_{\bm{n}_j}^0$ is the average energy of the ${}^3P_J$ manifold for ${}^{88}$Sr and of the combined ${}^3P_J$ and ${}^1P_1$ manifold for ${}^{174}$Yb. We then rewrite the F\"orster defect as $\Delta E_{\rm F}(\bm{n},\alpha,\beta) \equiv \Delta E_{\rm F}^0(\bm{n},\alpha,\beta)+\delta E_{\bm{n}_1}+\delta E_{\bm{n}_2}$, where $\Delta E_{\rm F}^0(\bm{n},\alpha,\beta)$ is the F\"orster defect evaluated using $E_{\bm{n}_j}^0$. From the above results, the range of $\delta E_{\bm{n}_1}+\delta E_{\bm{n}_2}$ in ${}^{174}$Yb is broader than that in ${}^{88}$Sr. In addition to the fact that the number of allowed intermediate pair states in ${}^{174}$Yb is larger than that in ${}^{88}$Sr, this suggests that ${}^{174}$Yb has more F\"orster resonance points than ${}^{88}$Sr.

\begin{figure}[t]
\centering
\includegraphics[width=8.6cm,clip]{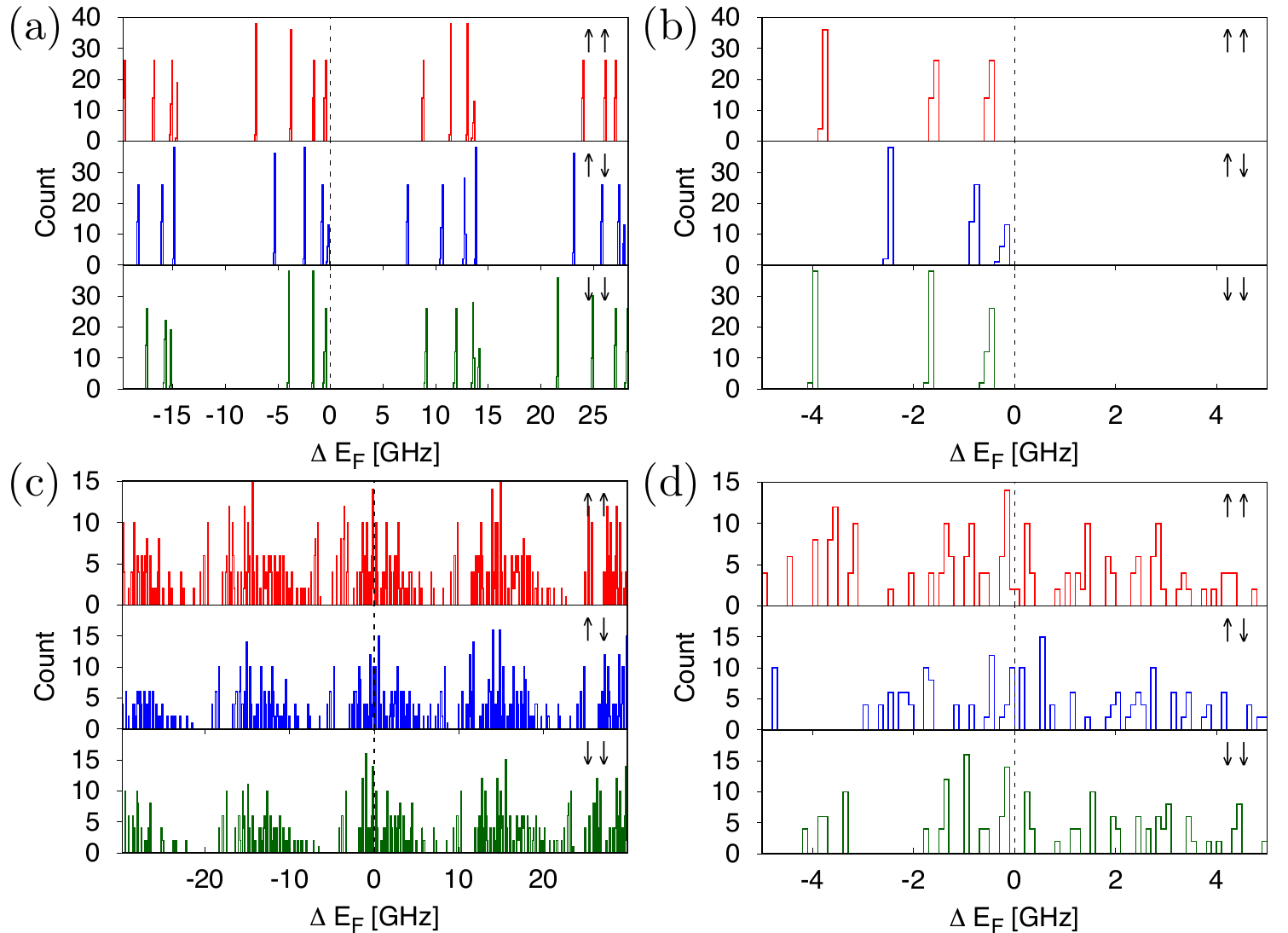}
\vspace{-1em}
\caption{Histograms of F\"orster defects for (a) ${}^{88}$Sr and (c) ${}^{174}$Yb at $n=80$ and $|m_J|=1$. [(b) and (d)] Magnified views of the near-resonant region around $\Delta E_{\mathrm{F}}=0$ in panels (a) and (c), respectively. Here, $\uparrow\uparrow$, $\uparrow\downarrow$, and $\downarrow\downarrow$ denote the F\"orster defects $\Delta E_{\mathrm{F}}(\bm{n},\uparrow,\uparrow)$, $\Delta E_{\mathrm{F}}(\bm{n},\uparrow,\downarrow)$, and $\Delta E_{\mathrm{F}}(\bm{n},\downarrow,\downarrow)$, respectively. Only intermediate states satisfying $R_{\rm{c}}>0.1~\mathrm{nm}$ are included to exclude narrow resonances. The three spin channels are shown in separate vertically stacked subpanels for each case. The bin width is $0.1~\mathrm{GHz}$. The vertical dashed line marks $\Delta E_{\mathrm{F}}=0$.
}
\label{fig:histgram_forster_defect}
\vspace{-0.75em}
\end{figure}

\begin{figure}[h]
\centering
\includegraphics[width=8.6cm,clip]{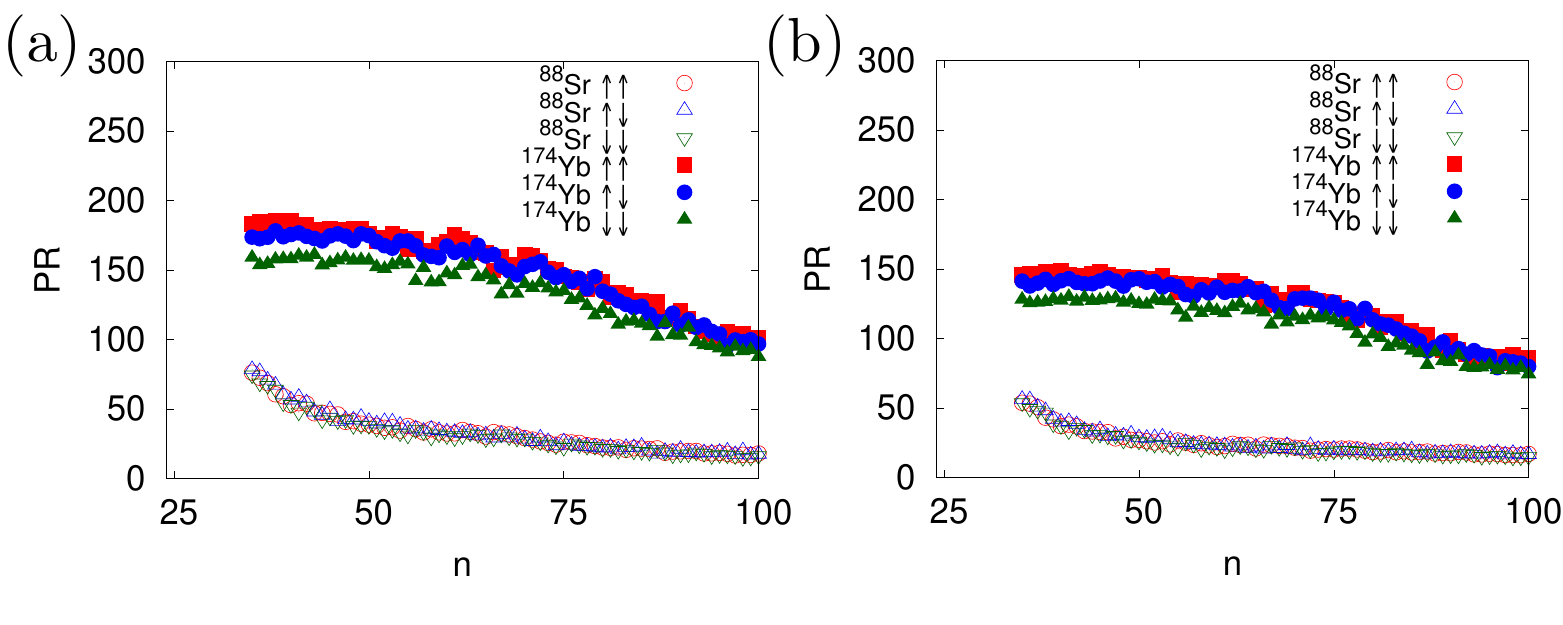}
\caption{PR of the distribution of the F\"orster defects for (a) $|m_J|=1$ and (b) $m_J=0$. Here, $\uparrow\uparrow$, $\uparrow\downarrow$, and $\downarrow\downarrow$ represent the F\"orster defects $\Delta E_{\rm F}(\bm{n},\uparrow,\uparrow)$, $\Delta E_{\rm F}(\bm{n},\uparrow,\downarrow)$, and $\Delta E_{\rm F}(\bm{n},\downarrow,\downarrow)$, respectively.
}
\label{fig:pr_forster_defect}
\vspace{-0.75em}
\end{figure}

Although the large energy splitting may account for the large number of F\"orster resonance points, it does not explain why $|C_6^{\uparrow\downarrow}|>|C_6|$. To clarify the origin of this behavior, we next examine the F\"orster defects. Figure~\ref{fig:histgram_forster_defect} shows histograms of the F\"orster defects for $n=80$ and $|m_J|=1$. We find that the distribution of the F\"orster defects in ${}^{174}$Yb is broader than that in ${}^{88}$Sr. To quantify this trend, we calculate the participation ratio (PR) of the distribution of F\"orster defects, which is larger for a broader distribution and smaller when the distribution is concentrated in a small number of bins. The PR is defined as
\begin{align}
\mathrm{PR} \equiv \frac{1}{\sum_j p_j^2},
\label{eq:definition_of_PR}
\end{align}
where $p_j$ is the normalized weight of the $j$th bin. The results are shown in Fig.~\ref{fig:pr_forster_defect}. We find that the PR of ${}^{174}$Yb is larger than that of ${}^{88}$Sr for all principal quantum numbers, suggesting that the F\"orster defects in ${}^{174}$Yb are more broadly distributed.  

Here, we focus on the distribution of the F\"orster defects around $\Delta E_{\rm F}\approx 0$. In this regime, the F\"orster defects take negative values for ${}^{88}$Sr [see Fig.~\ref{fig:histgram_forster_defect}(b)]. This behavior can be understood from the quantum defects. For the case of $\alpha=~\uparrow$ and $\beta=~\downarrow$, the intermediate pair states that give small F\"orster defects are $|n, {}^3P_J,m_J\rangle^{\otimes2}$. According to the MQDT model for ${}^{88}$Sr~\cite{robicheaux2019calculations}, the quantum defects for ${}^3S_1$ and ${}^3P_J$ have weak energy dependence and can be approximated by $\delta_{{\rm eff},{}^3S_1}\simeq 3.37$ and $\delta_{{\rm eff},{}^3P_J}\simeq 2.88$. The F\"orster defect can then be approximated as
\begin{align}
&\frac{\Delta E_{\rm F}(\bm{n},\uparrow,\downarrow)}{h c R_{\rm M}}\notag \\
&\simeq -\frac{2}{(n-2.88)^2}+\frac{1}{(n-2.37)^2}+\frac{1}{(n-3.37)^2},
\label{eq:Forster_defect_most_88Sr_small}
\end{align}
where $\bm{n}=|n,{}^3P_J,m_J\rangle^{\otimes2}$. One can show that $\Delta E_{\rm F}(\bm{n},\uparrow,\downarrow)<0$ for $n>40$, which is consistent with our numerical results. In addition, we find that the numerator in the expression for $C_6$, $\bra{\uparrow\downarrow}\hat{V}_{\rm dd}\cket{\bm{n}}\bra{\bm{n}}\hat{V}_{\rm dd}\cket{\downarrow\uparrow}$, is positive for the intermediate pair states with small $|\Delta E_{\rm F}|$. From these observations, we conclude that $C_6^{\uparrow\downarrow}$ and $C_6$ for ${}^{88}$Sr take almost the same values at zero magnetic field, because the dominant contributions to the expressions for $C_6^{\uparrow\downarrow}$ and $C_6$ have the same sign.

By contrast, for ${}^{174}$Yb, the F\"orster defects take both positive and negative values around $\Delta E_{\rm F}\approx 0$, as shown in Fig.~\ref{fig:histgram_forster_defect}(d). This redistribution of the F\"orster defects can be attributed to the large energy splittings and multichannel mixing of the intermediate $P$ states. In addition, we find that, for most of the dominant intermediate pair states, the matrix elements $\bra{\uparrow\downarrow}\hat{V}_{\rm dd}\cket{\bm{n}}\bra{\bm{n}}\hat{V}_{\rm dd}\cket{\downarrow\uparrow}$ are positive. These results imply that the dominant summands in the expressions for $C_6^{\uparrow\downarrow}$ and $C_6$ can have both signs. This property can lead to deviations between $C_6^{\uparrow\downarrow}$ and $C_6$ owing to cancellations among different terms. The trend $|C_6^{\uparrow\downarrow}|>|C_6|$ can be understood from the Cauchy-Schwarz inequality. See Appendix~\ref{app:trand_C6_parameters} for details. 

In summary, the large $|\delta|$ in ${}^{174}$Yb does not originate from spin-orbit coupling alone. Strong spin-orbit coupling induces singlet-triplet mixing and produces much larger splittings of the intermediate $P$ manifold than in ${}^{88}$Sr. Together with the multichannel character of the intermediate states, these effects redistribute the near-resonant F\"orster defects on both sides of $\Delta E_{\mathrm F}=0$, in contrast to the ${}^{88}$Sr case. Since the dominant matrix-element products are mostly positive, the signs of the perturbative summands are mainly controlled by the signs of the F\"orster defects. This leads to cancellations among different contributions and, together with the Cauchy-Schwarz argument in Appendix~A, explains the trend $|C_6^{\uparrow\downarrow}|>|C_6|$ and the resulting large $|\delta|$.

\subsection{In the presence of a magnetic field}\label{subsec:presence_of_magnetic_field}

Here, we discuss interactions between Rydberg atoms in the presence of a magnetic field. In this subsection, we focus on several representative points. The complete data are shown in Appendix~\ref{app:full_C6_parameters}.

\begin{figure}[h]
\centering
\includegraphics[width=9cm,clip]{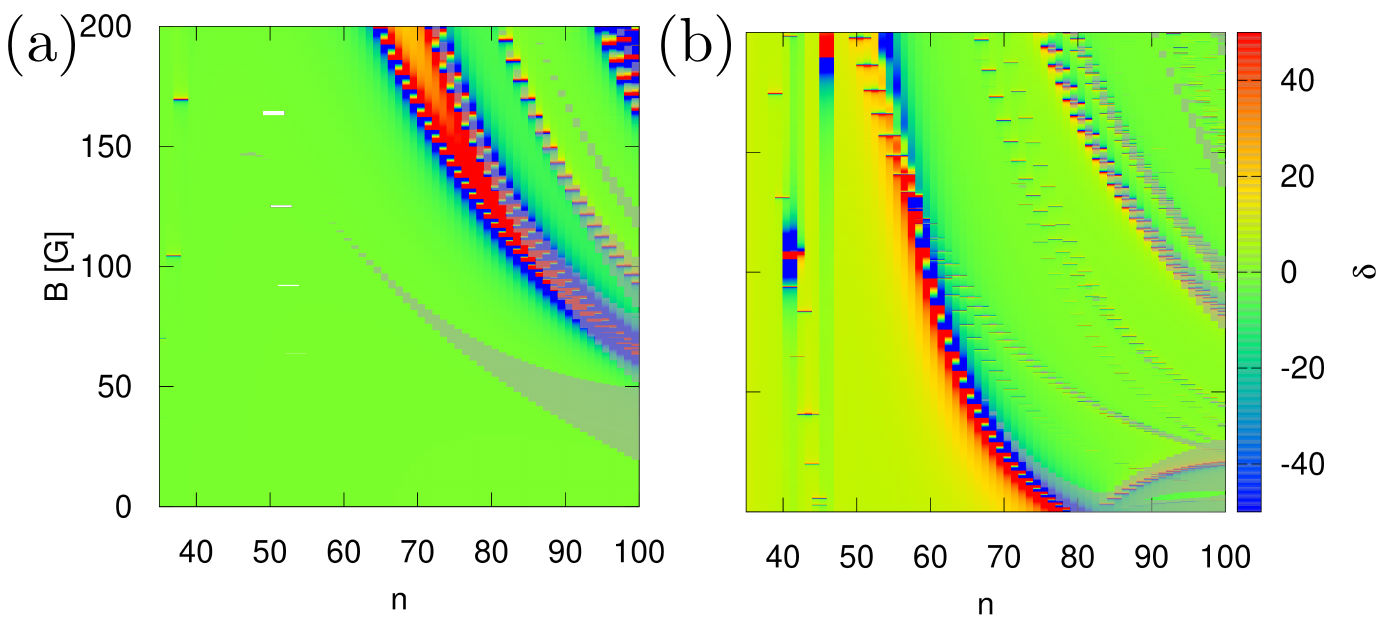}
\caption{Magnetic-field and principal quantum number dependence of the anisotropy parameter. (a) ${}^{88}$Sr with $m_J=1$ and (b) ${}^{174}$Yb with $m_J=1$. The white region in (a) indicates the absence of data due to strong $S$--$D$ mixing. The gray shaded regions indicate the parameter regions where $R_{\rm c}\ge 10~\mu{\rm m}$.
}
\label{fig:delta_nonzero_field}
\vspace{-0.75em}
\end{figure}

First, we consider ${}^{88}$Sr atoms. The magnetic-field and principal quantum number dependencies of the anisotropy parameter for $m_J=1$ are shown in Fig.~\ref{fig:delta_nonzero_field}(a). We find that the anisotropy parameter can be tuned continuously by changing the magnetic field as in the case of the alkali atoms~\cite{kunimi2025proposal}. The gray-shaded regions in Fig.~\ref{fig:delta_nonzero_field} indicate the parameter regions where $R_{\rm c}\ge 10~\mu{\rm m}$. Since the validity of the perturbative treatment depends on the interatomic distance through the condition $R^3\gg R_{\rm c}^3$, we use $R_{\rm c}=10~\mu{\rm m}$ as a practical threshold for flagging regions where the perturbative description may not be quantitatively reliable. In the shaded regions, the effective {\it XXZ} parameters should be regarded as qualitative indicators unless $R$ is sufficiently larger than $R_{\rm c}$.

Here, we focus on several representative points, such as $\delta\simeq 0, 1$, and $-5$, which correspond to the {\it XY} model, the Heisenberg model, and the large-anisotropy {\it XXZ} model, respectively. The parameters are summarized in Table~\ref{table:main_paremeter_list}. In addition to the system parameters, we evaluate $d\delta/dB$, which represents the sensitivity of $\delta$ to magnetic-field fluctuations~\cite{kunimi2025proposal}. For most of the representative points in Table~\ref{table:main_paremeter_list}, $|d\delta/dB|$ is of order $10^{-2}-10^{-1}~{\rm G}^{-1}$. A magnetic-field fluctuation of $0.1~{\rm G}$ therefore results in $|\Delta \delta|\lesssim 4\times 10^{-2}$ for most of these points. Several works~\cite{Fuchs2008binding,Inada2008collisional,Chang2020collisional,Xie2025Feshbach} have reported magnetic-field stability of order $0.01~{\rm G}$. Therefore, the required magnetic-field stability is achievable within current experimental techniques.

We note that, in the case of ${}^{88}$Sr with $m_J=\pm1$, some dressed ${}^3S_1$ states have a strong overlap with $D$ states, as indicated by the white regions in Figs.~\ref{fig:c6_and_delta_nonzero_field_sr_p} and \ref{fig:c6_and_delta_nonzero_field_sr_m}. As mentioned in Sec.~\ref{subsec:numerical_calculations}, we do not evaluate the $C_6$ coefficients in these cases because it is difficult to identify the dressed state uniquely. Here, we show only the data for which the overlaps between the dressed pair $S$ states and the corresponding bare pair $S$ states are all larger than 0.5.

Next, we consider ${}^{174}$Yb atoms. The magnetic-field and principal quantum number dependencies of the anisotropy parameter are shown in Fig.~\ref{fig:delta_nonzero_field}(b). As in the ${}^{88}$Sr case, we consider the cases $\delta\simeq 0, 1$, and $-5$. The results are summarized in Table~\ref{table:main_paremeter_list}. The characteristic interaction time, $\hbar/|J|=[2\pi |J|/h]^{-1}$, is on the order of $0.1~{\mu}{\rm s}$ for the representative points listed in Table I, sufficiently shorter than the experimentally reported $10$--$100~\mu{\rm s}$ coherence and lifetime scales of alkaline-earth ${}^3S_1$ Rydberg states~\cite{madjarov2020high,wilson2022trapping} to permit the observation of coherent interaction dynamics.

As in the case of alkali atoms, we find that magnetic fields can induce F\"orster resonances and continuously tune the anisotropy parameter. Alkaline-earth(-like) atoms tend to exhibit many more F\"orster resonance points than alkali atoms (see the Supplemental Material of Ref.~\cite{kunimi2025proposal}). This may be attributed to the larger number of possible intermediate pair states. By exploiting this rich F\"orster resonance structure, one can engineer many-body Hamiltonians over a wide parameter range.

\begin{table}[h]
\centering
\caption{List of parameters for representative values of $\delta$. $J$ is the exchange interaction evaluated at $R=2R_c$, for which the estimated fourth-order correction is about
$1.6\%$ of the second-order contribution~\cite{Note1}.
}
\begin{tabular}{cccccccc}\hline\hline
Species & $n$ & $B$ & $\delta$ &     $R_{\rm c}$     &          $J/h$       & $d\delta/dB$\\  
&           & (G)&            & $(\mu{\rm m})$ & $({\rm MHz})$ & $({\rm G}^{-1})$\\ \hline
${}^{88}$Sr $m_J=1$ & 84 & 195.75 & $-0.001$ & 6.75 & 2.00 & $-0.12$  \\ 
${}^{88}$Sr $m_J=0$ & 86 & 176.60 & 0.004 & 7.37 & 2.08 & $-0.17$  \\ 
${}^{88}$Sr $m_J=-1$ & 48 & 190.75 & 0.002 & 2.39 & 1.42 & $-0.16$  \\ 
${}^{88}$Sr $m_J=1$ & 38 & 182.40 & 1.001 & 1.38 & 1.10 & $-0.14$  \\ 
${}^{88}$Sr $m_J=0$ & 85 & 180.00 & 1.000 & 7.08 & 1.76 & $-0.37$  \\ 
${}^{88}$Sr $m_J=-1$ & 85 & 194.25 & 0.996 & 7.10 & 1.05 & $-0.33$  \\ 
${}^{88}$Sr $m_J=1$ & 80 & 95.35 & $-4.998$ & 5.67 & $-1.45$ & $-0.35$  \\ 
${}^{88}$Sr $m_J=0$ & 64 & 175.85 & $-5.001$ & 3.59 & $-1.30$ & $-0.21$  \\ 
${}^{88}$Sr $m_J=-1$ & 78 & 119.25 & $-5.004$ & 5.05 & $-2.13$ & $-0.40$  \\ \hline
${}^{174}$Yb $m_J=1$ & 81 & 42.40 & 0.005  & 5.64 & $-1.04$ & $0.25$  \\ 
${}^{174}$Yb $m_J=0$ & 69 & 148.50 & $-0.001$  & 3.15 & $-4.64$ & $0.03$  \\ 
${}^{174}$Yb $m_J=-1$ & 92 & 159.75 & 0.001  & 8.45 & $-1.92$ & $2.8$  \\ 
${}^{174}$Yb $m_J=1$ & 46 &5.35&  1.000 & 1.29 & $-8.82$ & $-0.027$  \\ 
${}^{174}$Yb $m_J=0$ & 80 &106.65&  0.999 & 5.48 & $-1.46$ & $-0.039$  \\ 
${}^{174}$Yb $m_J=-1$ & 46 &13.30&  0.996 & 1.71 & $-1.89$ & $0.2$  \\ 
${}^{174}$Yb $m_J=1$ & 60 &164.40&  $-5.001$ & 3.32 & $-0.71$ & $0.12$  \\ 
${}^{174}$Yb $m_J=0$ & 46 & 126.15& $-4.999$ & 1.82 & $-1.62$ & $-0.036$  \\ 
${}^{174}$Yb $m_J=-1$ & 80 & 162.40& $-4.998$ & 6.20 & $-1.16$ & $0.11$  \\ \hline\hline
\end{tabular}
\label{table:main_paremeter_list}
\end{table}

\begin{figure}[t]
\centering
\includegraphics[width=8.6cm,clip]{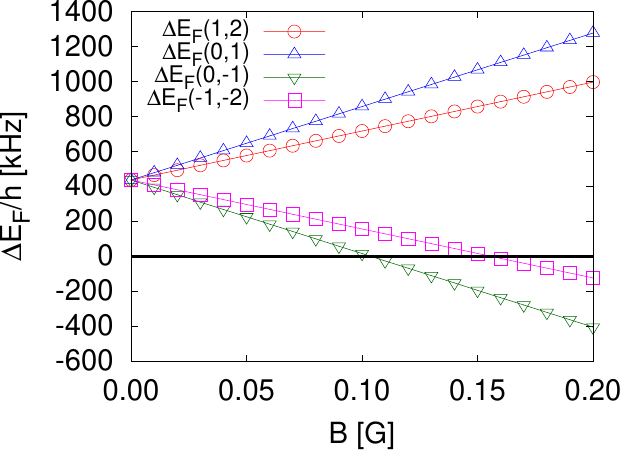}
\caption{Magnetic-field dependence of the F\"orster defect. Here, $\Delta E_{\rm F}(m_J,m_J')$ denotes the F\"orster defect for a pair of ${}^3S_1$ states with magnetic quantum number $m_J$ and a pair of ${}^3P_2$ states with magnetic quantum number $m_J'$. The solid black line indicates $\Delta E_{\rm F}(m_J,m_J')=0$.
}
\label{fig:forster_defect_n83}
\vspace{-0.75em}
\end{figure}

Here, we discuss the F\"orster resonances of ${}^{174}$Yb for $n=83$ in the presence of a magnetic field. As mentioned in Sec.~\ref{subsec:absence_of_magnetic_field}, the F\"orster resonances $|83, {}^3S_1,m_J=\pm1\rangle|84, {}^3S_1,m_J=\pm1\rangle \leftrightarrow |82, {}^3P_2,m_J=\pm2\rangle|84, {}^3P_2,m_J=\pm2\rangle$ for $m_J=\pm1$ and $|83, {}^3S_1,m_J=0\rangle|84, {}^3S_1,m_J=0\rangle \leftrightarrow |82, {}^3P_2,m_J=\pm1\rangle|84, {}^3P_2,m_J=\pm1\rangle$ for $m_J=0$ are already present at zero magnetic field. The F\"orster defects for these processes are extremely small but nonzero. We find that these F\"orster defects can be further reduced by applying a weak magnetic field. Figure \ref{fig:forster_defect_n83} shows the magnetic field dependence of the F\"orster defects. For $m_J=-1$, the F\"orster defect becomes smaller than $1~{\rm kHz}$ at $B\simeq 0.156~{\rm G}$, whereas for $m_J=0$ it becomes smaller than $1~{\rm kHz}$ at $B\simeq 0.103~{\rm G}$.

We now evaluate the $C_3$ coefficients for these resonances. We first consider the case $m_J=\pm1$. For notational simplicity, we define $\cket{1}\equiv |83,{}^3S_1,m_J=\pm 1\rangle$, $\cket{2}\equiv |84,{}^3S_1,m_J=\pm 1\rangle$, $\cket{3}\equiv |82,{}^3P_2,m_J=\pm 2\rangle$, and $\cket{4}\equiv |84,{}^3P_2,m_J=\pm2\rangle$. The interaction Hamiltonian for this four-level system can be written as
\begin{align}
\hat{H}_{\rm int}^{m_J=\pm1}&=\frac{h C_3^{\pm}}{R^3}\left(\cket{3}\otimes\cket{4}\bra{1}\otimes\bra{2}+\cket{1}\otimes\cket{2}\bra{3}\otimes\bra{4}\right.\notag \\
&\left.\hspace{3.0em}+\cket{4}\otimes\cket{3}\bra{2}\otimes\bra{1}+\cket{2}\otimes\cket{1}\bra{4}\otimes\bra{3}\right)\notag \\
&+\frac{h C_3^{'\pm}}{R^3}\left(\cket{4}\otimes\cket{3}\bra{1}\otimes\bra{2}+\cket{1}\otimes\cket{2}\bra{4}\otimes\bra{3}\right.\notag \\
&\left.\hspace{3.0em}+\cket{3}\otimes\cket{4}\bra{2}\otimes\bra{1}+\cket{2}\otimes\cket{1}\bra{3}\otimes\bra{4}\right),
\label{eq:definition_of_interaction_Hamiltonian_four_level_systems}\\
C_3^{\pm}&\equiv -\frac{3}{8\pi\epsilon_0h}\bra{3}\hat{d}^{\pm}\cket{1}\bra{4}\hat{d}^{\pm}\cket{2},
\label{eq:definition_of_C3_mJ_pm1}\\
C_3^{'\pm}&\equiv -\frac{3}{8\pi\epsilon_0h}\bra{4}\hat{d}^{\pm}\cket{1}\bra{3}\hat{d}^{\pm}\cket{2}.
\label{eq:definition_of_C3'_mJ_pm1}
\end{align}
We numerically evaluate $C_3^{\pm}$ and $C_3^{'\pm}$ and obtain $C_3^{\pm}=-23.30~{\rm GHz}\cdot\mu{\rm m}^3$ and $C_3^{'\pm}=-0.40~{\rm GHz}\cdot\mu{\rm m}^3$ at zero magnetic field. In the weak-magnetic-field regime, the magnetic-field dependence of the $C_3$ coefficients is very weak, and we therefore neglect it. 

We next consider the case $m_J=0$. We define $\cket{1}\equiv |83,{}^3S_1,m_J=0\rangle$, $\cket{2}\equiv |84,{}^3S_1,m_J=0\rangle$, $\cket{3}\equiv |82,{}^3P_2,m_J=1\rangle$, $\cket{4}\equiv |84,{}^3P_2,m_J=1\rangle$, $\cket{5}\equiv |82,{}^3P_2,m_J=-1\rangle$, and $\cket{6}\equiv |84,{}^3P_2,m_J=-1\rangle$. The interaction Hamiltonian for this six-level system is written as
\begin{align}
\hat{H}_{\rm int}^0&=\frac{h C_3^0}{R^3}(\cket{3}\otimes\cket{4}\bra{1}\otimes\bra{2}+\cket{1}\otimes\cket{2}\bra{3}\otimes\bra{4}\notag \\
&\hspace{3.0em}+\cket{4}\otimes\cket{3}\bra{2}\otimes\bra{1}+\cket{2}\otimes\cket{1}\bra{4}\otimes\bra{3}\notag\\
&\hspace{3.0em}+\cket{5}\otimes\cket{6}\bra{1}\otimes\bra{2}+\cket{1}\otimes\cket{2}\bra{5}\otimes\bra{6}\notag \\
&\hspace{3.0em}+\cket{6}\otimes\cket{5}\bra{2}\otimes\bra{1}+\cket{2}\otimes\cket{1}\bra{6}\otimes\bra{5})\notag \\
&+\frac{h C_3^{'0}}{R^3}(\cket{4}\otimes\cket{3}\bra{1}\otimes\bra{2}+\cket{1}\otimes\cket{2}\bra{4}\otimes\bra{3}\notag \\
&\hspace{3.0em}+\cket{3}\otimes\cket{4}\bra{2}\otimes\bra{1}+\cket{2}\otimes\cket{1}\bra{3}\otimes\bra{4}\notag \\
&\hspace{3.0em}+\cket{6}\otimes\cket{5}\bra{1}\otimes\bra{2}+\cket{1}\otimes\cket{2}\bra{6}\otimes\bra{5}\notag \\
&\hspace{3.0em}+\cket{5}\otimes\cket{6}\bra{2}\otimes\bra{1}+\cket{2}\otimes\cket{1}\bra{5}\otimes\bra{6}),
\label{eq:definition_interaction_Hamiltonian_for_six_level_systems}\\
C_3^0&\equiv -\frac{3}{8\pi\epsilon_0 h}\bra{3}\hat{d}^+\cket{1}\bra{4}\hat{d}^+\cket{2}\notag \\
&=-\frac{3}{8\pi\epsilon_0 h}\bra{5}\hat{d}^-\cket{1}\bra{6}\hat{d}^-\cket{2},
\label{eq:definition_of_C3_mJ=0}\\
C_3^{'0}&\equiv -\frac{3}{8\pi\epsilon_0 h}\bra{4}\hat{d}^+\cket{1}\bra{3}\hat{d}^+\cket{2}\notag \\
&=-\frac{3}{8\pi\epsilon_0 h}\bra{6}\hat{d}^-\cket{1}\bra{5}\hat{d}^-\cket{2},
\label{eq:definition_of_C3'_mJ=0}
\end{align}
where we have used time-reversal symmetry at zero magnetic field. The resulting $C_3$ coefficients are $C_3^0=-11.65~{\rm GHz}\cdot\mu{\rm m}^3$ and $C_3^{'0}=-0.20~{\rm GHz}\cdot\mu{\rm m}^3$.

\subsection{Application to one-dimensional chains}\label{subsec:application_to_one-dimensional}
Here, we apply our results to many-body problems. In this subsection, we consider $M$ Rydberg atoms arranged in a one-dimensional ring or an open chain. In our previous work~\cite{kunimi2025proposal}, we proposed realizations of the spin-1/2 $J_1$-$J_2$ Heisenberg chain and the spin-1 Heisenberg model by appropriately choosing the atomic positions and the magnetic-field strength. In the present work, we focus on the large-$|\delta|$ regime, because ${}^{174}$Yb can readily access this regime, as shown in the previous sections.

For the ring case, we place the $M$ Rydberg atoms on a ring in the $xy$ plane. The position of the $j$th atom is given by $\bm{r}_j \equiv r(\cos\phi_j\bm{e}_x+\sin\phi_j\bm{e}_y)$, where $r$ is the radius of the ring and $\phi_j \equiv 2\pi (j-1)/M$ $(j=1,2,\ldots,M)$. From Eq.~(\ref{eq:effective_Hamiltonian_spin_representation}), the many-body Hamiltonian for periodic boundary conditions (PBCs) is given by
{\small
\begin{align}
\hat{H}_{\rm PBC}&\equiv\frac{1}{2}\sum_{j,k,\,j\neq k}J_{jk}(\hat{S}_j^x\hat{S}_k^x+\hat{S}_j^y\hat{S}_k^y+\delta\hat{S}_j^z\hat{S}_k^z)+h^z\sum_{j=1}^M\hat{S}_j^z,
\label{eq:definition_of_XXZ_Hamiltonian_PBC}\\
J_{jk}&\equiv \frac{2h C_6}{|\bm{r}_j-\bm{r}_k|^6},
\label{eq:definition_of_Jjk_PBC}\\
h^z&\equiv E_{\uparrow}-E_{\downarrow}+h\frac{C_6^{\uparrow\uparrow}-C_6^{\downarrow\downarrow}}{2r_0^6}F_0,
\label{eq:definition_of_hj^z_PBC}\\
F_0&\equiv \frac{1}{8M^6}\sum_{j=2}^M\frac{1}{(1-\cos\phi_j)^3},
\label{eq:definition_of_f0}
\end{align}
}where we have defined $r_0 \equiv r/M$. For sufficiently large $M$, we obtain $F_0 \simeq 3.3\times10^{-5}$.

Here, we derive the effective Hamiltonian in the large-anisotropy regime $(|\delta|\gg 1)$. As discussed in Sec.~\ref{subsec:absence_of_magnetic_field}, ${}^{174}$Yb is ideal for exploring this regime. To this end, we define the Hilbert subspace as
\begin{align}
\mathcal{H}_P \equiv {\rm Span}\{ \cket{\bm{n}} \in \mathcal{H} ~|~ (n_j,n_{j+1}) \neq (\uparrow_j,\uparrow_{j+1})~\text{for all}~ j \},
\label{eq:definition_of_Hilbert_subspace_Hp}
\end{align}
where $\mathcal{H}$ is the spin-1/2 Hilbert space, $\cket{\bm{n}} \equiv \cket{n_1,n_2,\ldots,n_M}$ is a product basis in $\mathcal{H}$, $n_j=\uparrow_j, \downarrow_j$, and $n_{M+1}=n_1$ due to the periodic boundary conditions. In this subspace, no state contains the local configuration $\uparrow\uparrow$. The projection operator onto the subspace $\mathcal{H}_P$ is defined as
\begin{align}
\hat{\mathcal{P}}&\equiv \prod_{j=1}^{M}(1-\hat{P}_j'\hat{P}_{j+1}'),
\label{eq:definition_of_projection_operator_onto_Hp_OBC}\\
\hat{P}_j'&\equiv \frac{1}{2}+\hat{S}_j^z.
\label{eq:definition_of_Pj'}
\end{align}
We decompose the Hamiltonian into nonperturbative and perturbative parts as
\begin{align}
\hat{H}_{\rm PBC}&\equiv \hat{H}_0+\hat{V},
\label{eq:rewrite_Hamiltonian_perturbative_part_PBC}\\
\hat{H}_0&\equiv \delta J_1\sum_{j=1}^M\hat{S}_j^z\hat{S}_{j+1}^z+h^z\sum_{j=1}^M\hat{S}_j^z,
\label{eq:definition_of_H0_for_PBC}\\
\hat{V}&\equiv J_1\sum_{j=1}^M(\hat{S}_j^x\hat{S}_{j+1}^x+\hat{S}_{j}^y\hat{S}_{j+1}^y)\notag \\
&\quad+\delta J_2\sum_{j=1}^M\hat{S}_j^z\hat{S}_{j+2}^z+\hat{H}',
\label{eq:definition_of_V_for_PBC}
\end{align}
where $J_1\equiv 2hC_6/[2r\sin(\pi/M)]^6$ and $J_2\equiv 2hC_6/[2r\sin(2\pi/M)]^6$ denote the strengths of the nearest-neighbor and next-nearest-neighbor interactions, respectively. The effective Hamiltonian up to first order is given by
\begin{align}
\hat{H}^{\rm PBC}_{\rm eff}&=J_1\sum_{j=1}^M\hat{P}_{j-1}(\hat{S}_j^x\hat{S}_{j+1}^x+\hat{S}_j^y\hat{S}_{j+1}^y)\hat{P}_{j+2}
\notag \\
&+(h^z-\delta J_1)\sum_{j=1}^M\hat{S}_j^z+J_2\delta\sum_{j=1}^M\hat{\mathcal{P}}\hat{S}_j^z\hat{S}_{j+2}^z\hat{\mathcal{P}}+\hat{\mathcal{P}}\hat{H}'\hat{\mathcal{P}},
\label{eq:effective_Hamiltonian_upto_first_order_PBC}\\
\hat{P}_j&\equiv \frac{1}{2}-\hat{S}_j^z.
\label{eq:definition_of_Pj}
\end{align}
Here, we omit the constant terms arising from $\hat{\mathcal{P}}\hat{H}_0\hat{\mathcal{P}}$.

When the next-nearest-neighbor interaction terms and $\hat{H}'$ are neglected, this Hamiltonian reduces to the folded {\it XXZ} model~\cite{zadnik2021folded,zadnik2021folded2,pozsgay2021integrable}, which is integrable. In our setup, however, integrability is broken by the additional $J_2$ term and the longer-range interactions contained in $\hat{H}'$, both of which originate from the van der Waals $1/R^6$ interaction. If we neglect $\hat{H}'$ while retaining the next-nearest-neighbor $J_2$ $ZZ$ interaction, the Hamiltonian becomes identical to that discussed in Ref.~\cite{yang2020hilbert}. That work showed that this model exhibits Hilbert-space fragmentation, in which the Hilbert space splits into exponentially many dynamically disconnected sectors~\cite{Sala2020,Khemani2020}. In this case, the origin of the Hilbert-space fragmentation is the conservation of the domain-wall (DW) number:
\begin{align}
\hat{N}_{\rm DW}^{\rm PBC}\equiv 2\sum_{j=1}^M\left(\frac{1}{4}-\hat{S}_j^z\hat{S}_{j+1}^z\right).\label{eq:definition_of_domain_wall_number_operator}
\end{align}
Our proposal therefore provides a way to observe Hilbert-space fragmentation in the large-$|\delta|$ regime.

Other interesting phenomena have also been discussed in the folded {\it XXZ} model. Recently, quantum spin transport in this model has been investigated, and it has been shown analytically that some spin probability distributions are described by the Gaussian unitary ensemble Tracy-Widom distribution~\cite{fujimoto2024quantum}. The origin of this behavior may be related to the integrability of the folded {\it XXZ} model. We expect that this nontrivial distribution could be observed in our setup as a transient phenomenon.

\begin{figure}[t]
\centering
\includegraphics[width=8.6cm,clip]{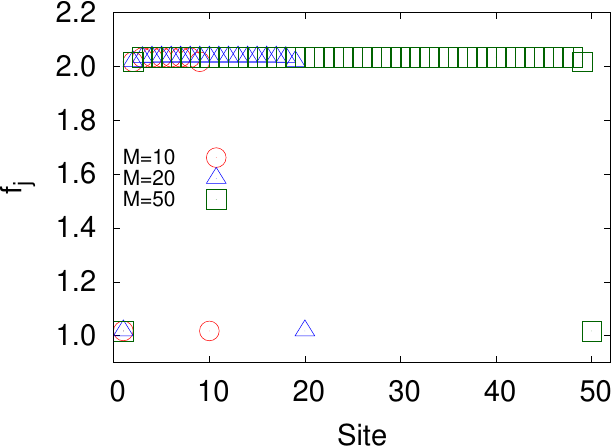}
\caption{Spatial dependence of $f_j$ for $M=10, 20,$ and $50$.
}
\label{fig:spatial_dependence_fj}
\vspace{-0.75em}
\end{figure}

For open chains, we place the $M$ Rydberg atoms in a one-dimensional array with a lattice constant $d$ in the $xy$-plane. From Eq.~(\ref{eq:effective_Hamiltonian_spin_representation}), the many-body Hamiltonian for open boundary conditions (OBCs) becomes{\small
\begin{align}
\hat{H}_{\rm OBC}&\equiv\frac{1}{2}\sum_{j,k,j\not=k}J_{jk}(\hat{S}_j^x\hat{S}_k^x+\hat{S}_j^y\hat{S}_k^y+\delta\hat{S}_j^z\hat{S}_k^z)+\sum_{j=1}^Mh_j^z\hat{S}_j^z,\label{eq:definition_of_XXZ_Hamiltonian_OBC}\\
J_{jk}&\equiv \frac{2h C_6}{d^6(j-k)^6},\label{eq:definition_of_Jjk_OBC}\\
h_j^z&\equiv E_{\uparrow}-E_{\downarrow}+h\frac{C_6^{\uparrow\uparrow}-C_6^{\downarrow\downarrow}}{2d^6}f_j,\label{eq:definition_of_hj^z}\\
f_j&\equiv \sum_{k\not=j}\frac{1}{(j-k)^6},~j=1,2,\ldots,M,\label{eq:definition_of_f_j}
\end{align}
}where $J_{jk}$ is the strength of the interaction between the sites $j$ and $k$, and $h_j^z$ is the strength of an effective magnetic field. Although the third term in Eq.~(\ref{eq:definition_of_hj^z}) has spatial dependence through $f_j$, which originates from interaction-induced Zeeman terms, we neglect it for simplicity. We can expect that the spatial dependence does not significantly affect the properties of the bulk because the nonuniformity of this term arises only on the edges of the system as shown in Fig.~\ref{fig:spatial_dependence_fj}. If one desires a more accurate Hamiltonian, the nonuniformity can be compensated by applying an additional AC Stark shift at the edges of the system. This procedure has been demonstrated experimentally~\cite{Bornet2024}. Hereafter, we use $h^z$ instead of $h_j^z$.

To derive the effective Hamiltonian for open boundary conditions, we define the Hilbert subspace as
\begin{align}
\mathcal{H}_P^{\rm OBC}&\equiv {\rm Span}\{\cket{\bm{n}}\in\mathcal{H}~|~(n_j,n_{j+1})\not=(\uparrow_j,\uparrow_{j+1})\notag \\
&\hspace{10.0em} \text{for }j=1,2,\ldots,M-1 \}.
\label{eq:definition_of_Hilbert_subspace_OBC_case}
\end{align}
We note that $\mathcal{H}_P^{\rm OBC}\neq\mathcal{H}_P$, because the configuration with $n_1=\uparrow_1$ and $n_M=\uparrow_M$ is allowed in the case of open boundary conditions. The projection operator onto the subspace $\mathcal{H}_P^{\rm OBC}$ is defined as
\begin{align}
\hat{\mathcal{P}}^{\rm OBC}\equiv \prod_{j=1}^{M-1}(1-\hat{P}_j'\hat{P}_{j+1}').
\label{eq:definition_of_projection_operator_for_OBC}
\end{align}
We decompose the Hamiltonian into nonperturbative and perturbative parts as
\begin{align}
\hat{H}_{\rm OBC}&\equiv \hat{H}_0^{\rm OBC}+\hat{V}^{\rm OBC},
\label{eq:rewrite_H_OBC_for_perturbation}\\
\hat{H}_0^{\rm OBC}&\equiv \delta J_1^{\rm OBC}\sum_{j=1}^{M-1}\hat{S}_j^z\hat{S}_{j+1}^z+h^z\sum_{j=1}^M\hat{S}_j^z,
\label{eq:definition_of_H0_OBC}\\
\hat{V}^{\rm OBC}&\equiv J_1^{\rm OBC}\sum_{j=1}^{M-1}(\hat{S}_j^x\hat{S}_{j+1}^x+\hat{S}_j^y\hat{S}_{j+1}^y)
\notag \\
&\quad +\delta J_2^{\rm OBC}\sum_{j=1}^{M-2}\hat{S}_j^z\hat{S}_{j+2}^z+\hat{H}'_{\rm OBC},
\label{eq:definition_of_V_OBC}
\end{align}
where $J_1^{\rm OBC}\equiv 2hC_6/d^6$ and $J_2^{\rm OBC}\equiv hC_6/(32d^6)=J_1^{\rm OBC}/64$. The effective Hamiltonian for open boundary conditions is given by{\footnotesize
\begin{align}
&\hat{H}_{\rm eff}^{\rm OBC}\notag \\
&=J_1^{\rm OBC}\sum_{j=1}^{M-1}\hat{P}_{j-1}(\hat{S}_j^x\hat{S}_{j+1}^x+\hat{S}_j^y\hat{S}_{j+1}^y)\hat{P}_{j+2}
\notag \\
&\quad +(h^z-\delta J_1^{\rm OBC})\sum_{j=1}^M\hat{S}_j^z+J_2^{\rm OBC}\delta\sum_{j=1}^{M-2}\hat{\mathcal{P}}^{\rm OBC}\hat{S}_j^z\hat{S}_{j+2}^z\hat{\mathcal{P}}^{\rm OBC}
\notag \\
&\quad +\hat{\mathcal{P}}^{\rm OBC}\hat{H}_{\rm OBC}'\hat{\mathcal{P}}^{\rm OBC}+\frac{J_1^{\rm OBC}\delta}{2}(\hat{S}_1^z+\hat{S}_M^z),
\label{eq:effctive_Hamiltonian_for_OBC_case}
\end{align}
}where $\hat{P}_0=1$ and $\hat{P}_{M+1}=1$. In contrast to the periodic boundary conditions, this Hamiltonian does not commute with the domain-wall-number operator defined as
\begin{align}
\hat{N}_{\rm DW}^{\rm OBC}\equiv 2\sum_{j=1}^{M-1}\left(\frac{1}{4}-\hat{S}_j^z\hat{S}_{j+1}^z\right).
\label{eq:definition_of_domain_number_operator_for_OBC}
\end{align}
The noncommuting contributions arise at the edges of the system. For example, consider the following process:
\begin{align}
(\hat{S}_1^x\hat{S}_2^x+\hat{S}_1^y\hat{S}_2^y)\cket{\uparrow_1\downarrow_2\downarrow_3}\propto \cket{\downarrow_1\uparrow_2\downarrow_3}.
\label{eq:process_breaks_domain_wall_number}
\end{align}
In this process, the domain-wall number changes from $1$ to $2$. To compensate for the edge effects, we modify the Hamiltonian and the domain-wall-number operator as
\begin{align}
\hat{H}_0'^{\rm OBC}&\equiv \hat{H}_0^{\rm OBC}-\frac{J_1^{\rm OBC}\delta}{2}(\hat{S}_1^z+\hat{S}_M^z),
\label{eq:modify_H0_OBC}\\
\hat{N}_{\rm DW}'^{\rm OBC}&\equiv 2\sum_{j=0}^{M}\left(\frac{1}{4}-\hat{S}_j^z\hat{S}_{j+1}^z\right)
\notag \\
&=\hat{N}_{\rm DW}^{\rm OBC}+\left(\frac{1}{2}+\hat{S}_1^z\right)+\left(\frac{1}{2}+\hat{S}_M^z\right),
\label{eq:definition_of_modified_domain_wall_number_operator}
\end{align}
where $\hat{S}_0^z=-1/2$ and $\hat{S}_{M+1}^z=-1/2$. A direct calculation shows that the modified domain-wall-number operator commutes with the modified effective Hamiltonian,
\begin{align}
\hat{H}_{\rm eff}'^{\rm OBC}\equiv \hat{H}_{\rm eff}^{\rm OBC}-\frac{J_1^{\rm OBC}\delta}{2}(\hat{S}_1^z+\hat{S}_M^z).
\label{eq:definition_of_modified_effective_Hamiltonian}
\end{align}
A similar modification has been used for the DH model~\cite{Kunimi2024,kunimi2025systematic}. The edge magnetic-field terms can be realized by applying the AC Stark shift at the edges of the system~\cite{Bornet2024} as discussed above.

\subsection{Application to two-dimensional square lattices}\label{subsec:application_to_two-dimensional_square_lattice}
We now consider applications of our results to two-dimensional systems. Since spin-1/2 systems can be mapped onto hard-core boson systems, the {\it XXZ} model can be regarded as a hard-core Bose-Hubbard model with $1/r^6$ hopping and density-density interactions. This model is similar to the Bose-Hubbard model with nearest-neighbor hopping and dipolar interactions ($1/r^3$), which is known to exhibit a supersolid phase~\cite{dang2008vacancy,danshita2010critical,Capogrosso-Sansone2010quantum,yamamoto2012quantum}. 

The supersolid phase is characterized by the coexistence of superfluidity and crystalline order~\cite{andreev1969quantum,chester1970speculations,Leggett1970can}. In continuum systems, the supersolid phase has been observed in ultracold dipolar bosonic systems~\cite{tanzi2019observation,Bottcher2019transient,Chomaz2019long,guo2019low,norcia2021two}. In lattice systems, the supersolid phase has been discussed extensively and is characterized by the spontaneous breaking of discrete translational symmetry~\cite{matsuda1970off}. Although the dipolar Bose-Hubbard model has been realized experimentally~\cite{baier2016extended,su2023dipolar}, the supersolid phase has not yet been observed in lattice systems. 

In this subsection, we explore the ground-state phase diagram of our {\it XXZ} model within the mean-field approximation. Our method is similar to that used in previous works~\cite{Murthy1997superfluids,Bernardet2002analytical,danshita2010critical,yamamoto2012quantum}.

For simplicity, we consider an infinite two-dimensional (2D) square lattice. The position of the $j$th lattice site is denoted by $\bm{R}_j \equiv d(n_j\bm{e}_x+m_j\bm{e}_y)$, where $d$ is the lattice constant and $n_j$ and $m_j$ are integers. The Hamiltonian is defined as
\begin{align}
\hat{H}_{\rm 2D}&\equiv \frac{1}{2}\sum_{j,k,\,j\neq k}J_{jk}(\hat{S}_j^x\hat{S}_k^x+\hat{S}_j^y\hat{S}_k^y+\delta\hat{S}_j^z\hat{S}_k^z)-h_0\sum_j\hat{S}_j^z,
\label{eq:Hamiltonian_2D_system}\\
J_{jk}&\equiv \frac{2hC_6}{|\bm{R}_j-\bm{R}_k|^6}\equiv -\frac{J_0d^6}{|\bm{R}_j-\bm{R}_k|^6},
\label{eq:definition_of_interaction_in_2D}\\
h_0&\equiv -(E_{\uparrow}-E_{\downarrow})-2h\frac{C_6^{\uparrow\uparrow}-C_6^{\downarrow\downarrow}}{d^6}C_+,
\label{eq:definition_of_hz_2D}\\
C_+&\equiv \sum_{n=1}^{\infty}\frac{1}{n^6}+\sum_{n,m=1}^{\infty}\frac{1}{(n^2+m^2)^3}\simeq 1.165.
\label{eq:definition_of_C+}
\end{align}
In the hard-core boson representation, we map the states $\cket{\uparrow}$ and $\cket{\downarrow}$ onto the empty and occupied states $\cket{0}$ and $\cket{1}$, respectively. We assume $J_0>0$ and $\delta\le 0$, corresponding to positive hopping and repulsive density-density interactions. For example, this parameter regime can be realized in ${}^{174}$Yb for $n\gtrsim 80$ at zero magnetic field (see Figs.~\ref{fig:C6_in_the_absence_of_magnetic_field} and \ref{fig:delta_in_the_absence_of_magnetic_field}). For nonzero magnetic fields, we find $C_6\simeq -185.33~{\rm GHz}\cdot\mu{\rm m}^6$ and $\delta\simeq -2.06$ for $n=70$ and $m_J=1$ at $B=90~{\rm G}$. 

Here, we perform a mean-field (MF) analysis. We assume that the ground-state wave function is given by
\begin{align}
&\cket{\Psi_{\rm MF}}\notag \\
&\equiv \prod_j\left[e^{-i\varphi_j/2}\cos(\theta_j/2)\cket{\uparrow_j}+e^{+i\varphi_j/2}\sin(\theta_j/2)\cket{\downarrow_j}\right],
\label{eq:mean-field_wave_function}
\end{align}
where $0\le \theta_j\le \pi$ and $0\le \varphi_j<2\pi$. We define the mean-field energy as
\begin{align}
E_{\rm MF}\equiv \bra{\Psi_{\rm MF}}\hat{H}_{\rm 2D}\cket{\Psi_{\rm MF}}.
\label{eq:definition_of_mean-field_energy}
\end{align}
Using Eq.~(\ref{eq:mean-field_wave_function}), the mean-field energy reduces to
\begin{align}
E_{\rm MF}&=\frac{1}{8}\sum_{j,k,\,j\neq k}J_{jk}\left\{\left[\cos(\varphi_j-\varphi_k)\right]\sin\theta_j\sin\theta_k\right.\notag\\
&\left.\quad +\delta\cos\theta_j\cos\theta_k\right\}-\frac{h_0}{2}\sum_j\cos\theta_j.
\label{eq:mean-field_energy}
\end{align}
Without loss of generality, we may set $\varphi_j=0$ for all $j$, because we consider the ground state and the system possesses $U(1)$ symmetry. The spins are then aligned in the $xz$ plane. In addition, we employ the two-sublattice ansatz~\cite{danshita2010critical,yamamoto2012quantum}
\begin{align}
\bm{S}_j&=
\begin{cases}
\vspace{0.5em}\displaystyle{\frac{1}{2}(\sin\theta_A\bm{e}_x+\cos\theta_A\bm{e}_z)},\quad j\in \text{sublattice }A\\
\displaystyle{\frac{1}{2}(\sin\theta_B\bm{e}_x+\cos\theta_B\bm{e}_z)},\quad j\in \text{sublattice }B,
\end{cases}
\label{eq:two-sublattice_ansatz_for_2D}
\end{align}
where sublattices $A$ and $B$ are defined by $n_j+m_j$ being even and odd, respectively, and $\bm{S}_j\equiv \bra{\Psi_{\rm MF}}\hat{\bm{S}}_j\cket{\Psi_{\rm MF}}$ is the mean-field spin expectation value. The mean-field energy density then reduces to
\begin{align}
\mathcal{E}&=-\frac{J_0}{8}(C_+-C_-)(\sin^2\theta_A+\sin^2\theta_B)\notag \\
&-\frac{J_0}{4}(C_++C_-)\sin\theta_A\sin\theta_B\notag \\
&-\frac{\delta J_0}{8}(C_+-C_-)(\cos^2\theta_A+\cos^2\theta_B)\notag \\
&-\frac{\delta J_0}{4}(C_++C_-)\cos\theta_A\cos\theta_B\notag \\
&-\frac{h_0}{4}(\cos\theta_A+\cos\theta_B),
\label{eq:mean-field_energy_two-sublattice_ansat}\\
C_-&\equiv \sum_{n=1}^{\infty}\frac{(-1)^{n-1}}{n^6}+\sum_{n,m=1}^{\infty}\frac{(-1)^{n+m-1}}{(n^2+m^2)^{3}}\simeq 0.874,
\label{eq:definition_of_C-}
\end{align}
where $\mathcal{E}$ is defined as the mean-field energy per site.

Within the two-sublattice ansatz, the following five phases appear in the ground state~\cite{danshita2010critical,yamamoto2012quantum}:
\begin{align}
\theta_A&=\theta_B=0\quad (\text{empty}),
\label{eq:definition_of_empty_state}\\
\theta_A&=\theta_B=\pi\quad (\text{MI}),
\label{eq:definition_of_Mott_insulator}\\
\theta_A&=\theta_B\equiv \theta,\quad (0<\theta<\pi) \quad (\text{SF}),
\label{eq:definition_of_superfluid}\\
\theta_A&=0 \text{ and }\theta_B=\pi, \text{ or }\theta_A=\pi \text{ and } \theta_B=0\; (\text{CS}),
\label{eq:definition_of_CS}\\
\theta_A&\neq\theta_B,\quad (0<\theta_A, \theta_B<\pi)\quad (\text{CSS}),
\label{eq:definition_of_CSS}
\end{align}
where MI, SF, CS, and CSS denote the Mott insulator, superfluid, checkerboard solid, and checkerboard supersolid, respectively. The energy densities of the empty, MI, and CS states are given by
\begin{align}
\mathcal{E}_{\rm empty}&=-\frac{h_0}{2}-\frac{\delta J_0C_+}{2},
\label{eq:energy_of_empty_state}\\
\mathcal{E}_{\rm MI}&=+\frac{h_0}{2}-\frac{\delta J_0C_+}{2},
\label{eq:energy_of_Mott_insulator_state}\\
\mathcal{E}_{\rm CS}&=\frac{\delta J_0C_-}{2}.
\label{eq:energy_of_CS}
\end{align}
Substituting Eq.~(\ref{eq:definition_of_superfluid}) into Eq.~(\ref{eq:mean-field_energy_two-sublattice_ansat}) and minimizing $\mathcal{E}$ with respect to $\theta$, we obtain the solution for the SF state,
\begin{align}
\theta&=\cos^{-1}\left[\frac{h_0}{2C_+(1-\delta)J_0}\right]
\notag \\
&\quad \text{for }-2C_+(1-\delta)\le\frac{h_0}{J_0}\le 2C_+(1-\delta),
\label{eq:theta_for_SF_State}
\end{align}
where we have used $1-\delta\ge 1$. Therefore, the energy density of the SF state is given by
\begin{align}
\mathcal{E}_{\rm SF}=-\frac{J_0C_+}{2}-\frac{h^2_0}{8C_+(1-\delta)J_0}.
\label{eq:energy_of_SF_state}
\end{align}
By minimizing Eq.~(\ref{eq:mean-field_energy_two-sublattice_ansat}) with respect to $\theta_A$ and $\theta_B$ under the condition $\theta_A\neq\theta_B$, we obtain the energy density of the CSS state as follows (see the derivation in Appendix~\ref{app:derivation_CSS}):
\begin{align}
\mathcal{E}_{\rm CSS\pm}&=\frac{\delta J_0C_-}{2}-\frac{[h_0\pm \sqrt{1-\alpha^2}(C_+-C_-+2\delta C_-)J_0]^2}{8(1-\delta)J_0(C_+-C_-)},\label{eq:energy_CSS}\\
\alpha&\equiv \frac{C_++C_-}{-(C_+-C_-)-2\delta C_-},
\label{eq:re_alpha_Rydberg}
\end{align}
where $+(-)$ represents the solution for $\theta_A+\theta_B<\pi\; (\theta_A+\theta_B>\pi)$.

\begin{figure}[t]
\centering
\includegraphics[width=8.6cm,clip]{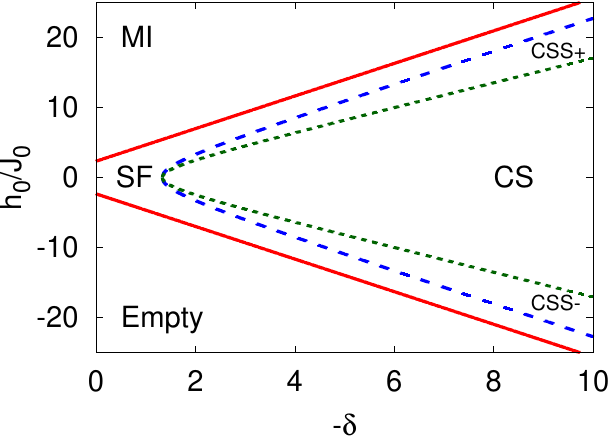}
\caption{Ground-state phase diagram. The solid red lines represent the MI-SF and empty-SF phase boundaries, the dashed blue line represents the SF-CSS phase boundary, and the dotted green line represents the CSS-CS phase boundary.
}
\label{fig:ground_state_phase_diagram}
\vspace{-0.75em}
\end{figure}

By comparing the energy densities, we can construct the mean-field ground-state phase diagram. The phase boundaries are obtained as
\begin{align}
\frac{h_0}{J_0}&=+2C_+(1+|\delta |)\quad \text{(SF-empty)},
\label{eq:boundary_SF-empty_Rydberg}\\
\frac{h_0}{J_0}&=-2C_+(1+|\delta|)\quad \text{(SF-MI)},
\label{eq:boundary_SF-MI_Rydberg}\\
\frac{h_0}{J_0}&=\sqrt{1-\alpha^2}\frac{C_+}{C_-}(-C_++C_-+2|\delta|C_-),
\notag \\
&\quad \text{for }|\delta|\ge C_+/C_-\quad \text{(SF-CSS$+$)},
\label{eq:boundary_SF-CSS+_Rydberg}\\
\frac{h_0}{J_0}&=-\sqrt{1-\alpha^2}\frac{C_+}{C_-}(-C_++C_-+2|\delta|C_-),
\notag \\
&\quad \text{for }|\delta|\ge C_+/C_-\quad \text{(SF-CSS$-$)},
\label{eq:boundary_SF-CSS-_Rydberg}\\
\frac{h_0}{J_0}&=+\sqrt{1-\alpha^2}(-C_++C_-+2|\delta|C_-),
\notag \\
&\quad \text{for }|\delta|\ge C_+/C_-\quad \text{(CSS$+$-CS)},
\label{eq:CSS+-CS_boundary_Rydberg}\\
\frac{h_0}{J_0}&=-\sqrt{1-\alpha^2}(-C_++C_-+2|\delta|C_-),
\notag \\
&\quad \text{for }|\delta|\ge C_+/C_-\quad \text{(CSS$-$-CS)}.
\label{eq:CSS--CS_boundary_Rydberg}
\end{align}
We note that all phase transitions are second order. The ground-state phase diagram is shown in Fig.~\ref{fig:ground_state_phase_diagram}. The supersolid phase appears in the region with large $|\delta|$ and intermediate $h_0/J_0$. This behavior is similar to that of dipolar hard-core bosons~\cite{danshita2010critical,yamamoto2012quantum}. As shown in those previous works, in the large-$|\delta|$ regime, long-period structures that cannot be captured by the two-sublattice ansatz emerge. Therefore, the CSS phase boundary may be modified in the intermediate-$|\delta|$ and intermediate-$h_0/J_0$ regions, which are readily accessible in ${}^{174}$Yb systems. 

Here, we discuss the stability of the CSS phase beyond the mean-field approximation. Previous work showed that the hard-core Bose-Hubbard model on the square lattice with next-nearest-neighbor interactions does not exhibit a supersolid phase based on quantum Monte Carlo calculations~\cite{Batrouni2000phase}, whereas the hard-core Bose-Hubbard model with dipolar interactions $(1/r^3)$ does exhibit supersolid phases~\cite{Capogrosso-Sansone2010quantum}. These results indicate that it is nontrivial whether our model, namely, hard-core bosons with $1/r^6$ hopping and density-density interactions, exhibits a supersolid phase beyond the mean-field approximation. Therefore, the stability of the CSS phase should be examined using beyond-mean-field methods, such as quantum Monte Carlo simulations~\cite{gubernatis2016quantum}.

Another possible route to realizing a supersolid phase is to consider hard-core bosons on triangular lattices. Previous works showed that supersolid phases of hard-core bosons on triangular lattices can survive beyond mean-field theory~\cite{Wessel2005supersolid,Yamamoto2012dipolar}. These results suggest that supersolid phases on triangular lattices are more robust against quantum fluctuations than those on square lattices. Since triangular lattices are feasible in current experiments, hard-core bosons on a triangular lattice may provide a promising candidate for realizing a supersolid phase in Rydberg-atom quantum simulators.

Finally, we remark on previous works discussing supersolidity in Rydberg tweezer arrays~\cite{liu2024supersolidity,homeier2025supersolidity}. Those works consider spin-1 systems with dipole-dipole interactions ($1/r^3$) encoded in three Rydberg states. In contrast, our system is a spin-1/2 system with van der Waals interactions ($1/r^6$).
\section{Summary}\label{sec:Summary}
In this paper, we investigated the interactions between the alkaline-earth(-like) Rydberg atoms ${}^{88}$Sr and ${}^{174}$Yb. We considered two Rydberg states, $|n,{}^3S_1,m_J\rangle$ and $|n+1,{}^3S_1,m_J\rangle$, and showed that the effective interaction Hamiltonian becomes an {\it XXZ} model as in the case of the alkali atoms. We found that the behavior of the anisotropy parameter of ${}^{174}$Yb at zero magnetic field is markedly different from that of ${}^{88}$Sr and alkali atoms. This difference arises from the interplay of strong spin-orbit coupling, multichannel mixing of the intermediate $P$ states, and the resulting redistribution of F\"orster defects. We also investigated the interaction properties in the presence of a magnetic field. As in the case of alkali atoms, the interaction parameters can be tuned by the magnetic field~\cite{kunimi2025proposal}. In addition, we found that ${}^{174}$Yb exhibits a F\"orster resonance at $n=83$ with an extremely small F\"orster defect.

As an application of the above results to quantum many-body problems, we considered two different setups. One is a one-dimensional ring or chain. In this setup, we showed that, in the large-$|\delta|$ regime, the effective Hamiltonian is reduced to the folded {\it XXZ} model with additional interaction terms. Because this model has a nontrivial conserved quantity, namely, the domain-wall-number operator, Hilbert-space fragmentation may emerge.

The other setup is a two-dimensional system. Since the spin-1/2 {\it XXZ} model can be mapped onto a hard-core boson model with long-range hopping and density-density interactions, a supersolid phase may arise in the corresponding hard-core boson system. Our mean-field analysis suggests that such a supersolid phase may appear in the regime of large  $|\delta|$ and intermediate $h_0/J_0$.

The present work suggests several interesting directions for future study. One possible extension is to generalize our analysis to dual-species or dual-isotope Rydberg systems~\cite{Zeng2017entangling,sheng2022defect,anand2024dual,nakamura2024hybrid,wei2024dual,white2026quantum,miles2026qubit,wang2026multi}. As we have shown, ${}^{174}$Yb exhibits interaction properties that differ from those of other atomic species. Exploiting this feature, one may be able to engineer more complex {\it XXZ} Hamiltonians. For example, ancillary atoms of a second species could be used as Rydberg mediators to engineer effective interactions between the primary Rydberg atoms~\cite{Kuznetsova2018effective,cesa2026engineering}. Several theoretical proposals have also explored quantum simulation with dual-species Rydberg-atom arrays~\cite{Li2024uncovering,wang2025tricritical,farouk2026quantum}. Another interesting direction is the study of three-dimensional systems. For example, in the large-anisotropy regime, the {\it XXZ} model on the pyrochlore lattice reduces to a ring-exchange Hamiltonian~\cite{Hermele2004pyrochlore}, which exhibits a spin-liquid ground state (we note that another route to realizing a ring-exchange Hamiltonian in Rydberg systems has also been proposed~\cite{shah2025quantum}). Three-dimensional atom arrays have also been developed~\cite{barredo2018synthetic,Kim2020quantum_Ising,Song2021quantum_simulation,kusano2025plane}. In the future, such exotic phases may be observed in Rydberg-atom quantum simulators.

\begin{acknowledgments}

The authors thank K. Fujimoto, Y. Takahashi, Y. Takasu, Y. Nakamura, R. Kuroda, S. Weber, J. M\"{o}gerle, H. Kuji, I. Danshita, T. Nikuni, and M. Shimizu for their useful discussions, and J. M\"{o}gerle in particular for a critical reading of the manuscript. We also acknowledge S. Era and Y. Imai for earlier contributions to the calculations of supersolid states. This work was supported by JSPS KAKENHI Grant No.~JP25K00215 (M.K.), JST ASPIRE Grant No.~JPMJAP24C2 (M.K.), MEXT Quantum Leap Flagship Program (MEXT Q-LEAP) Grant No. JPMXS0118069021 (T.T.), and JST Moonshot R\&D Program Grant No. JPMJMS2269 (T.T.).
\end{acknowledgments}

\section*{Data availability}
The data that support the findings of this article are openly available~\cite{kunimi_2026_21367229}.

\appendix
\section{Trend for $|C_6^{\uparrow\downarrow}|>|C_6|$}\label{app:trand_C6_parameters}
In this appendix, we discuss the origin of the trend $|C_6^{\uparrow\downarrow}|>|C_6|$ for ${}^{174}$Yb. Let us consider a set of intermediate pair states $\{\bm{n}\}$ that share the same F\"orster defect. We introduce the following quantities:
\begin{align}
F(\{\bm{n}\})&\equiv \sum_{\{\bm{n}\}}|\bra{\bm{n}}\hat{V}_{\rm dd}\cket{\uparrow\downarrow}|^2
=\sum_{\{\bm{n}\}}|\bra{\bm{n}}\hat{V}_{\rm dd}\cket{\downarrow\uparrow}|^2,
\label{eq:definition_of_f_of_n}\\
G(\{\bm{n}\})&\equiv \sum_{\{\bm{n}\}}\bra{\downarrow\uparrow}\hat{V}_{\rm dd}\cket{\bm{n}}\bra{\bm{n}}\hat{V}_{\rm dd}\cket{\uparrow\downarrow},
\label{eq:definition_of_g_of_n}
\end{align}
where in the second equality of Eq.~(\ref{eq:definition_of_f_of_n}) we have used the exchange symmetry of the dipole-dipole interaction. Here, we assume that $\bra{\downarrow\uparrow}\hat{V}_{\rm dd}\cket{\bm{n}}\bra{\bm{n}}\hat{V}_{\rm dd}\cket{\uparrow\downarrow}$ is positive for all states in $\{\bm{n}\}$. Using the Cauchy-Schwarz inequality, we obtain
\begin{align}
F(\{\bm{n}\})\ge G(\{\bm{n}\}).
\label{eq:inequality_for_f_and_g}
\end{align}
This result suggests that, for intermediate states with the same F\"orster defect, the absolute value of the corresponding contribution to $C_6^{\uparrow\downarrow}$ is no smaller than that of the corresponding contribution to $C_6$. This explains the trend $|C_6^{\uparrow\downarrow}|>|C_6|$.

\section{Magnetic-field and principal quantum number dependence of the physical quantities}\label{app:full_C6_parameters}
\begin{figure*}[t]
\centering
\includegraphics[width=17.0cm,clip]{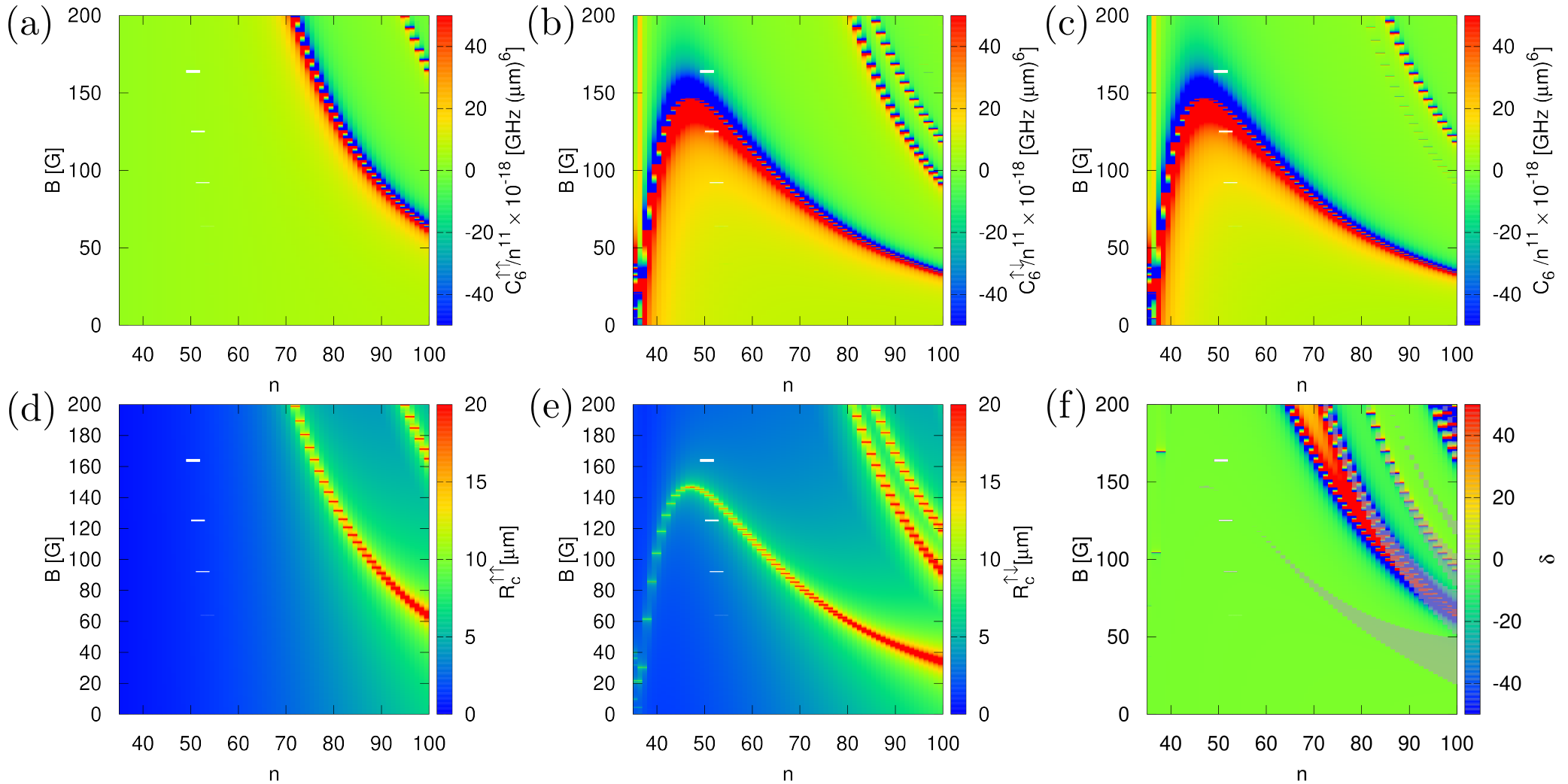}
\caption{Magnetic-field and principal quantum number dependence of various physical quantities for ${}^{88}$Sr with $m_J=+1$ at $\theta=\pi/2$: (a) $C_6^{\uparrow\uparrow}$, (b) $C_6^{\uparrow\downarrow}$, (c) $C_6$, (d) $R_{\rm c}^{\uparrow\uparrow}$, (e) $R_{\rm c}^{\uparrow\downarrow}$, and (f) anisotropy parameter $\delta$. The white region indicates the absence of data due to strong $S$-$D$ mixing. The gray-shaded region in panel (f) indicates the region where $R_{\rm c}\ge 10~\mu{\rm m}$. 
}
\label{fig:c6_and_delta_nonzero_field_sr_p}
\vspace{-0.75em}
\end{figure*}

\begin{figure*}[t]
\centering
\includegraphics[width=17.0cm,clip]{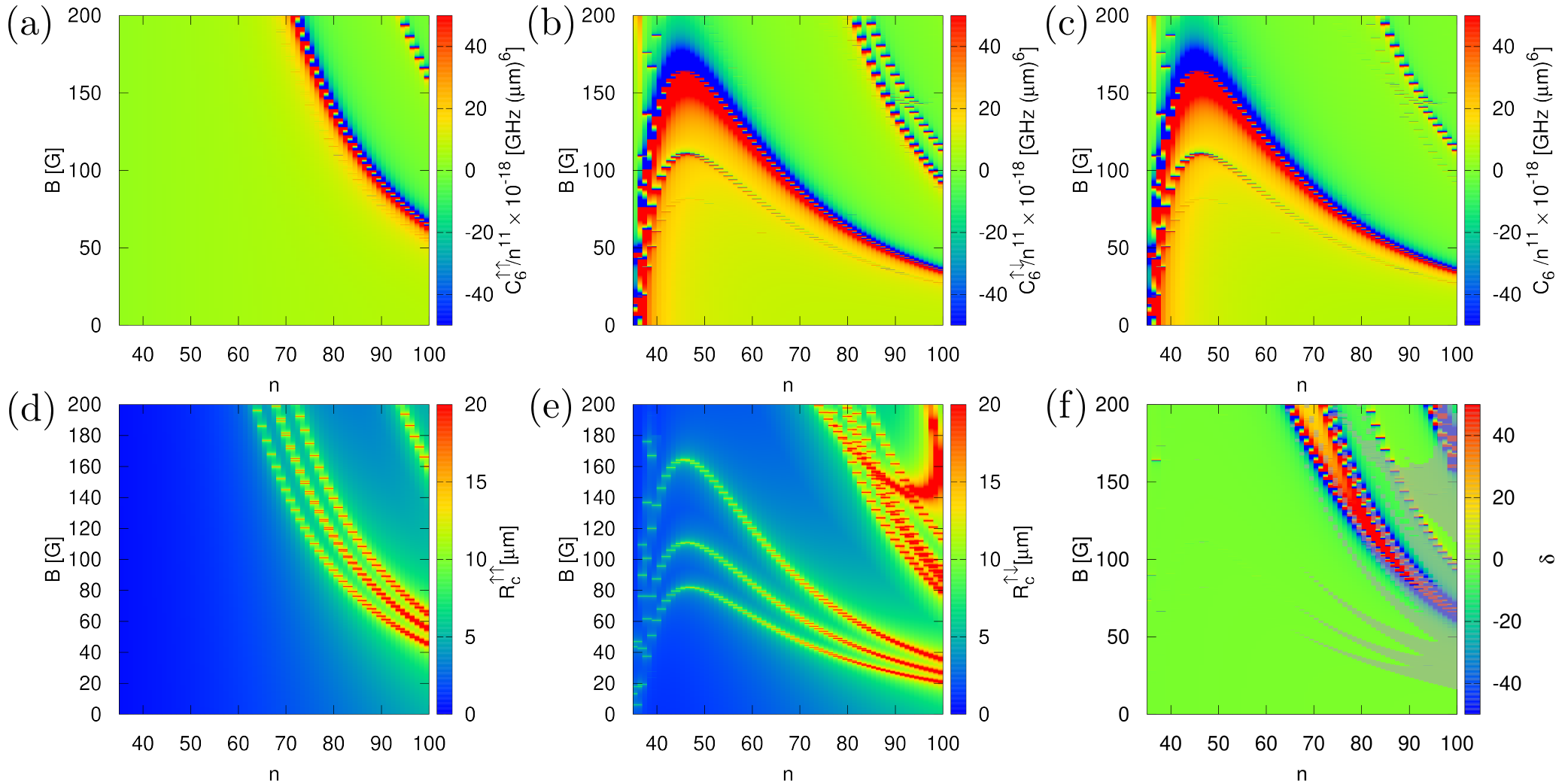}
\caption{Magnetic-field and principal quantum number dependence of various physical quantities for ${}^{88}$Sr with $m_J=0$ at $\theta=\pi/2$: (a) $C^{\uparrow\uparrow}_6$, (b) $C_6^{\uparrow\downarrow}$, (c) $C_6$, (d) $R_{\rm c}^{\uparrow\uparrow}$, (e) $R^{\uparrow\downarrow}_{\rm c}$, and (f) anisotropy parameter $\delta$. The gray-shaded region in panel (f) indicates the region where $R_{\rm c}\ge 10~\mu{\rm m}$.
}
\label{fig:c6_and_delta_nonzero_field_sr_z}
\vspace{-0.75em}
\end{figure*}

\begin{figure*}[t]
\centering
\includegraphics[width=17.0cm,clip]{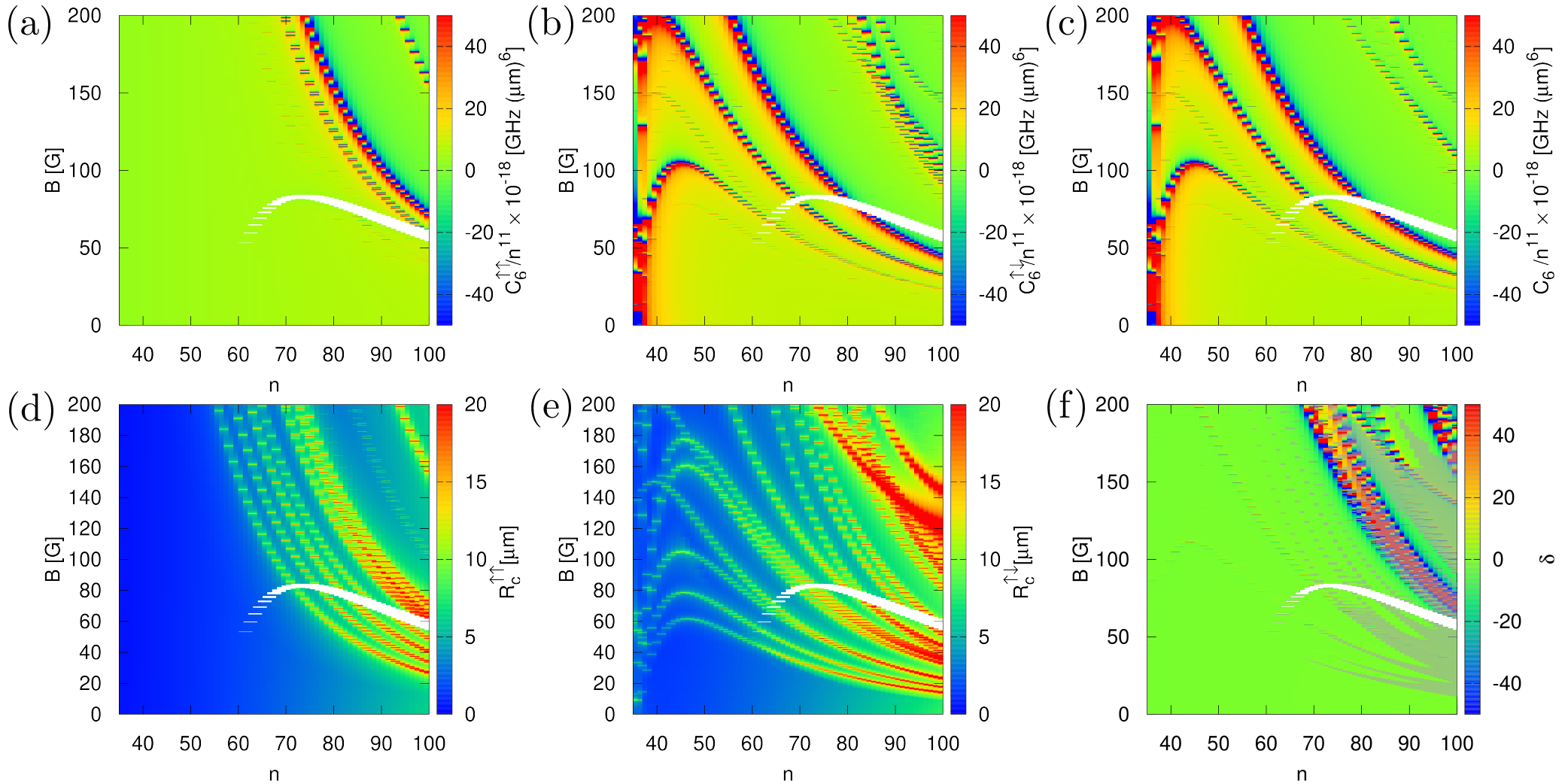}
\caption{Magnetic-field and principal quantum number dependence of various physical quantities for ${}^{88}$Sr with $m_J=-1$ at $\theta=\pi/2$: (a) $C^{\uparrow\uparrow}_6$, (b) $C_6^{\uparrow\downarrow}$, (c) $C_6$, (d) $R_{\rm c}^{\uparrow\uparrow}$, (e) $R^{\uparrow\downarrow}_{\rm c}$, and (f) anisotropy parameter $\delta$. The white region indicates the absence of data due to strong $S$-$D$ mixing. The gray-shaded region in panel (f) indicates the region where $R_{\rm c}\ge 10~\mu{\rm m}$.
}
\label{fig:c6_and_delta_nonzero_field_sr_m}
\vspace{-0.75em}
\end{figure*}

\begin{figure*}[t]
\centering
\includegraphics[width=17.0cm,clip]{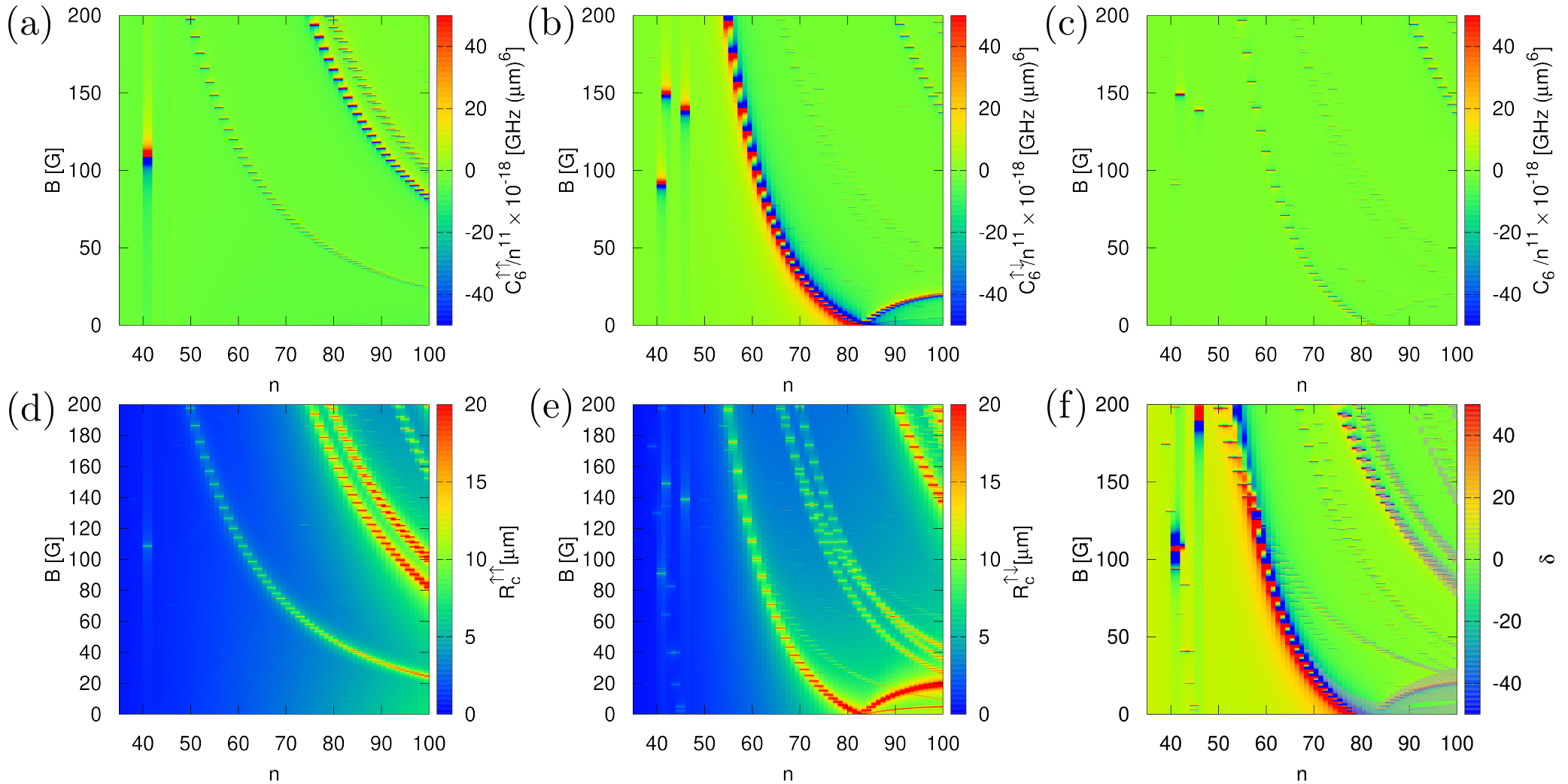}
\caption{Magnetic-field and principal quantum number dependence of various physical quantities for ${}^{174}$Yb with $m_J=+1$ at $\theta=\pi/2$: (a) $C^{\uparrow\uparrow}_6$, (b) $C_6^{\uparrow\downarrow}$, (c) $C_6$, (d) $R_{\rm c}^{\uparrow\uparrow}$, (e) $R^{\uparrow\downarrow}_{\rm c}$, and (f) anisotropy parameter $\delta$. The gray-shaded region in panel (f) indicates the region where $R_{\rm c}\ge 10~\mu{\rm m}$.
}
\label{fig:c6_and_delta_nonzero_field_yb_p}
\vspace{-0.75em}
\end{figure*}

\begin{figure*}[t]
\centering
\includegraphics[width=17.0cm,clip]{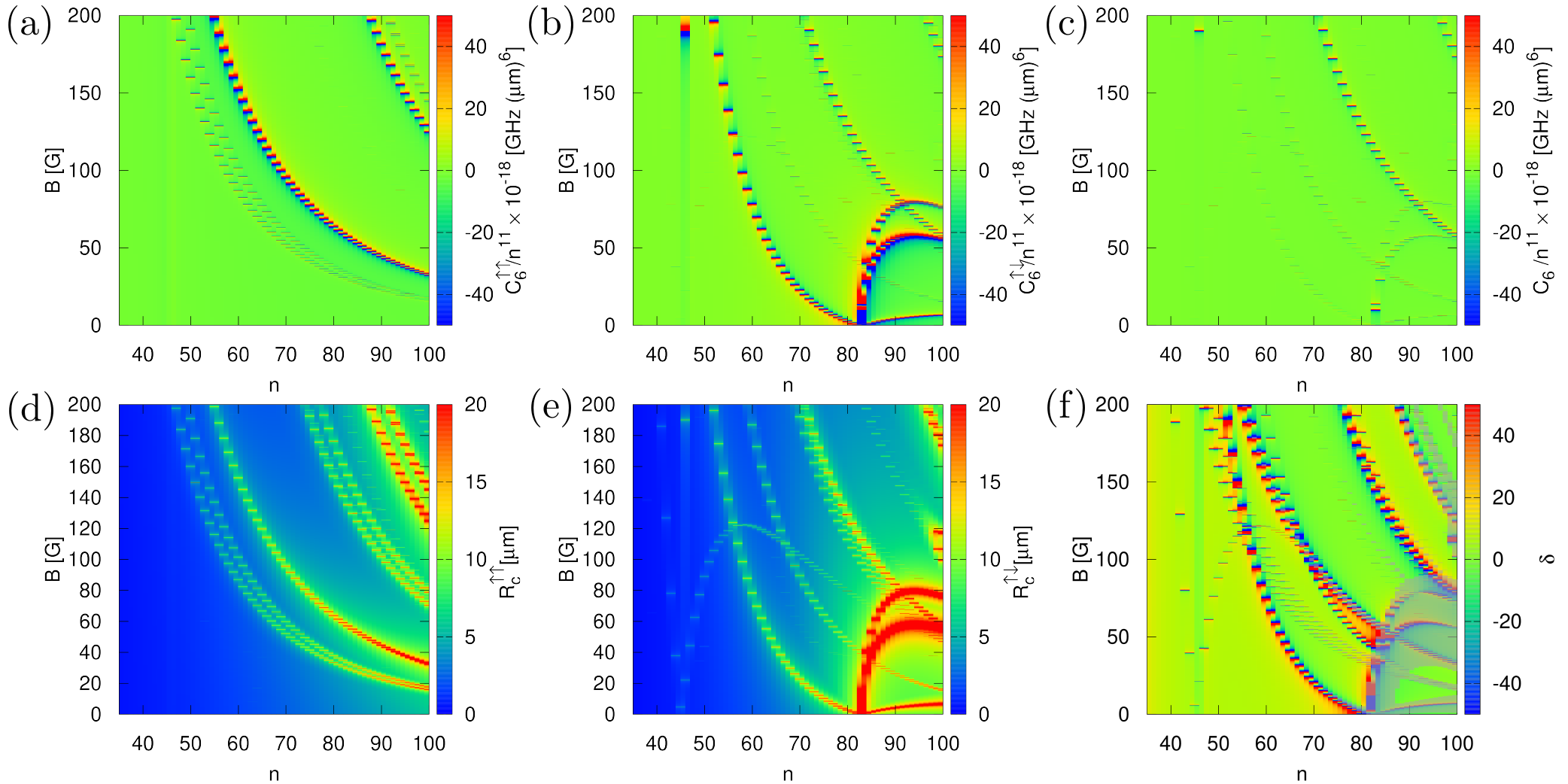}
\caption{Magnetic-field and principal quantum number dependence of various physical quantities for ${}^{174}$Yb with $m_J=0$ at $\theta=\pi/2$: (a) $C^{\uparrow\uparrow}_6$, (b) $C_6^{\uparrow\downarrow}$, (c) $C_6$, (d) $R_{\rm c}^{\uparrow\uparrow}$, (e) $R^{\uparrow\downarrow}_{\rm c}$, and (f) anisotropy parameter $\delta$. The gray-shaded region in panel (f) indicates the region where $R_{\rm c}\ge 10~\mu{\rm m}$.
}
\label{fig:c6_and_delta_nonzero_field_yb_z}
\vspace{-0.75em}
\end{figure*}

\begin{figure*}[t]
\centering
\includegraphics[width=17.0cm,clip]{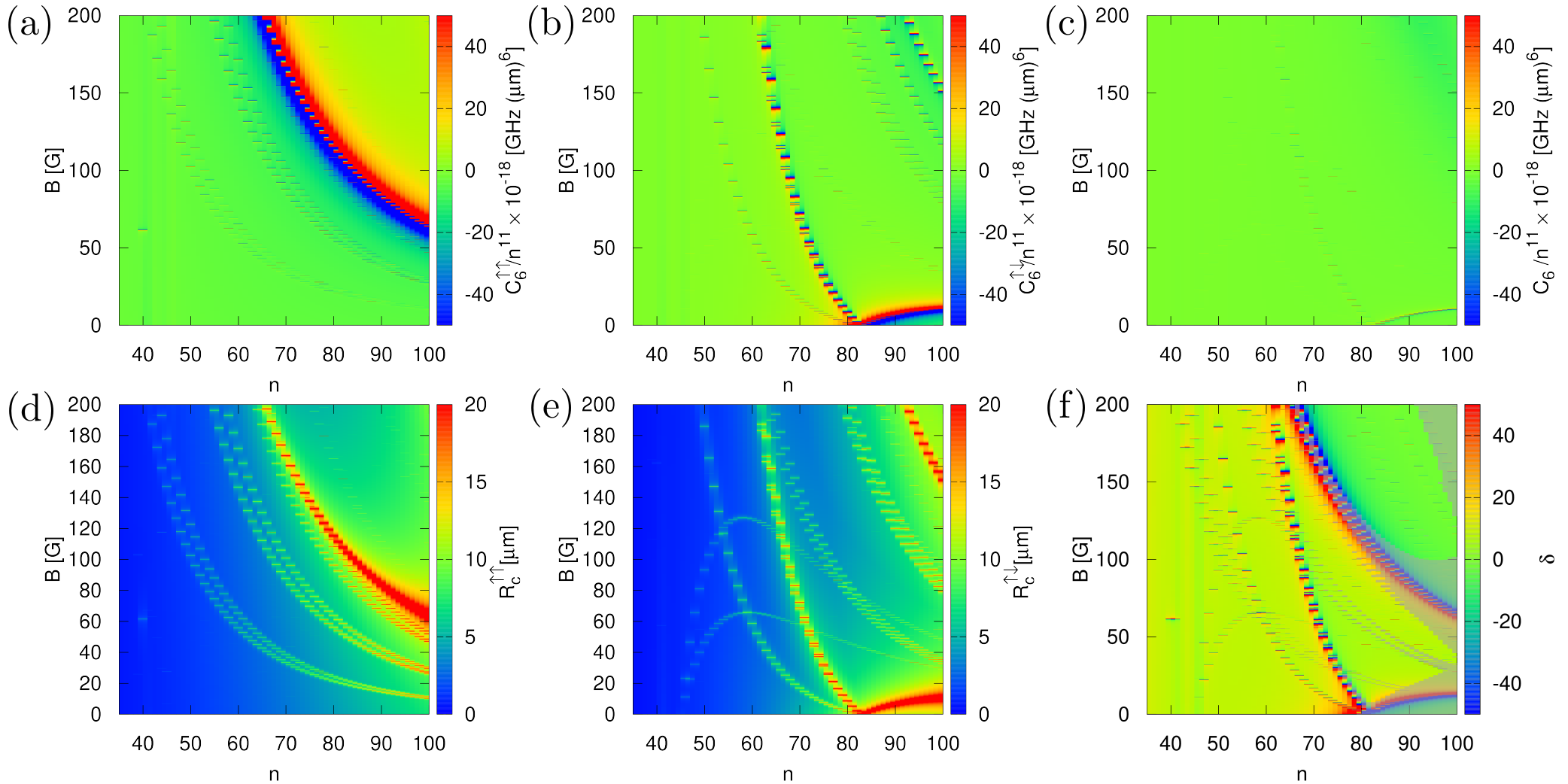}
\caption{Magnetic-field and principal quantum number dependence of various physical quantities for ${}^{174}$Yb with $m_J=-1$ at $\theta=\pi/2$: (a) $C^{\uparrow\uparrow}_6$, (b) $C_6^{\uparrow\downarrow}$, (c) $C_6$, (d) $R_{\rm c}^{\uparrow\uparrow}$, (e) $R^{\uparrow\downarrow}_{\rm c}$, and (f) anisotropy parameter $\delta$. The gray-shaded region in panel (f) indicates the region where $R_{\rm c}\ge 10~\mu{\rm m}$.
}
\label{fig:c6_and_delta_nonzero_field_yb_m}
\vspace{-0.75em}
\end{figure*}

In this appendix, we present the complete data for the $C_6$ coefficients, $R_{\rm c}$, and $\delta$.
Figures~\ref{fig:c6_and_delta_nonzero_field_sr_p}--\ref{fig:c6_and_delta_nonzero_field_yb_m} show the magnetic field and principal quantum number dependence of the physical quantities, including the $C_6$ coefficients, the critical radius, and the anisotropy parameter, for ${}^{88}$Sr and ${}^{174}$Yb. We do not show $C_6^{\downarrow\downarrow}$ or $R_{\rm c}^{\downarrow\downarrow}$, because for a given principal quantum number $n$ they are identical to $C_6^{\uparrow\uparrow}$ and $R_{\rm c}^{\uparrow\uparrow}$ evaluated at $n-1$. 

\clearpage
\section{Derivation of the energy density of the checkerboard supersolid state}\label{app:derivation_CSS}
In this appendix, we derive the expression for the energy density of the CSS state [Eq.~(\ref{eq:energy_CSS})]. To this end, we first rewrite Eq.~(\ref{eq:mean-field_energy_two-sublattice_ansat}) as
\begin{align}
\mathcal{E}(\theta_A,\theta_B)&=-\frac{J_0}{4}(C_+-C_-)-\frac{h_0}{4}(\cos\theta_A+\cos\theta_B)\notag \\
&-\frac{J_0}{4}(C_++C_-)\sin\theta_A\sin\theta_B\notag \\
&+\frac{J_0(C_+-C_-)(1-\delta)}{8}(\cos\theta_A+\cos\theta_B)^2\notag \\
&-\frac{J_0}{4}(C_+-C_-+2\delta C_-)\cos\theta_A\cos\theta_B.
\label{eq:rewrite_energy_density}
\end{align}
We then minimize $\mathcal{E}(\theta_A,\theta_B)$ with respect to $\theta_A$ and $\theta_B$:
{\small
\begin{align}
\frac{\partial\mathcal{E}(\theta_A,\theta_B)}{\partial\theta_A}&=\frac{h_0}{4}\sin\theta_A-\frac{J_0}{4}(C_++C_-)\cos\theta_A\sin\theta_B\notag \\
&-\frac{J_0(C_+-C_-)(1-\delta)}{4}(\cos\theta_A+\cos\theta_B)\sin\theta_A\notag \\
&+\frac{J_0}{4}(C_+-C_-+2\delta C_-)\sin\theta_A\cos\theta_B=0,
\label{eq:derivative_of_theta_A_energy_density}\\
\frac{\partial\mathcal{E}(\theta_A,\theta_B)}{\partial\theta_B}&=\frac{h_0}{4}\sin\theta_B-\frac{J_0}{4}(C_++C_-)\sin\theta_A\cos\theta_B\notag \\
&-\frac{J_0(C_+-C_-)(1-\delta)}{4}(\cos\theta_A+\cos\theta_B)\sin\theta_B\notag \\
&+\frac{J_0}{4}(C_+-C_-+2\delta C_-)\cos\theta_A\sin\theta_B=0.
\label{eq:derivative_of_theta_B_energy_density}
\end{align}
}By multiplying Eq.~(\ref{eq:derivative_of_theta_A_energy_density}) by $\sin\theta_B$ and subtracting Eq.~(\ref{eq:derivative_of_theta_B_energy_density}) multiplied by $\sin\theta_A$, we obtain the condition
\begin{align}
\sin\theta_A\sin\theta_B&=\alpha(1+\cos\theta_A\cos\theta_B),
\label{eq:condition1_for_CSS}\\
\alpha&\equiv \frac{C_+ +C_-}{-(C_+-C_-)-2\delta C_-}.
\label{eq:definition_of_alpha_for_CSS}
\end{align}
Similarly, by multiplying Eq.~(\ref{eq:derivative_of_theta_A_energy_density}) by $\sin\theta_B$ and adding Eq.~(\ref{eq:derivative_of_theta_B_energy_density}) multiplied by $\sin\theta_A$, we obtain the condition
\begin{align}
&h_0\sin\theta_A\sin\theta_B\notag \\
&-\frac{J_0}{2}(C_++C_-)(\cos\theta_A+\cos\theta_B)(1-\cos\theta_A\cos\theta_B)\notag \\
&-J_0(C_+-C_-)(1-\delta)(\cos\theta_A+\cos\theta_B)\sin\theta_A\sin\theta_B\notag \\
&+\frac{J_0}{2}(C_+-C_-+2\delta C_-)(\cos\theta_A+\cos\theta_B)\sin\theta_A\sin\theta_B\notag \\
&=0.\label{eq:codntion2_for_CSS}
\end{align}
For convenience, we introduce the variables~\cite{matsuda1970off}
\begin{align}
t&\equiv \cos\theta_A+\cos\theta_B,
\label{eq:definition_of_t_for_CSS}\\
u&\equiv \cos\theta_A\cos\theta_B.
\label{eq:definition_of_u_for_CSS}
\end{align}
Using Eqs.~(\ref{eq:definition_of_t_for_CSS}) and (\ref{eq:definition_of_u_for_CSS}), Eq.~(\ref{eq:condition1_for_CSS}) reduces to
\begin{align}
t=\pm\sqrt{1-\alpha^2}(1+u).
\label{eq:relation_t_and_u}
\end{align}
The sign $+(-)$ corresponds to the solution $\theta_A+\theta_B<\pi\;(\theta_A+\theta_B>\pi)$. Because $1-\alpha^2>0$, we obtain the condition
\begin{align}
\delta>1\text{ or }\delta<-\frac{C_+}{C_-}.
\label{eq:condition_for_delta_CSS_state}
\end{align}
Since we assume $\delta<0$, the condition for the CSS state is given by $\delta<-C_+/C_-$. Substituting Eq.~(\ref{eq:relation_t_and_u}) into Eq.~(\ref{eq:codntion2_for_CSS}), we obtain
\begin{align}
u&=\frac{\pm h_0/J_0+\sqrt{1-\alpha^2}\delta(C_+ +C_-)}{\sqrt{1-\alpha^2}(1-\delta)(C_+-C_-)}.
\label{eq:solution_for_u_CSS}
\end{align}
From the relation between the solutions and the coefficients of a quadratic equation, $\cos\theta_A$ and $\cos\theta_B$ are solutions of
\begin{align}
Z^2-tZ+u=0.
\label{eq:quadratic_equation_for_cos_theta_A_and_cos_theta_B}
\end{align}
Since $\cos\theta_A$ and $\cos\theta_B$ are real, the following condition must hold:
\begin{align}
t^2>4u.\label{eq:condition_reality_for_CSS}
\end{align}
Using Eqs.~(\ref{eq:relation_t_and_u}) and (\ref{eq:solution_for_u_CSS}), and $|u|<1$, inequality (\ref{eq:condition_reality_for_CSS}) reduces to
\begin{align}
u<\frac{1-\alpha}{1+\alpha}=\frac{(-C_+-\delta C_-)}{(1-\delta)C_-}.\label{eq:inequality_u_and_alpha}
\end{align}
Then, we obtain
\begin{align}
\pm\frac{h_0}{J_0}&<\sqrt{1-\alpha^2}\frac{C_+}{C_-}(-C_++C_--2\delta C_-).\label{eq:condition1_for_h}
\end{align}
From Eq.~(\ref{eq:solution_for_u_CSS}) and $|u|<1$, we can obtain another inequality:
\begin{widetext}
\begin{align}
-\sqrt{1-\alpha^2}(C_+-C_-+2\delta C_-)<\pm\frac{h_0}{J_0}<\sqrt{1-\alpha^2}(C_+-C_--2\delta C_+).\label{eq:condition2_for_h}
\end{align}
Therefore, we obtain the following inequalities for $h_0/J_0$:
\begin{align}
\sqrt{1-\alpha^2}(-C_++C_--2\delta C_-)<\pm\frac{h_0}{J_0}<\sqrt{1-\alpha^2}\frac{C_+}{C_-}(-C_++C_--2\delta C_-).\label{eq:reange_h_for_CSS}
\end{align}
\end{widetext}
Using the above results, we obtain expression (\ref{eq:energy_CSS}).

\newpage
\bibliography{references}

\begin{thebibliography}{183}%
\makeatletter
\providecommand \@ifxundefined [1]{%
 \@ifx{#1\undefined}
}%
\providecommand \@ifnum [1]{%
 \ifnum #1\expandafter \@firstoftwo
 \else \expandafter \@secondoftwo
 \fi
}%
\providecommand \@ifx [1]{%
 \ifx #1\expandafter \@firstoftwo
 \else \expandafter \@secondoftwo
 \fi
}%
\providecommand \natexlab [1]{#1}%
\providecommand \enquote  [1]{``#1''}%
\providecommand \bibnamefont  [1]{#1}%
\providecommand \bibfnamefont [1]{#1}%
\providecommand \citenamefont [1]{#1}%
\providecommand \href@noop [0]{\@secondoftwo}%
\providecommand \href [0]{\begingroup \@sanitize@url \@href}%
\providecommand \@href[1]{\@@startlink{#1}\@@href}%
\providecommand \@@href[1]{\endgroup#1\@@endlink}%
\providecommand \@sanitize@url [0]{\catcode `\\12\catcode `\$12\catcode
  `\&12\catcode `\#12\catcode `\^12\catcode `\_12\catcode `\%12\relax}%
\providecommand \@@startlink[1]{}%
\providecommand \@@endlink[0]{}%
\providecommand \url  [0]{\begingroup\@sanitize@url \@url }%
\providecommand \@url [1]{\endgroup\@href {#1}{\urlprefix }}%
\providecommand \urlprefix  [0]{URL }%
\providecommand \Eprint [0]{\href }%
\providecommand \doibase [0]{https://doi.org/}%
\providecommand \selectlanguage [0]{\@gobble}%
\providecommand \bibinfo  [0]{\@secondoftwo}%
\providecommand \bibfield  [0]{\@secondoftwo}%
\providecommand \translation [1]{[#1]}%
\providecommand \BibitemOpen [0]{}%
\providecommand \bibitemStop [0]{}%
\providecommand \bibitemNoStop [0]{.\EOS\space}%
\providecommand \EOS [0]{\spacefactor3000\relax}%
\providecommand \BibitemShut  [1]{\csname bibitem#1\endcsname}%
\let\auto@bib@innerbib\@empty
\bibitem [{\citenamefont {Bloch}\ \emph {et~al.}(2008)\citenamefont {Bloch},
  \citenamefont {Dalibard},\ and\ \citenamefont {Zwerger}}]{bloch2008many}%
  \BibitemOpen
  \bibfield  {author} {\bibinfo {author} {\bibfnamefont {I.}~\bibnamefont
  {Bloch}}, \bibinfo {author} {\bibfnamefont {J.}~\bibnamefont {Dalibard}},\
  and\ \bibinfo {author} {\bibfnamefont {W.}~\bibnamefont {Zwerger}},\
  }\bibfield  {title} {\bibinfo {title} {Many-body physics with ultracold
  gases},\ }\href@noop {} {\bibfield  {journal} {\bibinfo  {journal} {Rev. Mod.
  Phys.}\ }\textbf {\bibinfo {volume} {80}},\ \bibinfo {pages} {885} (\bibinfo
  {year} {2008})}\BibitemShut {NoStop}%
\bibitem [{\citenamefont {Gross}\ and\ \citenamefont
  {Bloch}(2017)}]{gross2017quantum}%
  \BibitemOpen
  \bibfield  {author} {\bibinfo {author} {\bibfnamefont {C.}~\bibnamefont
  {Gross}}\ and\ \bibinfo {author} {\bibfnamefont {I.}~\bibnamefont {Bloch}},\
  }\bibfield  {title} {\bibinfo {title} {Quantum simulations with ultracold
  atoms in optical lattices},\ }\href@noop {} {\bibfield  {journal} {\bibinfo
  {journal} {Science}\ }\textbf {\bibinfo {volume} {357}},\ \bibinfo {pages}
  {995} (\bibinfo {year} {2017})}\BibitemShut {NoStop}%
\bibitem [{\citenamefont {Sch{\"a}fer}\ \emph {et~al.}(2020)\citenamefont
  {Sch{\"a}fer}, \citenamefont {Fukuhara}, \citenamefont {Sugawa},
  \citenamefont {Takasu},\ and\ \citenamefont {Takahashi}}]{schafer2020tools}%
  \BibitemOpen
  \bibfield  {author} {\bibinfo {author} {\bibfnamefont {F.}~\bibnamefont
  {Sch{\"a}fer}}, \bibinfo {author} {\bibfnamefont {T.}~\bibnamefont
  {Fukuhara}}, \bibinfo {author} {\bibfnamefont {S.}~\bibnamefont {Sugawa}},
  \bibinfo {author} {\bibfnamefont {Y.}~\bibnamefont {Takasu}},\ and\ \bibinfo
  {author} {\bibfnamefont {Y.}~\bibnamefont {Takahashi}},\ }\bibfield  {title}
  {\bibinfo {title} {Tools for quantum simulation with ultracold atoms in
  optical lattices},\ }\href@noop {} {\bibfield  {journal} {\bibinfo  {journal}
  {Nat. Rev. Phys.}\ }\textbf {\bibinfo {volume} {2}},\ \bibinfo {pages} {411}
  (\bibinfo {year} {2020})}\BibitemShut {NoStop}%
\bibitem [{\citenamefont {Carr}\ \emph {et~al.}(2009)\citenamefont {Carr},
  \citenamefont {DeMille}, \citenamefont {Krems},\ and\ \citenamefont
  {Ye}}]{carr2009cold}%
  \BibitemOpen
  \bibfield  {author} {\bibinfo {author} {\bibfnamefont {L.~D.}\ \bibnamefont
  {Carr}}, \bibinfo {author} {\bibfnamefont {D.}~\bibnamefont {DeMille}},
  \bibinfo {author} {\bibfnamefont {R.~V.}\ \bibnamefont {Krems}},\ and\
  \bibinfo {author} {\bibfnamefont {J.}~\bibnamefont {Ye}},\ }\bibfield
  {title} {\bibinfo {title} {Cold and ultracold molecules: science, technology
  and applications},\ }\href@noop {} {\bibfield  {journal} {\bibinfo  {journal}
  {New J. Phys.}\ }\textbf {\bibinfo {volume} {11}},\ \bibinfo {pages} {055049}
  (\bibinfo {year} {2009})}\BibitemShut {NoStop}%
\bibitem [{\citenamefont {Cornish}\ \emph {et~al.}(2024)\citenamefont
  {Cornish}, \citenamefont {Tarbutt},\ and\ \citenamefont
  {Hazzard}}]{cornish2024quantum}%
  \BibitemOpen
  \bibfield  {author} {\bibinfo {author} {\bibfnamefont {S.~L.}\ \bibnamefont
  {Cornish}}, \bibinfo {author} {\bibfnamefont {M.~R.}\ \bibnamefont
  {Tarbutt}},\ and\ \bibinfo {author} {\bibfnamefont {K.~R.}\ \bibnamefont
  {Hazzard}},\ }\bibfield  {title} {\bibinfo {title} {Quantum computation and
  quantum simulation with ultracold molecules},\ }\href@noop {} {\bibfield
  {journal} {\bibinfo  {journal} {Nat. Phys.}\ }\textbf {\bibinfo {volume}
  {20}},\ \bibinfo {pages} {730} (\bibinfo {year} {2024})}\BibitemShut
  {NoStop}%
\bibitem [{\citenamefont {Blatt}\ and\ \citenamefont
  {Roos}(2012)}]{blatt2012quantum}%
  \BibitemOpen
  \bibfield  {author} {\bibinfo {author} {\bibfnamefont {R.}~\bibnamefont
  {Blatt}}\ and\ \bibinfo {author} {\bibfnamefont {C.~F.}\ \bibnamefont
  {Roos}},\ }\bibfield  {title} {\bibinfo {title} {Quantum simulations with
  trapped ions},\ }\href@noop {} {\bibfield  {journal} {\bibinfo  {journal}
  {Nat. Phys.}\ }\textbf {\bibinfo {volume} {8}},\ \bibinfo {pages} {277}
  (\bibinfo {year} {2012})}\BibitemShut {NoStop}%
\bibitem [{\citenamefont {Monroe}\ \emph {et~al.}(2021)\citenamefont {Monroe},
  \citenamefont {Campbell}, \citenamefont {Duan}, \citenamefont {Gong},
  \citenamefont {Gorshkov}, \citenamefont {Hess}, \citenamefont {Islam},
  \citenamefont {Kim}, \citenamefont {Linke}, \citenamefont {Pagano},
  \citenamefont {Richerme}, \citenamefont {Senko},\ and\ \citenamefont
  {Yao}}]{monroe2021programmable}%
  \BibitemOpen
  \bibfield  {author} {\bibinfo {author} {\bibfnamefont {C.}~\bibnamefont
  {Monroe}}, \bibinfo {author} {\bibfnamefont {W.~C.}\ \bibnamefont
  {Campbell}}, \bibinfo {author} {\bibfnamefont {L.-M.}\ \bibnamefont {Duan}},
  \bibinfo {author} {\bibfnamefont {Z.-X.}\ \bibnamefont {Gong}}, \bibinfo
  {author} {\bibfnamefont {A.~V.}\ \bibnamefont {Gorshkov}}, \bibinfo {author}
  {\bibfnamefont {P.~W.}\ \bibnamefont {Hess}}, \bibinfo {author}
  {\bibfnamefont {R.}~\bibnamefont {Islam}}, \bibinfo {author} {\bibfnamefont
  {K.}~\bibnamefont {Kim}}, \bibinfo {author} {\bibfnamefont {N.~M.}\
  \bibnamefont {Linke}}, \bibinfo {author} {\bibfnamefont {G.}~\bibnamefont
  {Pagano}}, \bibinfo {author} {\bibfnamefont {P.}~\bibnamefont {Richerme}},
  \bibinfo {author} {\bibfnamefont {C.}~\bibnamefont {Senko}},\ and\ \bibinfo
  {author} {\bibfnamefont {N.~Y.}\ \bibnamefont {Yao}},\ }\bibfield  {title}
  {\bibinfo {title} {Programmable quantum simulations of spin systems with
  trapped ions},\ }\href@noop {} {\bibfield  {journal} {\bibinfo  {journal}
  {Rev. Mod. Phys.}\ }\textbf {\bibinfo {volume} {93}},\ \bibinfo {pages}
  {025001} (\bibinfo {year} {2021})}\BibitemShut {NoStop}%
\bibitem [{\citenamefont {Browaeys}\ and\ \citenamefont
  {Lahaye}(2020)}]{Browaeys2020}%
  \BibitemOpen
  \bibfield  {author} {\bibinfo {author} {\bibfnamefont {A.}~\bibnamefont
  {Browaeys}}\ and\ \bibinfo {author} {\bibfnamefont {T.}~\bibnamefont
  {Lahaye}},\ }\bibfield  {title} {\bibinfo {title} {Many-body physics with
  individually controlled {Rydberg} atoms},\ }\href@noop {} {\bibfield
  {journal} {\bibinfo  {journal} {Nat. Phys.}\ }\textbf {\bibinfo {volume}
  {16}},\ \bibinfo {pages} {132} (\bibinfo {year} {2020})}\BibitemShut
  {NoStop}%
\bibitem [{\citenamefont {Morgado}\ and\ \citenamefont
  {Whitlock}(2021)}]{morgado2021quantum}%
  \BibitemOpen
  \bibfield  {author} {\bibinfo {author} {\bibfnamefont {M.}~\bibnamefont
  {Morgado}}\ and\ \bibinfo {author} {\bibfnamefont {S.}~\bibnamefont
  {Whitlock}},\ }\bibfield  {title} {\bibinfo {title} {Quantum simulation and
  computing with {Rydberg-interacting} qubits},\ }\href@noop {} {\bibfield
  {journal} {\bibinfo  {journal} {AVS Quantum Science}\ }\textbf {\bibinfo
  {volume} {3}},\ \bibinfo {pages} {023501} (\bibinfo {year}
  {2021})}\BibitemShut {NoStop}%
\bibitem [{\citenamefont {Wendin}(2017)}]{wendin2017quantum}%
  \BibitemOpen
  \bibfield  {author} {\bibinfo {author} {\bibfnamefont {G.}~\bibnamefont
  {Wendin}},\ }\bibfield  {title} {\bibinfo {title} {Quantum information
  processing with superconducting circuits: a review},\ }\href@noop {}
  {\bibfield  {journal} {\bibinfo  {journal} {Rep. Prog. Phys.}\ }\textbf
  {\bibinfo {volume} {80}},\ \bibinfo {pages} {106001} (\bibinfo {year}
  {2017})}\BibitemShut {NoStop}%
\bibitem [{\citenamefont {Kjaergaard}\ \emph {et~al.}(2020)\citenamefont
  {Kjaergaard}, \citenamefont {Schwartz}, \citenamefont {Braum{\"u}ller},
  \citenamefont {Krantz}, \citenamefont {Wang}, \citenamefont {Gustavsson},\
  and\ \citenamefont {Oliver}}]{kjaergaard2020superconducting}%
  \BibitemOpen
  \bibfield  {author} {\bibinfo {author} {\bibfnamefont {M.}~\bibnamefont
  {Kjaergaard}}, \bibinfo {author} {\bibfnamefont {M.~E.}\ \bibnamefont
  {Schwartz}}, \bibinfo {author} {\bibfnamefont {J.}~\bibnamefont
  {Braum{\"u}ller}}, \bibinfo {author} {\bibfnamefont {P.}~\bibnamefont
  {Krantz}}, \bibinfo {author} {\bibfnamefont {J.~I.-J.}\ \bibnamefont {Wang}},
  \bibinfo {author} {\bibfnamefont {S.}~\bibnamefont {Gustavsson}},\ and\
  \bibinfo {author} {\bibfnamefont {W.~D.}\ \bibnamefont {Oliver}},\ }\bibfield
   {title} {\bibinfo {title} {Superconducting qubits: Current state of play},\
  }\href@noop {} {\bibfield  {journal} {\bibinfo  {journal} {Annu. Rev.
  Condens. Matter Phys.}\ }\textbf {\bibinfo {volume} {11}},\ \bibinfo {pages}
  {369} (\bibinfo {year} {2020})}\BibitemShut {NoStop}%
\bibitem [{\citenamefont {Bernien}\ \emph {et~al.}(2017)\citenamefont
  {Bernien}, \citenamefont {Schwartz}, \citenamefont {Keesling}, \citenamefont
  {Levine}, \citenamefont {Omran}, \citenamefont {Pichler}, \citenamefont
  {Choi}, \citenamefont {Zibrov}, \citenamefont {Endres}, \citenamefont
  {Greiner}, \citenamefont {Vuleti\'c},\ and\ \citenamefont
  {Lukin}}]{Bernien2017}%
  \BibitemOpen
  \bibfield  {author} {\bibinfo {author} {\bibfnamefont {H.}~\bibnamefont
  {Bernien}}, \bibinfo {author} {\bibfnamefont {S.}~\bibnamefont {Schwartz}},
  \bibinfo {author} {\bibfnamefont {A.}~\bibnamefont {Keesling}}, \bibinfo
  {author} {\bibfnamefont {H.}~\bibnamefont {Levine}}, \bibinfo {author}
  {\bibfnamefont {A.}~\bibnamefont {Omran}}, \bibinfo {author} {\bibfnamefont
  {H.}~\bibnamefont {Pichler}}, \bibinfo {author} {\bibfnamefont
  {S.}~\bibnamefont {Choi}}, \bibinfo {author} {\bibfnamefont {A.~S.}\
  \bibnamefont {Zibrov}}, \bibinfo {author} {\bibfnamefont {M.}~\bibnamefont
  {Endres}}, \bibinfo {author} {\bibfnamefont {M.}~\bibnamefont {Greiner}},
  \bibinfo {author} {\bibfnamefont {V.}~\bibnamefont {Vuleti\'c}},\ and\
  \bibinfo {author} {\bibfnamefont {M.~D.}\ \bibnamefont {Lukin}},\ }\bibfield
  {title} {\bibinfo {title} {Probing many-body dynamics on a 51-atom quantum
  simulator},\ }\href@noop {} {\bibfield  {journal} {\bibinfo  {journal}
  {Nature}\ }\textbf {\bibinfo {volume} {551}},\ \bibinfo {pages} {579}
  (\bibinfo {year} {2017})}\BibitemShut {NoStop}%
\bibitem [{\citenamefont {de~L{\'e}s{\'e}leuc}\ \emph
  {et~al.}(2019)\citenamefont {de~L{\'e}s{\'e}leuc}, \citenamefont {Lienhard},
  \citenamefont {Scholl}, \citenamefont {Barredo}, \citenamefont {Weber},
  \citenamefont {Lang}, \citenamefont {B{\"u}chler}, \citenamefont {Lahaye},\
  and\ \citenamefont {Browaeys}}]{Leseleuc2019}%
  \BibitemOpen
  \bibfield  {author} {\bibinfo {author} {\bibfnamefont {S.}~\bibnamefont
  {de~L{\'e}s{\'e}leuc}}, \bibinfo {author} {\bibfnamefont {V.}~\bibnamefont
  {Lienhard}}, \bibinfo {author} {\bibfnamefont {P.}~\bibnamefont {Scholl}},
  \bibinfo {author} {\bibfnamefont {D.}~\bibnamefont {Barredo}}, \bibinfo
  {author} {\bibfnamefont {S.}~\bibnamefont {Weber}}, \bibinfo {author}
  {\bibfnamefont {N.}~\bibnamefont {Lang}}, \bibinfo {author} {\bibfnamefont
  {H.~P.}\ \bibnamefont {B{\"u}chler}}, \bibinfo {author} {\bibfnamefont
  {T.}~\bibnamefont {Lahaye}},\ and\ \bibinfo {author} {\bibfnamefont
  {A.}~\bibnamefont {Browaeys}},\ }\bibfield  {title} {\bibinfo {title}
  {Observation of a symmetry-protected topological phase of interacting bosons
  with {Rydberg} atoms},\ }\href@noop {} {\bibfield  {journal} {\bibinfo
  {journal} {Science}\ }\textbf {\bibinfo {volume} {365}},\ \bibinfo {pages}
  {775} (\bibinfo {year} {2019})}\BibitemShut {NoStop}%
\bibitem [{\citenamefont {Semeghini}\ \emph {et~al.}(2021)\citenamefont
  {Semeghini}, \citenamefont {Levine}, \citenamefont {Keesling}, \citenamefont
  {Ebadi}, \citenamefont {Wang}, \citenamefont {Bluvstein}, \citenamefont
  {Verresen}, \citenamefont {Pichler}, \citenamefont {Kalinowski},
  \citenamefont {Samajdar}, \citenamefont {Omran}, \citenamefont {Sachdev},
  \citenamefont {Vishwanath}, \citenamefont {Greiner}, \citenamefont
  {Vuleti\'c},\ and\ \citenamefont {Lukin}}]{Semeghini2021}%
  \BibitemOpen
  \bibfield  {author} {\bibinfo {author} {\bibfnamefont {G.}~\bibnamefont
  {Semeghini}}, \bibinfo {author} {\bibfnamefont {H.}~\bibnamefont {Levine}},
  \bibinfo {author} {\bibfnamefont {A.}~\bibnamefont {Keesling}}, \bibinfo
  {author} {\bibfnamefont {S.}~\bibnamefont {Ebadi}}, \bibinfo {author}
  {\bibfnamefont {T.~T.}\ \bibnamefont {Wang}}, \bibinfo {author}
  {\bibfnamefont {D.}~\bibnamefont {Bluvstein}}, \bibinfo {author}
  {\bibfnamefont {R.}~\bibnamefont {Verresen}}, \bibinfo {author}
  {\bibfnamefont {H.}~\bibnamefont {Pichler}}, \bibinfo {author} {\bibfnamefont
  {M.}~\bibnamefont {Kalinowski}}, \bibinfo {author} {\bibfnamefont
  {R.}~\bibnamefont {Samajdar}}, \bibinfo {author} {\bibfnamefont
  {A.}~\bibnamefont {Omran}}, \bibinfo {author} {\bibfnamefont
  {S.}~\bibnamefont {Sachdev}}, \bibinfo {author} {\bibfnamefont
  {A.}~\bibnamefont {Vishwanath}}, \bibinfo {author} {\bibfnamefont
  {M.}~\bibnamefont {Greiner}}, \bibinfo {author} {\bibfnamefont
  {V.}~\bibnamefont {Vuleti\'c}},\ and\ \bibinfo {author} {\bibfnamefont
  {M.~D.}\ \bibnamefont {Lukin}},\ }\bibfield  {title} {\bibinfo {title}
  {Probing topological spin liquids on a programmable quantum simulator},\
  }\href@noop {} {\bibfield  {journal} {\bibinfo  {journal} {Science}\ }\textbf
  {\bibinfo {volume} {374}},\ \bibinfo {pages} {1242} (\bibinfo {year}
  {2021})}\BibitemShut {NoStop}%
\bibitem [{\citenamefont {Chen}\ \emph {et~al.}(2023)\citenamefont {Chen},
  \citenamefont {Bornet}, \citenamefont {Bintz}, \citenamefont {Emperauger},
  \citenamefont {Leclerc}, \citenamefont {Liu}, \citenamefont {Scholl},
  \citenamefont {Barredo}, \citenamefont {Hauschild}, \citenamefont
  {Chatterjee}, \citenamefont {Schuler}, \citenamefont {L\"auchli},
  \citenamefont {Zaletel}, \citenamefont {Lahaye}, \citenamefont {Yao},\ and\
  \citenamefont {Browaeys}}]{Chen2023}%
  \BibitemOpen
  \bibfield  {author} {\bibinfo {author} {\bibfnamefont {C.}~\bibnamefont
  {Chen}}, \bibinfo {author} {\bibfnamefont {G.}~\bibnamefont {Bornet}},
  \bibinfo {author} {\bibfnamefont {M.}~\bibnamefont {Bintz}}, \bibinfo
  {author} {\bibfnamefont {G.}~\bibnamefont {Emperauger}}, \bibinfo {author}
  {\bibfnamefont {L.}~\bibnamefont {Leclerc}}, \bibinfo {author} {\bibfnamefont
  {V.~S.}\ \bibnamefont {Liu}}, \bibinfo {author} {\bibfnamefont
  {P.}~\bibnamefont {Scholl}}, \bibinfo {author} {\bibfnamefont
  {D.}~\bibnamefont {Barredo}}, \bibinfo {author} {\bibfnamefont
  {J.}~\bibnamefont {Hauschild}}, \bibinfo {author} {\bibfnamefont
  {S.}~\bibnamefont {Chatterjee}}, \bibinfo {author} {\bibfnamefont
  {M.}~\bibnamefont {Schuler}}, \bibinfo {author} {\bibfnamefont {A.~M.}\
  \bibnamefont {L\"auchli}}, \bibinfo {author} {\bibfnamefont {M.~P.}\
  \bibnamefont {Zaletel}}, \bibinfo {author} {\bibfnamefont {T.}~\bibnamefont
  {Lahaye}}, \bibinfo {author} {\bibfnamefont {N.~Y.}\ \bibnamefont {Yao}},\
  and\ \bibinfo {author} {\bibfnamefont {A.}~\bibnamefont {Browaeys}},\
  }\bibfield  {title} {\bibinfo {title} {Continuous symmetry breaking in a
  two-dimensional {Rydberg} array},\ }\href@noop {} {\bibfield  {journal}
  {\bibinfo  {journal} {Nature}\ }\textbf {\bibinfo {volume} {616}},\ \bibinfo
  {pages} {691} (\bibinfo {year} {2023})}\BibitemShut {NoStop}%
\bibitem [{\citenamefont {Labuhn}\ \emph {et~al.}(2016)\citenamefont {Labuhn},
  \citenamefont {Barredo}, \citenamefont {Ravets}, \citenamefont
  {de~L{\'e}s{\'e}leuc}, \citenamefont {Macr{\`\i}}, \citenamefont {Lahaye},\
  and\ \citenamefont {Browaeys}}]{Labuhn2016}%
  \BibitemOpen
  \bibfield  {author} {\bibinfo {author} {\bibfnamefont {H.}~\bibnamefont
  {Labuhn}}, \bibinfo {author} {\bibfnamefont {D.}~\bibnamefont {Barredo}},
  \bibinfo {author} {\bibfnamefont {S.}~\bibnamefont {Ravets}}, \bibinfo
  {author} {\bibfnamefont {S.}~\bibnamefont {de~L{\'e}s{\'e}leuc}}, \bibinfo
  {author} {\bibfnamefont {T.}~\bibnamefont {Macr{\`\i}}}, \bibinfo {author}
  {\bibfnamefont {T.}~\bibnamefont {Lahaye}},\ and\ \bibinfo {author}
  {\bibfnamefont {A.}~\bibnamefont {Browaeys}},\ }\bibfield  {title} {\bibinfo
  {title} {Tunable two-dimensional arrays of single {Rydberg} atoms for
  realizing quantum {Ising} models},\ }\href@noop {} {\bibfield  {journal}
  {\bibinfo  {journal} {Nature}\ }\textbf {\bibinfo {volume} {534}},\ \bibinfo
  {pages} {667} (\bibinfo {year} {2016})}\BibitemShut {NoStop}%
\bibitem [{\citenamefont {Zeiher}\ \emph {et~al.}(2016)\citenamefont {Zeiher},
  \citenamefont {van Bijnen}, \citenamefont {Schau{\ss}}, \citenamefont {Hild},
  \citenamefont {Choi}, \citenamefont {Pohl}, \citenamefont {Bloch},\ and\
  \citenamefont {Gross}}]{zeiher2016many}%
  \BibitemOpen
  \bibfield  {author} {\bibinfo {author} {\bibfnamefont {J.}~\bibnamefont
  {Zeiher}}, \bibinfo {author} {\bibfnamefont {R.}~\bibnamefont {van Bijnen}},
  \bibinfo {author} {\bibfnamefont {P.}~\bibnamefont {Schau{\ss}}}, \bibinfo
  {author} {\bibfnamefont {S.}~\bibnamefont {Hild}}, \bibinfo {author}
  {\bibfnamefont {J.-y.}\ \bibnamefont {Choi}}, \bibinfo {author}
  {\bibfnamefont {T.}~\bibnamefont {Pohl}}, \bibinfo {author} {\bibfnamefont
  {I.}~\bibnamefont {Bloch}},\ and\ \bibinfo {author} {\bibfnamefont
  {C.}~\bibnamefont {Gross}},\ }\bibfield  {title} {\bibinfo {title} {Many-body
  interferometry of a {Rydberg-dressed} spin lattice},\ }\href@noop {}
  {\bibfield  {journal} {\bibinfo  {journal} {Nat. Phys.}\ }\textbf {\bibinfo
  {volume} {12}},\ \bibinfo {pages} {1095} (\bibinfo {year}
  {2016})}\BibitemShut {NoStop}%
\bibitem [{\citenamefont {Zeiher}\ \emph {et~al.}(2017)\citenamefont {Zeiher},
  \citenamefont {Choi}, \citenamefont {Rubio-Abadal}, \citenamefont {Pohl},
  \citenamefont {van Bijnen}, \citenamefont {Bloch},\ and\ \citenamefont
  {Gross}}]{Zeiher2017}%
  \BibitemOpen
  \bibfield  {author} {\bibinfo {author} {\bibfnamefont {J.}~\bibnamefont
  {Zeiher}}, \bibinfo {author} {\bibfnamefont {J.-y.}\ \bibnamefont {Choi}},
  \bibinfo {author} {\bibfnamefont {A.}~\bibnamefont {Rubio-Abadal}}, \bibinfo
  {author} {\bibfnamefont {T.}~\bibnamefont {Pohl}}, \bibinfo {author}
  {\bibfnamefont {R.}~\bibnamefont {van Bijnen}}, \bibinfo {author}
  {\bibfnamefont {I.}~\bibnamefont {Bloch}},\ and\ \bibinfo {author}
  {\bibfnamefont {C.}~\bibnamefont {Gross}},\ }\bibfield  {title} {\bibinfo
  {title} {Coherent many-body spin dynamics in a long-range interacting {Ising}
  chain},\ }\href@noop {} {\bibfield  {journal} {\bibinfo  {journal} {Phys.
  Rev. X}\ }\textbf {\bibinfo {volume} {7}},\ \bibinfo {pages} {041063}
  (\bibinfo {year} {2017})}\BibitemShut {NoStop}%
\bibitem [{\citenamefont {de~L{\'e}s{\'e}leuc}\ \emph
  {et~al.}(2018)\citenamefont {de~L{\'e}s{\'e}leuc}, \citenamefont {Weber},
  \citenamefont {Lienhard}, \citenamefont {Barredo}, \citenamefont
  {B{\"u}chler}, \citenamefont {Lahaye},\ and\ \citenamefont
  {Browaeys}}]{Leseleuc2018}%
  \BibitemOpen
  \bibfield  {author} {\bibinfo {author} {\bibfnamefont {S.}~\bibnamefont
  {de~L{\'e}s{\'e}leuc}}, \bibinfo {author} {\bibfnamefont {S.}~\bibnamefont
  {Weber}}, \bibinfo {author} {\bibfnamefont {V.}~\bibnamefont {Lienhard}},
  \bibinfo {author} {\bibfnamefont {D.}~\bibnamefont {Barredo}}, \bibinfo
  {author} {\bibfnamefont {H.~P.}\ \bibnamefont {B{\"u}chler}}, \bibinfo
  {author} {\bibfnamefont {T.}~\bibnamefont {Lahaye}},\ and\ \bibinfo {author}
  {\bibfnamefont {A.}~\bibnamefont {Browaeys}},\ }\bibfield  {title} {\bibinfo
  {title} {Accurate mapping of multilevel {Rydberg} atoms on interacting
  spin-1/2 particles for the quantum simulation of {Ising} models},\
  }\href@noop {} {\bibfield  {journal} {\bibinfo  {journal} {Phys. Rev. Lett.}\
  }\textbf {\bibinfo {volume} {120}},\ \bibinfo {pages} {113602} (\bibinfo
  {year} {2018})}\BibitemShut {NoStop}%
\bibitem [{\citenamefont {Lienhard}\ \emph {et~al.}(2018)\citenamefont
  {Lienhard}, \citenamefont {de~L{\'e}s{\'e}leuc}, \citenamefont {Barredo},
  \citenamefont {Lahaye}, \citenamefont {Browaeys}, \citenamefont {Schuler},
  \citenamefont {Henry},\ and\ \citenamefont {L{\"a}uchli}}]{Lienhard2018}%
  \BibitemOpen
  \bibfield  {author} {\bibinfo {author} {\bibfnamefont {V.}~\bibnamefont
  {Lienhard}}, \bibinfo {author} {\bibfnamefont {S.}~\bibnamefont
  {de~L{\'e}s{\'e}leuc}}, \bibinfo {author} {\bibfnamefont {D.}~\bibnamefont
  {Barredo}}, \bibinfo {author} {\bibfnamefont {T.}~\bibnamefont {Lahaye}},
  \bibinfo {author} {\bibfnamefont {A.}~\bibnamefont {Browaeys}}, \bibinfo
  {author} {\bibfnamefont {M.}~\bibnamefont {Schuler}}, \bibinfo {author}
  {\bibfnamefont {L.-P.}\ \bibnamefont {Henry}},\ and\ \bibinfo {author}
  {\bibfnamefont {A.~M.}\ \bibnamefont {L{\"a}uchli}},\ }\bibfield  {title}
  {\bibinfo {title} {Observing the space-and time-dependent growth of
  correlations in dynamically tuned synthetic {Ising} models with
  antiferromagnetic interactions},\ }\href@noop {} {\bibfield  {journal}
  {\bibinfo  {journal} {Phys. Rev. X}\ }\textbf {\bibinfo {volume} {8}},\
  \bibinfo {pages} {021070} (\bibinfo {year} {2018})}\BibitemShut {NoStop}%
\bibitem [{\citenamefont {Guardado-Sanchez}\ \emph {et~al.}(2018)\citenamefont
  {Guardado-Sanchez}, \citenamefont {Brown}, \citenamefont {Mitra},
  \citenamefont {Devakul}, \citenamefont {Huse}, \citenamefont {Schau{\ss}},\
  and\ \citenamefont {Bakr}}]{Guardado2018}%
  \BibitemOpen
  \bibfield  {author} {\bibinfo {author} {\bibfnamefont {E.}~\bibnamefont
  {Guardado-Sanchez}}, \bibinfo {author} {\bibfnamefont {P.~T.}\ \bibnamefont
  {Brown}}, \bibinfo {author} {\bibfnamefont {D.}~\bibnamefont {Mitra}},
  \bibinfo {author} {\bibfnamefont {T.}~\bibnamefont {Devakul}}, \bibinfo
  {author} {\bibfnamefont {D.~A.}\ \bibnamefont {Huse}}, \bibinfo {author}
  {\bibfnamefont {P.}~\bibnamefont {Schau{\ss}}},\ and\ \bibinfo {author}
  {\bibfnamefont {W.~S.}\ \bibnamefont {Bakr}},\ }\bibfield  {title} {\bibinfo
  {title} {Probing the quench dynamics of antiferromagnetic correlations in a
  {2D} quantum {Ising} spin system},\ }\href@noop {} {\bibfield  {journal}
  {\bibinfo  {journal} {Phys. Rev. X}\ }\textbf {\bibinfo {volume} {8}},\
  \bibinfo {pages} {021069} (\bibinfo {year} {2018})}\BibitemShut {NoStop}%
\bibitem [{\citenamefont {Keesling}\ \emph {et~al.}(2019)\citenamefont
  {Keesling}, \citenamefont {Omran}, \citenamefont {Levine}, \citenamefont
  {Bernien}, \citenamefont {Pichler}, \citenamefont {Choi}, \citenamefont
  {Samajdar}, \citenamefont {Schwartz}, \citenamefont {Silvi}, \citenamefont
  {Sachdev}, \citenamefont {Zoller}, \citenamefont {Endres}, \citenamefont
  {Greiner}, \citenamefont {Vuleti\'c},\ and\ \citenamefont
  {Lukin}}]{Keesling2019}%
  \BibitemOpen
  \bibfield  {author} {\bibinfo {author} {\bibfnamefont {A.}~\bibnamefont
  {Keesling}}, \bibinfo {author} {\bibfnamefont {A.}~\bibnamefont {Omran}},
  \bibinfo {author} {\bibfnamefont {H.}~\bibnamefont {Levine}}, \bibinfo
  {author} {\bibfnamefont {H.}~\bibnamefont {Bernien}}, \bibinfo {author}
  {\bibfnamefont {H.}~\bibnamefont {Pichler}}, \bibinfo {author} {\bibfnamefont
  {S.}~\bibnamefont {Choi}}, \bibinfo {author} {\bibfnamefont {R.}~\bibnamefont
  {Samajdar}}, \bibinfo {author} {\bibfnamefont {S.}~\bibnamefont {Schwartz}},
  \bibinfo {author} {\bibfnamefont {P.}~\bibnamefont {Silvi}}, \bibinfo
  {author} {\bibfnamefont {S.}~\bibnamefont {Sachdev}}, \bibinfo {author}
  {\bibfnamefont {P.}~\bibnamefont {Zoller}}, \bibinfo {author} {\bibfnamefont
  {M.}~\bibnamefont {Endres}}, \bibinfo {author} {\bibfnamefont
  {M.}~\bibnamefont {Greiner}}, \bibinfo {author} {\bibfnamefont
  {V.}~\bibnamefont {Vuleti\'c}},\ and\ \bibinfo {author} {\bibfnamefont
  {M.~D.}\ \bibnamefont {Lukin}},\ }\bibfield  {title} {\bibinfo {title}
  {Quantum {Kibble--Zurek} mechanism and critical dynamics on a programmable
  {Rydberg} simulator},\ }\href@noop {} {\bibfield  {journal} {\bibinfo
  {journal} {Nature}\ }\textbf {\bibinfo {volume} {568}},\ \bibinfo {pages}
  {207} (\bibinfo {year} {2019})}\BibitemShut {NoStop}%
\bibitem [{\citenamefont {Tamura}\ \emph {et~al.}(2020)\citenamefont {Tamura},
  \citenamefont {Yamakoshi},\ and\ \citenamefont
  {Nakagawa}}]{tamura2020analysis}%
  \BibitemOpen
  \bibfield  {author} {\bibinfo {author} {\bibfnamefont {H.}~\bibnamefont
  {Tamura}}, \bibinfo {author} {\bibfnamefont {T.}~\bibnamefont {Yamakoshi}},\
  and\ \bibinfo {author} {\bibfnamefont {K.}~\bibnamefont {Nakagawa}},\
  }\bibfield  {title} {\bibinfo {title} {Analysis of coherent dynamics of a
  {Rydberg-atom} quantum simulator},\ }\href@noop {} {\bibfield  {journal}
  {\bibinfo  {journal} {Phys. Rev. A}\ }\textbf {\bibinfo {volume} {101}},\
  \bibinfo {pages} {043421} (\bibinfo {year} {2020})}\BibitemShut {NoStop}%
\bibitem [{\citenamefont {Kim}\ \emph {et~al.}(2020)\citenamefont {Kim},
  \citenamefont {Song}, \citenamefont {Kim},\ and\ \citenamefont
  {Ahn}}]{Kim2020quantum_Ising}%
  \BibitemOpen
  \bibfield  {author} {\bibinfo {author} {\bibfnamefont {M.}~\bibnamefont
  {Kim}}, \bibinfo {author} {\bibfnamefont {Y.}~\bibnamefont {Song}}, \bibinfo
  {author} {\bibfnamefont {J.}~\bibnamefont {Kim}},\ and\ \bibinfo {author}
  {\bibfnamefont {J.}~\bibnamefont {Ahn}},\ }\bibfield  {title} {\bibinfo
  {title} {Quantum {Ising} {Hamiltonian} programming in trio, quartet, and
  sextet qubit systems},\ }\href {https://doi.org/10.1103/PRXQuantum.1.020323}
  {\bibfield  {journal} {\bibinfo  {journal} {PRX Quantum}\ }\textbf {\bibinfo
  {volume} {1}},\ \bibinfo {pages} {020323} (\bibinfo {year}
  {2020})}\BibitemShut {NoStop}%
\bibitem [{\citenamefont {Borish}\ \emph {et~al.}(2020)\citenamefont {Borish},
  \citenamefont {Markovi\'c}, \citenamefont {Hines}, \citenamefont
  {Rajagopal},\ and\ \citenamefont {Schleier-Smith}}]{Borish2020transverse}%
  \BibitemOpen
  \bibfield  {author} {\bibinfo {author} {\bibfnamefont {V.}~\bibnamefont
  {Borish}}, \bibinfo {author} {\bibfnamefont {O.}~\bibnamefont {Markovi\'c}},
  \bibinfo {author} {\bibfnamefont {J.~A.}\ \bibnamefont {Hines}}, \bibinfo
  {author} {\bibfnamefont {S.~V.}\ \bibnamefont {Rajagopal}},\ and\ \bibinfo
  {author} {\bibfnamefont {M.}~\bibnamefont {Schleier-Smith}},\ }\bibfield
  {title} {\bibinfo {title} {Transverse-field {Ising} dynamics in a
  {Rydberg}-dressed atomic gas},\ }\href
  {https://doi.org/10.1103/PhysRevLett.124.063601} {\bibfield  {journal}
  {\bibinfo  {journal} {Phys. Rev. Lett.}\ }\textbf {\bibinfo {volume} {124}},\
  \bibinfo {pages} {063601} (\bibinfo {year} {2020})}\BibitemShut {NoStop}%
\bibitem [{\citenamefont {Song}\ \emph {et~al.}(2021)\citenamefont {Song},
  \citenamefont {Kim}, \citenamefont {Hwang}, \citenamefont {Lee},\ and\
  \citenamefont {Ahn}}]{Song2021quantum_simulation}%
  \BibitemOpen
  \bibfield  {author} {\bibinfo {author} {\bibfnamefont {Y.}~\bibnamefont
  {Song}}, \bibinfo {author} {\bibfnamefont {M.}~\bibnamefont {Kim}}, \bibinfo
  {author} {\bibfnamefont {H.}~\bibnamefont {Hwang}}, \bibinfo {author}
  {\bibfnamefont {W.}~\bibnamefont {Lee}},\ and\ \bibinfo {author}
  {\bibfnamefont {J.}~\bibnamefont {Ahn}},\ }\bibfield  {title} {\bibinfo
  {title} {Quantum simulation of {Cayley}-tree {Ising} {Hamiltonians} with
  three-dimensional {Rydberg} atoms},\ }\href
  {https://doi.org/10.1103/PhysRevResearch.3.013286} {\bibfield  {journal}
  {\bibinfo  {journal} {Phys. Rev. Res.}\ }\textbf {\bibinfo {volume} {3}},\
  \bibinfo {pages} {013286} (\bibinfo {year} {2021})}\BibitemShut {NoStop}%
\bibitem [{\citenamefont {Ebadi}\ \emph {et~al.}(2021)\citenamefont {Ebadi},
  \citenamefont {Wang}, \citenamefont {Levine}, \citenamefont {Keesling},
  \citenamefont {Semeghini}, \citenamefont {Omran}, \citenamefont {Bluvstein},
  \citenamefont {Samajdar}, \citenamefont {Pichler}, \citenamefont {Ho},
  \citenamefont {Choi}, \citenamefont {Sachdev}, \citenamefont {Greiner},
  \citenamefont {Vuleti\'c},\ and\ \citenamefont {Lukin}}]{Ebadi2021}%
  \BibitemOpen
  \bibfield  {author} {\bibinfo {author} {\bibfnamefont {S.}~\bibnamefont
  {Ebadi}}, \bibinfo {author} {\bibfnamefont {T.~T.}\ \bibnamefont {Wang}},
  \bibinfo {author} {\bibfnamefont {H.}~\bibnamefont {Levine}}, \bibinfo
  {author} {\bibfnamefont {A.}~\bibnamefont {Keesling}}, \bibinfo {author}
  {\bibfnamefont {G.}~\bibnamefont {Semeghini}}, \bibinfo {author}
  {\bibfnamefont {A.}~\bibnamefont {Omran}}, \bibinfo {author} {\bibfnamefont
  {D.}~\bibnamefont {Bluvstein}}, \bibinfo {author} {\bibfnamefont
  {R.}~\bibnamefont {Samajdar}}, \bibinfo {author} {\bibfnamefont
  {H.}~\bibnamefont {Pichler}}, \bibinfo {author} {\bibfnamefont {W.~W.}\
  \bibnamefont {Ho}}, \bibinfo {author} {\bibfnamefont {S.}~\bibnamefont
  {Choi}}, \bibinfo {author} {\bibfnamefont {S.}~\bibnamefont {Sachdev}},
  \bibinfo {author} {\bibfnamefont {M.}~\bibnamefont {Greiner}}, \bibinfo
  {author} {\bibfnamefont {V.}~\bibnamefont {Vuleti\'c}},\ and\ \bibinfo
  {author} {\bibfnamefont {M.~D.}\ \bibnamefont {Lukin}},\ }\bibfield  {title}
  {\bibinfo {title} {Quantum phases of matter on a 256-atom programmable
  quantum simulator},\ }\href@noop {} {\bibfield  {journal} {\bibinfo
  {journal} {Nature}\ }\textbf {\bibinfo {volume} {595}},\ \bibinfo {pages}
  {227} (\bibinfo {year} {2021})}\BibitemShut {NoStop}%
\bibitem [{\citenamefont {Bluvstein}\ \emph {et~al.}(2021)\citenamefont
  {Bluvstein}, \citenamefont {Omran}, \citenamefont {Levine}, \citenamefont
  {Keesling}, \citenamefont {Semeghini}, \citenamefont {Ebadi}, \citenamefont
  {Wang}, \citenamefont {Michailidis}, \citenamefont {Maskara}, \citenamefont
  {Ho}, \citenamefont {Choi}, \citenamefont {Serbyn}, \citenamefont {Greiner},
  \citenamefont {Vuleti\'c},\ and\ \citenamefont {Lukin}}]{Bluvstein2021}%
  \BibitemOpen
  \bibfield  {author} {\bibinfo {author} {\bibfnamefont {D.}~\bibnamefont
  {Bluvstein}}, \bibinfo {author} {\bibfnamefont {A.}~\bibnamefont {Omran}},
  \bibinfo {author} {\bibfnamefont {H.}~\bibnamefont {Levine}}, \bibinfo
  {author} {\bibfnamefont {A.}~\bibnamefont {Keesling}}, \bibinfo {author}
  {\bibfnamefont {G.}~\bibnamefont {Semeghini}}, \bibinfo {author}
  {\bibfnamefont {S.}~\bibnamefont {Ebadi}}, \bibinfo {author} {\bibfnamefont
  {T.~T.}\ \bibnamefont {Wang}}, \bibinfo {author} {\bibfnamefont {A.~A.}\
  \bibnamefont {Michailidis}}, \bibinfo {author} {\bibfnamefont
  {N.}~\bibnamefont {Maskara}}, \bibinfo {author} {\bibfnamefont {W.~W.}\
  \bibnamefont {Ho}}, \bibinfo {author} {\bibfnamefont {S.}~\bibnamefont
  {Choi}}, \bibinfo {author} {\bibfnamefont {M.}~\bibnamefont {Serbyn}},
  \bibinfo {author} {\bibfnamefont {M.}~\bibnamefont {Greiner}}, \bibinfo
  {author} {\bibfnamefont {V.}~\bibnamefont {Vuleti\'c}},\ and\ \bibinfo
  {author} {\bibfnamefont {M.~D.}\ \bibnamefont {Lukin}},\ }\bibfield  {title}
  {\bibinfo {title} {Controlling quantum many-body dynamics in driven {Rydberg}
  atom arrays},\ }\href@noop {} {\bibfield  {journal} {\bibinfo  {journal}
  {Science}\ }\textbf {\bibinfo {volume} {371}},\ \bibinfo {pages} {1355}
  (\bibinfo {year} {2021})}\BibitemShut {NoStop}%
\bibitem [{\citenamefont {Scholl}\ \emph {et~al.}(2021)\citenamefont {Scholl},
  \citenamefont {Schuler}, \citenamefont {Williams}, \citenamefont
  {Eberharter}, \citenamefont {Barredo}, \citenamefont {Schymik}, \citenamefont
  {Lienhard}, \citenamefont {Henry}, \citenamefont {Lang}, \citenamefont
  {Lahaye}, \citenamefont {L\"auchli},\ and\ \citenamefont
  {Browaeys}}]{Scholl2021}%
  \BibitemOpen
  \bibfield  {author} {\bibinfo {author} {\bibfnamefont {P.}~\bibnamefont
  {Scholl}}, \bibinfo {author} {\bibfnamefont {M.}~\bibnamefont {Schuler}},
  \bibinfo {author} {\bibfnamefont {H.~J.}\ \bibnamefont {Williams}}, \bibinfo
  {author} {\bibfnamefont {A.~A.}\ \bibnamefont {Eberharter}}, \bibinfo
  {author} {\bibfnamefont {D.}~\bibnamefont {Barredo}}, \bibinfo {author}
  {\bibfnamefont {K.-N.}\ \bibnamefont {Schymik}}, \bibinfo {author}
  {\bibfnamefont {V.}~\bibnamefont {Lienhard}}, \bibinfo {author}
  {\bibfnamefont {L.-P.}\ \bibnamefont {Henry}}, \bibinfo {author}
  {\bibfnamefont {T.~C.}\ \bibnamefont {Lang}}, \bibinfo {author}
  {\bibfnamefont {T.}~\bibnamefont {Lahaye}}, \bibinfo {author} {\bibfnamefont
  {A.~M.}\ \bibnamefont {L\"auchli}},\ and\ \bibinfo {author} {\bibfnamefont
  {A.}~\bibnamefont {Browaeys}},\ }\bibfield  {title} {\bibinfo {title}
  {Quantum simulation of {2D} antiferromagnets with hundreds of {Rydberg}
  atoms},\ }\href@noop {} {\bibfield  {journal} {\bibinfo  {journal} {Nature}\
  }\textbf {\bibinfo {volume} {595}},\ \bibinfo {pages} {233} (\bibinfo {year}
  {2021})}\BibitemShut {NoStop}%
\bibitem [{\citenamefont {Hollerith}\ \emph {et~al.}(2022)\citenamefont
  {Hollerith}, \citenamefont {Srakaew}, \citenamefont {Wei}, \citenamefont
  {Rubio-Abadal}, \citenamefont {Adler}, \citenamefont {Weckesser},
  \citenamefont {Kruckenhauser}, \citenamefont {Walther}, \citenamefont {van
  Bijnen}, \citenamefont {Rui}, \citenamefont {Gross}, \citenamefont {Bloch},\
  and\ \citenamefont {Zeiher}}]{Hollerith2022}%
  \BibitemOpen
  \bibfield  {author} {\bibinfo {author} {\bibfnamefont {S.}~\bibnamefont
  {Hollerith}}, \bibinfo {author} {\bibfnamefont {K.}~\bibnamefont {Srakaew}},
  \bibinfo {author} {\bibfnamefont {D.}~\bibnamefont {Wei}}, \bibinfo {author}
  {\bibfnamefont {A.}~\bibnamefont {Rubio-Abadal}}, \bibinfo {author}
  {\bibfnamefont {D.}~\bibnamefont {Adler}}, \bibinfo {author} {\bibfnamefont
  {P.}~\bibnamefont {Weckesser}}, \bibinfo {author} {\bibfnamefont
  {A.}~\bibnamefont {Kruckenhauser}}, \bibinfo {author} {\bibfnamefont
  {V.}~\bibnamefont {Walther}}, \bibinfo {author} {\bibfnamefont
  {R.}~\bibnamefont {van Bijnen}}, \bibinfo {author} {\bibfnamefont
  {J.}~\bibnamefont {Rui}}, \bibinfo {author} {\bibfnamefont {C.}~\bibnamefont
  {Gross}}, \bibinfo {author} {\bibfnamefont {I.}~\bibnamefont {Bloch}},\ and\
  \bibinfo {author} {\bibfnamefont {J.}~\bibnamefont {Zeiher}},\ }\bibfield
  {title} {\bibinfo {title} {Realizing distance-selective interactions in a
  {Rydberg-dressed} atom array},\ }\href@noop {} {\bibfield  {journal}
  {\bibinfo  {journal} {Phys. Rev. Lett.}\ }\textbf {\bibinfo {volume} {128}},\
  \bibinfo {pages} {113602} (\bibinfo {year} {2022})}\BibitemShut {NoStop}%
\bibitem [{\citenamefont {Zhao}\ \emph {et~al.}(2023)\citenamefont {Zhao},
  \citenamefont {Lee}, \citenamefont {Aliyu},\ and\ \citenamefont
  {Loh}}]{Zhao2023}%
  \BibitemOpen
  \bibfield  {author} {\bibinfo {author} {\bibfnamefont {L.}~\bibnamefont
  {Zhao}}, \bibinfo {author} {\bibfnamefont {M.~D.~K.}\ \bibnamefont {Lee}},
  \bibinfo {author} {\bibfnamefont {M.~M.}\ \bibnamefont {Aliyu}},\ and\
  \bibinfo {author} {\bibfnamefont {H.}~\bibnamefont {Loh}},\ }\bibfield
  {title} {\bibinfo {title} {Floquet-tailored {Rydberg} interactions},\
  }\href@noop {} {\bibfield  {journal} {\bibinfo  {journal} {Nat. Commun.}\
  }\textbf {\bibinfo {volume} {14}},\ \bibinfo {pages} {7128} (\bibinfo {year}
  {2023})}\BibitemShut {NoStop}%
\bibitem [{\citenamefont {Bharti}\ \emph {et~al.}(2023)\citenamefont {Bharti},
  \citenamefont {Sugawa}, \citenamefont {Mizoguchi}, \citenamefont {Kunimi},
  \citenamefont {Zhang}, \citenamefont {de~L{\'e}s{\'e}leuc}, \citenamefont
  {Tomita}, \citenamefont {Franz}, \citenamefont {Weidem{\"u}ller},\ and\
  \citenamefont {Ohmori}}]{Bharti2023}%
  \BibitemOpen
  \bibfield  {author} {\bibinfo {author} {\bibfnamefont {V.}~\bibnamefont
  {Bharti}}, \bibinfo {author} {\bibfnamefont {S.}~\bibnamefont {Sugawa}},
  \bibinfo {author} {\bibfnamefont {M.}~\bibnamefont {Mizoguchi}}, \bibinfo
  {author} {\bibfnamefont {M.}~\bibnamefont {Kunimi}}, \bibinfo {author}
  {\bibfnamefont {Y.}~\bibnamefont {Zhang}}, \bibinfo {author} {\bibfnamefont
  {S.}~\bibnamefont {de~L{\'e}s{\'e}leuc}}, \bibinfo {author} {\bibfnamefont
  {T.}~\bibnamefont {Tomita}}, \bibinfo {author} {\bibfnamefont
  {T.}~\bibnamefont {Franz}}, \bibinfo {author} {\bibfnamefont
  {M.}~\bibnamefont {Weidem{\"u}ller}},\ and\ \bibinfo {author} {\bibfnamefont
  {K.}~\bibnamefont {Ohmori}},\ }\bibfield  {title} {\bibinfo {title}
  {Picosecond-scale ultrafast many-body dynamics in an ultracold
  {Rydberg-excited} atomic {Mott} insulator},\ }\href@noop {} {\bibfield
  {journal} {\bibinfo  {journal} {Phys. Rev. Lett.}\ }\textbf {\bibinfo
  {volume} {131}},\ \bibinfo {pages} {123201} (\bibinfo {year}
  {2023})}\BibitemShut {NoStop}%
\bibitem [{\citenamefont {Franz}\ \emph {et~al.}(2024)\citenamefont {Franz},
  \citenamefont {Geier}, \citenamefont {Hainaut}, \citenamefont {Braemer},
  \citenamefont {Thaicharoen}, \citenamefont {Hornung}, \citenamefont {Braun},
  \citenamefont {G{\"a}rttner}, \citenamefont {Z{\"u}rn},\ and\ \citenamefont
  {Weidem{\"u}ller}}]{Franz2024}%
  \BibitemOpen
  \bibfield  {author} {\bibinfo {author} {\bibfnamefont {T.}~\bibnamefont
  {Franz}}, \bibinfo {author} {\bibfnamefont {S.}~\bibnamefont {Geier}},
  \bibinfo {author} {\bibfnamefont {C.}~\bibnamefont {Hainaut}}, \bibinfo
  {author} {\bibfnamefont {A.}~\bibnamefont {Braemer}}, \bibinfo {author}
  {\bibfnamefont {N.}~\bibnamefont {Thaicharoen}}, \bibinfo {author}
  {\bibfnamefont {M.}~\bibnamefont {Hornung}}, \bibinfo {author} {\bibfnamefont
  {E.}~\bibnamefont {Braun}}, \bibinfo {author} {\bibfnamefont
  {M.}~\bibnamefont {G{\"a}rttner}}, \bibinfo {author} {\bibfnamefont
  {G.}~\bibnamefont {Z{\"u}rn}},\ and\ \bibinfo {author} {\bibfnamefont
  {M.}~\bibnamefont {Weidem{\"u}ller}},\ }\bibfield  {title} {\bibinfo {title}
  {Observation of anisotropy-independent magnetization dynamics in spatially
  disordered {Heisenberg} spin systems},\ }\href@noop {} {\bibfield  {journal}
  {\bibinfo  {journal} {Phys. Rev. Res.}\ }\textbf {\bibinfo {volume} {6}},\
  \bibinfo {pages} {033131} (\bibinfo {year} {2024})}\BibitemShut {NoStop}%
\bibitem [{\citenamefont {Bharti}\ \emph {et~al.}(2024)\citenamefont {Bharti},
  \citenamefont {Sugawa}, \citenamefont {Kunimi}, \citenamefont {Chauhan},
  \citenamefont {Mahesh}, \citenamefont {Mizoguchi}, \citenamefont {Matsubara},
  \citenamefont {Tomita}, \citenamefont {de~L\'es\'eleuc},\ and\ \citenamefont
  {Ohmori}}]{Bharti2024}%
  \BibitemOpen
  \bibfield  {author} {\bibinfo {author} {\bibfnamefont {V.}~\bibnamefont
  {Bharti}}, \bibinfo {author} {\bibfnamefont {S.}~\bibnamefont {Sugawa}},
  \bibinfo {author} {\bibfnamefont {M.}~\bibnamefont {Kunimi}}, \bibinfo
  {author} {\bibfnamefont {V.~S.}\ \bibnamefont {Chauhan}}, \bibinfo {author}
  {\bibfnamefont {T.~P.}\ \bibnamefont {Mahesh}}, \bibinfo {author}
  {\bibfnamefont {M.}~\bibnamefont {Mizoguchi}}, \bibinfo {author}
  {\bibfnamefont {T.}~\bibnamefont {Matsubara}}, \bibinfo {author}
  {\bibfnamefont {T.}~\bibnamefont {Tomita}}, \bibinfo {author} {\bibfnamefont
  {S.}~\bibnamefont {de~L\'es\'eleuc}},\ and\ \bibinfo {author} {\bibfnamefont
  {K.}~\bibnamefont {Ohmori}},\ }\bibfield  {title} {\bibinfo {title} {Strong
  spin-motion coupling in the ultrafast dynamics of {Rydberg} atoms},\
  }\href@noop {} {\bibfield  {journal} {\bibinfo  {journal} {Phys. Rev. Lett.}\
  }\textbf {\bibinfo {volume} {133}},\ \bibinfo {pages} {093405} (\bibinfo
  {year} {2024})}\BibitemShut {NoStop}%
\bibitem [{\citenamefont {Kim}\ \emph {et~al.}(2024)\citenamefont {Kim},
  \citenamefont {Yang}, \citenamefont {M{\o}lmer},\ and\ \citenamefont
  {Ahn}}]{kim2024realization}%
  \BibitemOpen
  \bibfield  {author} {\bibinfo {author} {\bibfnamefont {K.}~\bibnamefont
  {Kim}}, \bibinfo {author} {\bibfnamefont {F.}~\bibnamefont {Yang}}, \bibinfo
  {author} {\bibfnamefont {K.}~\bibnamefont {M{\o}lmer}},\ and\ \bibinfo
  {author} {\bibfnamefont {J.}~\bibnamefont {Ahn}},\ }\bibfield  {title}
  {\bibinfo {title} {Realization of an extremely anisotropic {Heisenberg}
  magnet in {Rydberg} atom arrays},\ }\href@noop {} {\bibfield  {journal}
  {\bibinfo  {journal} {Phys. Rev. X}\ }\textbf {\bibinfo {volume} {14}},\
  \bibinfo {pages} {011025} (\bibinfo {year} {2024})}\BibitemShut {NoStop}%
\bibitem [{\citenamefont {Zhao}\ \emph {et~al.}(2025)\citenamefont {Zhao},
  \citenamefont {Datla}, \citenamefont {Tian}, \citenamefont {Aliyu},\ and\
  \citenamefont {Loh}}]{Zhao2025}%
  \BibitemOpen
  \bibfield  {author} {\bibinfo {author} {\bibfnamefont {L.}~\bibnamefont
  {Zhao}}, \bibinfo {author} {\bibfnamefont {P.~R.}\ \bibnamefont {Datla}},
  \bibinfo {author} {\bibfnamefont {W.}~\bibnamefont {Tian}}, \bibinfo {author}
  {\bibfnamefont {M.~M.}\ \bibnamefont {Aliyu}},\ and\ \bibinfo {author}
  {\bibfnamefont {H.}~\bibnamefont {Loh}},\ }\bibfield  {title} {\bibinfo
  {title} {Observation of quantum thermalization restricted to {Hilbert} space
  fragments and $\mathbb{Z}_{2k}$ scars},\ }\href@noop {} {\bibfield  {journal}
  {\bibinfo  {journal} {Phys. Rev. X}\ }\textbf {\bibinfo {volume} {15}},\
  \bibinfo {pages} {011035} (\bibinfo {year} {2025})}\BibitemShut {NoStop}%
\bibitem [{\citenamefont {Manovitz}\ \emph {et~al.}(2025)\citenamefont
  {Manovitz}, \citenamefont {Li}, \citenamefont {Ebadi}, \citenamefont
  {Samajdar}, \citenamefont {Geim}, \citenamefont {Evered}, \citenamefont
  {Bluvstein}, \citenamefont {Zhou}, \citenamefont {Koyluoglu}, \citenamefont
  {Feldmeier}, \citenamefont {Dolgirev}, \citenamefont {Maskara}, \citenamefont
  {Kalinowski}, \citenamefont {Sachdev}, \citenamefont {Huse}, \citenamefont
  {Greiner}, \citenamefont {Vuleti\'c},\ and\ \citenamefont
  {Lukin}}]{manovitz2025quantum}%
  \BibitemOpen
  \bibfield  {author} {\bibinfo {author} {\bibfnamefont {T.}~\bibnamefont
  {Manovitz}}, \bibinfo {author} {\bibfnamefont {S.~H.}\ \bibnamefont {Li}},
  \bibinfo {author} {\bibfnamefont {S.}~\bibnamefont {Ebadi}}, \bibinfo
  {author} {\bibfnamefont {R.}~\bibnamefont {Samajdar}}, \bibinfo {author}
  {\bibfnamefont {A.~A.}\ \bibnamefont {Geim}}, \bibinfo {author}
  {\bibfnamefont {S.~J.}\ \bibnamefont {Evered}}, \bibinfo {author}
  {\bibfnamefont {D.}~\bibnamefont {Bluvstein}}, \bibinfo {author}
  {\bibfnamefont {H.}~\bibnamefont {Zhou}}, \bibinfo {author} {\bibfnamefont
  {N.~U.}\ \bibnamefont {Koyluoglu}}, \bibinfo {author} {\bibfnamefont
  {J.}~\bibnamefont {Feldmeier}}, \bibinfo {author} {\bibfnamefont {P.~E.}\
  \bibnamefont {Dolgirev}}, \bibinfo {author} {\bibfnamefont {N.}~\bibnamefont
  {Maskara}}, \bibinfo {author} {\bibfnamefont {M.}~\bibnamefont {Kalinowski}},
  \bibinfo {author} {\bibfnamefont {S.}~\bibnamefont {Sachdev}}, \bibinfo
  {author} {\bibfnamefont {D.~A.}\ \bibnamefont {Huse}}, \bibinfo {author}
  {\bibfnamefont {M.}~\bibnamefont {Greiner}}, \bibinfo {author} {\bibfnamefont
  {V.}~\bibnamefont {Vuleti\'c}},\ and\ \bibinfo {author} {\bibfnamefont
  {M.~D.}\ \bibnamefont {Lukin}},\ }\bibfield  {title} {\bibinfo {title}
  {Quantum coarsening and collective dynamics on a programmable simulator},\
  }\href@noop {} {\bibfield  {journal} {\bibinfo  {journal} {Nature}\ }\textbf
  {\bibinfo {volume} {638}},\ \bibinfo {pages} {86} (\bibinfo {year}
  {2025})}\BibitemShut {NoStop}%
\bibitem [{\citenamefont {Gonzalez-Cuadra}\ \emph {et~al.}(2025)\citenamefont
  {Gonzalez-Cuadra}, \citenamefont {Hamdan}, \citenamefont {Zache},
  \citenamefont {Braverman}, \citenamefont {Kornja{\v{c}}a}, \citenamefont
  {Lukin}, \citenamefont {Cant{\'u}}, \citenamefont {Liu}, \citenamefont
  {Wang}, \citenamefont {Keesling}, \citenamefont {Lukin}, \citenamefont
  {Zoller},\ and\ \citenamefont {Bylinskii}}]{gonzalez2025observation}%
  \BibitemOpen
  \bibfield  {author} {\bibinfo {author} {\bibfnamefont {D.}~\bibnamefont
  {Gonzalez-Cuadra}}, \bibinfo {author} {\bibfnamefont {M.}~\bibnamefont
  {Hamdan}}, \bibinfo {author} {\bibfnamefont {T.~V.}\ \bibnamefont {Zache}},
  \bibinfo {author} {\bibfnamefont {B.}~\bibnamefont {Braverman}}, \bibinfo
  {author} {\bibfnamefont {M.}~\bibnamefont {Kornja{\v{c}}a}}, \bibinfo
  {author} {\bibfnamefont {A.}~\bibnamefont {Lukin}}, \bibinfo {author}
  {\bibfnamefont {S.~H.}\ \bibnamefont {Cant{\'u}}}, \bibinfo {author}
  {\bibfnamefont {F.}~\bibnamefont {Liu}}, \bibinfo {author} {\bibfnamefont
  {S.-T.}\ \bibnamefont {Wang}}, \bibinfo {author} {\bibfnamefont
  {A.}~\bibnamefont {Keesling}}, \bibinfo {author} {\bibfnamefont {M.~D.}\
  \bibnamefont {Lukin}}, \bibinfo {author} {\bibfnamefont {P.}~\bibnamefont
  {Zoller}},\ and\ \bibinfo {author} {\bibfnamefont {A.}~\bibnamefont
  {Bylinskii}},\ }\bibfield  {title} {\bibinfo {title} {Observation of string
  breaking on a (2+1){D} {Rydberg} quantum simulator},\ }\href@noop {}
  {\bibfield  {journal} {\bibinfo  {journal} {Nature}\ }\textbf {\bibinfo
  {volume} {642}},\ \bibinfo {pages} {321} (\bibinfo {year}
  {2025})}\BibitemShut {NoStop}%
\bibitem [{\citenamefont {Zhang}\ \emph
  {et~al.}(2025{\natexlab{a}})\citenamefont {Zhang}, \citenamefont {Wang},
  \citenamefont {Zhang}, \citenamefont {Wang}, \citenamefont {Du},
  \citenamefont {Li}, \citenamefont {Wu}, \citenamefont {Li}, \citenamefont
  {Hu}, \citenamefont {Zhai},\ and\ \citenamefont
  {Chen}}]{zhang2025observation}%
  \BibitemOpen
  \bibfield  {author} {\bibinfo {author} {\bibfnamefont {T.}~\bibnamefont
  {Zhang}}, \bibinfo {author} {\bibfnamefont {H.}~\bibnamefont {Wang}},
  \bibinfo {author} {\bibfnamefont {W.}~\bibnamefont {Zhang}}, \bibinfo
  {author} {\bibfnamefont {Y.}~\bibnamefont {Wang}}, \bibinfo {author}
  {\bibfnamefont {A.}~\bibnamefont {Du}}, \bibinfo {author} {\bibfnamefont
  {Z.}~\bibnamefont {Li}}, \bibinfo {author} {\bibfnamefont {Y.}~\bibnamefont
  {Wu}}, \bibinfo {author} {\bibfnamefont {C.}~\bibnamefont {Li}}, \bibinfo
  {author} {\bibfnamefont {J.}~\bibnamefont {Hu}}, \bibinfo {author}
  {\bibfnamefont {H.}~\bibnamefont {Zhai}},\ and\ \bibinfo {author}
  {\bibfnamefont {W.}~\bibnamefont {Chen}},\ }\bibfield  {title} {\bibinfo
  {title} {Observation of near-critical {Kibble-Zurek} scaling in {Rydberg}
  atom arrays},\ }\href@noop {} {\bibfield  {journal} {\bibinfo  {journal}
  {Phys. Rev. Lett.}\ }\textbf {\bibinfo {volume} {135}},\ \bibinfo {pages}
  {093403} (\bibinfo {year} {2025}{\natexlab{a}})}\BibitemShut {NoStop}%
\bibitem [{\citenamefont {Da\u{g}}\ \emph {et~al.}(2025)\citenamefont
  {Da\u{g}}, \citenamefont {Ma}, \citenamefont {Eugenio}, \citenamefont
  {Fang},\ and\ \citenamefont {Yelin}}]{Dag2025emergent}%
  \BibitemOpen
  \bibfield  {author} {\bibinfo {author} {\bibfnamefont {C.~B.}\ \bibnamefont
  {Da\u{g}}}, \bibinfo {author} {\bibfnamefont {H.}~\bibnamefont {Ma}},
  \bibinfo {author} {\bibfnamefont {P.~M.}\ \bibnamefont {Eugenio}}, \bibinfo
  {author} {\bibfnamefont {F.}~\bibnamefont {Fang}},\ and\ \bibinfo {author}
  {\bibfnamefont {S.~F.}\ \bibnamefont {Yelin}},\ }\bibfield  {title} {\bibinfo
  {title} {Emergent disorder and sub-ballistic dynamics in quantum simulations
  of the {Ising} model using {Rydberg} atom arrays},\ }\href
  {https://doi.org/10.1103/jr7l-2cfb} {\bibfield  {journal} {\bibinfo
  {journal} {Phys. Rev. Lett.}\ }\textbf {\bibinfo {volume} {135}},\ \bibinfo
  {pages} {250403} (\bibinfo {year} {2025})}\BibitemShut {NoStop}%
\bibitem [{\citenamefont {Zhang}\ \emph
  {et~al.}(2025{\natexlab{b}})\citenamefont {Zhang}, \citenamefont {Cant{\'u}},
  \citenamefont {Liu}, \citenamefont {Bylinskii}, \citenamefont {Braverman},
  \citenamefont {Huber}, \citenamefont {Amato-Grill}, \citenamefont {Lukin},
  \citenamefont {Gemelke}, \citenamefont {Keesling}, \citenamefont {Wang},
  \citenamefont {Meurice},\ and\ \citenamefont {Tsai}}]{zhang2025probing}%
  \BibitemOpen
  \bibfield  {author} {\bibinfo {author} {\bibfnamefont {J.}~\bibnamefont
  {Zhang}}, \bibinfo {author} {\bibfnamefont {S.~H.}\ \bibnamefont
  {Cant{\'u}}}, \bibinfo {author} {\bibfnamefont {F.}~\bibnamefont {Liu}},
  \bibinfo {author} {\bibfnamefont {A.}~\bibnamefont {Bylinskii}}, \bibinfo
  {author} {\bibfnamefont {B.}~\bibnamefont {Braverman}}, \bibinfo {author}
  {\bibfnamefont {F.}~\bibnamefont {Huber}}, \bibinfo {author} {\bibfnamefont
  {J.}~\bibnamefont {Amato-Grill}}, \bibinfo {author} {\bibfnamefont
  {A.}~\bibnamefont {Lukin}}, \bibinfo {author} {\bibfnamefont
  {N.}~\bibnamefont {Gemelke}}, \bibinfo {author} {\bibfnamefont
  {A.}~\bibnamefont {Keesling}}, \bibinfo {author} {\bibfnamefont {S.-T.}\
  \bibnamefont {Wang}}, \bibinfo {author} {\bibfnamefont {Y.}~\bibnamefont
  {Meurice}},\ and\ \bibinfo {author} {\bibfnamefont {S.-W.}\ \bibnamefont
  {Tsai}},\ }\bibfield  {title} {\bibinfo {title} {Probing quantum floating
  phases in {Rydberg} atom arrays},\ }\href@noop {} {\bibfield  {journal}
  {\bibinfo  {journal} {Nat. Commun.}\ }\textbf {\bibinfo {volume} {16}},\
  \bibinfo {pages} {712} (\bibinfo {year} {2025}{\natexlab{b}})}\BibitemShut
  {NoStop}%
\bibitem [{\citenamefont {Datla}\ \emph {et~al.}(2026)\citenamefont {Datla},
  \citenamefont {Zhao}, \citenamefont {Ho}, \citenamefont {Klco},\ and\
  \citenamefont {Loh}}]{datla2026statistical}%
  \BibitemOpen
  \bibfield  {author} {\bibinfo {author} {\bibfnamefont {P.~R.}\ \bibnamefont
  {Datla}}, \bibinfo {author} {\bibfnamefont {L.}~\bibnamefont {Zhao}},
  \bibinfo {author} {\bibfnamefont {W.~W.}\ \bibnamefont {Ho}}, \bibinfo
  {author} {\bibfnamefont {N.}~\bibnamefont {Klco}},\ and\ \bibinfo {author}
  {\bibfnamefont {H.}~\bibnamefont {Loh}},\ }\bibfield  {title} {\bibinfo
  {title} {Statistical localization of {U(1)} lattice gauge theory in a
  {Rydberg} simulator},\ }\href@noop {} {\bibfield  {journal} {\bibinfo
  {journal} {Nat. Phys.}\ }\textbf {\bibinfo {volume} {22}},\ \bibinfo {pages}
  {355} (\bibinfo {year} {2026})}\BibitemShut {NoStop}%
\bibitem [{\citenamefont {Geim}\ \emph {et~al.}(2026)\citenamefont {Geim},
  \citenamefont {Koyluoglu}, \citenamefont {Evered}, \citenamefont {Sahay},
  \citenamefont {Li}, \citenamefont {Xu}, \citenamefont {Bluvstein},
  \citenamefont {Gjonbalaj}, \citenamefont {Maskara}, \citenamefont
  {Kalinowski}, \citenamefont {Manovitz}, \citenamefont {Verresen},
  \citenamefont {Yelin}, \citenamefont {Feldmeier}, \citenamefont {Greiner},
  \citenamefont {Vuleti\'c},\ and\ \citenamefont
  {Lukin}}]{geim2026engineering}%
  \BibitemOpen
  \bibfield  {author} {\bibinfo {author} {\bibfnamefont {A.~A.}\ \bibnamefont
  {Geim}}, \bibinfo {author} {\bibfnamefont {N.~U.}\ \bibnamefont {Koyluoglu}},
  \bibinfo {author} {\bibfnamefont {S.~J.}\ \bibnamefont {Evered}}, \bibinfo
  {author} {\bibfnamefont {R.}~\bibnamefont {Sahay}}, \bibinfo {author}
  {\bibfnamefont {S.~H.}\ \bibnamefont {Li}}, \bibinfo {author} {\bibfnamefont
  {M.}~\bibnamefont {Xu}}, \bibinfo {author} {\bibfnamefont {D.}~\bibnamefont
  {Bluvstein}}, \bibinfo {author} {\bibfnamefont {N.~O.}\ \bibnamefont
  {Gjonbalaj}}, \bibinfo {author} {\bibfnamefont {N.}~\bibnamefont {Maskara}},
  \bibinfo {author} {\bibfnamefont {M.}~\bibnamefont {Kalinowski}}, \bibinfo
  {author} {\bibfnamefont {T.}~\bibnamefont {Manovitz}}, \bibinfo {author}
  {\bibfnamefont {R.}~\bibnamefont {Verresen}}, \bibinfo {author}
  {\bibfnamefont {S.~F.}\ \bibnamefont {Yelin}}, \bibinfo {author}
  {\bibfnamefont {J.}~\bibnamefont {Feldmeier}}, \bibinfo {author}
  {\bibfnamefont {M.}~\bibnamefont {Greiner}}, \bibinfo {author} {\bibfnamefont
  {V.}~\bibnamefont {Vuleti\'c}},\ and\ \bibinfo {author} {\bibfnamefont
  {M.~D.}\ \bibnamefont {Lukin}},\ }\bibfield  {title} {\bibinfo {title}
  {Engineering quantum criticality and dynamics on an analog-digital
  simulator},\ }\href@noop {} {\bibfield  {journal} {\bibinfo  {journal}
  {arXiv:2602.18555}\ } (\bibinfo {year} {2026})}\BibitemShut {NoStop}%
\bibitem [{\citenamefont {Leclerc}\ \emph {et~al.}(2026)\citenamefont
  {Leclerc}, \citenamefont {Juli\`a-Farr\'e}, \citenamefont {Freitas},
  \citenamefont {Villaret}, \citenamefont {Albrecht}, \citenamefont {B\'eguin},
  \citenamefont {Bourachot}, \citenamefont {Briosne-Frejaville}, \citenamefont
  {Claveau}, \citenamefont {Cornillot}, \citenamefont {de~Hond}, \citenamefont
  {Diallo}, \citenamefont {Dupays}, \citenamefont {Dupont}, \citenamefont
  {Eritzpokhoff}, \citenamefont {Gottlob}, \citenamefont {Henriet},
  \citenamefont {Kaicher}, \citenamefont {Lassabli\`ere}, \citenamefont
  {Lindberg}, \citenamefont {Machu}, \citenamefont {Mamann}, \citenamefont
  {Pansiot}, \citenamefont {Ripoll}, \citenamefont {Choi}, \citenamefont
  {Signoles}, \citenamefont {Vovrosh}, \citenamefont {Ximenez}, \citenamefont
  {Zapf}, \citenamefont {Zhang}, \citenamefont {Zhou}, \citenamefont {Lee},
  \citenamefont {Mendes-Santos}, \citenamefont {Dalyac}, \citenamefont
  {Browaeys},\ and\ \citenamefont {Dauphin}}]{leclerc2026one}%
  \BibitemOpen
  \bibfield  {author} {\bibinfo {author} {\bibfnamefont {L.}~\bibnamefont
  {Leclerc}}, \bibinfo {author} {\bibfnamefont {S.}~\bibnamefont
  {Juli\`a-Farr\'e}}, \bibinfo {author} {\bibfnamefont {G.~S.}\ \bibnamefont
  {Freitas}}, \bibinfo {author} {\bibfnamefont {G.}~\bibnamefont {Villaret}},
  \bibinfo {author} {\bibfnamefont {B.}~\bibnamefont {Albrecht}}, \bibinfo
  {author} {\bibfnamefont {L.}~\bibnamefont {B\'eguin}}, \bibinfo {author}
  {\bibfnamefont {L.}~\bibnamefont {Bourachot}}, \bibinfo {author}
  {\bibfnamefont {C.}~\bibnamefont {Briosne-Frejaville}}, \bibinfo {author}
  {\bibfnamefont {D.}~\bibnamefont {Claveau}}, \bibinfo {author} {\bibfnamefont
  {A.}~\bibnamefont {Cornillot}}, \bibinfo {author} {\bibfnamefont
  {J.}~\bibnamefont {de~Hond}}, \bibinfo {author} {\bibfnamefont
  {D.}~\bibnamefont {Diallo}}, \bibinfo {author} {\bibfnamefont
  {C.}~\bibnamefont {Dupays}}, \bibinfo {author} {\bibfnamefont
  {R.}~\bibnamefont {Dupont}}, \bibinfo {author} {\bibfnamefont
  {T.}~\bibnamefont {Eritzpokhoff}}, \bibinfo {author} {\bibfnamefont
  {E.}~\bibnamefont {Gottlob}}, \bibinfo {author} {\bibfnamefont
  {L.}~\bibnamefont {Henriet}}, \bibinfo {author} {\bibfnamefont
  {M.}~\bibnamefont {Kaicher}}, \bibinfo {author} {\bibfnamefont
  {L.}~\bibnamefont {Lassabli\`ere}}, \bibinfo {author} {\bibfnamefont
  {A.}~\bibnamefont {Lindberg}}, \bibinfo {author} {\bibfnamefont
  {Y.}~\bibnamefont {Machu}}, \bibinfo {author} {\bibfnamefont
  {H.}~\bibnamefont {Mamann}}, \bibinfo {author} {\bibfnamefont
  {T.}~\bibnamefont {Pansiot}}, \bibinfo {author} {\bibfnamefont
  {J.}~\bibnamefont {Ripoll}}, \bibinfo {author} {\bibfnamefont {E.~S.}\
  \bibnamefont {Choi}}, \bibinfo {author} {\bibfnamefont {A.}~\bibnamefont
  {Signoles}}, \bibinfo {author} {\bibfnamefont {J.}~\bibnamefont {Vovrosh}},
  \bibinfo {author} {\bibfnamefont {B.}~\bibnamefont {Ximenez}}, \bibinfo
  {author} {\bibfnamefont {V.}~\bibnamefont {Zapf}}, \bibinfo {author}
  {\bibfnamefont {S.}~\bibnamefont {Zhang}}, \bibinfo {author} {\bibfnamefont
  {H.}~\bibnamefont {Zhou}}, \bibinfo {author} {\bibfnamefont {M.}~\bibnamefont
  {Lee}}, \bibinfo {author} {\bibfnamefont {T.}~\bibnamefont {Mendes-Santos}},
  \bibinfo {author} {\bibfnamefont {C.}~\bibnamefont {Dalyac}}, \bibinfo
  {author} {\bibfnamefont {A.}~\bibnamefont {Browaeys}},\ and\ \bibinfo
  {author} {\bibfnamefont {A.}~\bibnamefont {Dauphin}},\ }\bibfield  {title}
  {\bibinfo {title} {One-to-one quantum simulation of a frustrated magnet with
  256 qubits},\ }\href@noop {} {\bibfield  {journal} {\bibinfo  {journal}
  {arXiv:2603.20372}\ } (\bibinfo {year} {2026})}\BibitemShut {NoStop}%
\bibitem [{\citenamefont {Ravets}\ \emph {et~al.}(2014)\citenamefont {Ravets},
  \citenamefont {Labuhn}, \citenamefont {Barredo}, \citenamefont {B{\'e}guin},
  \citenamefont {Lahaye},\ and\ \citenamefont {Browaeys}}]{Ravets2014}%
  \BibitemOpen
  \bibfield  {author} {\bibinfo {author} {\bibfnamefont {S.}~\bibnamefont
  {Ravets}}, \bibinfo {author} {\bibfnamefont {H.}~\bibnamefont {Labuhn}},
  \bibinfo {author} {\bibfnamefont {D.}~\bibnamefont {Barredo}}, \bibinfo
  {author} {\bibfnamefont {L.}~\bibnamefont {B{\'e}guin}}, \bibinfo {author}
  {\bibfnamefont {T.}~\bibnamefont {Lahaye}},\ and\ \bibinfo {author}
  {\bibfnamefont {A.}~\bibnamefont {Browaeys}},\ }\bibfield  {title} {\bibinfo
  {title} {Coherent dipole--dipole coupling between two single {Rydberg} atoms
  at an electrically-tuned {F{\"o}rster} resonance},\ }\href@noop {} {\bibfield
   {journal} {\bibinfo  {journal} {Nat. Phys.}\ }\textbf {\bibinfo {volume}
  {10}},\ \bibinfo {pages} {914} (\bibinfo {year} {2014})}\BibitemShut
  {NoStop}%
\bibitem [{\citenamefont {Barredo}\ \emph {et~al.}(2015)\citenamefont
  {Barredo}, \citenamefont {Labuhn}, \citenamefont {Ravets}, \citenamefont
  {Lahaye}, \citenamefont {Browaeys},\ and\ \citenamefont
  {Adams}}]{Barredo2015coherent}%
  \BibitemOpen
  \bibfield  {author} {\bibinfo {author} {\bibfnamefont {D.}~\bibnamefont
  {Barredo}}, \bibinfo {author} {\bibfnamefont {H.}~\bibnamefont {Labuhn}},
  \bibinfo {author} {\bibfnamefont {S.}~\bibnamefont {Ravets}}, \bibinfo
  {author} {\bibfnamefont {T.}~\bibnamefont {Lahaye}}, \bibinfo {author}
  {\bibfnamefont {A.}~\bibnamefont {Browaeys}},\ and\ \bibinfo {author}
  {\bibfnamefont {C.~S.}\ \bibnamefont {Adams}},\ }\bibfield  {title} {\bibinfo
  {title} {Coherent excitation transfer in a spin chain of three {Rydberg}
  atoms},\ }\href {https://doi.org/10.1103/PhysRevLett.114.113002} {\bibfield
  {journal} {\bibinfo  {journal} {Phys. Rev. Lett.}\ }\textbf {\bibinfo
  {volume} {114}},\ \bibinfo {pages} {113002} (\bibinfo {year}
  {2015})}\BibitemShut {NoStop}%
\bibitem [{\citenamefont {Ravets}\ \emph {et~al.}(2015)\citenamefont {Ravets},
  \citenamefont {Labuhn}, \citenamefont {Barredo}, \citenamefont {Lahaye},\
  and\ \citenamefont {Browaeys}}]{Ravets2015}%
  \BibitemOpen
  \bibfield  {author} {\bibinfo {author} {\bibfnamefont {S.}~\bibnamefont
  {Ravets}}, \bibinfo {author} {\bibfnamefont {H.}~\bibnamefont {Labuhn}},
  \bibinfo {author} {\bibfnamefont {D.}~\bibnamefont {Barredo}}, \bibinfo
  {author} {\bibfnamefont {T.}~\bibnamefont {Lahaye}},\ and\ \bibinfo {author}
  {\bibfnamefont {A.}~\bibnamefont {Browaeys}},\ }\bibfield  {title} {\bibinfo
  {title} {Measurement of the angular dependence of the dipole-dipole
  interaction between two individual {Rydberg} atoms at a {F{\"o}rster
  resonance}},\ }\href@noop {} {\bibfield  {journal} {\bibinfo  {journal}
  {Phys. Rev. A}\ }\textbf {\bibinfo {volume} {92}},\ \bibinfo {pages}
  {020701(R)} (\bibinfo {year} {2015})}\BibitemShut {NoStop}%
\bibitem [{\citenamefont {Orioli}\ \emph {et~al.}(2018)\citenamefont {Orioli},
  \citenamefont {Signoles}, \citenamefont {Wildhagen}, \citenamefont
  {G{\"u}nter}, \citenamefont {Berges}, \citenamefont {Whitlock},\ and\
  \citenamefont {Weidem{\"u}ller}}]{Orioli2018}%
  \BibitemOpen
  \bibfield  {author} {\bibinfo {author} {\bibfnamefont {A.~P.}\ \bibnamefont
  {Orioli}}, \bibinfo {author} {\bibfnamefont {A.}~\bibnamefont {Signoles}},
  \bibinfo {author} {\bibfnamefont {H.}~\bibnamefont {Wildhagen}}, \bibinfo
  {author} {\bibfnamefont {G.}~\bibnamefont {G{\"u}nter}}, \bibinfo {author}
  {\bibfnamefont {J.}~\bibnamefont {Berges}}, \bibinfo {author} {\bibfnamefont
  {S.}~\bibnamefont {Whitlock}},\ and\ \bibinfo {author} {\bibfnamefont
  {M.}~\bibnamefont {Weidem{\"u}ller}},\ }\bibfield  {title} {\bibinfo {title}
  {Relaxation of an isolated dipolar-interacting {Rydberg} quantum spin
  system},\ }\href@noop {} {\bibfield  {journal} {\bibinfo  {journal} {Phys.
  Rev. Lett.}\ }\textbf {\bibinfo {volume} {120}},\ \bibinfo {pages} {063601}
  (\bibinfo {year} {2018})}\BibitemShut {NoStop}%
\bibitem [{\citenamefont {Lippe}\ \emph {et~al.}(2021)\citenamefont {Lippe},
  \citenamefont {Klas}, \citenamefont {Bender}, \citenamefont {Mischke},
  \citenamefont {Niederpr\"um},\ and\ \citenamefont
  {Ott}}]{lippe2021experimental}%
  \BibitemOpen
  \bibfield  {author} {\bibinfo {author} {\bibfnamefont {C.}~\bibnamefont
  {Lippe}}, \bibinfo {author} {\bibfnamefont {T.}~\bibnamefont {Klas}},
  \bibinfo {author} {\bibfnamefont {J.}~\bibnamefont {Bender}}, \bibinfo
  {author} {\bibfnamefont {P.}~\bibnamefont {Mischke}}, \bibinfo {author}
  {\bibfnamefont {T.}~\bibnamefont {Niederpr\"um}},\ and\ \bibinfo {author}
  {\bibfnamefont {H.}~\bibnamefont {Ott}},\ }\bibfield  {title} {\bibinfo
  {title} {Experimental realization of a {3D} random hopping model},\
  }\href@noop {} {\bibfield  {journal} {\bibinfo  {journal} {Nat. Commun.}\
  }\textbf {\bibinfo {volume} {12}},\ \bibinfo {pages} {6976} (\bibinfo {year}
  {2021})}\BibitemShut {NoStop}%
\bibitem [{\citenamefont {Chew}\ \emph {et~al.}(2022)\citenamefont {Chew},
  \citenamefont {Tomita}, \citenamefont {Mahesh}, \citenamefont {Sugawa},
  \citenamefont {de~L{\'e}s{\'e}leuc},\ and\ \citenamefont
  {Ohmori}}]{Chew2022}%
  \BibitemOpen
  \bibfield  {author} {\bibinfo {author} {\bibfnamefont {Y.}~\bibnamefont
  {Chew}}, \bibinfo {author} {\bibfnamefont {T.}~\bibnamefont {Tomita}},
  \bibinfo {author} {\bibfnamefont {T.~P.}\ \bibnamefont {Mahesh}}, \bibinfo
  {author} {\bibfnamefont {S.}~\bibnamefont {Sugawa}}, \bibinfo {author}
  {\bibfnamefont {S.}~\bibnamefont {de~L{\'e}s{\'e}leuc}},\ and\ \bibinfo
  {author} {\bibfnamefont {K.}~\bibnamefont {Ohmori}},\ }\bibfield  {title}
  {\bibinfo {title} {Ultrafast energy exchange between two single {Rydberg}
  atoms on a nanosecond timescale},\ }\href@noop {} {\bibfield  {journal}
  {\bibinfo  {journal} {Nat. Photonics}\ }\textbf {\bibinfo {volume} {16}},\
  \bibinfo {pages} {724} (\bibinfo {year} {2022})}\BibitemShut {NoStop}%
\bibitem [{\citenamefont {Franz}\ \emph {et~al.}(2022)\citenamefont {Franz},
  \citenamefont {Geier}, \citenamefont {Braemer}, \citenamefont {Hainaut},
  \citenamefont {Signoles}, \citenamefont {Thaicharoen}, \citenamefont
  {Tebben}, \citenamefont {Salzinger}, \citenamefont {G{\"a}rttner},
  \citenamefont {Z{\"u}rn},\ and\ \citenamefont {Weidem\"uller}}]{Franz2022a}%
  \BibitemOpen
  \bibfield  {author} {\bibinfo {author} {\bibfnamefont {T.}~\bibnamefont
  {Franz}}, \bibinfo {author} {\bibfnamefont {S.}~\bibnamefont {Geier}},
  \bibinfo {author} {\bibfnamefont {A.}~\bibnamefont {Braemer}}, \bibinfo
  {author} {\bibfnamefont {C.}~\bibnamefont {Hainaut}}, \bibinfo {author}
  {\bibfnamefont {A.}~\bibnamefont {Signoles}}, \bibinfo {author}
  {\bibfnamefont {N.}~\bibnamefont {Thaicharoen}}, \bibinfo {author}
  {\bibfnamefont {A.}~\bibnamefont {Tebben}}, \bibinfo {author} {\bibfnamefont
  {A.}~\bibnamefont {Salzinger}}, \bibinfo {author} {\bibfnamefont
  {M.}~\bibnamefont {G{\"a}rttner}}, \bibinfo {author} {\bibfnamefont
  {G.}~\bibnamefont {Z{\"u}rn}},\ and\ \bibinfo {author} {\bibfnamefont
  {M.}~\bibnamefont {Weidem\"uller}},\ }\bibfield  {title} {\bibinfo {title}
  {Emergent pair localization in a many-body quantum spin system},\ }\href@noop
  {} {\bibfield  {journal} {\bibinfo  {journal} {arXiv:2207.14216}\ } (\bibinfo
  {year} {2022})}\BibitemShut {NoStop}%
\bibitem [{\citenamefont {Bornet}\ \emph {et~al.}(2023)\citenamefont {Bornet},
  \citenamefont {Emperauger}, \citenamefont {Chen}, \citenamefont {Ye},
  \citenamefont {Block}, \citenamefont {Bintz}, \citenamefont {Boyd},
  \citenamefont {Barredo}, \citenamefont {Comparin}, \citenamefont {Mezzacapo},
  \citenamefont {Roscilde}, \citenamefont {Lahaye}, \citenamefont {Yao},\ and\
  \citenamefont {Browaeys}}]{Bornet2023}%
  \BibitemOpen
  \bibfield  {author} {\bibinfo {author} {\bibfnamefont {G.}~\bibnamefont
  {Bornet}}, \bibinfo {author} {\bibfnamefont {G.}~\bibnamefont {Emperauger}},
  \bibinfo {author} {\bibfnamefont {C.}~\bibnamefont {Chen}}, \bibinfo {author}
  {\bibfnamefont {B.}~\bibnamefont {Ye}}, \bibinfo {author} {\bibfnamefont
  {M.}~\bibnamefont {Block}}, \bibinfo {author} {\bibfnamefont
  {M.}~\bibnamefont {Bintz}}, \bibinfo {author} {\bibfnamefont {J.~A.}\
  \bibnamefont {Boyd}}, \bibinfo {author} {\bibfnamefont {D.}~\bibnamefont
  {Barredo}}, \bibinfo {author} {\bibfnamefont {T.}~\bibnamefont {Comparin}},
  \bibinfo {author} {\bibfnamefont {F.}~\bibnamefont {Mezzacapo}}, \bibinfo
  {author} {\bibfnamefont {T.}~\bibnamefont {Roscilde}}, \bibinfo {author}
  {\bibfnamefont {T.}~\bibnamefont {Lahaye}}, \bibinfo {author} {\bibfnamefont
  {N.~Y.}\ \bibnamefont {Yao}},\ and\ \bibinfo {author} {\bibfnamefont
  {A.}~\bibnamefont {Browaeys}},\ }\bibfield  {title} {\bibinfo {title}
  {Scalable spin squeezing in a dipolar {Rydberg} atom array},\ }\href@noop {}
  {\bibfield  {journal} {\bibinfo  {journal} {Nature}\ }\textbf {\bibinfo
  {volume} {621}},\ \bibinfo {pages} {728} (\bibinfo {year}
  {2023})}\BibitemShut {NoStop}%
\bibitem [{\citenamefont {Bornet}\ \emph {et~al.}(2024)\citenamefont {Bornet},
  \citenamefont {Emperauger}, \citenamefont {Chen}, \citenamefont {Machado},
  \citenamefont {Chern}, \citenamefont {Leclerc}, \citenamefont {G{\'e}ly},
  \citenamefont {Chew}, \citenamefont {Barredo}, \citenamefont {Lahaye},
  \citenamefont {Yao},\ and\ \citenamefont {Browaeys}}]{Bornet2024}%
  \BibitemOpen
  \bibfield  {author} {\bibinfo {author} {\bibfnamefont {G.}~\bibnamefont
  {Bornet}}, \bibinfo {author} {\bibfnamefont {G.}~\bibnamefont {Emperauger}},
  \bibinfo {author} {\bibfnamefont {C.}~\bibnamefont {Chen}}, \bibinfo {author}
  {\bibfnamefont {F.}~\bibnamefont {Machado}}, \bibinfo {author} {\bibfnamefont
  {S.}~\bibnamefont {Chern}}, \bibinfo {author} {\bibfnamefont
  {L.}~\bibnamefont {Leclerc}}, \bibinfo {author} {\bibfnamefont
  {B.}~\bibnamefont {G{\'e}ly}}, \bibinfo {author} {\bibfnamefont {Y.~T.}\
  \bibnamefont {Chew}}, \bibinfo {author} {\bibfnamefont {D.}~\bibnamefont
  {Barredo}}, \bibinfo {author} {\bibfnamefont {T.}~\bibnamefont {Lahaye}},
  \bibinfo {author} {\bibfnamefont {N.~Y.}\ \bibnamefont {Yao}},\ and\ \bibinfo
  {author} {\bibfnamefont {A.}~\bibnamefont {Browaeys}},\ }\bibfield  {title}
  {\bibinfo {title} {Enhancing a many-body dipolar {Rydberg} tweezer array with
  arbitrary local controls},\ }\href@noop {} {\bibfield  {journal} {\bibinfo
  {journal} {Phys. Rev. Lett.}\ }\textbf {\bibinfo {volume} {132}},\ \bibinfo
  {pages} {263601} (\bibinfo {year} {2024})}\BibitemShut {NoStop}%
\bibitem [{\citenamefont {Emperauger}\ \emph
  {et~al.}(2025{\natexlab{a}})\citenamefont {Emperauger}, \citenamefont {Qiao},
  \citenamefont {Chen}, \citenamefont {Caleca}, \citenamefont {Bocini},
  \citenamefont {Bintz}, \citenamefont {Bornet}, \citenamefont {Martin},
  \citenamefont {G\'ely}, \citenamefont {Klein}, \citenamefont {Barredo},
  \citenamefont {Chatterjee}, \citenamefont {Yao}, \citenamefont {Mezzacapo},
  \citenamefont {Lahaye}, \citenamefont {Roscilde},\ and\ \citenamefont
  {Browaeys}}]{Emperauger2025}%
  \BibitemOpen
  \bibfield  {author} {\bibinfo {author} {\bibfnamefont {G.}~\bibnamefont
  {Emperauger}}, \bibinfo {author} {\bibfnamefont {M.}~\bibnamefont {Qiao}},
  \bibinfo {author} {\bibfnamefont {C.}~\bibnamefont {Chen}}, \bibinfo {author}
  {\bibfnamefont {F.}~\bibnamefont {Caleca}}, \bibinfo {author} {\bibfnamefont
  {S.}~\bibnamefont {Bocini}}, \bibinfo {author} {\bibfnamefont
  {M.}~\bibnamefont {Bintz}}, \bibinfo {author} {\bibfnamefont
  {G.}~\bibnamefont {Bornet}}, \bibinfo {author} {\bibfnamefont
  {R.}~\bibnamefont {Martin}}, \bibinfo {author} {\bibfnamefont
  {B.}~\bibnamefont {G\'ely}}, \bibinfo {author} {\bibfnamefont
  {L.}~\bibnamefont {Klein}}, \bibinfo {author} {\bibfnamefont
  {D.}~\bibnamefont {Barredo}}, \bibinfo {author} {\bibfnamefont
  {S.}~\bibnamefont {Chatterjee}}, \bibinfo {author} {\bibfnamefont {N.~Y.}\
  \bibnamefont {Yao}}, \bibinfo {author} {\bibfnamefont {F.}~\bibnamefont
  {Mezzacapo}}, \bibinfo {author} {\bibfnamefont {T.}~\bibnamefont {Lahaye}},
  \bibinfo {author} {\bibfnamefont {T.}~\bibnamefont {Roscilde}},\ and\
  \bibinfo {author} {\bibfnamefont {A.}~\bibnamefont {Browaeys}},\ }\bibfield
  {title} {\bibinfo {title} {{Tomonaga-Luttinger} liquid behavior in a
  {Rydberg-Encoded} spin chain},\ }\href {https://doi.org/10.1103/qfnp-6dpz}
  {\bibfield  {journal} {\bibinfo  {journal} {Phys. Rev. X}\ }\textbf {\bibinfo
  {volume} {15}},\ \bibinfo {pages} {031021} (\bibinfo {year}
  {2025}{\natexlab{a}})}\BibitemShut {NoStop}%
\bibitem [{\citenamefont {Emperauger}\ \emph
  {et~al.}(2025{\natexlab{b}})\citenamefont {Emperauger}, \citenamefont {Qiao},
  \citenamefont {Bornet}, \citenamefont {Chen}, \citenamefont {Martin},
  \citenamefont {Chew}, \citenamefont {G\'ely}, \citenamefont {Klein},
  \citenamefont {Barredo}, \citenamefont {Browaeys},\ and\ \citenamefont
  {Lahaye}}]{Emperauger2025benchmarking}%
  \BibitemOpen
  \bibfield  {author} {\bibinfo {author} {\bibfnamefont {G.}~\bibnamefont
  {Emperauger}}, \bibinfo {author} {\bibfnamefont {M.}~\bibnamefont {Qiao}},
  \bibinfo {author} {\bibfnamefont {G.}~\bibnamefont {Bornet}}, \bibinfo
  {author} {\bibfnamefont {C.}~\bibnamefont {Chen}}, \bibinfo {author}
  {\bibfnamefont {R.}~\bibnamefont {Martin}}, \bibinfo {author} {\bibfnamefont
  {Y.~T.}\ \bibnamefont {Chew}}, \bibinfo {author} {\bibfnamefont
  {B.}~\bibnamefont {G\'ely}}, \bibinfo {author} {\bibfnamefont
  {L.}~\bibnamefont {Klein}}, \bibinfo {author} {\bibfnamefont
  {D.}~\bibnamefont {Barredo}}, \bibinfo {author} {\bibfnamefont
  {A.}~\bibnamefont {Browaeys}},\ and\ \bibinfo {author} {\bibfnamefont
  {T.}~\bibnamefont {Lahaye}},\ }\bibfield  {title} {\bibinfo {title}
  {Benchmarking direct and indirect dipolar spin-exchange interactions between
  two {Rydberg} atoms},\ }\href {https://doi.org/10.1103/PhysRevA.111.062806}
  {\bibfield  {journal} {\bibinfo  {journal} {Phys. Rev. A}\ }\textbf {\bibinfo
  {volume} {111}},\ \bibinfo {pages} {062806} (\bibinfo {year}
  {2025}{\natexlab{b}})}\BibitemShut {NoStop}%
\bibitem [{\citenamefont {Chen}\ \emph {et~al.}(2025)\citenamefont {Chen},
  \citenamefont {Emperauger}, \citenamefont {Bornet}, \citenamefont {Caleca},
  \citenamefont {G{\'e}ly}, \citenamefont {Bintz}, \citenamefont {Chatterjee},
  \citenamefont {Liu}, \citenamefont {Barredo}, \citenamefont {Yao},
  \citenamefont {Lahaye}, \citenamefont {Mezzacapo}, \citenamefont {Roscilde},\
  and\ \citenamefont {Browaeys}}]{chen2025spectroscopy}%
  \BibitemOpen
  \bibfield  {author} {\bibinfo {author} {\bibfnamefont {C.}~\bibnamefont
  {Chen}}, \bibinfo {author} {\bibfnamefont {G.}~\bibnamefont {Emperauger}},
  \bibinfo {author} {\bibfnamefont {G.}~\bibnamefont {Bornet}}, \bibinfo
  {author} {\bibfnamefont {F.}~\bibnamefont {Caleca}}, \bibinfo {author}
  {\bibfnamefont {B.}~\bibnamefont {G{\'e}ly}}, \bibinfo {author}
  {\bibfnamefont {M.}~\bibnamefont {Bintz}}, \bibinfo {author} {\bibfnamefont
  {S.}~\bibnamefont {Chatterjee}}, \bibinfo {author} {\bibfnamefont
  {V.}~\bibnamefont {Liu}}, \bibinfo {author} {\bibfnamefont {D.}~\bibnamefont
  {Barredo}}, \bibinfo {author} {\bibfnamefont {N.~Y.}\ \bibnamefont {Yao}},
  \bibinfo {author} {\bibfnamefont {T.}~\bibnamefont {Lahaye}}, \bibinfo
  {author} {\bibfnamefont {F.}~\bibnamefont {Mezzacapo}}, \bibinfo {author}
  {\bibfnamefont {T.}~\bibnamefont {Roscilde}},\ and\ \bibinfo {author}
  {\bibfnamefont {A.}~\bibnamefont {Browaeys}},\ }\bibfield  {title} {\bibinfo
  {title} {Spectroscopy of elementary excitations from quench dynamics in a
  dipolar {XY} {Rydberg} simulator},\ }\href
  {https://doi.org/10.1126/science.adn0618} {\bibfield  {journal} {\bibinfo
  {journal} {Science}\ }\textbf {\bibinfo {volume} {389}},\ \bibinfo {pages}
  {483} (\bibinfo {year} {2025})}\BibitemShut {NoStop}%
\bibitem [{\citenamefont {Emperauger}\ \emph
  {et~al.}(2025{\natexlab{c}})\citenamefont {Emperauger}, \citenamefont {Qiao},
  \citenamefont {Bornet}, \citenamefont {Chew}, \citenamefont {Martin},
  \citenamefont {G\'ely}, \citenamefont {Klein}, \citenamefont {Barredo},
  \citenamefont {Lahaye},\ and\ \citenamefont
  {Browaeys}}]{emperauger2025probing}%
  \BibitemOpen
  \bibfield  {author} {\bibinfo {author} {\bibfnamefont {G.}~\bibnamefont
  {Emperauger}}, \bibinfo {author} {\bibfnamefont {M.}~\bibnamefont {Qiao}},
  \bibinfo {author} {\bibfnamefont {G.}~\bibnamefont {Bornet}}, \bibinfo
  {author} {\bibfnamefont {Y.~T.}\ \bibnamefont {Chew}}, \bibinfo {author}
  {\bibfnamefont {R.}~\bibnamefont {Martin}}, \bibinfo {author} {\bibfnamefont
  {B.}~\bibnamefont {G\'ely}}, \bibinfo {author} {\bibfnamefont
  {L.}~\bibnamefont {Klein}}, \bibinfo {author} {\bibfnamefont
  {D.}~\bibnamefont {Barredo}}, \bibinfo {author} {\bibfnamefont
  {T.}~\bibnamefont {Lahaye}},\ and\ \bibinfo {author} {\bibfnamefont
  {A.}~\bibnamefont {Browaeys}},\ }\bibfield  {title} {\bibinfo {title}
  {Probing spin-motion coupling of two {Rydberg} atoms by a
  {Stern-Gerlach-like} experiment},\ }\href {https://doi.org/10.1103/jd4l-qy2j}
  {\bibfield  {journal} {\bibinfo  {journal} {Phys. Rev. A}\ }\textbf {\bibinfo
  {volume} {112}},\ \bibinfo {pages} {053717} (\bibinfo {year}
  {2025}{\natexlab{c}})}\BibitemShut {NoStop}%
\bibitem [{\citenamefont {Hornung}\ \emph {et~al.}(2025)\citenamefont
  {Hornung}, \citenamefont {Braun}, \citenamefont {Geier}, \citenamefont
  {Franz}, \citenamefont {Z{\"u}rn},\ and\ \citenamefont
  {Weidem{\"u}ller}}]{hornung2025observation}%
  \BibitemOpen
  \bibfield  {author} {\bibinfo {author} {\bibfnamefont {M.}~\bibnamefont
  {Hornung}}, \bibinfo {author} {\bibfnamefont {E.~J.}\ \bibnamefont {Braun}},
  \bibinfo {author} {\bibfnamefont {S.}~\bibnamefont {Geier}}, \bibinfo
  {author} {\bibfnamefont {T.}~\bibnamefont {Franz}}, \bibinfo {author}
  {\bibfnamefont {G.}~\bibnamefont {Z{\"u}rn}},\ and\ \bibinfo {author}
  {\bibfnamefont {M.}~\bibnamefont {Weidem{\"u}ller}},\ }\bibfield  {title}
  {\bibinfo {title} {Observation of hysteresis in an isolated quantum system of
  disordered {Heisenberg} spins},\ }\href@noop {} {\bibfield  {journal}
  {\bibinfo  {journal} {arXiv:2508.18197}\ } (\bibinfo {year}
  {2025})}\BibitemShut {NoStop}%
\bibitem [{\citenamefont {Chen}\ \emph {et~al.}(2026)\citenamefont {Chen},
  \citenamefont {Capizzi}, \citenamefont {March{\'e}}, \citenamefont {Bornet},
  \citenamefont {Emperauger}, \citenamefont {Lahaye}, \citenamefont {Browaeys},
  \citenamefont {Fagotti},\ and\ \citenamefont {Mazza}}]{chen2026observing}%
  \BibitemOpen
  \bibfield  {author} {\bibinfo {author} {\bibfnamefont {C.}~\bibnamefont
  {Chen}}, \bibinfo {author} {\bibfnamefont {L.}~\bibnamefont {Capizzi}},
  \bibinfo {author} {\bibfnamefont {A.}~\bibnamefont {March{\'e}}}, \bibinfo
  {author} {\bibfnamefont {G.}~\bibnamefont {Bornet}}, \bibinfo {author}
  {\bibfnamefont {G.}~\bibnamefont {Emperauger}}, \bibinfo {author}
  {\bibfnamefont {T.}~\bibnamefont {Lahaye}}, \bibinfo {author} {\bibfnamefont
  {A.}~\bibnamefont {Browaeys}}, \bibinfo {author} {\bibfnamefont
  {M.}~\bibnamefont {Fagotti}},\ and\ \bibinfo {author} {\bibfnamefont
  {L.}~\bibnamefont {Mazza}},\ }\bibfield  {title} {\bibinfo {title} {Observing
  weakly broken conservation laws in a dipolar {Rydberg} quantum spin chain},\
  }\href@noop {} {\bibfield  {journal} {\bibinfo  {journal} {arXiv:2602.02251}\
  } (\bibinfo {year} {2026})}\BibitemShut {NoStop}%
\bibitem [{\citenamefont {Bornet}\ \emph {et~al.}(2026)\citenamefont {Bornet},
  \citenamefont {Bintz}, \citenamefont {Chen}, \citenamefont {Emperauger},
  \citenamefont {Qiao}, \citenamefont {Martin}, \citenamefont {Barredo},
  \citenamefont {Chatterjee}, \citenamefont {Liu}, \citenamefont {Lahaye},
  \citenamefont {Zaletel}, \citenamefont {Yao},\ and\ \citenamefont
  {Browaeys}}]{bornet2026dirac}%
  \BibitemOpen
  \bibfield  {author} {\bibinfo {author} {\bibfnamefont {G.}~\bibnamefont
  {Bornet}}, \bibinfo {author} {\bibfnamefont {M.}~\bibnamefont {Bintz}},
  \bibinfo {author} {\bibfnamefont {C.}~\bibnamefont {Chen}}, \bibinfo {author}
  {\bibfnamefont {G.}~\bibnamefont {Emperauger}}, \bibinfo {author}
  {\bibfnamefont {M.}~\bibnamefont {Qiao}}, \bibinfo {author} {\bibfnamefont
  {R.}~\bibnamefont {Martin}}, \bibinfo {author} {\bibfnamefont
  {D.}~\bibnamefont {Barredo}}, \bibinfo {author} {\bibfnamefont
  {S.}~\bibnamefont {Chatterjee}}, \bibinfo {author} {\bibfnamefont {V.~S.}\
  \bibnamefont {Liu}}, \bibinfo {author} {\bibfnamefont {T.}~\bibnamefont
  {Lahaye}}, \bibinfo {author} {\bibfnamefont {M.~P.}\ \bibnamefont {Zaletel}},
  \bibinfo {author} {\bibfnamefont {N.~Y.}\ \bibnamefont {Yao}},\ and\ \bibinfo
  {author} {\bibfnamefont {A.}~\bibnamefont {Browaeys}},\ }\bibfield  {title}
  {\bibinfo {title} {Dirac spin liquid candidate in a {Rydberg} quantum
  simulator},\ }\href@noop {} {\bibfield  {journal} {\bibinfo  {journal}
  {arXiv:2602.14323}\ } (\bibinfo {year} {2026})}\BibitemShut {NoStop}%
\bibitem [{\citenamefont {Signoles}\ \emph {et~al.}(2021)\citenamefont
  {Signoles}, \citenamefont {Franz}, \citenamefont {Ferracini~Alves},
  \citenamefont {G{\"a}rttner}, \citenamefont {Whitlock}, \citenamefont
  {Z{\"u}rn},\ and\ \citenamefont {Weidem{\"u}ller}}]{Signoles2021}%
  \BibitemOpen
  \bibfield  {author} {\bibinfo {author} {\bibfnamefont {A.}~\bibnamefont
  {Signoles}}, \bibinfo {author} {\bibfnamefont {T.}~\bibnamefont {Franz}},
  \bibinfo {author} {\bibfnamefont {R.}~\bibnamefont {Ferracini~Alves}},
  \bibinfo {author} {\bibfnamefont {M.}~\bibnamefont {G{\"a}rttner}}, \bibinfo
  {author} {\bibfnamefont {S.}~\bibnamefont {Whitlock}}, \bibinfo {author}
  {\bibfnamefont {G.}~\bibnamefont {Z{\"u}rn}},\ and\ \bibinfo {author}
  {\bibfnamefont {M.}~\bibnamefont {Weidem{\"u}ller}},\ }\bibfield  {title}
  {\bibinfo {title} {Glassy dynamics in a disordered {Heisenberg} quantum spin
  system},\ }\href@noop {} {\bibfield  {journal} {\bibinfo  {journal} {Phys.
  Rev. X}\ }\textbf {\bibinfo {volume} {11}},\ \bibinfo {pages} {011011}
  (\bibinfo {year} {2021})}\BibitemShut {NoStop}%
\bibitem [{\citenamefont {Geier}\ \emph {et~al.}(2021)\citenamefont {Geier},
  \citenamefont {Thaicharoen}, \citenamefont {Hainaut}, \citenamefont {Franz},
  \citenamefont {Salzinger}, \citenamefont {Tebben}, \citenamefont
  {Grimshandl}, \citenamefont {Z{\"u}rn},\ and\ \citenamefont
  {Weidem{\"u}ller}}]{Geier2021}%
  \BibitemOpen
  \bibfield  {author} {\bibinfo {author} {\bibfnamefont {S.}~\bibnamefont
  {Geier}}, \bibinfo {author} {\bibfnamefont {N.}~\bibnamefont {Thaicharoen}},
  \bibinfo {author} {\bibfnamefont {C.}~\bibnamefont {Hainaut}}, \bibinfo
  {author} {\bibfnamefont {T.}~\bibnamefont {Franz}}, \bibinfo {author}
  {\bibfnamefont {A.}~\bibnamefont {Salzinger}}, \bibinfo {author}
  {\bibfnamefont {A.}~\bibnamefont {Tebben}}, \bibinfo {author} {\bibfnamefont
  {D.}~\bibnamefont {Grimshandl}}, \bibinfo {author} {\bibfnamefont
  {G.}~\bibnamefont {Z{\"u}rn}},\ and\ \bibinfo {author} {\bibfnamefont
  {M.}~\bibnamefont {Weidem{\"u}ller}},\ }\bibfield  {title} {\bibinfo {title}
  {Floquet {Hamiltonian} engineering of an isolated many-body spin system},\
  }\href@noop {} {\bibfield  {journal} {\bibinfo  {journal} {Science}\ }\textbf
  {\bibinfo {volume} {374}},\ \bibinfo {pages} {1149} (\bibinfo {year}
  {2021})}\BibitemShut {NoStop}%
\bibitem [{\citenamefont {Scholl}\ \emph {et~al.}(2022)\citenamefont {Scholl},
  \citenamefont {Williams}, \citenamefont {Bornet}, \citenamefont {Wallner},
  \citenamefont {Barredo}, \citenamefont {Henriet}, \citenamefont {Signoles},
  \citenamefont {Hainaut}, \citenamefont {Franz}, \citenamefont {Geier},
  \citenamefont {Tebben}, \citenamefont {Salzinger}, \citenamefont {Z\"{u}rn},
  \citenamefont {Lahaye}, \citenamefont {Weidem\"{u}ller},\ and\ \citenamefont
  {Browaeys}}]{Scholl2022}%
  \BibitemOpen
  \bibfield  {author} {\bibinfo {author} {\bibfnamefont {P.}~\bibnamefont
  {Scholl}}, \bibinfo {author} {\bibfnamefont {H.~J.}\ \bibnamefont
  {Williams}}, \bibinfo {author} {\bibfnamefont {G.}~\bibnamefont {Bornet}},
  \bibinfo {author} {\bibfnamefont {F.}~\bibnamefont {Wallner}}, \bibinfo
  {author} {\bibfnamefont {D.}~\bibnamefont {Barredo}}, \bibinfo {author}
  {\bibfnamefont {L.}~\bibnamefont {Henriet}}, \bibinfo {author} {\bibfnamefont
  {A.}~\bibnamefont {Signoles}}, \bibinfo {author} {\bibfnamefont
  {C.}~\bibnamefont {Hainaut}}, \bibinfo {author} {\bibfnamefont
  {T.}~\bibnamefont {Franz}}, \bibinfo {author} {\bibfnamefont
  {S.}~\bibnamefont {Geier}}, \bibinfo {author} {\bibfnamefont
  {A.}~\bibnamefont {Tebben}}, \bibinfo {author} {\bibfnamefont
  {A.}~\bibnamefont {Salzinger}}, \bibinfo {author} {\bibfnamefont
  {G.}~\bibnamefont {Z\"{u}rn}}, \bibinfo {author} {\bibfnamefont
  {T.}~\bibnamefont {Lahaye}}, \bibinfo {author} {\bibfnamefont
  {M.}~\bibnamefont {Weidem\"{u}ller}},\ and\ \bibinfo {author} {\bibfnamefont
  {A.}~\bibnamefont {Browaeys}},\ }\bibfield  {title} {\bibinfo {title}
  {Microwave engineering of programmable {XXZ} {Hamiltonians} in arrays of
  {Rydberg} atoms},\ }\href@noop {} {\bibfield  {journal} {\bibinfo  {journal}
  {PRX Quantum}\ }\textbf {\bibinfo {volume} {3}},\ \bibinfo {pages} {020303}
  (\bibinfo {year} {2022})}\BibitemShut {NoStop}%
\bibitem [{\citenamefont {Steinert}\ \emph {et~al.}(2023)\citenamefont
  {Steinert}, \citenamefont {Osterholz}, \citenamefont {Eberhard},
  \citenamefont {Festa}, \citenamefont {Lorenz}, \citenamefont {Chen},
  \citenamefont {Trautmann},\ and\ \citenamefont {Gross}}]{Steinert2023}%
  \BibitemOpen
  \bibfield  {author} {\bibinfo {author} {\bibfnamefont {L.-M.}\ \bibnamefont
  {Steinert}}, \bibinfo {author} {\bibfnamefont {P.}~\bibnamefont {Osterholz}},
  \bibinfo {author} {\bibfnamefont {R.}~\bibnamefont {Eberhard}}, \bibinfo
  {author} {\bibfnamefont {L.}~\bibnamefont {Festa}}, \bibinfo {author}
  {\bibfnamefont {N.}~\bibnamefont {Lorenz}}, \bibinfo {author} {\bibfnamefont
  {Z.}~\bibnamefont {Chen}}, \bibinfo {author} {\bibfnamefont {A.}~\bibnamefont
  {Trautmann}},\ and\ \bibinfo {author} {\bibfnamefont {C.}~\bibnamefont
  {Gross}},\ }\bibfield  {title} {\bibinfo {title} {Spatially tunable spin
  interactions in neutral atom arrays},\ }\href@noop {} {\bibfield  {journal}
  {\bibinfo  {journal} {Phys. Rev. Lett.}\ }\textbf {\bibinfo {volume} {130}},\
  \bibinfo {pages} {243001} (\bibinfo {year} {2023})}\BibitemShut {NoStop}%
\bibitem [{\citenamefont {M{\"o}gerle}\ \emph {et~al.}(2025)\citenamefont
  {M{\"o}gerle}, \citenamefont {Brechtelsbauer}, \citenamefont {Gea-Caballero},
  \citenamefont {Prior}, \citenamefont {Emperauger}, \citenamefont {Bornet},
  \citenamefont {Chen}, \citenamefont {Lahaye}, \citenamefont {Browaeys},\ and\
  \citenamefont {B{\"u}chler}}]{mogerle2025spin}%
  \BibitemOpen
  \bibfield  {author} {\bibinfo {author} {\bibfnamefont {J.}~\bibnamefont
  {M{\"o}gerle}}, \bibinfo {author} {\bibfnamefont {K.}~\bibnamefont
  {Brechtelsbauer}}, \bibinfo {author} {\bibfnamefont {A.~T.}\ \bibnamefont
  {Gea-Caballero}}, \bibinfo {author} {\bibfnamefont {J.}~\bibnamefont
  {Prior}}, \bibinfo {author} {\bibfnamefont {G.}~\bibnamefont {Emperauger}},
  \bibinfo {author} {\bibfnamefont {G.}~\bibnamefont {Bornet}}, \bibinfo
  {author} {\bibfnamefont {C.}~\bibnamefont {Chen}}, \bibinfo {author}
  {\bibfnamefont {T.}~\bibnamefont {Lahaye}}, \bibinfo {author} {\bibfnamefont
  {A.}~\bibnamefont {Browaeys}},\ and\ \bibinfo {author} {\bibfnamefont
  {H.~P.}\ \bibnamefont {B{\"u}chler}},\ }\bibfield  {title} {\bibinfo {title}
  {Spin-1 {Haldane} phase in a chain of {Rydberg} atoms},\ }\href@noop {}
  {\bibfield  {journal} {\bibinfo  {journal} {PRX Quantum}\ }\textbf {\bibinfo
  {volume} {6}},\ \bibinfo {pages} {020332} (\bibinfo {year}
  {2025})}\BibitemShut {NoStop}%
\bibitem [{\citenamefont {Qiao}\ \emph
  {et~al.}(2025{\natexlab{a}})\citenamefont {Qiao}, \citenamefont {Emperauger},
  \citenamefont {Chen}, \citenamefont {Homeier}, \citenamefont {Hollerith},
  \citenamefont {Bornet}, \citenamefont {Martin}, \citenamefont {G{\'e}ly},
  \citenamefont {Klein}, \citenamefont {Barredo}, \citenamefont {Geier},
  \citenamefont {Chiu}, \citenamefont {Grusdt}, \citenamefont {Bohrdt},
  \citenamefont {Lahaye},\ and\ \citenamefont {Browaeys}}]{Qiao2025}%
  \BibitemOpen
  \bibfield  {author} {\bibinfo {author} {\bibfnamefont {M.}~\bibnamefont
  {Qiao}}, \bibinfo {author} {\bibfnamefont {G.}~\bibnamefont {Emperauger}},
  \bibinfo {author} {\bibfnamefont {C.}~\bibnamefont {Chen}}, \bibinfo {author}
  {\bibfnamefont {L.}~\bibnamefont {Homeier}}, \bibinfo {author} {\bibfnamefont
  {S.}~\bibnamefont {Hollerith}}, \bibinfo {author} {\bibfnamefont
  {G.}~\bibnamefont {Bornet}}, \bibinfo {author} {\bibfnamefont
  {R.}~\bibnamefont {Martin}}, \bibinfo {author} {\bibfnamefont
  {B.}~\bibnamefont {G{\'e}ly}}, \bibinfo {author} {\bibfnamefont
  {L.}~\bibnamefont {Klein}}, \bibinfo {author} {\bibfnamefont
  {D.}~\bibnamefont {Barredo}}, \bibinfo {author} {\bibfnamefont
  {S.}~\bibnamefont {Geier}}, \bibinfo {author} {\bibfnamefont {N.-C.}\
  \bibnamefont {Chiu}}, \bibinfo {author} {\bibfnamefont {F.}~\bibnamefont
  {Grusdt}}, \bibinfo {author} {\bibfnamefont {A.}~\bibnamefont {Bohrdt}},
  \bibinfo {author} {\bibfnamefont {T.}~\bibnamefont {Lahaye}},\ and\ \bibinfo
  {author} {\bibfnamefont {A.}~\bibnamefont {Browaeys}},\ }\bibfield  {title}
  {\bibinfo {title} {Realization of a doped quantum antiferromagnet in a
  {Rydberg} tweezer array},\ }\href@noop {} {\bibfield  {journal} {\bibinfo
  {journal} {Nature}\ }\textbf {\bibinfo {volume} {644}},\ \bibinfo {pages}
  {889} (\bibinfo {year} {2025}{\natexlab{a}})}\BibitemShut {NoStop}%
\bibitem [{\citenamefont {Qiao}\ \emph
  {et~al.}(2025{\natexlab{b}})\citenamefont {Qiao}, \citenamefont {Martin},
  \citenamefont {Homeier}, \citenamefont {Morera}, \citenamefont {G{\'e}ly},
  \citenamefont {Klein}, \citenamefont {Chew}, \citenamefont {Barredo},
  \citenamefont {Lahaye}, \citenamefont {Demler},\ and\ \citenamefont
  {Browaeys}}]{qiao2025kinetically}%
  \BibitemOpen
  \bibfield  {author} {\bibinfo {author} {\bibfnamefont {M.}~\bibnamefont
  {Qiao}}, \bibinfo {author} {\bibfnamefont {R.}~\bibnamefont {Martin}},
  \bibinfo {author} {\bibfnamefont {L.}~\bibnamefont {Homeier}}, \bibinfo
  {author} {\bibfnamefont {I.}~\bibnamefont {Morera}}, \bibinfo {author}
  {\bibfnamefont {B.}~\bibnamefont {G{\'e}ly}}, \bibinfo {author}
  {\bibfnamefont {L.}~\bibnamefont {Klein}}, \bibinfo {author} {\bibfnamefont
  {Y.~T.}\ \bibnamefont {Chew}}, \bibinfo {author} {\bibfnamefont
  {D.}~\bibnamefont {Barredo}}, \bibinfo {author} {\bibfnamefont
  {T.}~\bibnamefont {Lahaye}}, \bibinfo {author} {\bibfnamefont
  {E.}~\bibnamefont {Demler}},\ and\ \bibinfo {author} {\bibfnamefont
  {A.}~\bibnamefont {Browaeys}},\ }\bibfield  {title} {\bibinfo {title}
  {Kinetically-induced bound states in a frustrated {Rydberg} tweezer array},\
  }\href@noop {} {\bibfield  {journal} {\bibinfo  {journal} {arXiv:2510.17183}\
  } (\bibinfo {year} {2025}{\natexlab{b}})}\BibitemShut {NoStop}%
\bibitem [{\citenamefont {Martin}\ \emph {et~al.}(2026)\citenamefont {Martin},
  \citenamefont {Qiao}, \citenamefont {Morera}, \citenamefont {Homeier},
  \citenamefont {G\'ely}, \citenamefont {Klein}, \citenamefont {Chew},
  \citenamefont {Barredo}, \citenamefont {Lahaye}, \citenamefont {Demler},\
  and\ \citenamefont {Browaeys}}]{martin2026measuring}%
  \BibitemOpen
  \bibfield  {author} {\bibinfo {author} {\bibfnamefont {R.}~\bibnamefont
  {Martin}}, \bibinfo {author} {\bibfnamefont {M.}~\bibnamefont {Qiao}},
  \bibinfo {author} {\bibfnamefont {I.}~\bibnamefont {Morera}}, \bibinfo
  {author} {\bibfnamefont {L.}~\bibnamefont {Homeier}}, \bibinfo {author}
  {\bibfnamefont {B.}~\bibnamefont {G\'ely}}, \bibinfo {author} {\bibfnamefont
  {L.}~\bibnamefont {Klein}}, \bibinfo {author} {\bibfnamefont {Y.~T.}\
  \bibnamefont {Chew}}, \bibinfo {author} {\bibfnamefont {D.}~\bibnamefont
  {Barredo}}, \bibinfo {author} {\bibfnamefont {T.}~\bibnamefont {Lahaye}},
  \bibinfo {author} {\bibfnamefont {E.}~\bibnamefont {Demler}},\ and\ \bibinfo
  {author} {\bibfnamefont {A.}~\bibnamefont {Browaeys}},\ }\bibfield  {title}
  {\bibinfo {title} {Measuring spectral functions of doped magnets with
  {Rydberg} tweezer arrays},\ }\href@noop {} {\bibfield  {journal} {\bibinfo
  {journal} {arXiv:2602.17600}\ } (\bibinfo {year} {2026})}\BibitemShut
  {NoStop}%
\bibitem [{\citenamefont {Evered}\ \emph {et~al.}(2025)\citenamefont {Evered},
  \citenamefont {Kalinowski}, \citenamefont {Geim}, \citenamefont {Manovitz},
  \citenamefont {Bluvstein}, \citenamefont {Li}, \citenamefont {Maskara},
  \citenamefont {Zhou}, \citenamefont {Ebadi}, \citenamefont {Xu},
  \citenamefont {Campo}, \citenamefont {Cain}, \citenamefont {Ostermann},
  \citenamefont {Yelin}, \citenamefont {Sachdev}, \citenamefont {Greiner},
  \citenamefont {Vuleti\'{c}},\ and\ \citenamefont
  {Lukin}}]{evered2025probing}%
  \BibitemOpen
  \bibfield  {author} {\bibinfo {author} {\bibfnamefont {S.~J.}\ \bibnamefont
  {Evered}}, \bibinfo {author} {\bibfnamefont {M.}~\bibnamefont {Kalinowski}},
  \bibinfo {author} {\bibfnamefont {A.~A.}\ \bibnamefont {Geim}}, \bibinfo
  {author} {\bibfnamefont {T.}~\bibnamefont {Manovitz}}, \bibinfo {author}
  {\bibfnamefont {D.}~\bibnamefont {Bluvstein}}, \bibinfo {author}
  {\bibfnamefont {S.~H.}\ \bibnamefont {Li}}, \bibinfo {author} {\bibfnamefont
  {N.}~\bibnamefont {Maskara}}, \bibinfo {author} {\bibfnamefont
  {H.}~\bibnamefont {Zhou}}, \bibinfo {author} {\bibfnamefont {S.}~\bibnamefont
  {Ebadi}}, \bibinfo {author} {\bibfnamefont {M.}~\bibnamefont {Xu}}, \bibinfo
  {author} {\bibfnamefont {J.}~\bibnamefont {Campo}}, \bibinfo {author}
  {\bibfnamefont {M.}~\bibnamefont {Cain}}, \bibinfo {author} {\bibfnamefont
  {S.}~\bibnamefont {Ostermann}}, \bibinfo {author} {\bibfnamefont {S.~F.}\
  \bibnamefont {Yelin}}, \bibinfo {author} {\bibfnamefont {S.}~\bibnamefont
  {Sachdev}}, \bibinfo {author} {\bibfnamefont {M.}~\bibnamefont {Greiner}},
  \bibinfo {author} {\bibfnamefont {V.}~\bibnamefont {Vuleti\'{c}}},\ and\
  \bibinfo {author} {\bibfnamefont {M.~D.}\ \bibnamefont {Lukin}},\ }\bibfield
  {title} {\bibinfo {title} {Probing the {Kitaev} honeycomb model on a
  neutral-atom quantum computer},\ }\href@noop {} {\bibfield  {journal}
  {\bibinfo  {journal} {Nature}\ }\textbf {\bibinfo {volume} {645}},\ \bibinfo
  {pages} {341} (\bibinfo {year} {2025})}\BibitemShut {NoStop}%
\bibitem [{\citenamefont {Lienhard}\ \emph {et~al.}(2020)\citenamefont
  {Lienhard}, \citenamefont {Scholl}, \citenamefont {Weber}, \citenamefont
  {Barredo}, \citenamefont {de~L{\'e}s{\'e}leuc}, \citenamefont {Bai},
  \citenamefont {Lang}, \citenamefont {Fleischhauer}, \citenamefont
  {B{\"u}chler}, \citenamefont {Lahaye},\ and\ \citenamefont
  {Browaeys}}]{Lienhard2020}%
  \BibitemOpen
  \bibfield  {author} {\bibinfo {author} {\bibfnamefont {V.}~\bibnamefont
  {Lienhard}}, \bibinfo {author} {\bibfnamefont {P.}~\bibnamefont {Scholl}},
  \bibinfo {author} {\bibfnamefont {S.}~\bibnamefont {Weber}}, \bibinfo
  {author} {\bibfnamefont {D.}~\bibnamefont {Barredo}}, \bibinfo {author}
  {\bibfnamefont {S.}~\bibnamefont {de~L{\'e}s{\'e}leuc}}, \bibinfo {author}
  {\bibfnamefont {R.}~\bibnamefont {Bai}}, \bibinfo {author} {\bibfnamefont
  {N.}~\bibnamefont {Lang}}, \bibinfo {author} {\bibfnamefont {M.}~\bibnamefont
  {Fleischhauer}}, \bibinfo {author} {\bibfnamefont {H.~P.}\ \bibnamefont
  {B{\"u}chler}}, \bibinfo {author} {\bibfnamefont {T.}~\bibnamefont
  {Lahaye}},\ and\ \bibinfo {author} {\bibfnamefont {A.}~\bibnamefont
  {Browaeys}},\ }\bibfield  {title} {\bibinfo {title} {Realization of a
  density-dependent {Peierls} phase in a synthetic, spin-orbit coupled
  {Rydberg} system},\ }\href@noop {} {\bibfield  {journal} {\bibinfo  {journal}
  {Phys. Rev. X}\ }\textbf {\bibinfo {volume} {10}},\ \bibinfo {pages} {021031}
  (\bibinfo {year} {2020})}\BibitemShut {NoStop}%
\bibitem [{\citenamefont {Kanungo}\ \emph {et~al.}(2022)\citenamefont
  {Kanungo}, \citenamefont {Whalen}, \citenamefont {Lu}, \citenamefont {Yuan},
  \citenamefont {Dasgupta}, \citenamefont {Dunning}, \citenamefont {Hazzard},\
  and\ \citenamefont {Killian}}]{Kanungo2022}%
  \BibitemOpen
  \bibfield  {author} {\bibinfo {author} {\bibfnamefont {S.~K.}\ \bibnamefont
  {Kanungo}}, \bibinfo {author} {\bibfnamefont {J.~D.}\ \bibnamefont {Whalen}},
  \bibinfo {author} {\bibfnamefont {Y.}~\bibnamefont {Lu}}, \bibinfo {author}
  {\bibfnamefont {M.}~\bibnamefont {Yuan}}, \bibinfo {author} {\bibfnamefont
  {S.}~\bibnamefont {Dasgupta}}, \bibinfo {author} {\bibfnamefont {F.~B.}\
  \bibnamefont {Dunning}}, \bibinfo {author} {\bibfnamefont {K.~R.~A.}\
  \bibnamefont {Hazzard}},\ and\ \bibinfo {author} {\bibfnamefont {T.~C.}\
  \bibnamefont {Killian}},\ }\bibfield  {title} {\bibinfo {title} {Realizing
  topological edge states with {Rydberg-atom} synthetic dimensions},\
  }\href@noop {} {\bibfield  {journal} {\bibinfo  {journal} {Nat. Commun.}\
  }\textbf {\bibinfo {volume} {13}},\ \bibinfo {pages} {972} (\bibinfo {year}
  {2022})}\BibitemShut {NoStop}%
\bibitem [{\citenamefont {Wu}\ \emph {et~al.}(2022{\natexlab{a}})\citenamefont
  {Wu}, \citenamefont {Yang}, \citenamefont {Yang}, \citenamefont {M{\o}lmer},
  \citenamefont {Pohl}, \citenamefont {Tey},\ and\ \citenamefont
  {You}}]{wu2022manipulating}%
  \BibitemOpen
  \bibfield  {author} {\bibinfo {author} {\bibfnamefont {X.}~\bibnamefont
  {Wu}}, \bibinfo {author} {\bibfnamefont {F.}~\bibnamefont {Yang}}, \bibinfo
  {author} {\bibfnamefont {S.}~\bibnamefont {Yang}}, \bibinfo {author}
  {\bibfnamefont {K.}~\bibnamefont {M{\o}lmer}}, \bibinfo {author}
  {\bibfnamefont {T.}~\bibnamefont {Pohl}}, \bibinfo {author} {\bibfnamefont
  {M.~K.}\ \bibnamefont {Tey}},\ and\ \bibinfo {author} {\bibfnamefont
  {L.}~\bibnamefont {You}},\ }\bibfield  {title} {\bibinfo {title}
  {Manipulating synthetic gauge fluxes via multicolor dressing of
  {Rydberg-atom} arrays},\ }\href@noop {} {\bibfield  {journal} {\bibinfo
  {journal} {Phys. Rev. Res.}\ }\textbf {\bibinfo {volume} {4}},\ \bibinfo
  {pages} {L032046} (\bibinfo {year} {2022}{\natexlab{a}})}\BibitemShut
  {NoStop}%
\bibitem [{\citenamefont {Weber}\ \emph {et~al.}(2022)\citenamefont {Weber},
  \citenamefont {Bai}, \citenamefont {Makki}, \citenamefont {M{\"o}gerle},
  \citenamefont {Lahaye}, \citenamefont {Browaeys}, \citenamefont {Daghofer},
  \citenamefont {Lang},\ and\ \citenamefont
  {B{\"u}chler}}]{weber2022experimentally}%
  \BibitemOpen
  \bibfield  {author} {\bibinfo {author} {\bibfnamefont {S.}~\bibnamefont
  {Weber}}, \bibinfo {author} {\bibfnamefont {R.}~\bibnamefont {Bai}}, \bibinfo
  {author} {\bibfnamefont {N.}~\bibnamefont {Makki}}, \bibinfo {author}
  {\bibfnamefont {J.}~\bibnamefont {M{\"o}gerle}}, \bibinfo {author}
  {\bibfnamefont {T.}~\bibnamefont {Lahaye}}, \bibinfo {author} {\bibfnamefont
  {A.}~\bibnamefont {Browaeys}}, \bibinfo {author} {\bibfnamefont
  {M.}~\bibnamefont {Daghofer}}, \bibinfo {author} {\bibfnamefont
  {N.}~\bibnamefont {Lang}},\ and\ \bibinfo {author} {\bibfnamefont {H.~P.}\
  \bibnamefont {B{\"u}chler}},\ }\bibfield  {title} {\bibinfo {title}
  {Experimentally accessible scheme for a fractional {Chern} insulator in
  {Rydberg} atoms},\ }\href@noop {} {\bibfield  {journal} {\bibinfo  {journal}
  {PRX Quantum}\ }\textbf {\bibinfo {volume} {3}},\ \bibinfo {pages} {030302}
  (\bibinfo {year} {2022})}\BibitemShut {NoStop}%
\bibitem [{\citenamefont {Yang}\ \emph {et~al.}(2022)\citenamefont {Yang},
  \citenamefont {Wang}, \citenamefont {Zhou},\ and\ \citenamefont
  {Liu}}]{yang2022quantum_Hall}%
  \BibitemOpen
  \bibfield  {author} {\bibinfo {author} {\bibfnamefont {T.-H.}\ \bibnamefont
  {Yang}}, \bibinfo {author} {\bibfnamefont {B.-Z.}\ \bibnamefont {Wang}},
  \bibinfo {author} {\bibfnamefont {X.-C.}\ \bibnamefont {Zhou}},\ and\
  \bibinfo {author} {\bibfnamefont {X.-J.}\ \bibnamefont {Liu}},\ }\bibfield
  {title} {\bibinfo {title} {Quantum {Hall} states for {Rydberg} arrays with
  laser-assisted dipole-dipole interactions},\ }\href
  {https://doi.org/10.1103/PhysRevA.106.L021101} {\bibfield  {journal}
  {\bibinfo  {journal} {Phys. Rev. A}\ }\textbf {\bibinfo {volume} {106}},\
  \bibinfo {pages} {L021101} (\bibinfo {year} {2022})}\BibitemShut {NoStop}%
\bibitem [{\citenamefont {Zhao}\ and\ \citenamefont
  {Shi}(2023)}]{zhao2023fractional}%
  \BibitemOpen
  \bibfield  {author} {\bibinfo {author} {\bibfnamefont {Y.}~\bibnamefont
  {Zhao}}\ and\ \bibinfo {author} {\bibfnamefont {X.-F.}\ \bibnamefont {Shi}},\
  }\bibfield  {title} {\bibinfo {title} {Fractional {Chern} insulator with
  {Rydberg-dressed} neutral atoms},\ }\href@noop {} {\bibfield  {journal}
  {\bibinfo  {journal} {Phys. Rev. A}\ }\textbf {\bibinfo {volume} {108}},\
  \bibinfo {pages} {053107} (\bibinfo {year} {2023})}\BibitemShut {NoStop}%
\bibitem [{\citenamefont {Nishad}\ \emph {et~al.}(2023)\citenamefont {Nishad},
  \citenamefont {Keselman}, \citenamefont {Lahaye}, \citenamefont {Browaeys},\
  and\ \citenamefont {Tsesses}}]{nishad2023quantum}%
  \BibitemOpen
  \bibfield  {author} {\bibinfo {author} {\bibfnamefont {N.}~\bibnamefont
  {Nishad}}, \bibinfo {author} {\bibfnamefont {A.}~\bibnamefont {Keselman}},
  \bibinfo {author} {\bibfnamefont {T.}~\bibnamefont {Lahaye}}, \bibinfo
  {author} {\bibfnamefont {A.}~\bibnamefont {Browaeys}},\ and\ \bibinfo
  {author} {\bibfnamefont {S.}~\bibnamefont {Tsesses}},\ }\bibfield  {title}
  {\bibinfo {title} {Quantum simulation of generic spin-exchange models in
  {Floquet-engineered} {Rydberg-atom} arrays},\ }\href@noop {} {\bibfield
  {journal} {\bibinfo  {journal} {Phys. Rev. A}\ }\textbf {\bibinfo {volume}
  {108}},\ \bibinfo {pages} {053318} (\bibinfo {year} {2023})}\BibitemShut
  {NoStop}%
\bibitem [{\citenamefont {Kuznetsova}\ \emph {et~al.}(2023)\citenamefont
  {Kuznetsova}, \citenamefont {Mistakidis}, \citenamefont {Rittenhouse},
  \citenamefont {Yelin},\ and\ \citenamefont
  {Sadeghpour}}]{kuznetsova2023engineering}%
  \BibitemOpen
  \bibfield  {author} {\bibinfo {author} {\bibfnamefont {E.}~\bibnamefont
  {Kuznetsova}}, \bibinfo {author} {\bibfnamefont {S.~I.}\ \bibnamefont
  {Mistakidis}}, \bibinfo {author} {\bibfnamefont {S.~T.}\ \bibnamefont
  {Rittenhouse}}, \bibinfo {author} {\bibfnamefont {S.~F.}\ \bibnamefont
  {Yelin}},\ and\ \bibinfo {author} {\bibfnamefont {H.~R.}\ \bibnamefont
  {Sadeghpour}},\ }\bibfield  {title} {\bibinfo {title} {Engineering chiral
  spin interactions with {Rydberg} atoms},\ }\href@noop {} {\bibfield
  {journal} {\bibinfo  {journal} {arXiv:2309.08795}\ } (\bibinfo {year}
  {2023})}\BibitemShut {NoStop}%
\bibitem [{\citenamefont {Chen}\ \emph {et~al.}(2024)\citenamefont {Chen},
  \citenamefont {Wang}, \citenamefont {Poon}, \citenamefont {Zhou},
  \citenamefont {Liu},\ and\ \citenamefont {Liu}}]{chen2024proposal}%
  \BibitemOpen
  \bibfield  {author} {\bibinfo {author} {\bibfnamefont {Y.-H.}\ \bibnamefont
  {Chen}}, \bibinfo {author} {\bibfnamefont {B.-Z.}\ \bibnamefont {Wang}},
  \bibinfo {author} {\bibfnamefont {T.-F.~J.}\ \bibnamefont {Poon}}, \bibinfo
  {author} {\bibfnamefont {X.-C.}\ \bibnamefont {Zhou}}, \bibinfo {author}
  {\bibfnamefont {Z.-X.}\ \bibnamefont {Liu}},\ and\ \bibinfo {author}
  {\bibfnamefont {X.-J.}\ \bibnamefont {Liu}},\ }\bibfield  {title} {\bibinfo
  {title} {Proposal for realization and detection of {Kitaev} quantum spin
  liquid with {Rydberg} atoms},\ }\href@noop {} {\bibfield  {journal} {\bibinfo
   {journal} {Phys. Rev. Res.}\ }\textbf {\bibinfo {volume} {6}},\ \bibinfo
  {pages} {L042054} (\bibinfo {year} {2024})}\BibitemShut {NoStop}%
\bibitem [{\citenamefont {Valencia-Tortora}\ \emph {et~al.}(2024)\citenamefont
  {Valencia-Tortora}, \citenamefont {Pancotti}, \citenamefont {Fleischhauer},
  \citenamefont {Bernien},\ and\ \citenamefont
  {Marino}}]{Valencia-Tortora2024rydberg}%
  \BibitemOpen
  \bibfield  {author} {\bibinfo {author} {\bibfnamefont {R.~J.}\ \bibnamefont
  {Valencia-Tortora}}, \bibinfo {author} {\bibfnamefont {N.}~\bibnamefont
  {Pancotti}}, \bibinfo {author} {\bibfnamefont {M.}~\bibnamefont
  {Fleischhauer}}, \bibinfo {author} {\bibfnamefont {H.}~\bibnamefont
  {Bernien}},\ and\ \bibinfo {author} {\bibfnamefont {J.}~\bibnamefont
  {Marino}},\ }\bibfield  {title} {\bibinfo {title} {Rydberg platform for
  nonergodic chiral quantum dynamics},\ }\href
  {https://doi.org/10.1103/PhysRevLett.132.223201} {\bibfield  {journal}
  {\bibinfo  {journal} {Phys. Rev. Lett.}\ }\textbf {\bibinfo {volume} {132}},\
  \bibinfo {pages} {223201} (\bibinfo {year} {2024})}\BibitemShut {NoStop}%
\bibitem [{\citenamefont {Kunimi}\ \emph {et~al.}(2024)\citenamefont {Kunimi},
  \citenamefont {Tomita}, \citenamefont {Katsura},\ and\ \citenamefont
  {Kato}}]{Kunimi2024}%
  \BibitemOpen
  \bibfield  {author} {\bibinfo {author} {\bibfnamefont {M.}~\bibnamefont
  {Kunimi}}, \bibinfo {author} {\bibfnamefont {T.}~\bibnamefont {Tomita}},
  \bibinfo {author} {\bibfnamefont {H.}~\bibnamefont {Katsura}},\ and\ \bibinfo
  {author} {\bibfnamefont {Y.}~\bibnamefont {Kato}},\ }\bibfield  {title}
  {\bibinfo {title} {Proposal for simulating quantum spin models with the
  {Dzyaloshinskii-Moriya} interaction using {Rydberg} atoms and the
  construction of asymptotic quantum many-body scar states},\ }\href@noop {}
  {\bibfield  {journal} {\bibinfo  {journal} {Phys. Rev. A}\ }\textbf {\bibinfo
  {volume} {110}},\ \bibinfo {pages} {043312} (\bibinfo {year}
  {2024})}\BibitemShut {NoStop}%
\bibitem [{\citenamefont {Yoshida}\ \emph {et~al.}(2026)\citenamefont
  {Yoshida}, \citenamefont {Kunimi},\ and\ \citenamefont
  {Nikuni}}]{yoshida2024proposal}%
  \BibitemOpen
  \bibfield  {author} {\bibinfo {author} {\bibfnamefont {T.}~\bibnamefont
  {Yoshida}}, \bibinfo {author} {\bibfnamefont {M.}~\bibnamefont {Kunimi}},\
  and\ \bibinfo {author} {\bibfnamefont {T.}~\bibnamefont {Nikuni}},\
  }\bibfield  {title} {\bibinfo {title} {Proposal for experimental realization
  of quantum spin chains with quasiperiodic interactions using {Rydberg}
  atoms},\ }\href {https://doi.org/10.1103/gbqy-526p} {\bibfield  {journal}
  {\bibinfo  {journal} {Phys. Rev. A}\ }\textbf {\bibinfo {volume} {114}},\
  \bibinfo {pages} {023311} (\bibinfo {year} {2026})}\BibitemShut {NoStop}%
\bibitem [{\citenamefont {Kuji}\ \emph {et~al.}(2025)\citenamefont {Kuji},
  \citenamefont {Kunimi},\ and\ \citenamefont {Nikuni}}]{kuji2025proposal}%
  \BibitemOpen
  \bibfield  {author} {\bibinfo {author} {\bibfnamefont {H.}~\bibnamefont
  {Kuji}}, \bibinfo {author} {\bibfnamefont {M.}~\bibnamefont {Kunimi}},\ and\
  \bibinfo {author} {\bibfnamefont {T.}~\bibnamefont {Nikuni}},\ }\bibfield
  {title} {\bibinfo {title} {Proposal for realizing quantum-spin systems on a
  two-dimensional square lattice with {Dzyaloshinskii-Moriya} interaction by
  {Floquet} engineering using {Rydberg} atoms},\ }\href@noop {} {\bibfield
  {journal} {\bibinfo  {journal} {Phys. Rev. A}\ }\textbf {\bibinfo {volume}
  {112}},\ \bibinfo {pages} {022614} (\bibinfo {year} {2025})}\BibitemShut
  {NoStop}%
\bibitem [{\citenamefont {Tian}\ \emph {et~al.}(2025)\citenamefont {Tian},
  \citenamefont {Samajdar},\ and\ \citenamefont
  {Gadway}}]{tian2025engineering}%
  \BibitemOpen
  \bibfield  {author} {\bibinfo {author} {\bibfnamefont {M.}~\bibnamefont
  {Tian}}, \bibinfo {author} {\bibfnamefont {R.}~\bibnamefont {Samajdar}},\
  and\ \bibinfo {author} {\bibfnamefont {B.}~\bibnamefont {Gadway}},\
  }\bibfield  {title} {\bibinfo {title} {Engineering frustrated {Rydberg} spin
  models by graphical {Floquet} modulation},\ }\href@noop {} {\bibfield
  {journal} {\bibinfo  {journal} {Phys. Rev. Lett.}\ }\textbf {\bibinfo
  {volume} {135}},\ \bibinfo {pages} {253001} (\bibinfo {year}
  {2025})}\BibitemShut {NoStop}%
\bibitem [{\citenamefont {Nill}\ \emph {et~al.}(2025)\citenamefont {Nill},
  \citenamefont {de~L{\'e}s{\'e}leuc}, \citenamefont {Gro{\ss}},\ and\
  \citenamefont {Lesanovsky}}]{nill2025resonant}%
  \BibitemOpen
  \bibfield  {author} {\bibinfo {author} {\bibfnamefont {C.}~\bibnamefont
  {Nill}}, \bibinfo {author} {\bibfnamefont {S.}~\bibnamefont
  {de~L{\'e}s{\'e}leuc}}, \bibinfo {author} {\bibfnamefont {C.}~\bibnamefont
  {Gro{\ss}}},\ and\ \bibinfo {author} {\bibfnamefont {I.}~\bibnamefont
  {Lesanovsky}},\ }\bibfield  {title} {\bibinfo {title} {Resonant stroboscopic
  {Rydberg dressing}: Electron-motion coupling and multibody interactions},\
  }\href@noop {} {\bibfield  {journal} {\bibinfo  {journal} {Phys. Rev. A}\
  }\textbf {\bibinfo {volume} {111}},\ \bibinfo {pages} {L041104} (\bibinfo
  {year} {2025})}\BibitemShut {NoStop}%
\bibitem [{\citenamefont {Shah}\ \emph {et~al.}(2025)\citenamefont {Shah},
  \citenamefont {Nambiar}, \citenamefont {Gorshkov},\ and\ \citenamefont
  {Galitski}}]{shah2025quantum}%
  \BibitemOpen
  \bibfield  {author} {\bibinfo {author} {\bibfnamefont {J.}~\bibnamefont
  {Shah}}, \bibinfo {author} {\bibfnamefont {G.}~\bibnamefont {Nambiar}},
  \bibinfo {author} {\bibfnamefont {A.~V.}\ \bibnamefont {Gorshkov}},\ and\
  \bibinfo {author} {\bibfnamefont {V.}~\bibnamefont {Galitski}},\ }\bibfield
  {title} {\bibinfo {title} {Quantum spin ice in three-dimensional {Rydberg}
  atom arrays},\ }\href@noop {} {\bibfield  {journal} {\bibinfo  {journal}
  {Phys. Rev. X}\ }\textbf {\bibinfo {volume} {15}},\ \bibinfo {pages} {011025}
  (\bibinfo {year} {2025})}\BibitemShut {NoStop}%
\bibitem [{\citenamefont {Mukherjee}\ \emph {et~al.}(2026)\citenamefont
  {Mukherjee}, \citenamefont {Barghathi}, \citenamefont {Del~Maestro},\ and\
  \citenamefont {Mukherjee}}]{mukherjee2026quantum}%
  \BibitemOpen
  \bibfield  {author} {\bibinfo {author} {\bibfnamefont {K.}~\bibnamefont
  {Mukherjee}}, \bibinfo {author} {\bibfnamefont {H.}~\bibnamefont
  {Barghathi}}, \bibinfo {author} {\bibfnamefont {A.}~\bibnamefont
  {Del~Maestro}},\ and\ \bibinfo {author} {\bibfnamefont {R.}~\bibnamefont
  {Mukherjee}},\ }\bibfield  {title} {\bibinfo {title} {Quantum simulation of
  {Motzkin} spin chain with {Rydberg} atoms},\ }\href@noop {} {\bibfield
  {journal} {\bibinfo  {journal} {arXiv:2603.23422}\ } (\bibinfo {year}
  {2026})}\BibitemShut {NoStop}%
\bibitem [{\citenamefont {Samajdar}\ \emph {et~al.}(2026)\citenamefont
  {Samajdar}, \citenamefont {Lukin},\ and\ \citenamefont
  {Walther}}]{Samajdar2026ThreeBodyRydberg}%
  \BibitemOpen
  \bibfield  {author} {\bibinfo {author} {\bibfnamefont {R.}~\bibnamefont
  {Samajdar}}, \bibinfo {author} {\bibfnamefont {M.~D.}\ \bibnamefont
  {Lukin}},\ and\ \bibinfo {author} {\bibfnamefont {V.}~\bibnamefont
  {Walther}},\ }\bibfield  {title} {\bibinfo {title} {Three-body interactions
  in {Rydberg} lattices},\ }\href@noop {} {\bibfield  {journal} {\bibinfo
  {journal} {arXiv:2604.11870}\ } (\bibinfo {year} {2026})}\BibitemShut
  {NoStop}%
\bibitem [{\citenamefont {Soto-Garcia}\ and\ \citenamefont
  {Chepiga}(2026)}]{SotoGarcia2026LongLivedRevivals}%
  \BibitemOpen
  \bibfield  {author} {\bibinfo {author} {\bibfnamefont {J.}~\bibnamefont
  {Soto-Garcia}}\ and\ \bibinfo {author} {\bibfnamefont {N.}~\bibnamefont
  {Chepiga}},\ }\bibfield  {title} {\bibinfo {title} {Long-lived revivals and
  real-space fragmentation in chains of multispecies {Rydberg} atoms},\
  }\href@noop {} {\bibfield  {journal} {\bibinfo  {journal} {arXiv:2604.13257}\
  } (\bibinfo {year} {2026})}\BibitemShut {NoStop}%
\bibitem [{\citenamefont {Kunimi}\ and\ \citenamefont
  {Tomita}(2025)}]{kunimi2025proposal}%
  \BibitemOpen
  \bibfield  {author} {\bibinfo {author} {\bibfnamefont {M.}~\bibnamefont
  {Kunimi}}\ and\ \bibinfo {author} {\bibfnamefont {T.}~\bibnamefont
  {Tomita}},\ }\bibfield  {title} {\bibinfo {title} {Proposal for realizing
  {Heisenberg-type} quantum-spin models in {Rydberg-atom} quantum simulators},\
  }\href {https://doi.org/10.1103/c97b-my2w} {\bibfield  {journal} {\bibinfo
  {journal} {Phys. Rev. A}\ }\textbf {\bibinfo {volume} {112}},\ \bibinfo
  {pages} {L051301} (\bibinfo {year} {2025})}\BibitemShut {NoStop}%
\bibitem [{\citenamefont {Norcia}\ \emph {et~al.}(2019)\citenamefont {Norcia},
  \citenamefont {Young}, \citenamefont {Eckner}, \citenamefont {Oelker},
  \citenamefont {Ye},\ and\ \citenamefont {Kaufman}}]{norcia2019seconds}%
  \BibitemOpen
  \bibfield  {author} {\bibinfo {author} {\bibfnamefont {M.~A.}\ \bibnamefont
  {Norcia}}, \bibinfo {author} {\bibfnamefont {A.~W.}\ \bibnamefont {Young}},
  \bibinfo {author} {\bibfnamefont {W.~J.}\ \bibnamefont {Eckner}}, \bibinfo
  {author} {\bibfnamefont {E.}~\bibnamefont {Oelker}}, \bibinfo {author}
  {\bibfnamefont {J.}~\bibnamefont {Ye}},\ and\ \bibinfo {author}
  {\bibfnamefont {A.~M.}\ \bibnamefont {Kaufman}},\ }\bibfield  {title}
  {\bibinfo {title} {Seconds-scale coherence on an optical clock transition in
  a tweezer array},\ }\href@noop {} {\bibfield  {journal} {\bibinfo  {journal}
  {Science}\ }\textbf {\bibinfo {volume} {366}},\ \bibinfo {pages} {93}
  (\bibinfo {year} {2019})}\BibitemShut {NoStop}%
\bibitem [{\citenamefont {Madjarov}\ \emph {et~al.}(2020)\citenamefont
  {Madjarov}, \citenamefont {Covey}, \citenamefont {Shaw}, \citenamefont
  {Choi}, \citenamefont {Kale}, \citenamefont {Cooper}, \citenamefont
  {Pichler}, \citenamefont {Schkolnik}, \citenamefont {Williams},\ and\
  \citenamefont {Endres}}]{madjarov2020high}%
  \BibitemOpen
  \bibfield  {author} {\bibinfo {author} {\bibfnamefont {I.~S.}\ \bibnamefont
  {Madjarov}}, \bibinfo {author} {\bibfnamefont {J.~P.}\ \bibnamefont {Covey}},
  \bibinfo {author} {\bibfnamefont {A.~L.}\ \bibnamefont {Shaw}}, \bibinfo
  {author} {\bibfnamefont {J.}~\bibnamefont {Choi}}, \bibinfo {author}
  {\bibfnamefont {A.}~\bibnamefont {Kale}}, \bibinfo {author} {\bibfnamefont
  {A.}~\bibnamefont {Cooper}}, \bibinfo {author} {\bibfnamefont
  {H.}~\bibnamefont {Pichler}}, \bibinfo {author} {\bibfnamefont
  {V.}~\bibnamefont {Schkolnik}}, \bibinfo {author} {\bibfnamefont {J.~R.}\
  \bibnamefont {Williams}},\ and\ \bibinfo {author} {\bibfnamefont
  {M.}~\bibnamefont {Endres}},\ }\bibfield  {title} {\bibinfo {title}
  {High-fidelity entanglement and detection of alkaline-earth {Rydberg}
  atoms},\ }\href@noop {} {\bibfield  {journal} {\bibinfo  {journal} {Nat.
  Phys.}\ }\textbf {\bibinfo {volume} {16}},\ \bibinfo {pages} {857} (\bibinfo
  {year} {2020})}\BibitemShut {NoStop}%
\bibitem [{\citenamefont {Young}\ \emph {et~al.}(2020)\citenamefont {Young},
  \citenamefont {Eckner}, \citenamefont {Milner}, \citenamefont {Kedar},
  \citenamefont {Norcia}, \citenamefont {Oelker}, \citenamefont {Schine},
  \citenamefont {Ye},\ and\ \citenamefont {Kaufman}}]{young2020half}%
  \BibitemOpen
  \bibfield  {author} {\bibinfo {author} {\bibfnamefont {A.~W.}\ \bibnamefont
  {Young}}, \bibinfo {author} {\bibfnamefont {W.~J.}\ \bibnamefont {Eckner}},
  \bibinfo {author} {\bibfnamefont {W.~R.}\ \bibnamefont {Milner}}, \bibinfo
  {author} {\bibfnamefont {D.}~\bibnamefont {Kedar}}, \bibinfo {author}
  {\bibfnamefont {M.~A.}\ \bibnamefont {Norcia}}, \bibinfo {author}
  {\bibfnamefont {E.}~\bibnamefont {Oelker}}, \bibinfo {author} {\bibfnamefont
  {N.}~\bibnamefont {Schine}}, \bibinfo {author} {\bibfnamefont
  {J.}~\bibnamefont {Ye}},\ and\ \bibinfo {author} {\bibfnamefont {A.~M.}\
  \bibnamefont {Kaufman}},\ }\bibfield  {title} {\bibinfo {title}
  {Half-minute-scale atomic coherence and high relative stability in a tweezer
  clock},\ }\href@noop {} {\bibfield  {journal} {\bibinfo  {journal} {Nature}\
  }\textbf {\bibinfo {volume} {588}},\ \bibinfo {pages} {408} (\bibinfo {year}
  {2020})}\BibitemShut {NoStop}%
\bibitem [{\citenamefont {Burgers}\ \emph {et~al.}(2022)\citenamefont
  {Burgers}, \citenamefont {Ma}, \citenamefont {Saskin}, \citenamefont
  {Wilson}, \citenamefont {Alarc{\'o}n}, \citenamefont {Greene},\ and\
  \citenamefont {Thompson}}]{burgers2022controlling}%
  \BibitemOpen
  \bibfield  {author} {\bibinfo {author} {\bibfnamefont {A.~P.}\ \bibnamefont
  {Burgers}}, \bibinfo {author} {\bibfnamefont {S.}~\bibnamefont {Ma}},
  \bibinfo {author} {\bibfnamefont {S.}~\bibnamefont {Saskin}}, \bibinfo
  {author} {\bibfnamefont {J.}~\bibnamefont {Wilson}}, \bibinfo {author}
  {\bibfnamefont {M.~A.}\ \bibnamefont {Alarc{\'o}n}}, \bibinfo {author}
  {\bibfnamefont {C.~H.}\ \bibnamefont {Greene}},\ and\ \bibinfo {author}
  {\bibfnamefont {J.~D.}\ \bibnamefont {Thompson}},\ }\bibfield  {title}
  {\bibinfo {title} {Controlling {Rydberg} excitations using ion-core
  transitions in alkaline-earth atom-tweezer arrays},\ }\href@noop {}
  {\bibfield  {journal} {\bibinfo  {journal} {PRX Quantum}\ }\textbf {\bibinfo
  {volume} {3}},\ \bibinfo {pages} {020326} (\bibinfo {year}
  {2022})}\BibitemShut {NoStop}%
\bibitem [{\citenamefont {Wilson}\ \emph {et~al.}(2022)\citenamefont {Wilson},
  \citenamefont {Saskin}, \citenamefont {Meng}, \citenamefont {Ma},
  \citenamefont {Dilip}, \citenamefont {Burgers},\ and\ \citenamefont
  {Thompson}}]{wilson2022trapping}%
  \BibitemOpen
  \bibfield  {author} {\bibinfo {author} {\bibfnamefont {J.~T.}\ \bibnamefont
  {Wilson}}, \bibinfo {author} {\bibfnamefont {S.}~\bibnamefont {Saskin}},
  \bibinfo {author} {\bibfnamefont {Y.}~\bibnamefont {Meng}}, \bibinfo {author}
  {\bibfnamefont {S.}~\bibnamefont {Ma}}, \bibinfo {author} {\bibfnamefont
  {R.}~\bibnamefont {Dilip}}, \bibinfo {author} {\bibfnamefont {A.~P.}\
  \bibnamefont {Burgers}},\ and\ \bibinfo {author} {\bibfnamefont {J.~D.}\
  \bibnamefont {Thompson}},\ }\bibfield  {title} {\bibinfo {title} {Trapping
  alkaline earth {Rydberg} atoms optical tweezer arrays},\ }\href@noop {}
  {\bibfield  {journal} {\bibinfo  {journal} {Phys. Rev. Lett.}\ }\textbf
  {\bibinfo {volume} {128}},\ \bibinfo {pages} {033201} (\bibinfo {year}
  {2022})}\BibitemShut {NoStop}%
\bibitem [{\citenamefont {Wu}\ \emph {et~al.}(2022{\natexlab{b}})\citenamefont
  {Wu}, \citenamefont {Kolkowitz}, \citenamefont {Puri},\ and\ \citenamefont
  {Thompson}}]{wu2022erasure}%
  \BibitemOpen
  \bibfield  {author} {\bibinfo {author} {\bibfnamefont {Y.}~\bibnamefont
  {Wu}}, \bibinfo {author} {\bibfnamefont {S.}~\bibnamefont {Kolkowitz}},
  \bibinfo {author} {\bibfnamefont {S.}~\bibnamefont {Puri}},\ and\ \bibinfo
  {author} {\bibfnamefont {J.~D.}\ \bibnamefont {Thompson}},\ }\bibfield
  {title} {\bibinfo {title} {Erasure conversion for fault-tolerant quantum
  computing in alkaline earth {Rydberg} atom arrays},\ }\href@noop {}
  {\bibfield  {journal} {\bibinfo  {journal} {Nat. Commun.}\ }\textbf {\bibinfo
  {volume} {13}},\ \bibinfo {pages} {4657} (\bibinfo {year}
  {2022}{\natexlab{b}})}\BibitemShut {NoStop}%
\bibitem [{\citenamefont {Jenkins}\ \emph {et~al.}(2022)\citenamefont
  {Jenkins}, \citenamefont {Lis}, \citenamefont {Senoo}, \citenamefont
  {McGrew},\ and\ \citenamefont {Kaufman}}]{jenkins2022ytterbium}%
  \BibitemOpen
  \bibfield  {author} {\bibinfo {author} {\bibfnamefont {A.}~\bibnamefont
  {Jenkins}}, \bibinfo {author} {\bibfnamefont {J.~W.}\ \bibnamefont {Lis}},
  \bibinfo {author} {\bibfnamefont {A.}~\bibnamefont {Senoo}}, \bibinfo
  {author} {\bibfnamefont {W.~F.}\ \bibnamefont {McGrew}},\ and\ \bibinfo
  {author} {\bibfnamefont {A.~M.}\ \bibnamefont {Kaufman}},\ }\bibfield
  {title} {\bibinfo {title} {Ytterbium nuclear-spin qubits in an optical
  tweezer array},\ }\href@noop {} {\bibfield  {journal} {\bibinfo  {journal}
  {Phys. Rev. X}\ }\textbf {\bibinfo {volume} {12}},\ \bibinfo {pages} {021027}
  (\bibinfo {year} {2022})}\BibitemShut {NoStop}%
\bibitem [{\citenamefont {Ma}\ \emph {et~al.}(2022)\citenamefont {Ma},
  \citenamefont {Burgers}, \citenamefont {Liu}, \citenamefont {Wilson},
  \citenamefont {Zhang},\ and\ \citenamefont {Thompson}}]{ma2022universal}%
  \BibitemOpen
  \bibfield  {author} {\bibinfo {author} {\bibfnamefont {S.}~\bibnamefont
  {Ma}}, \bibinfo {author} {\bibfnamefont {A.~P.}\ \bibnamefont {Burgers}},
  \bibinfo {author} {\bibfnamefont {G.}~\bibnamefont {Liu}}, \bibinfo {author}
  {\bibfnamefont {J.}~\bibnamefont {Wilson}}, \bibinfo {author} {\bibfnamefont
  {B.}~\bibnamefont {Zhang}},\ and\ \bibinfo {author} {\bibfnamefont {J.~D.}\
  \bibnamefont {Thompson}},\ }\bibfield  {title} {\bibinfo {title} {Universal
  gate operations on nuclear spin qubits in an optical tweezer array of
  ${}^{171}${Yb} atoms},\ }\href@noop {} {\bibfield  {journal} {\bibinfo
  {journal} {Phys. Rev. X}\ }\textbf {\bibinfo {volume} {12}},\ \bibinfo
  {pages} {021028} (\bibinfo {year} {2022})}\BibitemShut {NoStop}%
\bibitem [{\citenamefont {Lis}\ \emph {et~al.}(2023)\citenamefont {Lis},
  \citenamefont {Senoo}, \citenamefont {McGrew}, \citenamefont {R{\"o}nchen},
  \citenamefont {Jenkins},\ and\ \citenamefont {Kaufman}}]{lis2023midcircuit}%
  \BibitemOpen
  \bibfield  {author} {\bibinfo {author} {\bibfnamefont {J.~W.}\ \bibnamefont
  {Lis}}, \bibinfo {author} {\bibfnamefont {A.}~\bibnamefont {Senoo}}, \bibinfo
  {author} {\bibfnamefont {W.~F.}\ \bibnamefont {McGrew}}, \bibinfo {author}
  {\bibfnamefont {F.}~\bibnamefont {R{\"o}nchen}}, \bibinfo {author}
  {\bibfnamefont {A.}~\bibnamefont {Jenkins}},\ and\ \bibinfo {author}
  {\bibfnamefont {A.~M.}\ \bibnamefont {Kaufman}},\ }\bibfield  {title}
  {\bibinfo {title} {Midcircuit operations using the {\it omg} architecture in
  neutral atom arrays},\ }\href@noop {} {\bibfield  {journal} {\bibinfo
  {journal} {Phys. Rev. X}\ }\textbf {\bibinfo {volume} {13}},\ \bibinfo
  {pages} {041035} (\bibinfo {year} {2023})}\BibitemShut {NoStop}%
\bibitem [{\citenamefont {Ma}\ \emph {et~al.}(2023)\citenamefont {Ma},
  \citenamefont {Liu}, \citenamefont {Peng}, \citenamefont {Zhang},
  \citenamefont {Jandura}, \citenamefont {Claes}, \citenamefont {Burgers},
  \citenamefont {Pupillo}, \citenamefont {Puri},\ and\ \citenamefont
  {Thompson}}]{ma2023high}%
  \BibitemOpen
  \bibfield  {author} {\bibinfo {author} {\bibfnamefont {S.}~\bibnamefont
  {Ma}}, \bibinfo {author} {\bibfnamefont {G.}~\bibnamefont {Liu}}, \bibinfo
  {author} {\bibfnamefont {P.}~\bibnamefont {Peng}}, \bibinfo {author}
  {\bibfnamefont {B.}~\bibnamefont {Zhang}}, \bibinfo {author} {\bibfnamefont
  {S.}~\bibnamefont {Jandura}}, \bibinfo {author} {\bibfnamefont
  {J.}~\bibnamefont {Claes}}, \bibinfo {author} {\bibfnamefont {A.~P.}\
  \bibnamefont {Burgers}}, \bibinfo {author} {\bibfnamefont {G.}~\bibnamefont
  {Pupillo}}, \bibinfo {author} {\bibfnamefont {S.}~\bibnamefont {Puri}},\ and\
  \bibinfo {author} {\bibfnamefont {J.~D.}\ \bibnamefont {Thompson}},\
  }\bibfield  {title} {\bibinfo {title} {High-fidelity gates and mid-circuit
  erasure conversion in an atomic qubit},\ }\href@noop {} {\bibfield  {journal}
  {\bibinfo  {journal} {Nature}\ }\textbf {\bibinfo {volume} {622}},\ \bibinfo
  {pages} {279} (\bibinfo {year} {2023})}\BibitemShut {NoStop}%
\bibitem [{\citenamefont {Cao}\ \emph {et~al.}(2024)\citenamefont {Cao},
  \citenamefont {Eckner}, \citenamefont {Lukin~Yelin}, \citenamefont {Young},
  \citenamefont {Jandura}, \citenamefont {Yan}, \citenamefont {Kim},
  \citenamefont {Pupillo}, \citenamefont {Ye}, \citenamefont {Darkwah~Oppong},\
  and\ \citenamefont {Kaufman}}]{cao2024multi}%
  \BibitemOpen
  \bibfield  {author} {\bibinfo {author} {\bibfnamefont {A.}~\bibnamefont
  {Cao}}, \bibinfo {author} {\bibfnamefont {W.~J.}\ \bibnamefont {Eckner}},
  \bibinfo {author} {\bibfnamefont {T.}~\bibnamefont {Lukin~Yelin}}, \bibinfo
  {author} {\bibfnamefont {A.~W.}\ \bibnamefont {Young}}, \bibinfo {author}
  {\bibfnamefont {S.}~\bibnamefont {Jandura}}, \bibinfo {author} {\bibfnamefont
  {L.}~\bibnamefont {Yan}}, \bibinfo {author} {\bibfnamefont {K.}~\bibnamefont
  {Kim}}, \bibinfo {author} {\bibfnamefont {G.}~\bibnamefont {Pupillo}},
  \bibinfo {author} {\bibfnamefont {J.}~\bibnamefont {Ye}}, \bibinfo {author}
  {\bibfnamefont {N.}~\bibnamefont {Darkwah~Oppong}},\ and\ \bibinfo {author}
  {\bibfnamefont {A.~M.}\ \bibnamefont {Kaufman}},\ }\bibfield  {title}
  {\bibinfo {title} {Multi-qubit gates and {Schr{\"o}dinger} cat states in an
  optical clock},\ }\href@noop {} {\bibfield  {journal} {\bibinfo  {journal}
  {Nature}\ }\textbf {\bibinfo {volume} {634}},\ \bibinfo {pages} {315}
  (\bibinfo {year} {2024})}\BibitemShut {NoStop}%
\bibitem [{\citenamefont {Shaw}\ \emph {et~al.}(2024)\citenamefont {Shaw},
  \citenamefont {Chen}, \citenamefont {Choi}, \citenamefont {Mark},
  \citenamefont {Scholl}, \citenamefont {Finkelstein}, \citenamefont {Elben},
  \citenamefont {Choi},\ and\ \citenamefont {Endres}}]{shaw2024benchmarking}%
  \BibitemOpen
  \bibfield  {author} {\bibinfo {author} {\bibfnamefont {A.~L.}\ \bibnamefont
  {Shaw}}, \bibinfo {author} {\bibfnamefont {Z.}~\bibnamefont {Chen}}, \bibinfo
  {author} {\bibfnamefont {J.}~\bibnamefont {Choi}}, \bibinfo {author}
  {\bibfnamefont {D.~K.}\ \bibnamefont {Mark}}, \bibinfo {author}
  {\bibfnamefont {P.}~\bibnamefont {Scholl}}, \bibinfo {author} {\bibfnamefont
  {R.}~\bibnamefont {Finkelstein}}, \bibinfo {author} {\bibfnamefont
  {A.}~\bibnamefont {Elben}}, \bibinfo {author} {\bibfnamefont
  {S.}~\bibnamefont {Choi}},\ and\ \bibinfo {author} {\bibfnamefont
  {M.}~\bibnamefont {Endres}},\ }\bibfield  {title} {\bibinfo {title}
  {Benchmarking highly entangled states on a 60-atom analogue quantum
  simulator},\ }\href@noop {} {\bibfield  {journal} {\bibinfo  {journal}
  {Nature}\ }\textbf {\bibinfo {volume} {628}},\ \bibinfo {pages} {71}
  (\bibinfo {year} {2024})}\BibitemShut {NoStop}%
\bibitem [{\citenamefont {Finkelstein}\ \emph {et~al.}(2024)\citenamefont
  {Finkelstein}, \citenamefont {Tsai}, \citenamefont {Sun}, \citenamefont
  {Scholl}, \citenamefont {Direkci}, \citenamefont {Gefen}, \citenamefont
  {Choi}, \citenamefont {Shaw},\ and\ \citenamefont
  {Endres}}]{finkelstein2024universal}%
  \BibitemOpen
  \bibfield  {author} {\bibinfo {author} {\bibfnamefont {R.}~\bibnamefont
  {Finkelstein}}, \bibinfo {author} {\bibfnamefont {R.~B.-S.}\ \bibnamefont
  {Tsai}}, \bibinfo {author} {\bibfnamefont {X.}~\bibnamefont {Sun}}, \bibinfo
  {author} {\bibfnamefont {P.}~\bibnamefont {Scholl}}, \bibinfo {author}
  {\bibfnamefont {S.}~\bibnamefont {Direkci}}, \bibinfo {author} {\bibfnamefont
  {T.}~\bibnamefont {Gefen}}, \bibinfo {author} {\bibfnamefont
  {J.}~\bibnamefont {Choi}}, \bibinfo {author} {\bibfnamefont {A.~L.}\
  \bibnamefont {Shaw}},\ and\ \bibinfo {author} {\bibfnamefont
  {M.}~\bibnamefont {Endres}},\ }\bibfield  {title} {\bibinfo {title}
  {Universal quantum operations and ancilla-based read-out for tweezer
  clocks},\ }\href@noop {} {\bibfield  {journal} {\bibinfo  {journal} {Nature}\
  }\textbf {\bibinfo {volume} {634}},\ \bibinfo {pages} {321} (\bibinfo {year}
  {2024})}\BibitemShut {NoStop}%
\bibitem [{\citenamefont {Muniz}\ \emph {et~al.}(2025)\citenamefont {Muniz},
  \citenamefont {Stone}, \citenamefont {Stack}, \citenamefont {Jaffe},
  \citenamefont {Kindem}, \citenamefont {Wadleigh}, \citenamefont
  {Zalys-Geller}, \citenamefont {Zhang}, \citenamefont {Chen}, \citenamefont
  {Norcia}, \citenamefont {Epstein}, \citenamefont {Halperin}, \citenamefont
  {Hummel}, \citenamefont {Wilkason}, \citenamefont {Li}, \citenamefont
  {Barnes}, \citenamefont {Battaglino}, \citenamefont {Bohdanowicz},
  \citenamefont {Booth}, \citenamefont {Brown}, \citenamefont {Brown},
  \citenamefont {Cairncross}, \citenamefont {Cassella}, \citenamefont {Coxe},
  \citenamefont {Crow}, \citenamefont {Feldkamp}, \citenamefont {Griger},
  \citenamefont {Heinz}, \citenamefont {Jones}, \citenamefont {Kim},
  \citenamefont {King}, \citenamefont {Kotru}, \citenamefont {Lauigan},
  \citenamefont {Marjanovic}, \citenamefont {Megidish}, \citenamefont
  {Meredith}, \citenamefont {McDonald}, \citenamefont {Morshead}, \citenamefont
  {Narayanaswami}, \citenamefont {Nishiguchi}, \citenamefont {Paule},
  \citenamefont {Pawlak}, \citenamefont {Pudenz}, \citenamefont {P\'erez},
  \citenamefont {Ryou}, \citenamefont {Simon}, \citenamefont {Smull},
  \citenamefont {Urbanek}, \citenamefont {van~de Veerdonk}, \citenamefont
  {Vendeiro}, \citenamefont {Wu}, \citenamefont {Xie},\ and\ \citenamefont
  {Bloom}}]{Muniz2025high}%
  \BibitemOpen
  \bibfield  {author} {\bibinfo {author} {\bibfnamefont {J.~A.}\ \bibnamefont
  {Muniz}}, \bibinfo {author} {\bibfnamefont {M.}~\bibnamefont {Stone}},
  \bibinfo {author} {\bibfnamefont {D.~T.}\ \bibnamefont {Stack}}, \bibinfo
  {author} {\bibfnamefont {M.}~\bibnamefont {Jaffe}}, \bibinfo {author}
  {\bibfnamefont {J.~M.}\ \bibnamefont {Kindem}}, \bibinfo {author}
  {\bibfnamefont {L.}~\bibnamefont {Wadleigh}}, \bibinfo {author}
  {\bibfnamefont {E.}~\bibnamefont {Zalys-Geller}}, \bibinfo {author}
  {\bibfnamefont {X.}~\bibnamefont {Zhang}}, \bibinfo {author} {\bibfnamefont
  {C.-A.}\ \bibnamefont {Chen}}, \bibinfo {author} {\bibfnamefont {M.~A.}\
  \bibnamefont {Norcia}}, \bibinfo {author} {\bibfnamefont {J.}~\bibnamefont
  {Epstein}}, \bibinfo {author} {\bibfnamefont {E.}~\bibnamefont {Halperin}},
  \bibinfo {author} {\bibfnamefont {F.}~\bibnamefont {Hummel}}, \bibinfo
  {author} {\bibfnamefont {T.}~\bibnamefont {Wilkason}}, \bibinfo {author}
  {\bibfnamefont {M.}~\bibnamefont {Li}}, \bibinfo {author} {\bibfnamefont
  {K.}~\bibnamefont {Barnes}}, \bibinfo {author} {\bibfnamefont
  {P.}~\bibnamefont {Battaglino}}, \bibinfo {author} {\bibfnamefont {T.~C.}\
  \bibnamefont {Bohdanowicz}}, \bibinfo {author} {\bibfnamefont
  {G.}~\bibnamefont {Booth}}, \bibinfo {author} {\bibfnamefont
  {A.}~\bibnamefont {Brown}}, \bibinfo {author} {\bibfnamefont {M.~O.}\
  \bibnamefont {Brown}}, \bibinfo {author} {\bibfnamefont {W.~B.}\ \bibnamefont
  {Cairncross}}, \bibinfo {author} {\bibfnamefont {K.}~\bibnamefont
  {Cassella}}, \bibinfo {author} {\bibfnamefont {R.}~\bibnamefont {Coxe}},
  \bibinfo {author} {\bibfnamefont {D.}~\bibnamefont {Crow}}, \bibinfo {author}
  {\bibfnamefont {M.}~\bibnamefont {Feldkamp}}, \bibinfo {author}
  {\bibfnamefont {C.}~\bibnamefont {Griger}}, \bibinfo {author} {\bibfnamefont
  {A.}~\bibnamefont {Heinz}}, \bibinfo {author} {\bibfnamefont {A.~M.~W.}\
  \bibnamefont {Jones}}, \bibinfo {author} {\bibfnamefont {H.}~\bibnamefont
  {Kim}}, \bibinfo {author} {\bibfnamefont {J.}~\bibnamefont {King}}, \bibinfo
  {author} {\bibfnamefont {K.}~\bibnamefont {Kotru}}, \bibinfo {author}
  {\bibfnamefont {J.}~\bibnamefont {Lauigan}}, \bibinfo {author} {\bibfnamefont
  {J.}~\bibnamefont {Marjanovic}}, \bibinfo {author} {\bibfnamefont
  {E.}~\bibnamefont {Megidish}}, \bibinfo {author} {\bibfnamefont
  {M.}~\bibnamefont {Meredith}}, \bibinfo {author} {\bibfnamefont
  {M.}~\bibnamefont {McDonald}}, \bibinfo {author} {\bibfnamefont
  {R.}~\bibnamefont {Morshead}}, \bibinfo {author} {\bibfnamefont
  {S.}~\bibnamefont {Narayanaswami}}, \bibinfo {author} {\bibfnamefont
  {C.}~\bibnamefont {Nishiguchi}}, \bibinfo {author} {\bibfnamefont
  {T.}~\bibnamefont {Paule}}, \bibinfo {author} {\bibfnamefont {K.~A.}\
  \bibnamefont {Pawlak}}, \bibinfo {author} {\bibfnamefont {K.~L.}\
  \bibnamefont {Pudenz}}, \bibinfo {author} {\bibfnamefont {D.~R.}\
  \bibnamefont {P\'erez}}, \bibinfo {author} {\bibfnamefont {A.}~\bibnamefont
  {Ryou}}, \bibinfo {author} {\bibfnamefont {J.}~\bibnamefont {Simon}},
  \bibinfo {author} {\bibfnamefont {A.}~\bibnamefont {Smull}}, \bibinfo
  {author} {\bibfnamefont {M.}~\bibnamefont {Urbanek}}, \bibinfo {author}
  {\bibfnamefont {R.~J.~M.}\ \bibnamefont {van~de Veerdonk}}, \bibinfo {author}
  {\bibfnamefont {Z.}~\bibnamefont {Vendeiro}}, \bibinfo {author}
  {\bibfnamefont {T.-Y.}\ \bibnamefont {Wu}}, \bibinfo {author} {\bibfnamefont
  {X.}~\bibnamefont {Xie}},\ and\ \bibinfo {author} {\bibfnamefont {B.~J.}\
  \bibnamefont {Bloom}},\ }\bibfield  {title} {\bibinfo {title} {High-fidelity
  universal gates in the ${}^{171}$$\mathrm{Yb}$ ground-state nuclear-spin
  qubit},\ }\href {https://doi.org/10.1103/PRXQuantum.6.020334} {\bibfield
  {journal} {\bibinfo  {journal} {PRX Quantum}\ }\textbf {\bibinfo {volume}
  {6}},\ \bibinfo {pages} {020334} (\bibinfo {year} {2025})}\BibitemShut
  {NoStop}%
\bibitem [{\citenamefont {Seaton}(1966)}]{seaton1966quantum}%
  \BibitemOpen
  \bibfield  {author} {\bibinfo {author} {\bibfnamefont {M.~J.}\ \bibnamefont
  {Seaton}},\ }\bibfield  {title} {\bibinfo {title} {Quantum defect theory {I}.
  {General} formulation},\ }\href@noop {} {\bibfield  {journal} {\bibinfo
  {journal} {Proc. Phys. Soc.}\ }\textbf {\bibinfo {volume} {88}},\ \bibinfo
  {pages} {801} (\bibinfo {year} {1966})}\BibitemShut {NoStop}%
\bibitem [{\citenamefont {Fano}(1970)}]{fano1970quantum}%
  \BibitemOpen
  \bibfield  {author} {\bibinfo {author} {\bibfnamefont {U.}~\bibnamefont
  {Fano}},\ }\bibfield  {title} {\bibinfo {title} {Quantum defect theory of $l$
  uncoupling in {${\mathrm{H}}_{2}$} as an example of channel-interaction
  treatment},\ }\href {https://doi.org/10.1103/PhysRevA.2.353} {\bibfield
  {journal} {\bibinfo  {journal} {Phys. Rev. A}\ }\textbf {\bibinfo {volume}
  {2}},\ \bibinfo {pages} {353} (\bibinfo {year} {1970})}\BibitemShut {NoStop}%
\bibitem [{\citenamefont {Lu}(1971)}]{lu1971spectroscopy}%
  \BibitemOpen
  \bibfield  {author} {\bibinfo {author} {\bibfnamefont {K.~T.}\ \bibnamefont
  {Lu}},\ }\bibfield  {title} {\bibinfo {title} {Spectroscopy and collision
  theory. {The} {Xe} absorption spectrum},\ }\href@noop {} {\bibfield
  {journal} {\bibinfo  {journal} {Phys. Rev. A}\ }\textbf {\bibinfo {volume}
  {4}},\ \bibinfo {pages} {579} (\bibinfo {year} {1971})}\BibitemShut {NoStop}%
\bibitem [{\citenamefont {Lee}\ and\ \citenamefont
  {Lu}(1973)}]{lee1973spectroscopy}%
  \BibitemOpen
  \bibfield  {author} {\bibinfo {author} {\bibfnamefont {C.-M.}\ \bibnamefont
  {Lee}}\ and\ \bibinfo {author} {\bibfnamefont {K.~T.}\ \bibnamefont {Lu}},\
  }\bibfield  {title} {\bibinfo {title} {Spectroscopy and collision theory.
  {II}. {The} {Ar} absorption spectrum},\ }\href@noop {} {\bibfield  {journal}
  {\bibinfo  {journal} {Phys. Rev. A}\ }\textbf {\bibinfo {volume} {8}},\
  \bibinfo {pages} {1241} (\bibinfo {year} {1973})}\BibitemShut {NoStop}%
\bibitem [{\citenamefont {Seaton}(1983)}]{seaton1983quantum}%
  \BibitemOpen
  \bibfield  {author} {\bibinfo {author} {\bibfnamefont {M.~J.}\ \bibnamefont
  {Seaton}},\ }\bibfield  {title} {\bibinfo {title} {Quantum defect theory},\
  }\href@noop {} {\bibfield  {journal} {\bibinfo  {journal} {Rep. Prog. Phys.}\
  }\textbf {\bibinfo {volume} {46}},\ \bibinfo {pages} {167} (\bibinfo {year}
  {1983})}\BibitemShut {NoStop}%
\bibitem [{\citenamefont {Cooke}\ and\ \citenamefont
  {Cromer}(1985)}]{cooke1985multichannel}%
  \BibitemOpen
  \bibfield  {author} {\bibinfo {author} {\bibfnamefont {W.~E.}\ \bibnamefont
  {Cooke}}\ and\ \bibinfo {author} {\bibfnamefont {C.~L.}\ \bibnamefont
  {Cromer}},\ }\bibfield  {title} {\bibinfo {title} {Multichannel
  quantum-defect theory and an equivalent {N}-level system},\ }\href@noop {}
  {\bibfield  {journal} {\bibinfo  {journal} {Phys. Rev. A}\ }\textbf {\bibinfo
  {volume} {32}},\ \bibinfo {pages} {2725} (\bibinfo {year}
  {1985})}\BibitemShut {NoStop}%
\bibitem [{\citenamefont {Potvliege}(2024)}]{potvliege2024mqdtfit}%
  \BibitemOpen
  \bibfield  {author} {\bibinfo {author} {\bibfnamefont {R.~M.}\ \bibnamefont
  {Potvliege}},\ }\bibfield  {title} {\bibinfo {title} {mqdtfit: {A} collection
  of {Python} functions for empirical multichannel quantum defect
  calculations},\ }\href@noop {} {\bibfield  {journal} {\bibinfo  {journal}
  {Comput. Phys. Commun.}\ }\textbf {\bibinfo {volume} {300}},\ \bibinfo
  {pages} {109172} (\bibinfo {year} {2024})}\BibitemShut {NoStop}%
\bibitem [{\citenamefont {Seaton}(1958)}]{seaton1958quantum}%
  \BibitemOpen
  \bibfield  {author} {\bibinfo {author} {\bibfnamefont {M.~J.}\ \bibnamefont
  {Seaton}},\ }\bibfield  {title} {\bibinfo {title} {The quantum defect
  method},\ }\href@noop {} {\bibfield  {journal} {\bibinfo  {journal} {Mon.
  Not. R. Astron. Soc.}\ }\textbf {\bibinfo {volume} {118}},\ \bibinfo {pages}
  {504} (\bibinfo {year} {1958})}\BibitemShut {NoStop}%
\bibitem [{\citenamefont {Vaillant}\ \emph {et~al.}(2014)\citenamefont
  {Vaillant}, \citenamefont {Jones},\ and\ \citenamefont
  {Potvliege}}]{vaillant2014multichannel}%
  \BibitemOpen
  \bibfield  {author} {\bibinfo {author} {\bibfnamefont {C.}~\bibnamefont
  {Vaillant}}, \bibinfo {author} {\bibfnamefont {M.}~\bibnamefont {Jones}},\
  and\ \bibinfo {author} {\bibfnamefont {R.}~\bibnamefont {Potvliege}},\
  }\bibfield  {title} {\bibinfo {title} {Multichannel quantum defect theory of
  strontium bound {Rydberg} states},\ }\href@noop {} {\bibfield  {journal}
  {\bibinfo  {journal} {J. Phys. B: At. Mol. Opt. Phys.}\ }\textbf {\bibinfo
  {volume} {47}},\ \bibinfo {pages} {155001} (\bibinfo {year}
  {2014})}\BibitemShut {NoStop}%
\bibitem [{\citenamefont {Robicheaux}\ \emph {et~al.}(2018)\citenamefont
  {Robicheaux}, \citenamefont {Booth},\ and\ \citenamefont
  {Saffman}}]{robicheaux2018theory}%
  \BibitemOpen
  \bibfield  {author} {\bibinfo {author} {\bibfnamefont {F.}~\bibnamefont
  {Robicheaux}}, \bibinfo {author} {\bibfnamefont {D.~W.}\ \bibnamefont
  {Booth}},\ and\ \bibinfo {author} {\bibfnamefont {M.}~\bibnamefont
  {Saffman}},\ }\bibfield  {title} {\bibinfo {title} {Theory of long-range
  interactions for {Rydberg} states attached to hyperfine-split cores},\
  }\href@noop {} {\bibfield  {journal} {\bibinfo  {journal} {Phys. Rev. A}\
  }\textbf {\bibinfo {volume} {97}},\ \bibinfo {pages} {022508} (\bibinfo
  {year} {2018})}\BibitemShut {NoStop}%
\bibitem [{\citenamefont {Robicheaux}(2019)}]{robicheaux2019calculations}%
  \BibitemOpen
  \bibfield  {author} {\bibinfo {author} {\bibfnamefont {F.}~\bibnamefont
  {Robicheaux}},\ }\bibfield  {title} {\bibinfo {title} {Calculations of long
  range interactions for ${}^{87}${Sr} {Rydberg} states},\ }\href@noop {}
  {\bibfield  {journal} {\bibinfo  {journal} {J. Phys. B: At. Mol. Opt. Phys.}\
  }\textbf {\bibinfo {volume} {52}},\ \bibinfo {pages} {244001} (\bibinfo
  {year} {2019})}\BibitemShut {NoStop}%
\bibitem [{\citenamefont {Hummel}\ \emph {et~al.}(2024)\citenamefont {Hummel},
  \citenamefont {Weber}, \citenamefont {M{\"o}gerle}, \citenamefont {Menke},
  \citenamefont {King}, \citenamefont {Bloom}, \citenamefont {Hofferberth},\
  and\ \citenamefont {Li}}]{hummel2024engineering}%
  \BibitemOpen
  \bibfield  {author} {\bibinfo {author} {\bibfnamefont {F.}~\bibnamefont
  {Hummel}}, \bibinfo {author} {\bibfnamefont {S.}~\bibnamefont {Weber}},
  \bibinfo {author} {\bibfnamefont {J.}~\bibnamefont {M{\"o}gerle}}, \bibinfo
  {author} {\bibfnamefont {H.}~\bibnamefont {Menke}}, \bibinfo {author}
  {\bibfnamefont {J.}~\bibnamefont {King}}, \bibinfo {author} {\bibfnamefont
  {B.}~\bibnamefont {Bloom}}, \bibinfo {author} {\bibfnamefont
  {S.}~\bibnamefont {Hofferberth}},\ and\ \bibinfo {author} {\bibfnamefont
  {M.}~\bibnamefont {Li}},\ }\bibfield  {title} {\bibinfo {title} {Engineering
  {Rydberg-pair} interactions in divalent atoms with hyperfine-split ionization
  thresholds},\ }\href@noop {} {\bibfield  {journal} {\bibinfo  {journal}
  {Phys. Rev. A}\ }\textbf {\bibinfo {volume} {110}},\ \bibinfo {pages}
  {042821} (\bibinfo {year} {2024})}\BibitemShut {NoStop}%
\bibitem [{\citenamefont {Ding}\ \emph {et~al.}(2018)\citenamefont {Ding},
  \citenamefont {Whalen}, \citenamefont {Kanungo}, \citenamefont {Killian},
  \citenamefont {Dunning}, \citenamefont {Yoshida},\ and\ \citenamefont
  {Burgd\"orfer}}]{ding2018spectroscopy}%
  \BibitemOpen
  \bibfield  {author} {\bibinfo {author} {\bibfnamefont {R.}~\bibnamefont
  {Ding}}, \bibinfo {author} {\bibfnamefont {J.~D.}\ \bibnamefont {Whalen}},
  \bibinfo {author} {\bibfnamefont {S.~K.}\ \bibnamefont {Kanungo}}, \bibinfo
  {author} {\bibfnamefont {T.~C.}\ \bibnamefont {Killian}}, \bibinfo {author}
  {\bibfnamefont {F.~B.}\ \bibnamefont {Dunning}}, \bibinfo {author}
  {\bibfnamefont {S.}~\bibnamefont {Yoshida}},\ and\ \bibinfo {author}
  {\bibfnamefont {J.}~\bibnamefont {Burgd\"orfer}},\ }\bibfield  {title}
  {\bibinfo {title} {Spectroscopy of $^{87}\mathrm{Sr}$ triplet {Rydberg}
  states},\ }\href {https://doi.org/10.1103/PhysRevA.98.042505} {\bibfield
  {journal} {\bibinfo  {journal} {Phys. Rev. A}\ }\textbf {\bibinfo {volume}
  {98}},\ \bibinfo {pages} {042505} (\bibinfo {year} {2018})}\BibitemShut
  {NoStop}%
\bibitem [{\citenamefont {Lehec}\ \emph {et~al.}(2018)\citenamefont {Lehec},
  \citenamefont {Zuliani}, \citenamefont {Maineult}, \citenamefont
  {Luc-Koenig}, \citenamefont {Pillet}, \citenamefont {Cheinet}, \citenamefont
  {Niyaz},\ and\ \citenamefont {Gallagher}}]{lehec2018laser}%
  \BibitemOpen
  \bibfield  {author} {\bibinfo {author} {\bibfnamefont {H.}~\bibnamefont
  {Lehec}}, \bibinfo {author} {\bibfnamefont {A.}~\bibnamefont {Zuliani}},
  \bibinfo {author} {\bibfnamefont {W.}~\bibnamefont {Maineult}}, \bibinfo
  {author} {\bibfnamefont {E.}~\bibnamefont {Luc-Koenig}}, \bibinfo {author}
  {\bibfnamefont {P.}~\bibnamefont {Pillet}}, \bibinfo {author} {\bibfnamefont
  {P.}~\bibnamefont {Cheinet}}, \bibinfo {author} {\bibfnamefont
  {F.}~\bibnamefont {Niyaz}},\ and\ \bibinfo {author} {\bibfnamefont {T.~F.}\
  \bibnamefont {Gallagher}},\ }\bibfield  {title} {\bibinfo {title} {Laser and
  microwave spectroscopy of even-parity {Rydberg} states of neutral ytterbium
  and multichannel-quantum-defect-theory analysis},\ }\href@noop {} {\bibfield
  {journal} {\bibinfo  {journal} {Phys. Rev. A}\ }\textbf {\bibinfo {volume}
  {98}},\ \bibinfo {pages} {062506} (\bibinfo {year} {2018})}\BibitemShut
  {NoStop}%
\bibitem [{\citenamefont {Okuno}\ \emph {et~al.}(2022)\citenamefont {Okuno},
  \citenamefont {Nakamura}, \citenamefont {Kusano}, \citenamefont {Takasu},
  \citenamefont {Takei}, \citenamefont {Konishi},\ and\ \citenamefont
  {Takahashi}}]{okuno2022high}%
  \BibitemOpen
  \bibfield  {author} {\bibinfo {author} {\bibfnamefont {D.}~\bibnamefont
  {Okuno}}, \bibinfo {author} {\bibfnamefont {Y.}~\bibnamefont {Nakamura}},
  \bibinfo {author} {\bibfnamefont {T.}~\bibnamefont {Kusano}}, \bibinfo
  {author} {\bibfnamefont {Y.}~\bibnamefont {Takasu}}, \bibinfo {author}
  {\bibfnamefont {N.}~\bibnamefont {Takei}}, \bibinfo {author} {\bibfnamefont
  {H.}~\bibnamefont {Konishi}},\ and\ \bibinfo {author} {\bibfnamefont
  {Y.}~\bibnamefont {Takahashi}},\ }\bibfield  {title} {\bibinfo {title}
  {High-resolution spectroscopy and single-photon {Rydberg} excitation of
  reconfigurable ytterbium atom tweezer arrays utilizing a metastable state},\
  }\href@noop {} {\bibfield  {journal} {\bibinfo  {journal} {J. Phys. Soc.
  Jpn.}\ }\textbf {\bibinfo {volume} {91}},\ \bibinfo {pages} {084301}
  (\bibinfo {year} {2022})}\BibitemShut {NoStop}%
\bibitem [{\citenamefont {Peper}\ \emph {et~al.}(2025)\citenamefont {Peper},
  \citenamefont {Li}, \citenamefont {Knapp}, \citenamefont {Bileska},
  \citenamefont {Ma}, \citenamefont {Liu}, \citenamefont {Peng}, \citenamefont
  {Zhang}, \citenamefont {Horvath}, \citenamefont {Burgers},\ and\
  \citenamefont {Thompson}}]{peper2025spectroscopy}%
  \BibitemOpen
  \bibfield  {author} {\bibinfo {author} {\bibfnamefont {M.}~\bibnamefont
  {Peper}}, \bibinfo {author} {\bibfnamefont {Y.}~\bibnamefont {Li}}, \bibinfo
  {author} {\bibfnamefont {D.~Y.}\ \bibnamefont {Knapp}}, \bibinfo {author}
  {\bibfnamefont {M.}~\bibnamefont {Bileska}}, \bibinfo {author} {\bibfnamefont
  {S.}~\bibnamefont {Ma}}, \bibinfo {author} {\bibfnamefont {G.}~\bibnamefont
  {Liu}}, \bibinfo {author} {\bibfnamefont {P.}~\bibnamefont {Peng}}, \bibinfo
  {author} {\bibfnamefont {B.}~\bibnamefont {Zhang}}, \bibinfo {author}
  {\bibfnamefont {S.~P.}\ \bibnamefont {Horvath}}, \bibinfo {author}
  {\bibfnamefont {A.~P.}\ \bibnamefont {Burgers}},\ and\ \bibinfo {author}
  {\bibfnamefont {J.~D.}\ \bibnamefont {Thompson}},\ }\bibfield  {title}
  {\bibinfo {title} {Spectroscopy and modeling of ${}^{171}${Yb} {Rydberg}
  states for high-fidelity two-qubit gates},\ }\href@noop {} {\bibfield
  {journal} {\bibinfo  {journal} {Phys. Rev. X}\ }\textbf {\bibinfo {volume}
  {15}},\ \bibinfo {pages} {011009} (\bibinfo {year} {2025})}\BibitemShut
  {NoStop}%
\bibitem [{\citenamefont {Kuroda}\ \emph {et~al.}(2025)\citenamefont {Kuroda},
  \citenamefont {Hughes}, \citenamefont {Poitrinal}, \citenamefont {Peper},\
  and\ \citenamefont {Thompson}}]{kuroda2025microwave}%
  \BibitemOpen
  \bibfield  {author} {\bibinfo {author} {\bibfnamefont {R.}~\bibnamefont
  {Kuroda}}, \bibinfo {author} {\bibfnamefont {V.~M.}\ \bibnamefont {Hughes}},
  \bibinfo {author} {\bibfnamefont {M.}~\bibnamefont {Poitrinal}}, \bibinfo
  {author} {\bibfnamefont {M.}~\bibnamefont {Peper}},\ and\ \bibinfo {author}
  {\bibfnamefont {J.~D.}\ \bibnamefont {Thompson}},\ }\bibfield  {title}
  {\bibinfo {title} {Microwave spectroscopy and multichannel quantum defect
  analysis of ytterbium $6 snp$, $6 snf$, and $6 sng$ {Rydberg} states},\
  }\href@noop {} {\bibfield  {journal} {\bibinfo  {journal} {Phys. Rev. A}\
  }\textbf {\bibinfo {volume} {112}},\ \bibinfo {pages} {042817} (\bibinfo
  {year} {2025})}\BibitemShut {NoStop}%
\bibitem [{\citenamefont {Weber}\ \emph {et~al.}(2017)\citenamefont {Weber},
  \citenamefont {Tresp}, \citenamefont {Menke}, \citenamefont {Urvoy},
  \citenamefont {Firstenberg}, \citenamefont {B{\"u}chler},\ and\ \citenamefont
  {Hofferberth}}]{Weber2017}%
  \BibitemOpen
  \bibfield  {author} {\bibinfo {author} {\bibfnamefont {S.}~\bibnamefont
  {Weber}}, \bibinfo {author} {\bibfnamefont {C.}~\bibnamefont {Tresp}},
  \bibinfo {author} {\bibfnamefont {H.}~\bibnamefont {Menke}}, \bibinfo
  {author} {\bibfnamefont {A.}~\bibnamefont {Urvoy}}, \bibinfo {author}
  {\bibfnamefont {O.}~\bibnamefont {Firstenberg}}, \bibinfo {author}
  {\bibfnamefont {H.~P.}\ \bibnamefont {B{\"u}chler}},\ and\ \bibinfo {author}
  {\bibfnamefont {S.}~\bibnamefont {Hofferberth}},\ }\bibfield  {title}
  {\bibinfo {title} {Calculation of {Rydberg} interaction potentials},\
  }\href@noop {} {\bibfield  {journal} {\bibinfo  {journal} {J. Phys. B: At.
  Mol. Opt. Phys.}\ }\textbf {\bibinfo {volume} {50}},\ \bibinfo {pages}
  {133001} (\bibinfo {year} {2017})}\BibitemShut {NoStop}%
\bibitem [{\citenamefont {{\v{S}}ibali{\'c}}\ \emph {et~al.}(2017)\citenamefont
  {{\v{S}}ibali{\'c}}, \citenamefont {Pritchard}, \citenamefont {Adams},\ and\
  \citenamefont {Weatherill}}]{Sibalic2017}%
  \BibitemOpen
  \bibfield  {author} {\bibinfo {author} {\bibfnamefont {N.}~\bibnamefont
  {{\v{S}}ibali{\'c}}}, \bibinfo {author} {\bibfnamefont {J.~D.}\ \bibnamefont
  {Pritchard}}, \bibinfo {author} {\bibfnamefont {C.~S.}\ \bibnamefont
  {Adams}},\ and\ \bibinfo {author} {\bibfnamefont {K.~J.}\ \bibnamefont
  {Weatherill}},\ }\bibfield  {title} {\bibinfo {title} {{ARC}: {An}
  open-source library for calculating properties of alkali {Rydberg} atoms},\
  }\href@noop {} {\bibfield  {journal} {\bibinfo  {journal} {Comput. Phys.
  Commun.}\ }\textbf {\bibinfo {volume} {220}},\ \bibinfo {pages} {319}
  (\bibinfo {year} {2017})}\BibitemShut {NoStop}%
\bibitem [{\citenamefont {Robertson}\ \emph {et~al.}(2021)\citenamefont
  {Robertson}, \citenamefont {{\v{S}}ibali{\'c}}, \citenamefont {Potvliege},\
  and\ \citenamefont {Jones}}]{robertson2021arc}%
  \BibitemOpen
  \bibfield  {author} {\bibinfo {author} {\bibfnamefont {E.~J.}\ \bibnamefont
  {Robertson}}, \bibinfo {author} {\bibfnamefont {N.}~\bibnamefont
  {{\v{S}}ibali{\'c}}}, \bibinfo {author} {\bibfnamefont {R.~M.}\ \bibnamefont
  {Potvliege}},\ and\ \bibinfo {author} {\bibfnamefont {M.~P.}\ \bibnamefont
  {Jones}},\ }\bibfield  {title} {\bibinfo {title} {{ARC} 3.0: {An} expanded
  {Python} toolbox for atomic physics calculations},\ }\href@noop {} {\bibfield
   {journal} {\bibinfo  {journal} {Comput. Phys. Commun.}\ }\textbf {\bibinfo
  {volume} {261}},\ \bibinfo {pages} {107814} (\bibinfo {year}
  {2021})}\BibitemShut {NoStop}%
\bibitem [{\citenamefont {M{\"o}gerle}\ \emph {et~al.}(2026)\citenamefont
  {M{\"o}gerle}, \citenamefont {Hummel}, \citenamefont {Keil}, \citenamefont
  {Legrand}, \citenamefont {Braun}, \citenamefont {Menke}, \citenamefont
  {King}, \citenamefont {Olmos}, \citenamefont {Hofferberth}, \citenamefont
  {B{\"u}chler},\ and\ \citenamefont {Weber}}]{mogerle2026accurate}%
  \BibitemOpen
  \bibfield  {author} {\bibinfo {author} {\bibfnamefont {J.}~\bibnamefont
  {M{\"o}gerle}}, \bibinfo {author} {\bibfnamefont {F.}~\bibnamefont {Hummel}},
  \bibinfo {author} {\bibfnamefont {A.}~\bibnamefont {Keil}}, \bibinfo {author}
  {\bibfnamefont {T.}~\bibnamefont {Legrand}}, \bibinfo {author} {\bibfnamefont
  {E.~J.}\ \bibnamefont {Braun}}, \bibinfo {author} {\bibfnamefont
  {H.}~\bibnamefont {Menke}}, \bibinfo {author} {\bibfnamefont
  {J.}~\bibnamefont {King}}, \bibinfo {author} {\bibfnamefont {B.}~\bibnamefont
  {Olmos}}, \bibinfo {author} {\bibfnamefont {S.}~\bibnamefont {Hofferberth}},
  \bibinfo {author} {\bibfnamefont {H.~P.}\ \bibnamefont {B{\"u}chler}},\ and\
  \bibinfo {author} {\bibfnamefont {S.}~\bibnamefont {Weber}},\ }\bibfield
  {title} {\bibinfo {title} {Accurate modeling of {Rydberg} atoms and their
  interactions: {Theory} and implementation in {PairInteraction}},\ }\href@noop
  {} {\bibfield  {journal} {\bibinfo  {journal} {arXiv:2605.14993}\ } (\bibinfo
  {year} {2026})}\BibitemShut {NoStop}%
\bibitem [{\citenamefont {Zadnik}\ and\ \citenamefont
  {Fagotti}(2021)}]{zadnik2021folded}%
  \BibitemOpen
  \bibfield  {author} {\bibinfo {author} {\bibfnamefont {L.}~\bibnamefont
  {Zadnik}}\ and\ \bibinfo {author} {\bibfnamefont {M.}~\bibnamefont
  {Fagotti}},\ }\bibfield  {title} {\bibinfo {title} {The folded spin-1/2 {XXZ}
  model: {I}. {Diagonalisation}, jamming, and ground state properties},\
  }\href@noop {} {\bibfield  {journal} {\bibinfo  {journal} {SciPost Phys.
  Core}\ }\textbf {\bibinfo {volume} {4}},\ \bibinfo {pages} {010} (\bibinfo
  {year} {2021})}\BibitemShut {NoStop}%
\bibitem [{\citenamefont {Zadnik}\ \emph {et~al.}(2021)\citenamefont {Zadnik},
  \citenamefont {Bidzhiev},\ and\ \citenamefont {Fagotti}}]{zadnik2021folded2}%
  \BibitemOpen
  \bibfield  {author} {\bibinfo {author} {\bibfnamefont {L.}~\bibnamefont
  {Zadnik}}, \bibinfo {author} {\bibfnamefont {K.}~\bibnamefont {Bidzhiev}},\
  and\ \bibinfo {author} {\bibfnamefont {M.}~\bibnamefont {Fagotti}},\
  }\bibfield  {title} {\bibinfo {title} {The folded spin-1/2 {XXZ} model: {II}.
  {Thermodynamics} and hydrodynamics with a minimal set of charges},\
  }\href@noop {} {\bibfield  {journal} {\bibinfo  {journal} {SciPost Phys.}\
  }\textbf {\bibinfo {volume} {10}},\ \bibinfo {pages} {099} (\bibinfo {year}
  {2021})}\BibitemShut {NoStop}%
\bibitem [{\citenamefont {Pozsgay}\ \emph {et~al.}(2021)\citenamefont
  {Pozsgay}, \citenamefont {Gombor}, \citenamefont {Hutsalyuk}, \citenamefont
  {Jiang}, \citenamefont {Pristy{\'a}k},\ and\ \citenamefont
  {Vernier}}]{pozsgay2021integrable}%
  \BibitemOpen
  \bibfield  {author} {\bibinfo {author} {\bibfnamefont {B.}~\bibnamefont
  {Pozsgay}}, \bibinfo {author} {\bibfnamefont {T.}~\bibnamefont {Gombor}},
  \bibinfo {author} {\bibfnamefont {A.}~\bibnamefont {Hutsalyuk}}, \bibinfo
  {author} {\bibfnamefont {Y.}~\bibnamefont {Jiang}}, \bibinfo {author}
  {\bibfnamefont {L.}~\bibnamefont {Pristy{\'a}k}},\ and\ \bibinfo {author}
  {\bibfnamefont {E.}~\bibnamefont {Vernier}},\ }\bibfield  {title} {\bibinfo
  {title} {Integrable spin chain with {Hilbert} space fragmentation and
  solvable real-time dynamics},\ }\href@noop {} {\bibfield  {journal} {\bibinfo
   {journal} {Phys. Rev. E}\ }\textbf {\bibinfo {volume} {104}},\ \bibinfo
  {pages} {044106} (\bibinfo {year} {2021})}\BibitemShut {NoStop}%
\bibitem [{\citenamefont {Dang}\ \emph {et~al.}(2008)\citenamefont {Dang},
  \citenamefont {Boninsegni},\ and\ \citenamefont {Pollet}}]{dang2008vacancy}%
  \BibitemOpen
  \bibfield  {author} {\bibinfo {author} {\bibfnamefont {L.}~\bibnamefont
  {Dang}}, \bibinfo {author} {\bibfnamefont {M.}~\bibnamefont {Boninsegni}},\
  and\ \bibinfo {author} {\bibfnamefont {L.}~\bibnamefont {Pollet}},\
  }\bibfield  {title} {\bibinfo {title} {Vacancy supersolid of hard-core bosons
  on the square lattice},\ }\href {https://doi.org/10.1103/PhysRevB.78.132512}
  {\bibfield  {journal} {\bibinfo  {journal} {Phys. Rev. B}\ }\textbf {\bibinfo
  {volume} {78}},\ \bibinfo {pages} {132512} (\bibinfo {year}
  {2008})}\BibitemShut {NoStop}%
\bibitem [{\citenamefont {Danshita}\ and\ \citenamefont
  {Yamamoto}(2010)}]{danshita2010critical}%
  \BibitemOpen
  \bibfield  {author} {\bibinfo {author} {\bibfnamefont {I.}~\bibnamefont
  {Danshita}}\ and\ \bibinfo {author} {\bibfnamefont {D.}~\bibnamefont
  {Yamamoto}},\ }\bibfield  {title} {\bibinfo {title} {Critical velocity of
  flowing supersolids of dipolar {Bose} gases in optical lattices},\
  }\href@noop {} {\bibfield  {journal} {\bibinfo  {journal} {Phys. Rev. A}\
  }\textbf {\bibinfo {volume} {82}},\ \bibinfo {pages} {013645} (\bibinfo
  {year} {2010})}\BibitemShut {NoStop}%
\bibitem [{\citenamefont {Capogrosso-Sansone}\ \emph
  {et~al.}(2010)\citenamefont {Capogrosso-Sansone}, \citenamefont {Trefzger},
  \citenamefont {Lewenstein}, \citenamefont {Zoller},\ and\ \citenamefont
  {Pupillo}}]{Capogrosso-Sansone2010quantum}%
  \BibitemOpen
  \bibfield  {author} {\bibinfo {author} {\bibfnamefont {B.}~\bibnamefont
  {Capogrosso-Sansone}}, \bibinfo {author} {\bibfnamefont {C.}~\bibnamefont
  {Trefzger}}, \bibinfo {author} {\bibfnamefont {M.}~\bibnamefont
  {Lewenstein}}, \bibinfo {author} {\bibfnamefont {P.}~\bibnamefont {Zoller}},\
  and\ \bibinfo {author} {\bibfnamefont {G.}~\bibnamefont {Pupillo}},\
  }\bibfield  {title} {\bibinfo {title} {Quantum phases of cold polar molecules
  in {2D} optical lattices},\ }\href
  {https://doi.org/10.1103/PhysRevLett.104.125301} {\bibfield  {journal}
  {\bibinfo  {journal} {Phys. Rev. Lett.}\ }\textbf {\bibinfo {volume} {104}},\
  \bibinfo {pages} {125301} (\bibinfo {year} {2010})}\BibitemShut {NoStop}%
\bibitem [{\citenamefont {Yamamoto}\ \emph
  {et~al.}(2012{\natexlab{a}})\citenamefont {Yamamoto}, \citenamefont
  {Masaki},\ and\ \citenamefont {Danshita}}]{yamamoto2012quantum}%
  \BibitemOpen
  \bibfield  {author} {\bibinfo {author} {\bibfnamefont {D.}~\bibnamefont
  {Yamamoto}}, \bibinfo {author} {\bibfnamefont {A.}~\bibnamefont {Masaki}},\
  and\ \bibinfo {author} {\bibfnamefont {I.}~\bibnamefont {Danshita}},\
  }\bibfield  {title} {\bibinfo {title} {Quantum phases of hardcore bosons with
  long-range interactions on a square lattice},\ }\href@noop {} {\bibfield
  {journal} {\bibinfo  {journal} {Phys. Rev. B}\ }\textbf {\bibinfo {volume}
  {86}},\ \bibinfo {pages} {054516} (\bibinfo {year}
  {2012}{\natexlab{a}})}\BibitemShut {NoStop}%
\bibitem [{\citenamefont {van Bijnen}(2013)}]{Bijnen_PhD_thesis}%
  \BibitemOpen
  \bibfield  {author} {\bibinfo {author} {\bibfnamefont {R.~M.~W.}\
  \bibnamefont {van Bijnen}},\ }\emph {\bibinfo {title} {Quantum engineering
  with ultracold atoms}},\ \href@noop {} {\bibinfo {type} {Ph.{D}. thesis}},\
  \bibinfo  {school} {Eindhoven University of Technology} (\bibinfo {year}
  {2013})\BibitemShut {NoStop}%
\bibitem [{\citenamefont {Whitlock}\ \emph {et~al.}(2017)\citenamefont
  {Whitlock}, \citenamefont {Glaetzle},\ and\ \citenamefont
  {Hannaford}}]{Whitlock2017}%
  \BibitemOpen
  \bibfield  {author} {\bibinfo {author} {\bibfnamefont {S.}~\bibnamefont
  {Whitlock}}, \bibinfo {author} {\bibfnamefont {A.~W.}\ \bibnamefont
  {Glaetzle}},\ and\ \bibinfo {author} {\bibfnamefont {P.}~\bibnamefont
  {Hannaford}},\ }\bibfield  {title} {\bibinfo {title} {Simulating quantum spin
  models using {Rydberg-excited} atomic ensembles in magnetic microtrap
  arrays},\ }\href@noop {} {\bibfield  {journal} {\bibinfo  {journal} {J. Phys.
  B: At. Mol. Opt. Phys.}\ }\textbf {\bibinfo {volume} {50}},\ \bibinfo {pages}
  {074001} (\bibinfo {year} {2017})}\BibitemShut {NoStop}%
\bibitem [{\citenamefont {Wadenpfuhl}\ and\ \citenamefont
  {Adams}(2025)}]{Wadenpfuhl2025unraveling}%
  \BibitemOpen
  \bibfield  {author} {\bibinfo {author} {\bibfnamefont {K.}~\bibnamefont
  {Wadenpfuhl}}\ and\ \bibinfo {author} {\bibfnamefont {C.~S.}\ \bibnamefont
  {Adams}},\ }\bibfield  {title} {\bibinfo {title} {Unraveling the structures
  in the van der {Waals} interactions of alkali-metal {Rydberg} atoms},\ }\href
  {https://doi.org/10.1103/PhysRevA.111.062803} {\bibfield  {journal} {\bibinfo
   {journal} {Phys. Rev. A}\ }\textbf {\bibinfo {volume} {111}},\ \bibinfo
  {pages} {062803} (\bibinfo {year} {2025})}\BibitemShut {NoStop}%
\bibitem [{\citenamefont {Dobrzyniecki}\ \emph {et~al.}(2025)\citenamefont
  {Dobrzyniecki}, \citenamefont {Heim},\ and\ \citenamefont
  {Tomza}}]{dobrzyniecki2025tunable}%
  \BibitemOpen
  \bibfield  {author} {\bibinfo {author} {\bibfnamefont {J.}~\bibnamefont
  {Dobrzyniecki}}, \bibinfo {author} {\bibfnamefont {P.}~\bibnamefont {Heim}},\
  and\ \bibinfo {author} {\bibfnamefont {M.}~\bibnamefont {Tomza}},\ }\bibfield
   {title} {\bibinfo {title} {Tunable two-species spin models with {Rydberg}
  atoms in circular and elliptical states},\ }\href@noop {} {\bibfield
  {journal} {\bibinfo  {journal} {Phys. Rev. Res.}\ }\textbf {\bibinfo {volume}
  {7}},\ \bibinfo {pages} {013321} (\bibinfo {year} {2025})}\BibitemShut
  {NoStop}%
\bibitem [{Note1()}]{Note1}%
  \BibitemOpen
  \bibinfo {note} {As discussed in Ref.~\cite {kunimi2025proposal}, we can
  estimate the contribution of the next-to-leading-order correction. Because of
  the dipole selection rules, this correction arises at fourth order in
  perturbation theory. The ratio of the fourth-order correction to the
  second-order term scales as $(R_{\protect \rm c}/R)^6$. For $R=2R_{\protect
  \rm c}$, this ratio is $1/64\simeq 1.6\%$, indicating that the fourth-order
  correction is only of order $1\%$ of the second-order
  contribution.}\BibitemShut {Stop}%
\bibitem [{\citenamefont {Jiao}\ \emph {et~al.}(2022)\citenamefont {Jiao},
  \citenamefont {Bai}, \citenamefont {Song}, \citenamefont {Bao}, \citenamefont
  {Zhao},\ and\ \citenamefont {Jia}}]{jiao2022electric}%
  \BibitemOpen
  \bibfield  {author} {\bibinfo {author} {\bibfnamefont {Y.}~\bibnamefont
  {Jiao}}, \bibinfo {author} {\bibfnamefont {J.}~\bibnamefont {Bai}}, \bibinfo
  {author} {\bibfnamefont {R.}~\bibnamefont {Song}}, \bibinfo {author}
  {\bibfnamefont {S.}~\bibnamefont {Bao}}, \bibinfo {author} {\bibfnamefont
  {J.}~\bibnamefont {Zhao}},\ and\ \bibinfo {author} {\bibfnamefont
  {S.}~\bibnamefont {Jia}},\ }\bibfield  {title} {\bibinfo {title} {Electric
  field tuned dipolar interaction between {Rydberg} atoms},\ }\href@noop {}
  {\bibfield  {journal} {\bibinfo  {journal} {Front. Phys.}\ }\textbf {\bibinfo
  {volume} {10}},\ \bibinfo {pages} {892542} (\bibinfo {year}
  {2022})}\BibitemShut {NoStop}%
\bibitem [{\citenamefont {Manetsch}\ \emph {et~al.}(2025)\citenamefont
  {Manetsch}, \citenamefont {Nomura}, \citenamefont {Bataille}, \citenamefont
  {Lv}, \citenamefont {Leung},\ and\ \citenamefont
  {Endres}}]{manetsch2025tweezer}%
  \BibitemOpen
  \bibfield  {author} {\bibinfo {author} {\bibfnamefont {H.~J.}\ \bibnamefont
  {Manetsch}}, \bibinfo {author} {\bibfnamefont {G.}~\bibnamefont {Nomura}},
  \bibinfo {author} {\bibfnamefont {E.}~\bibnamefont {Bataille}}, \bibinfo
  {author} {\bibfnamefont {X.}~\bibnamefont {Lv}}, \bibinfo {author}
  {\bibfnamefont {K.~H.}\ \bibnamefont {Leung}},\ and\ \bibinfo {author}
  {\bibfnamefont {M.}~\bibnamefont {Endres}},\ }\bibfield  {title} {\bibinfo
  {title} {A tweezer array with 6100 highly coherent atomic qubits},\
  }\href@noop {} {\bibfield  {journal} {\bibinfo  {journal} {Nature}\ }\textbf
  {\bibinfo {volume} {647}},\ \bibinfo {pages} {60} (\bibinfo {year}
  {2025})}\BibitemShut {NoStop}%
\bibitem [{\citenamefont {Fuchs}\ \emph {et~al.}(2008)\citenamefont {Fuchs},
  \citenamefont {Ticknor}, \citenamefont {Dyke}, \citenamefont {Veeravalli},
  \citenamefont {Kuhnle}, \citenamefont {Rowlands}, \citenamefont {Hannaford},\
  and\ \citenamefont {Vale}}]{Fuchs2008binding}%
  \BibitemOpen
  \bibfield  {author} {\bibinfo {author} {\bibfnamefont {J.}~\bibnamefont
  {Fuchs}}, \bibinfo {author} {\bibfnamefont {C.}~\bibnamefont {Ticknor}},
  \bibinfo {author} {\bibfnamefont {P.}~\bibnamefont {Dyke}}, \bibinfo {author}
  {\bibfnamefont {G.}~\bibnamefont {Veeravalli}}, \bibinfo {author}
  {\bibfnamefont {E.}~\bibnamefont {Kuhnle}}, \bibinfo {author} {\bibfnamefont
  {W.}~\bibnamefont {Rowlands}}, \bibinfo {author} {\bibfnamefont
  {P.}~\bibnamefont {Hannaford}},\ and\ \bibinfo {author} {\bibfnamefont
  {C.~J.}\ \bibnamefont {Vale}},\ }\bibfield  {title} {\bibinfo {title}
  {Binding energies of $^{6}\text{L}\text{i}$ $p$-wave {Feshbach} molecules},\
  }\href {https://doi.org/10.1103/PhysRevA.77.053616} {\bibfield  {journal}
  {\bibinfo  {journal} {Phys. Rev. A}\ }\textbf {\bibinfo {volume} {77}},\
  \bibinfo {pages} {053616} (\bibinfo {year} {2008})}\BibitemShut {NoStop}%
\bibitem [{\citenamefont {Inada}\ \emph {et~al.}(2008)\citenamefont {Inada},
  \citenamefont {Horikoshi}, \citenamefont {Nakajima}, \citenamefont
  {Kuwata-Gonokami}, \citenamefont {Ueda},\ and\ \citenamefont
  {Mukaiyama}}]{Inada2008collisional}%
  \BibitemOpen
  \bibfield  {author} {\bibinfo {author} {\bibfnamefont {Y.}~\bibnamefont
  {Inada}}, \bibinfo {author} {\bibfnamefont {M.}~\bibnamefont {Horikoshi}},
  \bibinfo {author} {\bibfnamefont {S.}~\bibnamefont {Nakajima}}, \bibinfo
  {author} {\bibfnamefont {M.}~\bibnamefont {Kuwata-Gonokami}}, \bibinfo
  {author} {\bibfnamefont {M.}~\bibnamefont {Ueda}},\ and\ \bibinfo {author}
  {\bibfnamefont {T.}~\bibnamefont {Mukaiyama}},\ }\bibfield  {title} {\bibinfo
  {title} {Collisional properties of $p$-wave {Feshbach} molecules},\ }\href
  {https://doi.org/10.1103/PhysRevLett.101.100401} {\bibfield  {journal}
  {\bibinfo  {journal} {Phys. Rev. Lett.}\ }\textbf {\bibinfo {volume} {101}},\
  \bibinfo {pages} {100401} (\bibinfo {year} {2008})}\BibitemShut {NoStop}%
\bibitem [{\citenamefont {Chang}\ \emph {et~al.}(2020)\citenamefont {Chang},
  \citenamefont {Senaratne}, \citenamefont {Cavazos-Cavazos},\ and\
  \citenamefont {Hulet}}]{Chang2020collisional}%
  \BibitemOpen
  \bibfield  {author} {\bibinfo {author} {\bibfnamefont {Y.-T.}\ \bibnamefont
  {Chang}}, \bibinfo {author} {\bibfnamefont {R.}~\bibnamefont {Senaratne}},
  \bibinfo {author} {\bibfnamefont {D.}~\bibnamefont {Cavazos-Cavazos}},\ and\
  \bibinfo {author} {\bibfnamefont {R.~G.}\ \bibnamefont {Hulet}},\ }\bibfield
  {title} {\bibinfo {title} {Collisional loss of one-dimensional fermions near
  a $p$-wave {Feshbach} resonance},\ }\href
  {https://doi.org/10.1103/PhysRevLett.125.263402} {\bibfield  {journal}
  {\bibinfo  {journal} {Phys. Rev. Lett.}\ }\textbf {\bibinfo {volume} {125}},\
  \bibinfo {pages} {263402} (\bibinfo {year} {2020})}\BibitemShut {NoStop}%
\bibitem [{\citenamefont {Xie}\ \emph {et~al.}(2025)\citenamefont {Xie},
  \citenamefont {Li}, \citenamefont {Zhou}, \citenamefont {Luo}, \citenamefont
  {Wang}, \citenamefont {Nie}, \citenamefont {Shen}, \citenamefont {Chen},
  \citenamefont {Yao},\ and\ \citenamefont {Pan}}]{Xie2025Feshbach}%
  \BibitemOpen
  \bibfield  {author} {\bibinfo {author} {\bibfnamefont {K.}~\bibnamefont
  {Xie}}, \bibinfo {author} {\bibfnamefont {X.}~\bibnamefont {Li}}, \bibinfo
  {author} {\bibfnamefont {Y.-Y.}\ \bibnamefont {Zhou}}, \bibinfo {author}
  {\bibfnamefont {J.-H.}\ \bibnamefont {Luo}}, \bibinfo {author} {\bibfnamefont
  {S.}~\bibnamefont {Wang}}, \bibinfo {author} {\bibfnamefont {Y.-Z.}\
  \bibnamefont {Nie}}, \bibinfo {author} {\bibfnamefont {H.-C.}\ \bibnamefont
  {Shen}}, \bibinfo {author} {\bibfnamefont {Y.-A.}\ \bibnamefont {Chen}},
  \bibinfo {author} {\bibfnamefont {X.-C.}\ \bibnamefont {Yao}},\ and\ \bibinfo
  {author} {\bibfnamefont {J.-W.}\ \bibnamefont {Pan}},\ }\bibfield  {title}
  {\bibinfo {title} {Feshbach spectroscopy of ultracold mixtures of
  $^{6}\mathrm{Li}$ and $^{164}\mathrm{Dy}$ atoms},\ }\href
  {https://doi.org/10.1103/PhysRevA.111.023327} {\bibfield  {journal} {\bibinfo
   {journal} {Phys. Rev. A}\ }\textbf {\bibinfo {volume} {111}},\ \bibinfo
  {pages} {023327} (\bibinfo {year} {2025})}\BibitemShut {NoStop}%
\bibitem [{\citenamefont {Yang}\ \emph {et~al.}(2020)\citenamefont {Yang},
  \citenamefont {Liu}, \citenamefont {Gorshkov},\ and\ \citenamefont
  {Iadecola}}]{yang2020hilbert}%
  \BibitemOpen
  \bibfield  {author} {\bibinfo {author} {\bibfnamefont {Z.-C.}\ \bibnamefont
  {Yang}}, \bibinfo {author} {\bibfnamefont {F.}~\bibnamefont {Liu}}, \bibinfo
  {author} {\bibfnamefont {A.~V.}\ \bibnamefont {Gorshkov}},\ and\ \bibinfo
  {author} {\bibfnamefont {T.}~\bibnamefont {Iadecola}},\ }\bibfield  {title}
  {\bibinfo {title} {Hilbert-space fragmentation from strict confinement},\
  }\href@noop {} {\bibfield  {journal} {\bibinfo  {journal} {Phys. Rev. Lett.}\
  }\textbf {\bibinfo {volume} {124}},\ \bibinfo {pages} {207602} (\bibinfo
  {year} {2020})}\BibitemShut {NoStop}%
\bibitem [{\citenamefont {Sala}\ \emph {et~al.}(2020)\citenamefont {Sala},
  \citenamefont {Rakovszky}, \citenamefont {Verresen}, \citenamefont {Knap},\
  and\ \citenamefont {Pollmann}}]{Sala2020}%
  \BibitemOpen
  \bibfield  {author} {\bibinfo {author} {\bibfnamefont {P.}~\bibnamefont
  {Sala}}, \bibinfo {author} {\bibfnamefont {T.}~\bibnamefont {Rakovszky}},
  \bibinfo {author} {\bibfnamefont {R.}~\bibnamefont {Verresen}}, \bibinfo
  {author} {\bibfnamefont {M.}~\bibnamefont {Knap}},\ and\ \bibinfo {author}
  {\bibfnamefont {F.}~\bibnamefont {Pollmann}},\ }\bibfield  {title} {\bibinfo
  {title} {Ergodicity breaking arising from {Hilbert} space fragmentation in
  dipole-conserving {Hamiltonians}},\ }\href@noop {} {\bibfield  {journal}
  {\bibinfo  {journal} {Phys. Rev. X}\ }\textbf {\bibinfo {volume} {10}},\
  \bibinfo {pages} {011047} (\bibinfo {year} {2020})}\BibitemShut {NoStop}%
\bibitem [{\citenamefont {Khemani}\ \emph {et~al.}(2020)\citenamefont
  {Khemani}, \citenamefont {Hermele},\ and\ \citenamefont
  {Nandkishore}}]{Khemani2020}%
  \BibitemOpen
  \bibfield  {author} {\bibinfo {author} {\bibfnamefont {V.}~\bibnamefont
  {Khemani}}, \bibinfo {author} {\bibfnamefont {M.}~\bibnamefont {Hermele}},\
  and\ \bibinfo {author} {\bibfnamefont {R.}~\bibnamefont {Nandkishore}},\
  }\bibfield  {title} {\bibinfo {title} {Localization from {Hilbert} space
  shattering: From theory to physical realizations},\ }\href@noop {} {\bibfield
   {journal} {\bibinfo  {journal} {Phys. Rev. B}\ }\textbf {\bibinfo {volume}
  {101}},\ \bibinfo {pages} {174204} (\bibinfo {year} {2020})}\BibitemShut
  {NoStop}%
\bibitem [{\citenamefont {Fujimoto}\ and\ \citenamefont
  {Sasamoto}(2026)}]{fujimoto2024quantum}%
  \BibitemOpen
  \bibfield  {author} {\bibinfo {author} {\bibfnamefont {K.}~\bibnamefont
  {Fujimoto}}\ and\ \bibinfo {author} {\bibfnamefont {T.}~\bibnamefont
  {Sasamoto}},\ }\bibfield  {title} {\bibinfo {title} {Quantum transport in
  interacting spin chains: Exact derivation of the {Tracy-Widom}
  distribution},\ }\href {https://doi.org/10.1103/lbl9-n9ws} {\bibfield
  {journal} {\bibinfo  {journal} {Phys. Rev. Lett.}\ }\textbf {\bibinfo
  {volume} {136}},\ \bibinfo {pages} {087102} (\bibinfo {year}
  {2026})}\BibitemShut {NoStop}%
\bibitem [{\citenamefont {Kunimi}\ \emph {et~al.}(2025)\citenamefont {Kunimi},
  \citenamefont {Kato},\ and\ \citenamefont {Katsura}}]{kunimi2025systematic}%
  \BibitemOpen
  \bibfield  {author} {\bibinfo {author} {\bibfnamefont {M.}~\bibnamefont
  {Kunimi}}, \bibinfo {author} {\bibfnamefont {Y.}~\bibnamefont {Kato}},\ and\
  \bibinfo {author} {\bibfnamefont {H.}~\bibnamefont {Katsura}},\ }\bibfield
  {title} {\bibinfo {title} {Systematic construction of asymptotic quantum
  many-body scar states and their relation to supersymmetric quantum
  mechanics},\ }\href {https://doi.org/10.1103/zstl-s1y2} {\bibfield  {journal}
  {\bibinfo  {journal} {Phys. Rev. Res.}\ }\textbf {\bibinfo {volume} {7}},\
  \bibinfo {pages} {043107} (\bibinfo {year} {2025})}\BibitemShut {NoStop}%
\bibitem [{\citenamefont {Andreev}\ and\ \citenamefont
  {Lifshitz}(1969)}]{andreev1969quantum}%
  \BibitemOpen
  \bibfield  {author} {\bibinfo {author} {\bibfnamefont {A.~F.}\ \bibnamefont
  {Andreev}}\ and\ \bibinfo {author} {\bibfnamefont {I.~M.}\ \bibnamefont
  {Lifshitz}},\ }\bibfield  {title} {\bibinfo {title} {Quantum theory of
  defects in crystals},\ }\href@noop {} {\bibfield  {journal} {\bibinfo
  {journal} {Sov. Phys. JETP}\ }\textbf {\bibinfo {volume} {29}},\ \bibinfo
  {pages} {1107} (\bibinfo {year} {1969})}\BibitemShut {NoStop}%
\bibitem [{\citenamefont {Chester}(1970)}]{chester1970speculations}%
  \BibitemOpen
  \bibfield  {author} {\bibinfo {author} {\bibfnamefont {G.~V.}\ \bibnamefont
  {Chester}},\ }\bibfield  {title} {\bibinfo {title} {Speculations on
  {Bose-Einstein} condensation and quantum crystals},\ }\href
  {https://doi.org/10.1103/PhysRevA.2.256} {\bibfield  {journal} {\bibinfo
  {journal} {Phys. Rev. A}\ }\textbf {\bibinfo {volume} {2}},\ \bibinfo {pages}
  {256} (\bibinfo {year} {1970})}\BibitemShut {NoStop}%
\bibitem [{\citenamefont {Leggett}(1970)}]{Leggett1970can}%
  \BibitemOpen
  \bibfield  {author} {\bibinfo {author} {\bibfnamefont {A.~J.}\ \bibnamefont
  {Leggett}},\ }\bibfield  {title} {\bibinfo {title} {Can a solid be
  "superfluid"?},\ }\href {https://doi.org/10.1103/PhysRevLett.25.1543}
  {\bibfield  {journal} {\bibinfo  {journal} {Phys. Rev. Lett.}\ }\textbf
  {\bibinfo {volume} {25}},\ \bibinfo {pages} {1543} (\bibinfo {year}
  {1970})}\BibitemShut {NoStop}%
\bibitem [{\citenamefont {Tanzi}\ \emph {et~al.}(2019)\citenamefont {Tanzi},
  \citenamefont {Lucioni}, \citenamefont {Fam{\`a}}, \citenamefont {Catani},
  \citenamefont {Fioretti}, \citenamefont {Gabbanini}, \citenamefont {Bisset},
  \citenamefont {Santos},\ and\ \citenamefont
  {Modugno}}]{tanzi2019observation}%
  \BibitemOpen
  \bibfield  {author} {\bibinfo {author} {\bibfnamefont {L.}~\bibnamefont
  {Tanzi}}, \bibinfo {author} {\bibfnamefont {E.}~\bibnamefont {Lucioni}},
  \bibinfo {author} {\bibfnamefont {F.}~\bibnamefont {Fam{\`a}}}, \bibinfo
  {author} {\bibfnamefont {J.}~\bibnamefont {Catani}}, \bibinfo {author}
  {\bibfnamefont {A.}~\bibnamefont {Fioretti}}, \bibinfo {author}
  {\bibfnamefont {C.}~\bibnamefont {Gabbanini}}, \bibinfo {author}
  {\bibfnamefont {R.~N.}\ \bibnamefont {Bisset}}, \bibinfo {author}
  {\bibfnamefont {L.}~\bibnamefont {Santos}},\ and\ \bibinfo {author}
  {\bibfnamefont {G.}~\bibnamefont {Modugno}},\ }\bibfield  {title} {\bibinfo
  {title} {Observation of a dipolar quantum gas with metastable supersolid
  properties},\ }\href@noop {} {\bibfield  {journal} {\bibinfo  {journal}
  {Phys. Rev. Lett.}\ }\textbf {\bibinfo {volume} {122}},\ \bibinfo {pages}
  {130405} (\bibinfo {year} {2019})}\BibitemShut {NoStop}%
\bibitem [{\citenamefont {B\"ottcher}\ \emph {et~al.}(2019)\citenamefont
  {B\"ottcher}, \citenamefont {Schmidt}, \citenamefont {Wenzel}, \citenamefont
  {Hertkorn}, \citenamefont {Guo}, \citenamefont {Langen},\ and\ \citenamefont
  {Pfau}}]{Bottcher2019transient}%
  \BibitemOpen
  \bibfield  {author} {\bibinfo {author} {\bibfnamefont {F.}~\bibnamefont
  {B\"ottcher}}, \bibinfo {author} {\bibfnamefont {J.-N.}\ \bibnamefont
  {Schmidt}}, \bibinfo {author} {\bibfnamefont {M.}~\bibnamefont {Wenzel}},
  \bibinfo {author} {\bibfnamefont {J.}~\bibnamefont {Hertkorn}}, \bibinfo
  {author} {\bibfnamefont {M.}~\bibnamefont {Guo}}, \bibinfo {author}
  {\bibfnamefont {T.}~\bibnamefont {Langen}},\ and\ \bibinfo {author}
  {\bibfnamefont {T.}~\bibnamefont {Pfau}},\ }\bibfield  {title} {\bibinfo
  {title} {Transient supersolid properties in an array of dipolar quantum
  droplets},\ }\href {https://doi.org/10.1103/PhysRevX.9.011051} {\bibfield
  {journal} {\bibinfo  {journal} {Phys. Rev. X}\ }\textbf {\bibinfo {volume}
  {9}},\ \bibinfo {pages} {011051} (\bibinfo {year} {2019})}\BibitemShut
  {NoStop}%
\bibitem [{\citenamefont {Chomaz}\ \emph {et~al.}(2019)\citenamefont {Chomaz},
  \citenamefont {Petter}, \citenamefont {Ilzh\"ofer}, \citenamefont {Natale},
  \citenamefont {Trautmann}, \citenamefont {Politi}, \citenamefont
  {Durastante}, \citenamefont {van Bijnen}, \citenamefont {Patscheider},
  \citenamefont {Sohmen}, \citenamefont {Mark},\ and\ \citenamefont
  {Ferlaino}}]{Chomaz2019long}%
  \BibitemOpen
  \bibfield  {author} {\bibinfo {author} {\bibfnamefont {L.}~\bibnamefont
  {Chomaz}}, \bibinfo {author} {\bibfnamefont {D.}~\bibnamefont {Petter}},
  \bibinfo {author} {\bibfnamefont {P.}~\bibnamefont {Ilzh\"ofer}}, \bibinfo
  {author} {\bibfnamefont {G.}~\bibnamefont {Natale}}, \bibinfo {author}
  {\bibfnamefont {A.}~\bibnamefont {Trautmann}}, \bibinfo {author}
  {\bibfnamefont {C.}~\bibnamefont {Politi}}, \bibinfo {author} {\bibfnamefont
  {G.}~\bibnamefont {Durastante}}, \bibinfo {author} {\bibfnamefont {R.~M.~W.}\
  \bibnamefont {van Bijnen}}, \bibinfo {author} {\bibfnamefont
  {A.}~\bibnamefont {Patscheider}}, \bibinfo {author} {\bibfnamefont
  {M.}~\bibnamefont {Sohmen}}, \bibinfo {author} {\bibfnamefont {M.~J.}\
  \bibnamefont {Mark}},\ and\ \bibinfo {author} {\bibfnamefont
  {F.}~\bibnamefont {Ferlaino}},\ }\bibfield  {title} {\bibinfo {title}
  {Long-lived and transient supersolid behaviors in dipolar quantum gases},\
  }\href {https://doi.org/10.1103/PhysRevX.9.021012} {\bibfield  {journal}
  {\bibinfo  {journal} {Phys. Rev. X}\ }\textbf {\bibinfo {volume} {9}},\
  \bibinfo {pages} {021012} (\bibinfo {year} {2019})}\BibitemShut {NoStop}%
\bibitem [{\citenamefont {Guo}\ \emph {et~al.}(2019)\citenamefont {Guo},
  \citenamefont {B{\"o}ttcher}, \citenamefont {Hertkorn}, \citenamefont
  {Schmidt}, \citenamefont {Wenzel}, \citenamefont {B{\"u}chler}, \citenamefont
  {Langen},\ and\ \citenamefont {Pfau}}]{guo2019low}%
  \BibitemOpen
  \bibfield  {author} {\bibinfo {author} {\bibfnamefont {M.}~\bibnamefont
  {Guo}}, \bibinfo {author} {\bibfnamefont {F.}~\bibnamefont {B{\"o}ttcher}},
  \bibinfo {author} {\bibfnamefont {J.}~\bibnamefont {Hertkorn}}, \bibinfo
  {author} {\bibfnamefont {J.-N.}\ \bibnamefont {Schmidt}}, \bibinfo {author}
  {\bibfnamefont {M.}~\bibnamefont {Wenzel}}, \bibinfo {author} {\bibfnamefont
  {H.~P.}\ \bibnamefont {B{\"u}chler}}, \bibinfo {author} {\bibfnamefont
  {T.}~\bibnamefont {Langen}},\ and\ \bibinfo {author} {\bibfnamefont
  {T.}~\bibnamefont {Pfau}},\ }\bibfield  {title} {\bibinfo {title} {The
  low-energy {Goldstone} mode in a trapped dipolar supersolid},\ }\href@noop {}
  {\bibfield  {journal} {\bibinfo  {journal} {Nature}\ }\textbf {\bibinfo
  {volume} {574}},\ \bibinfo {pages} {386} (\bibinfo {year}
  {2019})}\BibitemShut {NoStop}%
\bibitem [{\citenamefont {Norcia}\ \emph {et~al.}(2021)\citenamefont {Norcia},
  \citenamefont {Politi}, \citenamefont {Klaus}, \citenamefont {Poli},
  \citenamefont {Sohmen}, \citenamefont {Mark}, \citenamefont {Bisset},
  \citenamefont {Santos},\ and\ \citenamefont {Ferlaino}}]{norcia2021two}%
  \BibitemOpen
  \bibfield  {author} {\bibinfo {author} {\bibfnamefont {M.~A.}\ \bibnamefont
  {Norcia}}, \bibinfo {author} {\bibfnamefont {C.}~\bibnamefont {Politi}},
  \bibinfo {author} {\bibfnamefont {L.}~\bibnamefont {Klaus}}, \bibinfo
  {author} {\bibfnamefont {E.}~\bibnamefont {Poli}}, \bibinfo {author}
  {\bibfnamefont {M.}~\bibnamefont {Sohmen}}, \bibinfo {author} {\bibfnamefont
  {M.~J.}\ \bibnamefont {Mark}}, \bibinfo {author} {\bibfnamefont {R.~N.}\
  \bibnamefont {Bisset}}, \bibinfo {author} {\bibfnamefont {L.}~\bibnamefont
  {Santos}},\ and\ \bibinfo {author} {\bibfnamefont {F.}~\bibnamefont
  {Ferlaino}},\ }\bibfield  {title} {\bibinfo {title} {Two-dimensional
  supersolidity in a dipolar quantum gas},\ }\href@noop {} {\bibfield
  {journal} {\bibinfo  {journal} {Nature}\ }\textbf {\bibinfo {volume} {596}},\
  \bibinfo {pages} {357} (\bibinfo {year} {2021})}\BibitemShut {NoStop}%
\bibitem [{\citenamefont {Matsuda}\ and\ \citenamefont
  {Tsuneto}(1970)}]{matsuda1970off}%
  \BibitemOpen
  \bibfield  {author} {\bibinfo {author} {\bibfnamefont {H.}~\bibnamefont
  {Matsuda}}\ and\ \bibinfo {author} {\bibfnamefont {T.}~\bibnamefont
  {Tsuneto}},\ }\bibfield  {title} {\bibinfo {title} {Off-diagonal long-range
  order in solids},\ }\href@noop {} {\bibfield  {journal} {\bibinfo  {journal}
  {Suppl. Prog. Theor. Phys.}\ }\textbf {\bibinfo {volume} {46}},\ \bibinfo
  {pages} {411} (\bibinfo {year} {1970})}\BibitemShut {NoStop}%
\bibitem [{\citenamefont {Baier}\ \emph {et~al.}(2016)\citenamefont {Baier},
  \citenamefont {Mark}, \citenamefont {Petter}, \citenamefont {Aikawa},
  \citenamefont {Chomaz}, \citenamefont {Cai}, \citenamefont {Baranov},
  \citenamefont {Zoller},\ and\ \citenamefont {Ferlaino}}]{baier2016extended}%
  \BibitemOpen
  \bibfield  {author} {\bibinfo {author} {\bibfnamefont {S.}~\bibnamefont
  {Baier}}, \bibinfo {author} {\bibfnamefont {M.~J.}\ \bibnamefont {Mark}},
  \bibinfo {author} {\bibfnamefont {D.}~\bibnamefont {Petter}}, \bibinfo
  {author} {\bibfnamefont {K.}~\bibnamefont {Aikawa}}, \bibinfo {author}
  {\bibfnamefont {L.}~\bibnamefont {Chomaz}}, \bibinfo {author} {\bibfnamefont
  {Z.}~\bibnamefont {Cai}}, \bibinfo {author} {\bibfnamefont {M.}~\bibnamefont
  {Baranov}}, \bibinfo {author} {\bibfnamefont {P.}~\bibnamefont {Zoller}},\
  and\ \bibinfo {author} {\bibfnamefont {F.}~\bibnamefont {Ferlaino}},\
  }\bibfield  {title} {\bibinfo {title} {Extended {Bose-Hubbard} models with
  ultracold magnetic atoms},\ }\href@noop {} {\bibfield  {journal} {\bibinfo
  {journal} {Science}\ }\textbf {\bibinfo {volume} {352}},\ \bibinfo {pages}
  {201} (\bibinfo {year} {2016})}\BibitemShut {NoStop}%
\bibitem [{\citenamefont {Su}\ \emph {et~al.}(2023)\citenamefont {Su},
  \citenamefont {Douglas}, \citenamefont {Szurek}, \citenamefont {Groth},
  \citenamefont {Ozturk}, \citenamefont {Krahn}, \citenamefont {H{\'e}bert},
  \citenamefont {Phelps}, \citenamefont {Ebadi}, \citenamefont {Dickerson},
  \citenamefont {Ferlaino}, \citenamefont {Markovi\'c},\ and\ \citenamefont
  {Greiner}}]{su2023dipolar}%
  \BibitemOpen
  \bibfield  {author} {\bibinfo {author} {\bibfnamefont {L.}~\bibnamefont
  {Su}}, \bibinfo {author} {\bibfnamefont {A.}~\bibnamefont {Douglas}},
  \bibinfo {author} {\bibfnamefont {M.}~\bibnamefont {Szurek}}, \bibinfo
  {author} {\bibfnamefont {R.}~\bibnamefont {Groth}}, \bibinfo {author}
  {\bibfnamefont {S.~F.}\ \bibnamefont {Ozturk}}, \bibinfo {author}
  {\bibfnamefont {A.}~\bibnamefont {Krahn}}, \bibinfo {author} {\bibfnamefont
  {A.~H.}\ \bibnamefont {H{\'e}bert}}, \bibinfo {author} {\bibfnamefont
  {G.~A.}\ \bibnamefont {Phelps}}, \bibinfo {author} {\bibfnamefont
  {S.}~\bibnamefont {Ebadi}}, \bibinfo {author} {\bibfnamefont
  {S.}~\bibnamefont {Dickerson}}, \bibinfo {author} {\bibfnamefont
  {F.}~\bibnamefont {Ferlaino}}, \bibinfo {author} {\bibfnamefont
  {O.}~\bibnamefont {Markovi\'c}},\ and\ \bibinfo {author} {\bibfnamefont
  {M.}~\bibnamefont {Greiner}},\ }\bibfield  {title} {\bibinfo {title} {Dipolar
  quantum solids emerging in a {Hubbard} quantum simulator},\ }\href@noop {}
  {\bibfield  {journal} {\bibinfo  {journal} {Nature}\ }\textbf {\bibinfo
  {volume} {622}},\ \bibinfo {pages} {724} (\bibinfo {year}
  {2023})}\BibitemShut {NoStop}%
\bibitem [{\citenamefont {Murthy}\ \emph {et~al.}(1997)\citenamefont {Murthy},
  \citenamefont {Arovas},\ and\ \citenamefont
  {Auerbach}}]{Murthy1997superfluids}%
  \BibitemOpen
  \bibfield  {author} {\bibinfo {author} {\bibfnamefont {G.}~\bibnamefont
  {Murthy}}, \bibinfo {author} {\bibfnamefont {D.}~\bibnamefont {Arovas}},\
  and\ \bibinfo {author} {\bibfnamefont {A.}~\bibnamefont {Auerbach}},\
  }\bibfield  {title} {\bibinfo {title} {Superfluids and supersolids on
  frustrated two-dimensional lattices},\ }\href
  {https://doi.org/10.1103/PhysRevB.55.3104} {\bibfield  {journal} {\bibinfo
  {journal} {Phys. Rev. B}\ }\textbf {\bibinfo {volume} {55}},\ \bibinfo
  {pages} {3104} (\bibinfo {year} {1997})}\BibitemShut {NoStop}%
\bibitem [{\citenamefont {Bernardet}\ \emph {et~al.}(2002)\citenamefont
  {Bernardet}, \citenamefont {Batrouni}, \citenamefont {Meunier}, \citenamefont
  {Schmid}, \citenamefont {Troyer},\ and\ \citenamefont
  {Dorneich}}]{Bernardet2002analytical}%
  \BibitemOpen
  \bibfield  {author} {\bibinfo {author} {\bibfnamefont {K.}~\bibnamefont
  {Bernardet}}, \bibinfo {author} {\bibfnamefont {G.~G.}\ \bibnamefont
  {Batrouni}}, \bibinfo {author} {\bibfnamefont {J.-L.}\ \bibnamefont
  {Meunier}}, \bibinfo {author} {\bibfnamefont {G.}~\bibnamefont {Schmid}},
  \bibinfo {author} {\bibfnamefont {M.}~\bibnamefont {Troyer}},\ and\ \bibinfo
  {author} {\bibfnamefont {A.}~\bibnamefont {Dorneich}},\ }\bibfield  {title}
  {\bibinfo {title} {Analytical and numerical study of hardcore bosons in two
  dimensions},\ }\href {https://doi.org/10.1103/PhysRevB.65.104519} {\bibfield
  {journal} {\bibinfo  {journal} {Phys. Rev. B}\ }\textbf {\bibinfo {volume}
  {65}},\ \bibinfo {pages} {104519} (\bibinfo {year} {2002})}\BibitemShut
  {NoStop}%
\bibitem [{\citenamefont {Batrouni}\ and\ \citenamefont
  {Scalettar}(2000)}]{Batrouni2000phase}%
  \BibitemOpen
  \bibfield  {author} {\bibinfo {author} {\bibfnamefont {G.~G.}\ \bibnamefont
  {Batrouni}}\ and\ \bibinfo {author} {\bibfnamefont {R.~T.}\ \bibnamefont
  {Scalettar}},\ }\bibfield  {title} {\bibinfo {title} {Phase separation in
  supersolids},\ }\href {https://doi.org/10.1103/PhysRevLett.84.1599}
  {\bibfield  {journal} {\bibinfo  {journal} {Phys. Rev. Lett.}\ }\textbf
  {\bibinfo {volume} {84}},\ \bibinfo {pages} {1599} (\bibinfo {year}
  {2000})}\BibitemShut {NoStop}%
\bibitem [{\citenamefont {Gubernatis}\ \emph {et~al.}(2016)\citenamefont
  {Gubernatis}, \citenamefont {Kawashima},\ and\ \citenamefont
  {Werner}}]{gubernatis2016quantum}%
  \BibitemOpen
  \bibfield  {author} {\bibinfo {author} {\bibfnamefont {J.}~\bibnamefont
  {Gubernatis}}, \bibinfo {author} {\bibfnamefont {N.}~\bibnamefont
  {Kawashima}},\ and\ \bibinfo {author} {\bibfnamefont {P.}~\bibnamefont
  {Werner}},\ }\href@noop {} {\emph {\bibinfo {title} {Quantum Monte Carlo
  Methods}}}\ (\bibinfo  {publisher} {Cambridge University Press},\ \bibinfo
  {year} {2016})\BibitemShut {NoStop}%
\bibitem [{\citenamefont {Wessel}\ and\ \citenamefont
  {Troyer}(2005)}]{Wessel2005supersolid}%
  \BibitemOpen
  \bibfield  {author} {\bibinfo {author} {\bibfnamefont {S.}~\bibnamefont
  {Wessel}}\ and\ \bibinfo {author} {\bibfnamefont {M.}~\bibnamefont
  {Troyer}},\ }\bibfield  {title} {\bibinfo {title} {Supersolid hard-core
  bosons on the triangular lattice},\ }\href
  {https://doi.org/10.1103/PhysRevLett.95.127205} {\bibfield  {journal}
  {\bibinfo  {journal} {Phys. Rev. Lett.}\ }\textbf {\bibinfo {volume} {95}},\
  \bibinfo {pages} {127205} (\bibinfo {year} {2005})}\BibitemShut {NoStop}%
\bibitem [{\citenamefont {Yamamoto}\ \emph
  {et~al.}(2012{\natexlab{b}})\citenamefont {Yamamoto}, \citenamefont
  {Danshita},\ and\ \citenamefont {S\'a~de Melo}}]{Yamamoto2012dipolar}%
  \BibitemOpen
  \bibfield  {author} {\bibinfo {author} {\bibfnamefont {D.}~\bibnamefont
  {Yamamoto}}, \bibinfo {author} {\bibfnamefont {I.}~\bibnamefont {Danshita}},\
  and\ \bibinfo {author} {\bibfnamefont {C.~A.~R.}\ \bibnamefont {S\'a~de
  Melo}},\ }\bibfield  {title} {\bibinfo {title} {Dipolar bosons in triangular
  optical lattices: Quantum phase transitions and anomalous hysteresis},\
  }\href {https://doi.org/10.1103/PhysRevA.85.021601} {\bibfield  {journal}
  {\bibinfo  {journal} {Phys. Rev. A}\ }\textbf {\bibinfo {volume} {85}},\
  \bibinfo {pages} {021601(R)} (\bibinfo {year}
  {2012}{\natexlab{b}})}\BibitemShut {NoStop}%
\bibitem [{\citenamefont {Liu}\ \emph {et~al.}(2024)\citenamefont {Liu},
  \citenamefont {Bintz}, \citenamefont {Block}, \citenamefont {Samajdar},
  \citenamefont {Kemp},\ and\ \citenamefont {Yao}}]{liu2024supersolidity}%
  \BibitemOpen
  \bibfield  {author} {\bibinfo {author} {\bibfnamefont {V.~S.}\ \bibnamefont
  {Liu}}, \bibinfo {author} {\bibfnamefont {M.}~\bibnamefont {Bintz}}, \bibinfo
  {author} {\bibfnamefont {M.}~\bibnamefont {Block}}, \bibinfo {author}
  {\bibfnamefont {R.}~\bibnamefont {Samajdar}}, \bibinfo {author}
  {\bibfnamefont {J.}~\bibnamefont {Kemp}},\ and\ \bibinfo {author}
  {\bibfnamefont {N.~Y.}\ \bibnamefont {Yao}},\ }\bibfield  {title} {\bibinfo
  {title} {Supersolidity and simplex phases in spin-1 {Rydberg} atom arrays},\
  }\href@noop {} {\bibfield  {journal} {\bibinfo  {journal} {arXiv:2407.17554}\
  } (\bibinfo {year} {2024})}\BibitemShut {NoStop}%
\bibitem [{\citenamefont {Homeier}\ \emph {et~al.}(2025)\citenamefont
  {Homeier}, \citenamefont {Hollerith}, \citenamefont {Geier}, \citenamefont
  {Chiu}, \citenamefont {Browaeys},\ and\ \citenamefont
  {Pollet}}]{homeier2025supersolidity}%
  \BibitemOpen
  \bibfield  {author} {\bibinfo {author} {\bibfnamefont {L.}~\bibnamefont
  {Homeier}}, \bibinfo {author} {\bibfnamefont {S.}~\bibnamefont {Hollerith}},
  \bibinfo {author} {\bibfnamefont {S.}~\bibnamefont {Geier}}, \bibinfo
  {author} {\bibfnamefont {N.-C.}\ \bibnamefont {Chiu}}, \bibinfo {author}
  {\bibfnamefont {A.}~\bibnamefont {Browaeys}},\ and\ \bibinfo {author}
  {\bibfnamefont {L.}~\bibnamefont {Pollet}},\ }\bibfield  {title} {\bibinfo
  {title} {Supersolidity in {Rydberg} tweezer arrays},\ }\href@noop {}
  {\bibfield  {journal} {\bibinfo  {journal} {Phys. Rev. A}\ }\textbf {\bibinfo
  {volume} {111}},\ \bibinfo {pages} {L011305} (\bibinfo {year}
  {2025})}\BibitemShut {NoStop}%
\bibitem [{\citenamefont {Zeng}\ \emph {et~al.}(2017)\citenamefont {Zeng},
  \citenamefont {Xu}, \citenamefont {He}, \citenamefont {Liu}, \citenamefont
  {Liu}, \citenamefont {Wang}, \citenamefont {Papoular}, \citenamefont
  {Shlyapnikov},\ and\ \citenamefont {Zhan}}]{Zeng2017entangling}%
  \BibitemOpen
  \bibfield  {author} {\bibinfo {author} {\bibfnamefont {Y.}~\bibnamefont
  {Zeng}}, \bibinfo {author} {\bibfnamefont {P.}~\bibnamefont {Xu}}, \bibinfo
  {author} {\bibfnamefont {X.}~\bibnamefont {He}}, \bibinfo {author}
  {\bibfnamefont {Y.}~\bibnamefont {Liu}}, \bibinfo {author} {\bibfnamefont
  {M.}~\bibnamefont {Liu}}, \bibinfo {author} {\bibfnamefont {J.}~\bibnamefont
  {Wang}}, \bibinfo {author} {\bibfnamefont {D.~J.}\ \bibnamefont {Papoular}},
  \bibinfo {author} {\bibfnamefont {G.~V.}\ \bibnamefont {Shlyapnikov}},\ and\
  \bibinfo {author} {\bibfnamefont {M.}~\bibnamefont {Zhan}},\ }\bibfield
  {title} {\bibinfo {title} {Entangling two individual atoms of different
  isotopes via {Rydberg} blockade},\ }\href
  {https://doi.org/10.1103/PhysRevLett.119.160502} {\bibfield  {journal}
  {\bibinfo  {journal} {Phys. Rev. Lett.}\ }\textbf {\bibinfo {volume} {119}},\
  \bibinfo {pages} {160502} (\bibinfo {year} {2017})}\BibitemShut {NoStop}%
\bibitem [{\citenamefont {Sheng}\ \emph {et~al.}(2022)\citenamefont {Sheng},
  \citenamefont {Hou}, \citenamefont {He}, \citenamefont {Wang}, \citenamefont
  {Guo}, \citenamefont {Zhuang}, \citenamefont {Mamat}, \citenamefont {Xu},
  \citenamefont {Liu}, \citenamefont {Wang},\ and\ \citenamefont
  {Zhan}}]{sheng2022defect}%
  \BibitemOpen
  \bibfield  {author} {\bibinfo {author} {\bibfnamefont {C.}~\bibnamefont
  {Sheng}}, \bibinfo {author} {\bibfnamefont {J.}~\bibnamefont {Hou}}, \bibinfo
  {author} {\bibfnamefont {X.}~\bibnamefont {He}}, \bibinfo {author}
  {\bibfnamefont {K.}~\bibnamefont {Wang}}, \bibinfo {author} {\bibfnamefont
  {R.}~\bibnamefont {Guo}}, \bibinfo {author} {\bibfnamefont {J.}~\bibnamefont
  {Zhuang}}, \bibinfo {author} {\bibfnamefont {B.}~\bibnamefont {Mamat}},
  \bibinfo {author} {\bibfnamefont {P.}~\bibnamefont {Xu}}, \bibinfo {author}
  {\bibfnamefont {M.}~\bibnamefont {Liu}}, \bibinfo {author} {\bibfnamefont
  {J.}~\bibnamefont {Wang}},\ and\ \bibinfo {author} {\bibfnamefont
  {M.}~\bibnamefont {Zhan}},\ }\bibfield  {title} {\bibinfo {title}
  {Defect-free arbitrary-geometry assembly of mixed-species atom arrays},\
  }\href@noop {} {\bibfield  {journal} {\bibinfo  {journal} {Phys. Rev. Lett.}\
  }\textbf {\bibinfo {volume} {128}},\ \bibinfo {pages} {083202} (\bibinfo
  {year} {2022})}\BibitemShut {NoStop}%
\bibitem [{\citenamefont {Anand}\ \emph {et~al.}(2024)\citenamefont {Anand},
  \citenamefont {Bradley}, \citenamefont {White}, \citenamefont {Ramesh},
  \citenamefont {Singh},\ and\ \citenamefont {Bernien}}]{anand2024dual}%
  \BibitemOpen
  \bibfield  {author} {\bibinfo {author} {\bibfnamefont {S.}~\bibnamefont
  {Anand}}, \bibinfo {author} {\bibfnamefont {C.~E.}\ \bibnamefont {Bradley}},
  \bibinfo {author} {\bibfnamefont {R.}~\bibnamefont {White}}, \bibinfo
  {author} {\bibfnamefont {V.}~\bibnamefont {Ramesh}}, \bibinfo {author}
  {\bibfnamefont {K.}~\bibnamefont {Singh}},\ and\ \bibinfo {author}
  {\bibfnamefont {H.}~\bibnamefont {Bernien}},\ }\bibfield  {title} {\bibinfo
  {title} {A dual-species {Rydberg} array},\ }\href@noop {} {\bibfield
  {journal} {\bibinfo  {journal} {Nat. Phys.}\ }\textbf {\bibinfo {volume}
  {20}},\ \bibinfo {pages} {1744} (\bibinfo {year} {2024})}\BibitemShut
  {NoStop}%
\bibitem [{\citenamefont {Nakamura}\ \emph {et~al.}(2024)\citenamefont
  {Nakamura}, \citenamefont {Kusano}, \citenamefont {Yokoyama}, \citenamefont
  {Saito}, \citenamefont {Higashi}, \citenamefont {Ozawa}, \citenamefont
  {Takano}, \citenamefont {Takasu},\ and\ \citenamefont
  {Takahashi}}]{nakamura2024hybrid}%
  \BibitemOpen
  \bibfield  {author} {\bibinfo {author} {\bibfnamefont {Y.}~\bibnamefont
  {Nakamura}}, \bibinfo {author} {\bibfnamefont {T.}~\bibnamefont {Kusano}},
  \bibinfo {author} {\bibfnamefont {R.}~\bibnamefont {Yokoyama}}, \bibinfo
  {author} {\bibfnamefont {K.}~\bibnamefont {Saito}}, \bibinfo {author}
  {\bibfnamefont {K.}~\bibnamefont {Higashi}}, \bibinfo {author} {\bibfnamefont
  {N.}~\bibnamefont {Ozawa}}, \bibinfo {author} {\bibfnamefont
  {T.}~\bibnamefont {Takano}}, \bibinfo {author} {\bibfnamefont
  {Y.}~\bibnamefont {Takasu}},\ and\ \bibinfo {author} {\bibfnamefont
  {Y.}~\bibnamefont {Takahashi}},\ }\bibfield  {title} {\bibinfo {title}
  {Hybrid atom tweezer array of nuclear spin and optical clock qubits},\
  }\href@noop {} {\bibfield  {journal} {\bibinfo  {journal} {Phys. Rev. X}\
  }\textbf {\bibinfo {volume} {14}},\ \bibinfo {pages} {041062} (\bibinfo
  {year} {2024})}\BibitemShut {NoStop}%
\bibitem [{\citenamefont {Wei}\ \emph {et~al.}(2024)\citenamefont {Wei},
  \citenamefont {Wei}, \citenamefont {Li},\ and\ \citenamefont
  {Yan}}]{wei2024dual}%
  \BibitemOpen
  \bibfield  {author} {\bibinfo {author} {\bibfnamefont {Y.}~\bibnamefont
  {Wei}}, \bibinfo {author} {\bibfnamefont {K.}~\bibnamefont {Wei}}, \bibinfo
  {author} {\bibfnamefont {S.}~\bibnamefont {Li}},\ and\ \bibinfo {author}
  {\bibfnamefont {B.}~\bibnamefont {Yan}},\ }\bibfield  {title} {\bibinfo
  {title} {Dual-species optical tweezer for {Rb} and {K} atoms},\ }\href@noop
  {} {\bibfield  {journal} {\bibinfo  {journal} {Phys. Rev. A}\ }\textbf
  {\bibinfo {volume} {110}},\ \bibinfo {pages} {043118} (\bibinfo {year}
  {2024})}\BibitemShut {NoStop}%
\bibitem [{\citenamefont {White}\ \emph {et~al.}(2026)\citenamefont {White},
  \citenamefont {Ramesh}, \citenamefont {Impertro}, \citenamefont {Anand},
  \citenamefont {Cesa}, \citenamefont {Giudici}, \citenamefont {Iadecola},
  \citenamefont {Pichler},\ and\ \citenamefont {Bernien}}]{white2026quantum}%
  \BibitemOpen
  \bibfield  {author} {\bibinfo {author} {\bibfnamefont {R.}~\bibnamefont
  {White}}, \bibinfo {author} {\bibfnamefont {V.}~\bibnamefont {Ramesh}},
  \bibinfo {author} {\bibfnamefont {A.}~\bibnamefont {Impertro}}, \bibinfo
  {author} {\bibfnamefont {S.}~\bibnamefont {Anand}}, \bibinfo {author}
  {\bibfnamefont {F.}~\bibnamefont {Cesa}}, \bibinfo {author} {\bibfnamefont
  {G.}~\bibnamefont {Giudici}}, \bibinfo {author} {\bibfnamefont
  {T.}~\bibnamefont {Iadecola}}, \bibinfo {author} {\bibfnamefont
  {H.}~\bibnamefont {Pichler}},\ and\ \bibinfo {author} {\bibfnamefont
  {H.}~\bibnamefont {Bernien}},\ }\bibfield  {title} {\bibinfo {title} {Quantum
  cellular automata on a dual-species {Rydberg} processor},\ }\href@noop {}
  {\bibfield  {journal} {\bibinfo  {journal} {arXiv:2601.16257}\ } (\bibinfo
  {year} {2026})}\BibitemShut {NoStop}%
\bibitem [{\citenamefont {Miles}\ \emph {et~al.}(2026)\citenamefont {Miles},
  \citenamefont {Lichtman}, \citenamefont {Scott}, \citenamefont {Scott},
  \citenamefont {Norrell}, \citenamefont {Bedalov}, \citenamefont {Belknap},
  \citenamefont {Cole}, \citenamefont {Eubanks}, \citenamefont {Gillette},
  \citenamefont {Gokhale}, \citenamefont {Goldwin}, \citenamefont {Iliev},
  \citenamefont {Jones}, \citenamefont {Kuper}, \citenamefont {Mason},
  \citenamefont {Mitchell}, \citenamefont {Murphree}, \citenamefont
  {Neff-Mallon}, \citenamefont {Noel}, \citenamefont {Radnaev}, \citenamefont
  {Vinogradov},\ and\ \citenamefont {Saffman}}]{miles2026qubit}%
  \BibitemOpen
  \bibfield  {author} {\bibinfo {author} {\bibfnamefont {J.}~\bibnamefont
  {Miles}}, \bibinfo {author} {\bibfnamefont {M.~T.}\ \bibnamefont {Lichtman}},
  \bibinfo {author} {\bibfnamefont {A.~M.}\ \bibnamefont {Scott}}, \bibinfo
  {author} {\bibfnamefont {J.}~\bibnamefont {Scott}}, \bibinfo {author}
  {\bibfnamefont {S.~A.}\ \bibnamefont {Norrell}}, \bibinfo {author}
  {\bibfnamefont {M.~J.}\ \bibnamefont {Bedalov}}, \bibinfo {author}
  {\bibfnamefont {D.~A.}\ \bibnamefont {Belknap}}, \bibinfo {author}
  {\bibfnamefont {D.~C.}\ \bibnamefont {Cole}}, \bibinfo {author}
  {\bibfnamefont {S.~Y.}\ \bibnamefont {Eubanks}}, \bibinfo {author}
  {\bibfnamefont {M.}~\bibnamefont {Gillette}}, \bibinfo {author}
  {\bibfnamefont {P.}~\bibnamefont {Gokhale}}, \bibinfo {author} {\bibfnamefont
  {J.}~\bibnamefont {Goldwin}}, \bibinfo {author} {\bibfnamefont
  {M.}~\bibnamefont {Iliev}}, \bibinfo {author} {\bibfnamefont {R.~A.}\
  \bibnamefont {Jones}}, \bibinfo {author} {\bibfnamefont {K.~W.}\ \bibnamefont
  {Kuper}}, \bibinfo {author} {\bibfnamefont {D.}~\bibnamefont {Mason}},
  \bibinfo {author} {\bibfnamefont {P.~T.}\ \bibnamefont {Mitchell}}, \bibinfo
  {author} {\bibfnamefont {J.~D.}\ \bibnamefont {Murphree}}, \bibinfo {author}
  {\bibfnamefont {N.~A.}\ \bibnamefont {Neff-Mallon}}, \bibinfo {author}
  {\bibfnamefont {T.~W.}\ \bibnamefont {Noel}}, \bibinfo {author}
  {\bibfnamefont {A.~G.}\ \bibnamefont {Radnaev}}, \bibinfo {author}
  {\bibfnamefont {I.~V.}\ \bibnamefont {Vinogradov}},\ and\ \bibinfo {author}
  {\bibfnamefont {M.}~\bibnamefont {Saffman}},\ }\bibfield  {title} {\bibinfo
  {title} {Qubit syndrome measurements with a high fidelity {Rb-Cs} {Rydberg}
  gate},\ }\href@noop {} {\bibfield  {journal} {\bibinfo  {journal}
  {arXiv:2603.13492}\ } (\bibinfo {year} {2026})}\BibitemShut {NoStop}%
\bibitem [{\citenamefont {Wang}\ \emph {et~al.}(2026)\citenamefont {Wang},
  \citenamefont {Cimmino}, \citenamefont {Wang}, \citenamefont {Lopez},
  \citenamefont {Li}, \citenamefont {Koh}, \citenamefont {Hall{\'e}n},
  \citenamefont {Matthies}, \citenamefont {Yao},\ and\ \citenamefont
  {Ni}}]{wang2026multi}%
  \BibitemOpen
  \bibfield  {author} {\bibinfo {author} {\bibfnamefont {Y.}~\bibnamefont
  {Wang}}, \bibinfo {author} {\bibfnamefont {R.}~\bibnamefont {Cimmino}},
  \bibinfo {author} {\bibfnamefont {K.}~\bibnamefont {Wang}}, \bibinfo {author}
  {\bibfnamefont {S.}~\bibnamefont {Lopez}}, \bibinfo {author} {\bibfnamefont
  {J.}~\bibnamefont {Li}}, \bibinfo {author} {\bibfnamefont {J.~M.}\
  \bibnamefont {Koh}}, \bibinfo {author} {\bibfnamefont {J.~N.}\ \bibnamefont
  {Hall{\'e}n}}, \bibinfo {author} {\bibfnamefont {A.}~\bibnamefont
  {Matthies}}, \bibinfo {author} {\bibfnamefont {N.~Y.}\ \bibnamefont {Yao}},\
  and\ \bibinfo {author} {\bibfnamefont {K.-K.}\ \bibnamefont {Ni}},\
  }\bibfield  {title} {\bibinfo {title} {Multi-qubit stabilizer readout on a
  dual-species {Rydberg} array},\ }\href@noop {} {\bibfield  {journal}
  {\bibinfo  {journal} {arXiv:2605.10924}\ } (\bibinfo {year}
  {2026})}\BibitemShut {NoStop}%
\bibitem [{\citenamefont {Kuznetsova}\ \emph {et~al.}(2018)\citenamefont
  {Kuznetsova}, \citenamefont {Rittenhouse}, \citenamefont {Beterov},
  \citenamefont {Scully}, \citenamefont {Yelin},\ and\ \citenamefont
  {Sadeghpour}}]{Kuznetsova2018effective}%
  \BibitemOpen
  \bibfield  {author} {\bibinfo {author} {\bibfnamefont {E.}~\bibnamefont
  {Kuznetsova}}, \bibinfo {author} {\bibfnamefont {S.~T.}\ \bibnamefont
  {Rittenhouse}}, \bibinfo {author} {\bibfnamefont {I.~I.}\ \bibnamefont
  {Beterov}}, \bibinfo {author} {\bibfnamefont {M.~O.}\ \bibnamefont {Scully}},
  \bibinfo {author} {\bibfnamefont {S.~F.}\ \bibnamefont {Yelin}},\ and\
  \bibinfo {author} {\bibfnamefont {H.~R.}\ \bibnamefont {Sadeghpour}},\
  }\bibfield  {title} {\bibinfo {title} {Effective spin-spin interactions in
  bilayers of {Rydberg} atoms and polar molecules},\ }\href
  {https://doi.org/10.1103/PhysRevA.98.043609} {\bibfield  {journal} {\bibinfo
  {journal} {Phys. Rev. A}\ }\textbf {\bibinfo {volume} {98}},\ \bibinfo
  {pages} {043609} (\bibinfo {year} {2018})}\BibitemShut {NoStop}%
\bibitem [{\citenamefont {Cesa}\ \emph {et~al.}(2026)\citenamefont {Cesa},
  \citenamefont {Di~Fini}, \citenamefont {Korbany}, \citenamefont {Tricarico},
  \citenamefont {Bernien}, \citenamefont {Pichler},\ and\ \citenamefont
  {Piroli}}]{cesa2026engineering}%
  \BibitemOpen
  \bibfield  {author} {\bibinfo {author} {\bibfnamefont {F.}~\bibnamefont
  {Cesa}}, \bibinfo {author} {\bibfnamefont {A.}~\bibnamefont {Di~Fini}},
  \bibinfo {author} {\bibfnamefont {D.~A.}\ \bibnamefont {Korbany}}, \bibinfo
  {author} {\bibfnamefont {R.}~\bibnamefont {Tricarico}}, \bibinfo {author}
  {\bibfnamefont {H.}~\bibnamefont {Bernien}}, \bibinfo {author} {\bibfnamefont
  {H.}~\bibnamefont {Pichler}},\ and\ \bibinfo {author} {\bibfnamefont
  {L.}~\bibnamefont {Piroli}},\ }\bibfield  {title} {\bibinfo {title}
  {Engineering discrete local dynamics in globally driven dual-species atom
  arrays},\ }\href@noop {} {\bibfield  {journal} {\bibinfo  {journal}
  {arXiv:2601.16961}\ } (\bibinfo {year} {2026})}\BibitemShut {NoStop}%
\bibitem [{\citenamefont {Li}\ \emph {et~al.}(2024)\citenamefont {Li},
  \citenamefont {Liu}, \citenamefont {Wang}, \citenamefont {Zhang},
  \citenamefont {Li}, \citenamefont {Zhai},\ and\ \citenamefont
  {Gu}}]{Li2024uncovering}%
  \BibitemOpen
  \bibfield  {author} {\bibinfo {author} {\bibfnamefont {C.}~\bibnamefont
  {Li}}, \bibinfo {author} {\bibfnamefont {S.}~\bibnamefont {Liu}}, \bibinfo
  {author} {\bibfnamefont {H.}~\bibnamefont {Wang}}, \bibinfo {author}
  {\bibfnamefont {W.}~\bibnamefont {Zhang}}, \bibinfo {author} {\bibfnamefont
  {Z.-X.}\ \bibnamefont {Li}}, \bibinfo {author} {\bibfnamefont
  {H.}~\bibnamefont {Zhai}},\ and\ \bibinfo {author} {\bibfnamefont
  {Y.}~\bibnamefont {Gu}},\ }\bibfield  {title} {\bibinfo {title} {Uncovering
  emergent spacetime supersymmetry with {Rydberg} atom arrays},\ }\href
  {https://doi.org/10.1103/PhysRevLett.133.223401} {\bibfield  {journal}
  {\bibinfo  {journal} {Phys. Rev. Lett.}\ }\textbf {\bibinfo {volume} {133}},\
  \bibinfo {pages} {223401} (\bibinfo {year} {2024})}\BibitemShut {NoStop}%
\bibitem [{\citenamefont {Wang}\ \emph {et~al.}(2025)\citenamefont {Wang},
  \citenamefont {Li},\ and\ \citenamefont {Li}}]{wang2025tricritical}%
  \BibitemOpen
  \bibfield  {author} {\bibinfo {author} {\bibfnamefont {H.}~\bibnamefont
  {Wang}}, \bibinfo {author} {\bibfnamefont {X.}~\bibnamefont {Li}},\ and\
  \bibinfo {author} {\bibfnamefont {C.}~\bibnamefont {Li}},\ }\bibfield
  {title} {\bibinfo {title} {Tricritical {Kibble-Zurek} scaling in {Rydberg}
  atom ladders},\ }\href {https://doi.org/10.1038/s41467-025-65652-9}
  {\bibfield  {journal} {\bibinfo  {journal} {Nat. Commun.}\ }\textbf {\bibinfo
  {volume} {16}},\ \bibinfo {pages} {10584} (\bibinfo {year}
  {2025})}\BibitemShut {NoStop}%
\bibitem [{\citenamefont {Farouk}\ \emph {et~al.}(2026)\citenamefont {Farouk},
  \citenamefont {Beterov}, \citenamefont {Suliman}, \citenamefont {Chen},\ and\
  \citenamefont {Ryabtsev}}]{farouk2026quantum}%
  \BibitemOpen
  \bibfield  {author} {\bibinfo {author} {\bibfnamefont {A.~M.}\ \bibnamefont
  {Farouk}}, \bibinfo {author} {\bibfnamefont {I.~I.}\ \bibnamefont {Beterov}},
  \bibinfo {author} {\bibfnamefont {G.}~\bibnamefont {Suliman}}, \bibinfo
  {author} {\bibfnamefont {J.}~\bibnamefont {Chen}},\ and\ \bibinfo {author}
  {\bibfnamefont {I.~I.}\ \bibnamefont {Ryabtsev}},\ }\bibfield  {title}
  {\bibinfo {title} {Quantum spin liquid state of a dual-species atomic array
  on kagome lattice},\ }\href@noop {} {\bibfield  {journal} {\bibinfo
  {journal} {arXiv:2605.03579}\ } (\bibinfo {year} {2026})}\BibitemShut
  {NoStop}%
\bibitem [{\citenamefont {Hermele}\ \emph {et~al.}(2004)\citenamefont
  {Hermele}, \citenamefont {Fisher},\ and\ \citenamefont
  {Balents}}]{Hermele2004pyrochlore}%
  \BibitemOpen
  \bibfield  {author} {\bibinfo {author} {\bibfnamefont {M.}~\bibnamefont
  {Hermele}}, \bibinfo {author} {\bibfnamefont {M.~P.~A.}\ \bibnamefont
  {Fisher}},\ and\ \bibinfo {author} {\bibfnamefont {L.}~\bibnamefont
  {Balents}},\ }\bibfield  {title} {\bibinfo {title} {Pyrochlore photons: The
  {$U(1)$} spin liquid in a {$S=\frac{1}{2}$} three-dimensional frustrated
  magnet},\ }\href {https://doi.org/10.1103/PhysRevB.69.064404} {\bibfield
  {journal} {\bibinfo  {journal} {Phys. Rev. B}\ }\textbf {\bibinfo {volume}
  {69}},\ \bibinfo {pages} {064404} (\bibinfo {year} {2004})}\BibitemShut
  {NoStop}%
\bibitem [{\citenamefont {Barredo}\ \emph {et~al.}(2018)\citenamefont
  {Barredo}, \citenamefont {Lienhard}, \citenamefont {de~L\'es\'eleuc},
  \citenamefont {Lahaye},\ and\ \citenamefont
  {Browaeys}}]{barredo2018synthetic}%
  \BibitemOpen
  \bibfield  {author} {\bibinfo {author} {\bibfnamefont {D.}~\bibnamefont
  {Barredo}}, \bibinfo {author} {\bibfnamefont {V.}~\bibnamefont {Lienhard}},
  \bibinfo {author} {\bibfnamefont {S.}~\bibnamefont {de~L\'es\'eleuc}},
  \bibinfo {author} {\bibfnamefont {T.}~\bibnamefont {Lahaye}},\ and\ \bibinfo
  {author} {\bibfnamefont {A.}~\bibnamefont {Browaeys}},\ }\bibfield  {title}
  {\bibinfo {title} {Synthetic three-dimensional atomic structures assembled
  atom by atom},\ }\href@noop {} {\bibfield  {journal} {\bibinfo  {journal}
  {Nature}\ }\textbf {\bibinfo {volume} {561}},\ \bibinfo {pages} {79}
  (\bibinfo {year} {2018})}\BibitemShut {NoStop}%
\bibitem [{\citenamefont {Kusano}\ \emph {et~al.}(2025)\citenamefont {Kusano},
  \citenamefont {Nakamura}, \citenamefont {Yokoyama}, \citenamefont {Ozawa},
  \citenamefont {Shibata}, \citenamefont {Takano}, \citenamefont {Takasu},\
  and\ \citenamefont {Takahashi}}]{kusano2025plane}%
  \BibitemOpen
  \bibfield  {author} {\bibinfo {author} {\bibfnamefont {T.}~\bibnamefont
  {Kusano}}, \bibinfo {author} {\bibfnamefont {Y.}~\bibnamefont {Nakamura}},
  \bibinfo {author} {\bibfnamefont {R.}~\bibnamefont {Yokoyama}}, \bibinfo
  {author} {\bibfnamefont {N.}~\bibnamefont {Ozawa}}, \bibinfo {author}
  {\bibfnamefont {K.}~\bibnamefont {Shibata}}, \bibinfo {author} {\bibfnamefont
  {T.}~\bibnamefont {Takano}}, \bibinfo {author} {\bibfnamefont
  {Y.}~\bibnamefont {Takasu}},\ and\ \bibinfo {author} {\bibfnamefont
  {Y.}~\bibnamefont {Takahashi}},\ }\bibfield  {title} {\bibinfo {title}
  {Plane-selective manipulations of nuclear spin qubits in a three-dimensional
  optical tweezer array},\ }\href
  {https://doi.org/10.1103/PhysRevResearch.7.L022045} {\bibfield  {journal}
  {\bibinfo  {journal} {Phys. Rev. Res.}\ }\textbf {\bibinfo {volume} {7}},\
  \bibinfo {pages} {L022045} (\bibinfo {year} {2025})}\BibitemShut {NoStop}%
\bibitem [{\citenamefont {Kunimi}\ and\ \citenamefont
  {Tomita}(2026)}]{kunimi_2026_21367229}%
  \BibitemOpen
  \bibfield  {author} {\bibinfo {author} {\bibfnamefont {M.}~\bibnamefont
  {Kunimi}}\ and\ \bibinfo {author} {\bibfnamefont {T.}~\bibnamefont
  {Tomita}},\ }\bibfield  {title} {\bibinfo {title} {Dataset for
  "{Magnetic-field} control of interactions in alkaline-earth {Rydberg} atoms
  and applications to {XXZ} models"},\ }\href
  {https://doi.org/10.5281/zenodo.21367229} {10.5281/zenodo.21367229} (\bibinfo
  {year} {2026})\BibitemShut {NoStop}%
\end{thebibliography}%
\end{document}